\documentclass[11pt,a4paper,twoside,openright]{report}
\usepackage{amsmath}
\usepackage{amsfonts}
\usepackage{amssymb}
\allowdisplaybreaks[1]    
\usepackage[a4paper,includehead,
   marginratio={1:1,10:13}, scale={0.65,0.82}]{geometry}
\usepackage[dvips]{graphicx}
\usepackage{slashed}
\usepackage{fancyhdr}
\usepackage{axodraw}
\usepackage{pstricks}
\usepackage{color}

\renewcommand{\chaptermark}[1]%
  {\markboth{\myheaderpart{} \thechapter\hspace{2em}#1}{}}
\fancypagestyle{plain}{%
  \fancyhf{}
  \fancyhead[LE]{{\bfseries \thepage}}
  \fancyhead[RO]{{\bfseries \thepage}}	

}

\def\as{\alpha_s}
\def\gl{\tilde{g}}
\def\sq{\tilde{q}}
\def\sqb{\bar{\tilde{q}}}
\def\qb{\bar{q}}

\def\ms{m_{\tilde q}}
\def\mg{m_{\tilde g}}

\def\md{m_{-}}

\def\ghat{\hat{g}_s}

\def\bs{\beta_{\sq}}

\def\bg{\beta_{\gl}}

\usepackage[square,comma,numbers,sort&compress]{natbib}

\newcommand{\myheaderpart}{\chaptername}
\newcommand{\mythesistitle}{%
A Decay Chain Spin Analysis for SUSY and UED at the LHC
}
\newcommand{\mythesisauthor}{%
   Sebastian Johannes Reinartz
}
\newcommand{\mythesisdate}{%
   December 2007
}

\usepackage[ps2pdf,bookmarksnumbered,
   pdftitle={\mythesistitle},%
   pdfauthor={\mythesisauthor}]{hyperref}

\begin{document}
  \pagenumbering{roman}
\thispagestyle{empty}
\vspace*{\stretch{3}}
\noindent
\begin{flushright}
   \rule{\textwidth}{3pt}\\
   \vspace*{\stretch{1}}
   {\sffamily\bfseries\Huge\mythesistitle\\}
   \vspace*{\stretch{1}}
   {\sffamily\bfseries\Large\mythesisauthor\\}
   \vspace*{\stretch{1}}
   \rule{\textwidth}{3pt}\\
\end{flushright}
\begin{center}
\vspace*{\stretch{20}}
\includegraphics[width=1.4in]{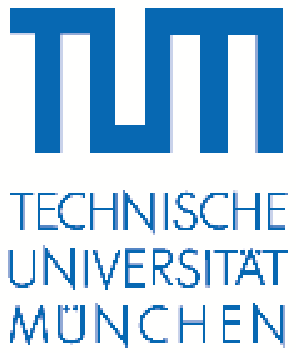}
\hspace*{-20pt}
\includegraphics[width=1.3in]{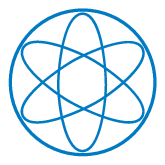}
\hspace{\stretch{1}}
\includegraphics[width=1.8in]{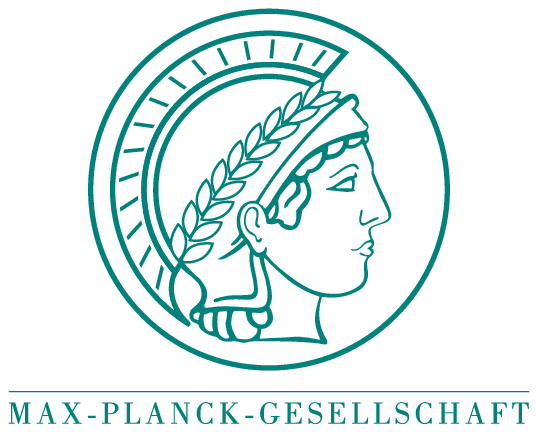}
\end{center}
\clearpage{\pagestyle{empty}\cleardoublepage}

\pagenumbering{roman}
\thispagestyle{empty}
\vspace*{\stretch{3}}
\noindent
\begin{flushright}
   \rule{\textwidth}{3pt}\\
   \vspace*{\stretch{1}}
   {\sffamily\bfseries\Huge\mythesistitle\\}
   \vspace*{\stretch{1}}
   {\sffamily\bfseries\Large\mythesisauthor\\}
   \vspace*{\stretch{1}}
   \rule{\textwidth}{3pt}\\
\end{flushright}
\vspace*{\stretch{4}}
\begin{center}
   {\Large
      Diploma Thesis\\
      by\\
      \mythesisauthor\\
      \vspace*{\stretch{4}}
      Technische Universit\"at M\"unchen\\
      Department f\"ur Physik\\
      \vspace*{\stretch{1}}
      Max-Planck-Institut f\"ur Physik\\
      (Werner-Heisenberg-Institut)\\
      \vspace*{\stretch{8.8}}  
      \mythesisdate
   }
\end{center}
\clearpage{\pagestyle{empty}\cleardoublepage}

\tableofcontents


  \clearpage{\pagestyle{empty}\cleardoublepage}

\pagenumbering{arabic}
  \chapter{Introduction}
In 2008, the Large Hadron Collider (LHC) at the Conseil Europ\'{e}en pour la Recherche Nucl\'{e}aire (CERN) will start operating. The LHC is the next step in discovering the mysteries of particle physics and the fundamental forces. Reaching energies at the level of $14 \; \text{TeV}$, the discovery of physics beyond the SM and new particles will become possible. Since the Standard Model (SM) of particle physics is most probably not the final theory, searches for theoretically motivated models have already begun and physicists wait for their confirmation.

But making sure that one really observes a specific model is far from trivial. Two very promising candidates for physics beyond the Standard Model are Supersymmetry (SUSY)~\cite{Wess} and Universal Extra Dimensions (UED)~\cite{appelquist-2001-64}. An important difference between them is their difference in the spin of the particle fields. In Supersymmetry, the usual space-time symmetry of the SM is enlarged towards a more general supersymmetry algebra by adding operators that change the spin of the particles. There are theoretical bounds like the Higgs mass, which allow to exclude or confirm SUSY at the LHC to a high certainty level. The model of Universal Extra Dimensions is a string theory inspired, extended version of the well known Standard model. Allowing all particles to propagate in five or more dimensions produces new particles. In the theory they appear after performing compactification of the extra dimensions. 

To show that SUSY is indeed the theory which is realized by nature, one has to measure the spin of all new particles. Since SUSY particles are not stable, except for the lightest supersymmetric particle, they are treated as parts of long decay chains. In this thesis we concentrate on the spin of the gluino, which is a Majorana fermion in SUSY. Majorana fermions are interesting anyway, since they can produce like-sign dilepton signatures which are a clear sign for new physics, as explained in~\cite{barnett}. Using like-sign dileptons in searches for new physics yields a much clearer signature than the one from missing energy measurements. But since like-sign dileptons can also be produced from the decay of a boson with an adjoint color charge, measuring the spin is crucial. What has to be done is to compare SUSY to a model with a ``bosonic gluino``, assuming equal masses of the particles. We therefore consider UED, where the gluino is substituted by a heavy Kaluza-Klein-gluon (KK-gluon) with bosonic spin-statistics. 

In recent publications the topic of spin measurements in decay chains gained higher popularity. In~\cite{smillie-2005-0510}, it is shown that measuring the spin of the squark is possible by comparing angular and charge asymmetries of the decay products in UED and SUSY.

It is demonstrated in~\cite{alves-2006-74}, that the spin of the gluino has a measurable impact on kinematic distributions and angular correlations of the decay products in gluon-gluon and gluon-quark collisions. Therefore leptonic and hadronic correlations of the decay products are investigated in UED and SUSY. It is presumed that the difference in the boost of the gluino and the KK-gluon is the main reason for the differences in kinematics of the decay products. As a second influence the coupling structure is mentioned. 

In~\cite{csaki-2007}, it is studied if UED can perfectly imitate the gluino decay, not necessarily assuming equal masses for the particle spectrum of both theories. It is shown for the decay of the gluino into two jets and a neutralino, that UED and SUSY can be differentiated, assuming that the gluino is lighter than all squarks. The influence of the coupling structure is investigated in a toy model and a comparison of the invariant masses for SUSY and UED is performed. The number of events needed to discriminate both theories is calculated. 

In this thesis we investigate the process of gluon-quark collision, producing two $b$-jets, a light quark jet and two neutralinos in SUSY. An equivalent process is calculated for UED and all cross sections are normalized to make a comparison possible. Kinematic distributions of the decay products are compared in SUSY and UED, aiming to explain the difference in the angular distributions between the outgoing $b$-jets. We study whether the influence from the different boost of the gluino and KK-gluon or the different coupling structure is more relevant to the correlations of the final state bottom jets. Therefore we vary the Kaluza-Klein mixing angle and numerically eliminate the effect from the difference in the boost distributions. In comparison to~\cite{csaki-2007}, we assume that the gluino is heavier than all squarks as it is e.g.\ in the SPS 1a scenario.
\bigskip

This thesis is structured as follows: In the second chapter, we review the definition of the cross section and sum up a few important concepts. 

In the third chapter, we shortly introduce SUSY, by giving an impression of how a supersymmetric Lagrangian in general looks like in the spinor notation. We do not extensively review the complete particle spectrum of SUSY here. The theory of UED is introduced in somewhat more detail and all Feynman rules and Lagrangians are calculated completely, since literature on this is more rare.

In the fourth chapter, we present our results for squark and gluino production cross-sections. The partonic cross sections are calculated analytically for all SUSY-QCD processes. Hadronic cross sections and kinematic distributions for the LHC are presented numerically for the case of gluino-squark pair production. The partonic cross-section for the production of a Kaluza-Klein gluon-quark pair in UED is also calculated analytically. This process is compared to the corresponding SUSY process at the production threshold.

In the fifth chapter, the general kinematics of decay chains are discussed and the concept of our program is explained. We explain how the decay phase space is tested.

In the sixth chapter, calculations are performed for the case of a SUSY and UED decay chain. We investigate the $2 \rightarrow 5$ process of quark-gluon collision in the two models and compare them. Since mass spectra are equal and cross sections are normalized, their only differences come from angular correlations, deriving from the spin of the particles. Results are compared for SUSY and UED. The influence of a varying mixing angle in the UED couplings is investigated. We also discuss the origin of the differences in the angular distributions which were claimed to be due to the different boosts of KK-gluon and gluino or to the different coupling structure of the heavy quark partners.

  \clearpage{\pagestyle{empty}\cleardoublepage}

  \chapter{Basic Concepts}
\label{ch:2}
\section{Calculation of Cross Sections}
The total cross-section is an important theoretical result of high-energy particle physics since it gives an observable quantity which can be compared to experiments. Differential distributions of the cross section can equally well be calculated and observed at colliders.
	
\subsection{Definition of the Cross Section}
In general the differential cross section is defined as the product of transition rate per scattering center and the number of final states reachable, divided by the incident flux. The transition rate from an initial state $\vert i \rangle $ to a final state $\langle f \vert$ is given by 
\begin{equation}
\omega=\frac{\vert \langle f \vert S \vert i \rangle \vert ^2}{T}=\frac{\vert S_{fi}\vert ^2}{T} \;,
\end{equation}
where $T$ is a finite time interval and $S_{fi}$ is called the $S$-matrix element. In a process where two initial particles with momenta $p^{\mu}_i=(E_i,\vec{p}_i)$, $i=1,2$ collide and N particles with the momenta $p'^{\mu}_f=(E'_f,\vec{p}\,'_f)$, $f=1,\dots, N$ come out, the matrix element $\mathcal{M}$ is connected to the S-matrix element by
\begin{small}
\begin{eqnarray}
S_{fi}=\delta _{fi}+(2 \pi)^4 \delta ^{(4)} \left ( \sum p'_f- \sum p_i \right) \prod \limits_{i} \left ( \frac{1}{2 V E_i} \right)^\frac{1}{2} \prod \limits_{f} \left ( \frac{1}{2 V E'_f} \right)^\frac{1}{2}  \mathcal {M} \; .
\end{eqnarray}
\end{small}
\hspace*{-3pt}To get from these relations to the transition rate to final states with momenta in the intervals $(\vec{p}\,'_f,\vec{p}\,'_f+d \vec{p}\,'_f)$, one has to multiply by 
\begin{small}
\begin{equation}
\prod \limits_{f} \frac{V d^3 \vec{p}\,'_f}{(2 \pi)^3}\;,
\end{equation}
\end{small}
\hspace*{-3.5pt}where the volume $V$ is considered to contain exactly one scattering center. The incoming flux is $v_{rel}/V$, where $v_{rel}$ is the relative velocity of colliding particles.
Combining everything one finds
\begin{small}
\begin{eqnarray}
\begin{aligned}
d &\sigma = \omega \frac{V}{v_{rel}} \prod \limits_{f} \frac{V d^3 \vec{p}\,'_f}{(2 \pi)^3}= \\
&\hphantom{\sigma} = (2 \pi)^4 \delta ^{(4)} \left ( \sum p'_f- \sum p_i \right) \frac{1}{4 E_1 E_2 v_{rel}} \left( \prod \frac{d^3 \vec{p}\,'_f}{(2 \pi)^3 2 E'_f} \right )\vert \mathcal{M} \vert ^2 \;.
\end{aligned}
\end{eqnarray}
\end{small}
\subsection{Parametrization for $2 \rightarrow 2$ Processes}
For kinematical analysis it will be important to find the differential distributions of cross sections in variables convenient for observation. Therefore distributions of transverse momentum $p_t$ and rapidity $y$ will be used. We show how they can be derived in the case of two massless colliding particles with momenta $k_1, k_2$ producing two massive particles with momenta $p_1, p_2$ as in the case of SUSY.
The usual Mandelstam variables are given by
\begin{equation}
\label{eq:Mandelstams}
s=(k_1+k_2)^2, \quad t=(k_2-p_2)^2, \quad u=(k_1-p_2)^2 \; .
\end{equation}
In~\cite{Byckling}, the differential cross section is given as
\begin{equation}
\frac{d \sigma}{d t}= \frac{1}{16 \pi s^2}\vert \mathcal{M}(s,t)\vert ^2 \; .
\end{equation}
When the beam axis is parallel to the z-axis, the variables $p_t=(p_x^2+p_y^2)^{1/2}$ and $y$ for an outgoing particle with momentum $p_2$ and mass $m_2$ are given by
\begin{eqnarray}
\label{eq:def_theta}
{p}_{t}^{2}&=& \left[(p_2^{0})^2-m_2^2 \right]  \phantom{t} \text{sin}^2 \theta^{\ast} \label{eq:p_t_definition} \;, \\[2 mm]
y&=&\frac{1}{2} \text{ln} \left [ \frac{p_2^0+p_{2,L}}{p_2^0-p_{2,L}} \right] \; ,
\end{eqnarray}
where $\theta^{\ast}$ and the longitudinal momentum $p_L= \vert \vec{p} \vert \; \text{cos} \theta^{\ast}$ are defined in the center of momentum system (CMS). A very comfortable feature of $p_t$ for practical calculations is its independence of boosts parallel to the beam axis. The advantage of the rapidity is its additivity under boosts. In the CMS energy and momentum of the outgoing particles are given by
\begin{equation}
E^{\ast}_{1}=\frac{s+m_1^2-m_2^2}{2 \sqrt{s}} \;\;, \qquad p^{\ast}_1=p^{\ast}_2=\frac{\lambda^{\frac{1}{2}}(s,m_1^2,m_2^2)}{2 \sqrt{s}} \;,
\end{equation}
where 
\begin{equation}
\label{eq:lambda_kinematic}
\lambda(x,y,z)=\left( x-(\sqrt{y}+\sqrt{z})^2\right)\left( x-(\sqrt{y}-\sqrt{z})^2\right)
\end{equation}
and $E_2^{\ast}$ is obtained by exchanging the masses of the outgoing particles. 
For massless incoming and massive outgoing particles with masses $m_1$ and $m_2$ one finds
\begin{equation}
t=m_1^2-\frac{1}{2}(s+m_1^2-m_2^2)+\frac{1}{2}(s+m_2^2-m_1^2) \left ( \frac{e^{2y}-1}{e^{2y}+1} \right )\; ,
\end{equation}
which leads to 
\begin{eqnarray}
\frac{d t}{d y}=\frac{(s+m_2^2-m_1^2)}{2 \phantom{t} \text{cosh}^2 y}\; .
\end{eqnarray}
Finally the differential rapidity distribution for the particle with momentum $p_2$ is given by
\begin{equation}
\frac{d \sigma}{d y}= \frac{(s+m_2^2-m_1^2)}{32 \pi s^2 \text{cosh}^2 y } \vert \mathcal{M}(s,y) \vert ^2 \; .
\end{equation}
To derive the differential $p_t$ distribution we use 
\begin{equation}
t=m_1^2-\frac{1}{2}(s+m_1^2-m_2^2)+ \frac{1}{2} \; \lambda^{\frac{1}{2}}(s,m_1^2,m_2^2) \; \text{cos} \theta^{\ast} \; ,
\end{equation}
which is easily derived in the CMS, and with~(\ref{eq:p_t_definition}) we find
\begin{equation}
t_{1/2}=m_1^2-\frac{1}{2}(s+m_1^2-m_2^2) \pm \frac{1}{2} \lambda^{\frac{1}{2}}(s,m_1^2,m_2^2) \left[1-\frac{p_{t}^2}{\lambda(s,m_1^2,m_2^2)} \right] ^{\frac{1}{2}}\; ,
\end{equation}
where $t_{1}$ corresponds to the case of the ``+`` sign with $0< \Theta < \frac{\pi}{2}$ and $t_{2}$ corresponds to the ``-`` sign with $\frac{\pi}{2}<  \Theta< \pi$.
The resulting differential cross section of the transverse momentum is then given by 
\begin{eqnarray} 
\frac{d \sigma}{d p_t} = \left \vert \frac{d t}{d p_t} \right \vert \frac{d \sigma}{d t}= \frac{1}{32 \pi s^2} \frac{p_t}{\sqrt{\lambda(s,m_1^2,m_2^2)-p_t^2}} \Big[ \vert \mathcal{M}(s,t=t_{1}) \vert ^2+\vert \mathcal{M}(s,t=t_{2}) \vert ^2 \Big]
\nonumber \; .
\end{eqnarray}
Later we use histograms to calculate differential cross sections for any variable within just one run of our program. A parametrization for differential cross sections of hadronic $2 \rightarrow 2$ processes is given in section~\ref{sec:hadronic_param}. An overview of particle kinematics can be found in~\cite{Byckling}.

\section{The Parton Model}
\label{the_parton_model}
In this section the parton model is introduced. The parton model links the cross-sections on the hadronic level to the collisions of the constituents of the hadrons, quarks and gluons, also called partons. 

This is a fundamental task for making predictions for experiments at hadron colliders. Effectively, the parton model describes the collision of hadrons by the collisions of single partons while the so-called spectator partons are not involved in the reactions. The energies of the colliding hadrons are assumed to be very high while the transverse momentum of the partons is low. They all propagate in the same direction. When calculating cross sections the colliding particles are treated as free particles. This approximation is only valid since the interaction between partons is weak for high momentum transfers, i.e.\ the coupling $\alpha _s(Q^2)$ decreases with increasing momentum transfer $Q^2$. This behavior of $\alpha _s(Q^2)$ is called \textit{asymptotic freedom}. A short discussion on that can be found in~\cite{Peskin}.

If calculations are performed in the \textit{infinite-momentum-frame} partons carry the momentum fraction \textit{x} of the hadronic momentum. In this frame transverse momenta and masses of the partons can be neglected, since they are small compared to the partonic momenta. Therefore from momentum conservation one obtains
\begin{equation}
\sum \limits_{i} x_{a_i}=\sum \limits_{i} \frac{p_{a_i}}{P_{A }}=1
\end{equation}
for both hadrons, where $A$ denotes the hadron including parton $a_i$.
Then for the kinematics on parton level one finds that 
\begin{equation}
s=(p_a+p_b)^2=(x_a P_A+x_b P_B)^2 \approx 2 x_a x_b (P_A P_B) \;,
\end{equation}
where the parton masses were neglected.

\begin{figure}[t!]
\begin{center}
	\includegraphics[width=0.7\textwidth]{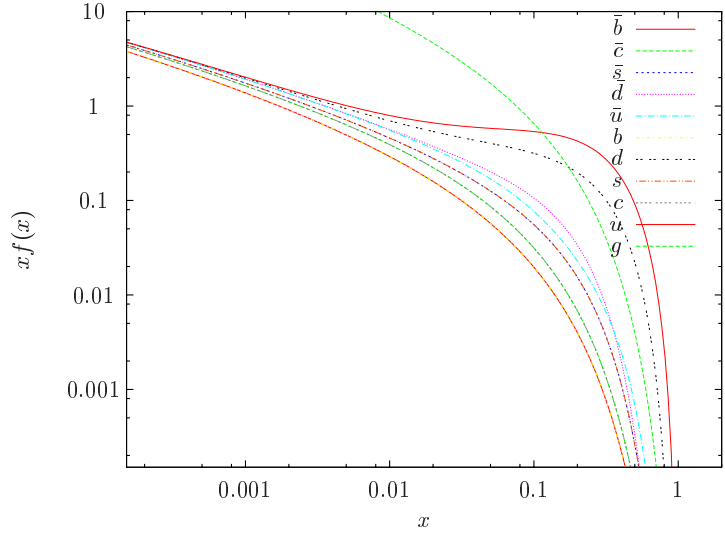}\\
\end{center}
\vspace*{-5mm}
\caption{The parton distribution functions of the proton as measured by the CTEQ collaboration in LO (CTEQ6L1)~\cite{Pumplin:hep-ph0512167}.
  \label{fig:pdfs}}
\end{figure}

For calculations of hadronic cross sections it is crucial to know the distribution of partons in the colliding hadrons. Since there is no way to find them by theoretical considerations or perturbative calculations, due to confinement of the quarks and gluons, the Parton Distribution Functions (PDFs) have to be measured. In this thesis we use the parton distribution functions for the proton from the CTEQ collaboration~\cite{Pumplin:hep-ph0512167}, shown in fig.~\ref{fig:pdfs}. One finds that u- and d-quarks have a very characteristic shape with a larger amount of higher momentum fractions $x$. The distributions of all other flavors and the antiquarks have a shape more similar to the gluon PDF, since they arise from next-to-leading-order effects in loops attached to the gluons as so-called sea-quarks.

Hadronic cross sections are finally obtained by convoluting the partonic cross-sections with the PDFs:
\begin{equation}
\vspace*{-1.5mm}
\label{eq:hadro_cross}
\sigma(S,Q^2)=\sum \limits_{i,j=g,q,\bar q} \int \limits_{x_1^{-}}^1 dx_{1} \int \limits_{x_2^{-}}^1 dx_{2} f_{i}^{h_{1}}(x_1,Q^2) f_{j}^{h_{2}}(x_2,Q^2) \hat \sigma_{ij}(x_1 x_2 S,Q^2) \; .
\end{equation}
Here the partons $i$ and $j$ carry momentum fraction $x_1$ and $x_2$ of the incoming hadrons $h_1$ and $h_2$. The sum is taken over all parton flavors, except the top quark which is not included in the PDFs of the proton. $Q$ denotes the factorization scale, usually chosen equal to the renormalization scale. We choose $Q$ to be the average mass of the final state particles. While we are able to do the partonic integration in an analytic way, all integrations on the hadronic level obviously have to be performed numerically. Therefore we make use of the Monte-Carlo integration routine Vegas~\cite{Lepage:1980dq}. 
More details about the parton model can be found in~\cite{RevModPhys.67.157}. Details on our hadronic calculations can be found in the chapters~\ref{ch:4} and~\ref{ch:6}.

To calculate the PDFs at any given scale, the running of $\alpha_s$ has to be considered. To be consistent with our choice of the leading-order PDFs, we consider the one loop running of $\alpha_s$ with five quarks, and a given value of 
\begin{equation}
\Lambda_{\text{QCD}}^{(5)}=0.165 \; \text{GeV} \; .
\end{equation}
The dependence of the renormalized strong coupling constant $\alpha_s(\mu)$ on the renormalization parameter $\mu$ in the $\overline{\textrm{MS}}$-scheme is described by the renormalization group equation 
\begin{equation}
  \label{eq:alpha_s_running}
  \frac{\partial\alpha_s (\mu)}{\partial\ln \mu^2}=\beta(\alpha_s) \; ,
\end{equation}
with
\begin{equation}
  \beta(\alpha_s) = -\frac{\alpha_s^2}{4\pi} \beta_0
\end{equation}
and
\begin{align}
  \beta_0 &= 11-\frac{2}{3}N_f \; .
\end{align}
$N_f$ denotes the effective number of quark flavors, i.e.\ the number of quarks with a mass much smaller than the energy scale $\mu$ of the process. The solution of Eq.~(\ref{eq:alpha_s_running}) at one-loop order reads
\begin{equation}
  \alpha_s(\mu)=\frac{\alpha_s(\mu_0)}{1+\frac{\beta_0}{4\pi}
  \alpha_s(\mu_0)\ln\frac{\mu^2}{\mu_0^2}}\, ,
\end{equation}
often also noted as
\begin{equation}
\frac{\alpha_s}{4 \pi}=\frac{1}{\beta_0 \text{ln} \frac{\mu^2}{\Lambda^{2}_{n_{f}}}} \; ,
\end{equation}
with $\Lambda_{n_{f}}$ accordingly defined.
If one wants to match the top-quark and other very massive particles from SUSY to the running of $\alpha_s$, one usually requires
\begin{equation}
\alpha_s^{(5)}(m_{\text{top}})=\alpha_s^{(6)}(m_{\text{top}})
\end{equation}
which ensures steadiness of the running $\alpha_s$.
At a scale of $580 \; \text{GeV}$, which is the average mass of the final state particles for our process under investigation, and including five quark flavors and the given $\Lambda_{\text{QCD}}^{(5)}$, one finds $\alpha_s=0.100375$.
Other heavier particles from supersymmetry could, in principle, also appear in loops. Due to their higher mass their effect on $\alpha_s$ should be small at LHC energies. If SUSY is studied at energies much higher than the masses of all supersymmetric particles, the whole particle content of SUSY should be included in the running, as it is explained in~\cite{Drees}. In this case one finds $\beta_0=3$ for the running of the QCD coupling.

  \clearpage{\pagestyle{empty}\cleardoublepage}

  \chapter{Standard Model Extensions}
\label{ch:3}
Until today the Standard Model of Elementary Particle Physics (SM) is very successfully in describing the observed phenomena in particle physics. But from theoretical considerations we expect that the SM is not the fundamental theory of particle physics. 

The SM describes the electroweak and strong interactions of elementary particles based on the gauge group $SU(3)_C \times SU(2)_L \times U(1)_Y$. The gauge group $SU (3)_C$ describes the strong interaction between quarks and gluons. This theory is called Quantum Chromodynamics (QCD). Electroweak interacting fields are described by the $SU(2)_L \times U(1)_Y$ gauge group. The SM contains 21 parameters which can not be derived from fundamental principles and have to be measured. Reviews of the Standard Model can be found in e.g.~\cite{Peskin},~\cite{Mandl}. An obvious weakness of the SM is that gravity is not included as one of the fundamental forces. String theories seem to be the best candidates for this.

Though the SM agrees with the observed phenomena and all predicted particles, except for Higgs boson, are found, it can only be an effective low-energy-theory since at the Planck scale $M_P\sim 10^{19} \;\text{GeV}$ quantum gravitational effects play an important role. The question why the ratio $\frac{M_P}{m_W}\sim10^{17}$ is so large can not be answered by the Standard Model itself and appears quite unnatural. This is called the \textit{hierachy problem}. It results in the instability of the energy scale of the Higgs boson when higher order corrections of the renormalizable Standard Model are calculated. The mass of the Higgs boson
\begin{equation}
m_H=v \sqrt{\frac{\lambda}{2}}
\end{equation}
depends on the vacuum expectation value $v=246 \; \text{GeV}$ and the strength of the Higgs self-coupling $\lambda$ from the Higgs potential
\begin{equation}
V=-\mu^2 \phi^{\dagger} \phi+\frac{\lambda}{4}(\phi^{\dagger} \phi)^2\; ,
\label{eq:Higgs_pot}
\end{equation}
with $\lambda > 0$ and $\mu^2 > 0$, which is essential for the spontaneous symmetry breaking mechanism. 
At one loop order, the 4-boson interaction in eq.~(\ref{eq:Higgs_pot}) yields
\begin{equation}
 \lambda \Lambda^2 \phi^{\dagger} \phi 
\end{equation}
in addition to the $-\mu^2 \phi^{\dagger} \phi$ term in the potential V, where $\Lambda$ is the cut-off parameter. The parameter $\Lambda$ represents the energy-scale where new physics appear and the Standard Model looses its validity.
These corrections to the mass appear in all renormalizable theories, i.e.\ a theory without divergences after renormalization. They are worst for scalar particles. Here one has to choose the parameters $\lambda$, $\Lambda$ and $\mu$ consistent with the phenomenologically fixed masses and the vacuum expectation value. This is usually called fine tuning. 
A more natural way to stabilize the scale of the Higgs mass would be to find a symmetry that makes the corrections for scalar particles disappear a priori. This is achieved by Supersymmetry (SUSY) which provides a fermionic partner to every boson and the other way round. Therefore fermions and bosons propagate in each loop and contribute to the mass corrections with different signs, due to the relative minus sign for closed fermion loops. Terms which are quadratic in $\Lambda$ cancel if the coupling of fermions to the Higgs boson is equal to the Higgs self-coupling $\lambda$. The corrections of the Higgs self-energy diagrams, including a fermion and the Higgs itself, yield a term 
\begin{eqnarray}
\vspace*{-15mm}
\lambda (m_H^2-m_f^2) \; \text{ln} \left(\frac{\Lambda}{m_H} \right)\; ,
\end{eqnarray}
which can be of the order of the $m^2_H$ if the fermion masses are not too large. This means that SUSY solves the hierachy problem and stabilizes the scale $m_H$ such that the corrections do not push it towards the Planck Scale $M_P$. To achieve the desired effect, the new supersymmetric partners can not be much heavier than $1-10\; \text{TeV}$~\cite{aitchison-2005}. This makes SUSY most likely observable at the LHC and leaves less room for SUSY if there are no new particles found.

Today physicists are searching for a theory including all forces and it is generally expected that at a very high scale electromagnetic, weak, strong and in the end also gravitational forces have the same value for the coupling constant. All particles are then arranged in a large multiplet, described by a larger symmetry group, interacting by only one force. This can be explicitly calculated by using renormalization group equations. While unification of the coupling constants does not work for the SM, the agreement with the idea of unification is much better for SUSY, due to supersymmetric loop contributions.

A further problem of the SM is that it does not solve the cosmological dark matter problem. Today it is known from various cosmological and astronomical observations, e.g.\ investigations of rotation curves of galaxies, that there has to be a large amount of dark matter. A candidate for this dark matter could be the lightest supersymmetric particle (LSP) if R parity, a newly introduced SUSY-quantum-number, is conserved. As a consequence the LSP does not decay to lighter SM particles.

But SUSY is not the only candidate for physics beyond the Standard Model. The dark matter problem could also be solved by Kaluza-Klein-number conserving extra dimensional theories. There one assumes that the particles known from the Standard Model can also propagate into a fifth or even more dimensions, inspired by higher dimensional string theory. In this thesis we will always use the simplest of all possible extensions to the Standard Model. In the case of Supersymmetry this is the Minimal Supersymmetric Standard Model (MSSM). As an extra dimensional model, we refer to a model with only one universal extra dimension.
\newpage
\section{Supersymmetry}
In this section we want to describe how a supersymmetric Lagrangian is constructed. In SUSY every particle from the SM receives equally many supersymmetric partner-degrees-of-freedom with the same quantum numbers except for the spin. The generators of a supersymmetric theory must turn fermionic states into bosonic states and vice versa:
\begin{equation}
\label{eq:SUSY_alg}
Q \vert \text{boson} \rangle =\vert \text{fermion} \rangle\; , \qquad  \qquad Q \vert \text{fermion} \rangle=\vert \text{boson} \rangle \; .
\end{equation}

In general there can be more than $N=1$ independent supersymmetric operator but for phenomenological studies we will assume the easiest case of $N=1$. If SUSY is an exact symmetry, the masses of particles and their supersymmetric partners would be equal. But since the SUSY partners of the known SM particles are not yet found, SUSY has to be broken.

The generators of the SUSY transformations, as given in eq.~(\ref{eq:SUSY_alg}), satisfy the following formulae:
\begin{eqnarray}
\lbrace Q_a,Q_b \rbrace &=&0 \;, \nonumber\\
\lbrace Q_a ^{\dagger},Q_b^{\dagger} \rbrace &=&0\;,\\
\lbrace Q_a,Q_b^{\dagger} \rbrace &=&\sigma^{\mu}_{ab}P_{\mu} \nonumber \; .
\end{eqnarray}
At this point one can already see that SUSY is even more than just a symmetry between fermionic and bosonic degrees of freedom but also an extension of the usual space time symmetry, since the momentum four vector is linked to the spin of the particles.

In SUSY all particles and their partners are arranged in two different kinds of supermultiplets. One is the matter or chiral multiplet that includes a two-component Weyl fermion and two real scalars called sfermions. The other is the gauge or vector multiplet including a spin 1 vector boson and a spin $\frac{1}{2}$ Weyl fermion called gaugino. Formulating a supersymmetric theory is possible by using the superfield notation or the more familiar spinor notation.

Following the discussion of~\cite{aitchison-2005}, we introduce the easiest ``supersymmetric'' Lagrangian in the two-component spinor notation, including only one complex scalar spin-0 field $\phi$ as well as one L-type, i.e. a left-chiral spinor field $\chi$:
\begin{eqnarray}
\mathcal{L}= \partial_{\mu} \phi^{\dagger} \partial^{\mu} \phi + \chi^{\dagger} i \bar{\sigma}^{\mu} \partial_{\mu} \chi \; .
\end{eqnarray}
The fields $\phi$ and $\chi$ are linked by the following SUSY transformations:
\begin{eqnarray}
\delta_{\xi} \phi &=& \xi^{T}(-i \sigma_2) \chi\; , \nonumber\\
\delta_{\xi} \chi &=& - [i \sigma^{\mu}(i \sigma_{2} \xi^{\ast})] \partial_{\mu} \phi \; .
\end{eqnarray}
Here $\xi$ is not a field, because it is independent of $x$, but a constant L-type spinor parametrizing the SUSY transformation. The transformations for their hermitian conjugates are found to be 
\begin{eqnarray}
\delta_{\xi} \phi^{\dagger} &=& \chi^{\dagger}(i \sigma_2) \xi^{\ast} \; , \nonumber\\
\delta_{\xi} \chi^{\dagger} &=& - \partial_{\mu} \phi^{\dagger} \xi^{T} i \sigma_{2} i \sigma^{\mu} \; .
\end{eqnarray}
Under these transformations the Lagrangian $\mathcal{L}$ changes only by a total derivative, leaving the action and the equations of motion invariant.
But in order to show that the theory is supersymmetric, one also has to show that the SUSY algebra closes, i.e.\ that a term like $(\delta_{\epsilon} \delta_{\xi}-\delta_{\xi} \delta_{\epsilon}) \phi$ with $\phi$ being a fermionic or bosonic field of our chiral multiplet vanishes. Unfortunately this does not hold for off-shell particles since the Weyl-equation does hold and therefore does not take out two fermionic degrees of freedom of the L-type spinor. As a result one has two bosonic and four fermionic degrees of freedom in the complex fields $\phi$ and $\chi$. Due to the fact that in every multiplet the number of degrees of freedom has to be equal, we have to introduce two more by a second scalar field $F$. 
The Lagrangian then becomes 
\begin{equation}
\mathcal{L}_{\text{chiral}}=\partial_{\mu} \phi ^{\dagger} \partial^{\mu} \phi+ \chi^{\dagger} i \bar \sigma^{\mu} \partial \chi +F^{\dagger} F \; .
\end{equation}
The SUSY transformations leaving this Lagrangian invariant are:
\begin{eqnarray}
\delta_{\xi} F &=& -i \xi^{\dagger} \bar \sigma^{\mu} \partial_{\mu} \chi \nonumber \; ,\\
\delta_{\xi} F^{\dagger} &=& \hphantom{-} i \partial_{\mu} \chi^{\dagger} \bar \sigma^{\mu} \xi \nonumber\; ,\\
\delta_{\xi} \chi &=& - [i \sigma^{\mu}(i \sigma_{2} \xi^{\ast})] \partial_{\mu} \phi+ F^{\dagger} \xi^{\dagger}\; , \\
\delta_{\xi} \chi^{\dagger} &=& - \partial_{\mu} \phi^{\dagger} \xi^{T} i \sigma_{2} i \sigma^{\mu}+\xi F \nonumber \; .
\end{eqnarray}
Under these transformations the supersymmetry algebra is closed. The first supersymmetric theory with interacting particles was formulated by Wess and Zumino~\cite{Wess}. They added the most general and renormalizable term 
\begin{equation}
\mathcal{L}_{\text{int}}=\sum_{ij} W_i(\phi,\phi^{\dagger}) F_i - \frac{1}{2} W_{ij}(\phi,\phi^{\dagger}) \chi_i \cdot \chi_j + h.c. \; .
\end{equation}
Certain conditions for the superpotential $W$ can be derived and after eliminating the non-physical degrees of freedom, i.e.\ the scalar field $F$, one finds
\begin{equation}
\mathcal{L}_{\text{WZ}}=\mathcal{L}_{\text{free}}-\vert W_i \vert ^2-\frac{1}{2} \lbrace W_{ij} \; \chi_i \cdot \chi_j +h.c. \rbrace \; ,\\
\end{equation}
with 
\begin{eqnarray}
W &=& \frac{1}{2}M_{ij} \, \phi_i \phi_j+ \frac{1}{6} y_{ijk} \, \phi_i \phi_j \phi_k \; ,\\
W_i &=& \frac{\partial W}{\partial \phi_i} \; ,\\
W_{ij} &=& \frac{\partial^2 W}{\partial \phi_i \partial \phi_j} \; .
\end{eqnarray}
$W$ is called \textit{superpotential}. Only taking into account terms which are quadratic in the fields and calculating their equations of motion one finds that for the fields $\phi$ and $\chi$ the usual Klein-Gordon-equation is satisfied for each component. Both fields have the same mass $M$ coming from the interaction term in the Wess-Zumino model. This shows that in unbroken SUSY masses are equal within the same supermultiplet.

Beside the chiral multiplets one also has to take into account the vector or gauge multiplets of the different gauge groups. Here we meet the same problems concerning the degrees of freedom as before and we have to introduce a second auxiliary field. For the $SU(2)$ and $SU(3)$ gauge groups the only difference is located in the different generators of the gauge groups and therefore in the number of corresponding gaugino fields. The resulting supersymmetric Lagrangian for the gauge multiplet is given by
\begin{eqnarray}
 {\cal{L}}_{{\rm gauge}}= -\frac{1}{4}F^\alpha_{\mu \nu}F^{\mu \nu \alpha} + 
    {\rm i} \lambda^{\alpha \dagger} {\bar{\sigma}}^\mu (D_\mu \lambda)^\alpha +
     \frac{1}{2}D^\alpha D^\alpha \; .
\end{eqnarray}
Here $D^{\alpha}$ is a scalar auxiliary field, contributing the additional degrees of freedom, $F^{\mu \nu}$ is the usual Maxwell field strength tensor and $\lambda$ is a L-type spinor, the gaugino. In the case of an SU(2) gauge group a triplet of gauginos is needed while for the SU(3) gauge group a gaugino octet appears. The same applies to the auxiliary field $D^{\alpha}$.
For the SUSY transformations one finds
\begin{eqnarray}
\delta_\xi W^{\mu \alpha}&=&\xi^\dagger {\bar{\sigma}}^\mu \lambda^\alpha + \lambda^{\alpha \dagger} 
    {\bar{\sigma}}^\mu \xi \;, \nonumber \\
    \delta_\xi\lambda^\alpha &=& \frac{1}{2}{\rm i} \sigma^\mu {\bar{\sigma}}^\nu \xi F^\alpha_{\mu \nu} 
    + \xi D^\alpha \;, \nonumber \\
    \delta_\xi D^\alpha &=& - {\rm i} (\xi^\dagger {\bar{\sigma}}^\mu (D_\mu \lambda)^\alpha 
    -(D_\mu \lambda)^{\alpha \dagger}{\bar{\sigma}}^\mu \xi) \;.
\end{eqnarray}
In order to retain gauge invariance of the supersymmetric Lagrangian one has to replace the usual derivative by the covariant derivative. 

Since we now have a SUSY invariant gauge and chiral supermultiplet Lagrangian we can ask for interactions between particles from those two multiplets. After dimensional considerations, inserting all possible interaction terms and checking SUSY invariance and gauge invariance, one obtains the following Lagrangian for the combined gauge and chiral multiplets in the non-abelian case:
\begin{align}
\mathcal{L}_{\text{gauge} \; \& \; \text{chiral}}=\;&\mathcal{L}_{\text{gauge}}+\mathcal{L}_{\text{chiral}}+\mathcal{L}_{\text{int.\! gauge} \; \& \; \text{chiral}}\nonumber\\
=&-\frac{1}{4}F^\alpha_{\mu \nu}F^{\mu \nu \alpha} + 
    {\rm i} \lambda^{\alpha \dagger} {\bar{\sigma}}^\mu (D_\mu \lambda)^\alpha +
     \frac{1}{2}D^\alpha D^\alpha+ \nonumber\\
&+D_\mu \phi_i^\dagger D^\mu \phi_i + \chi_i^\dagger {\rm i} {\bar{\sigma}}^\mu D_\mu \chi_i + F_i^\dagger F_i+\\ 
&+\left[ \frac{\partial W}{\partial \phi_i} F_i
-\frac{1}{2} \frac{\partial^2 W}{\partial \phi_i 
     \phi_j} \chi_i \cdot \chi_j + {\rm h.c.}\right]+ \nonumber\\ 
&-\sqrt{2} \; g \left[ (\phi_i^\dagger T^\alpha \chi_i) \cdot \lambda^\alpha + \lambda^{\alpha \dagger} \cdot 
       (\chi_i^\dagger T^\alpha \phi_i) \right] -g(\phi_i^\dagger T^\alpha \phi_i)D^\alpha \;. \nonumber
\end{align}
Both parameters $\xi$ parametrizing the SUSY transformations of chiral and gauge multiplet do not necessarily agree. In fact one finds for the gauge multiplet that the parameter $\xi$ has to be changed into $-\xi/\sqrt{2}$. The transformation of $F_i^{\dagger}$ for the combined Lagrangian has to be changed and is now given by
\begin{eqnarray}
\delta_\xi F_i^\dagger =-\sqrt{2}g \phi_i^\dagger T^\alpha \xi \cdot \lambda^\alpha + i \partial_{\mu} \chi^{\dagger} \bar \sigma^{\mu} \xi\; .
\end{eqnarray}
Since the fields $F$ and $D$ can be replaced in the given Lagrangian, one obtains the whole scalar potential consisting of the so-called $F$- and $D$-term, as
\begin{equation}
{{V}}(\phi_i, \phi_i^\dagger)=|W_i|^2 + 
     \frac{1}{2}\sum_{{\rm G}}\sum_\alpha \sum_{i, j} g_{\rm G}^2 
     \left( \phi_i^\dagger  T_{\rm G}^\alpha \phi_i \right) \left(\phi_j^\dagger T_{\rm G}^\alpha \phi_j \right)\; ,
\end{equation}
where in general the sum is taken over different gauge groups, their generators and the arising fields.

There are even some more possible interaction terms which keep the Lagrangian invariant under SUSY transformations and are renormalizable. But they either violate lepton number $L$ or baryon number $B$. Since one did not observe $L$ and $B$ violating processes, e.g.\ proton decay, one has to exclude these terms from the Lagrangian by claiming a symmetry to hold. This is the so-called $R$-parity, given by
\begin{equation}
R=(-1)^{3B+L+2s} \;,
\end{equation}
where $b$ is $1/3$ and $-1/3$ for quarks and antiquarks, $L$ equals $1$ and $-1$ for leptons and antileptons and $s$ corresponds to the spin of the particle. Therefore $R$ equals $+1$ for conventional particles and $-1$ for their SUSY-partners. As a consequence SUSY particles can only appear in pairs at a given vertex. In our calculations we assume $R$-parity to be always conserved. So the lightest supersymmetric particle (LSP) is stable and therefore a serious candidate for dark matter. This LSP is usually the lightest neutralino. When SUSY searches are performed at the LHC, two times the mass of the LSP is the minimum amount of missing energy. 

\subsection{The MSSM}
Since we now have an understanding of how the multiplets are set up, we present the particle spectrum of the MSSM in table~\ref{tab:all_particles}. Due to its length we do not want to introduce the Lagrangian for the physical fields. It can be found in~\cite{Haber:1984rc} and~\cite{PhysRevD.41.3464}. 
In the last section we mentioned the superpotential $W$ which has not been specified yet. The MSSM is given by the following choice for the superpotential
\begin{equation}
W= y_{\rm u}^{ij}{\bar{ u}}_i Q_j  H_{\rm u} - 
     y_{\rm d}^{ij}{\bar{ d}}_i Q_j H_{\rm d} 
      - y_{\rm e}^{ij} {\bar{ e}}_i L_j  H_{\rm d} + 
     \mu H_{\rm u} H_{\rm d} \; ,
\end{equation}
where the contraction over gauge indices is implicit. A new free parameter $\mu$, the Higgs self-coupling is introduced in this superpotential. $Q$, $L$, $e$, $u$ and $d$ denote the superfields as they are given in table~\ref{tab:all_particles}. $H_u$ and $H_d$ have to be understood as 
\begin{equation}
H_u=\left (H^+_u \atop H^0_u \right ) \; , \qquad \qquad  H_u\left (H^0_d \atop H^-_d \right ).
\end{equation}
For anomaly cancellation it is important to have two Higgs fields, both having opposite hypercharge. In order to break SUSY one introduces so-called \textit{soft SUSY-breaking terms} and breaks the symmetry explicitly. These terms are called ``soft`` because they do not destroy the solution of the hierachy problem by reintroducing new divergences. 

The origin of the soft SUSY-breaking terms is far from clear. Explaining the different possibilities for their existence is non-trivial and goes far beyond a short introduction. Including the SUSY breaking terms more than 100 parameters are added to SUSY, making concrete predictions quite difficult. When a special SUSY breaking mechanism is assumed this number is drastically reduced. The mechanisms most often discussed are minimal supergravity (mSUGRA), gauge-mediated SUSY breaking (GMSB) and anomaly-mediated SUSY breaking (AMSB). SUSY is broken in a hidden sector in all the scenarios and is mediated to the visible sector by gravitational or gauge interactions or by the Super-Weyl anomaly. A short overview of these SUSY breaking schemes is given in~\cite{martin-1997}.

In the mSUGRA scenario the whole parameter space is described by four parameters and a sign. These are the scalar mass parameter $m_0$, the gaugino mass parameter $m_{1/2}$, the trilinear coupling $A_0$ , the ratio of the Higgs vacuum expectation values $\text{tan} \beta$ and the sign of the Higgs sector parameter $\mu$.  From these parameters one obtains the mass spectra of the SUSY particles by renormalization group running from the SUSY breaking high-energy scale to the weak scale where they can be used for calculations of cross sections or decay branching ratios.

But even for a number of five parameters, instead of more than 100, scanning the whole parameter space for simulations of SUSY signatures is much too time-consuming. Therefore one uses selected parameter points~\cite{allanach-2001-010630}, which are consistent with all constraints from phenomenological studies, like the dark matter content of the universe and SUSY searches at Tevatron. For example, a region with $\text{tan} \beta \lesssim 3$ is excluded by LEP experiments.

As our standard scenario we choose the parameter point SPS 1a, which is a typical mSUGRA scenario. It is specified by 
\begin{equation}
m_0 = 100 \text{GeV}, \quad m_{1/2} = 250 \text{GeV}, \quad A_0 = -100 \text{GeV}, 
\quad \tan\beta = 10, \quad \mu > 0 \nonumber \; .
\end{equation}
\begin{table}
\begin{center}
	\includegraphics[width=1.0\textwidth]{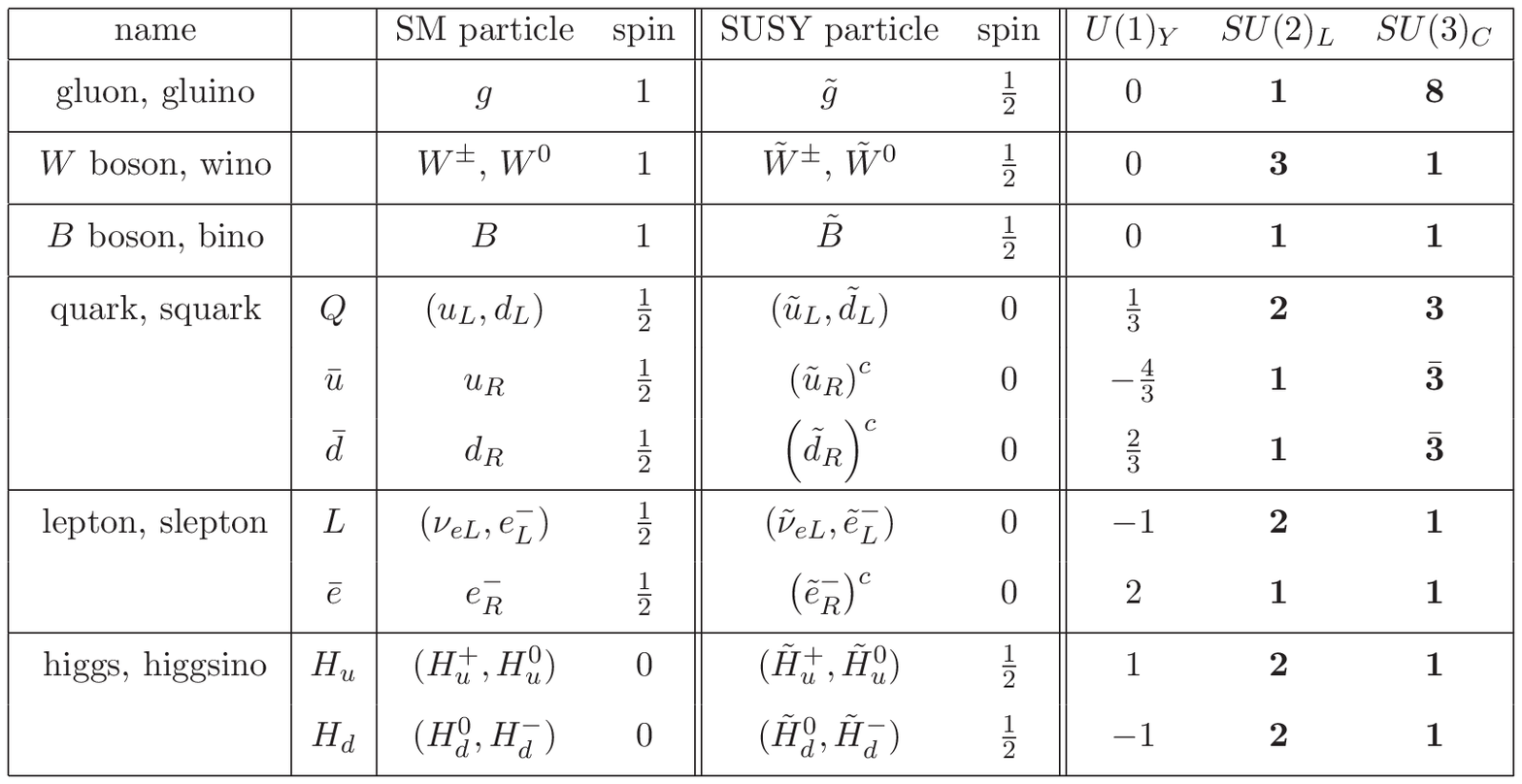}\\
\end{center}
\vspace*{-5mm}
\caption{Particle content of the MSSM with three families for quarks and leptons.
  \label{tab:all_particles}}
\end{table}

 \subsection{How to treat Majorana Particles}

Unlike in the SM, in SUSY various Majorana particles appear. Majorana particles have the interesting feature that they decay equally often into particles and antiparticles. Therefore processes like the gluino pair production can be used to find the helpful like-sign dilepton signature in order to claim that SUSY is observed. This is especially interesting at hadron colliders, since the production cross section for gluino pairs is quite large~\cite{barnett}.

Compared to off-shell fermions, the Dirac equation reduces the number of degrees of freedom of on-shell fermions by a factor of two. Off-shell Majorana fermions have, instead of eight degrees of freedom as usual fermions, a reduced number of four degrees of freedom. This is obvious from the general condition for Majorana fermions
\begin{equation}
\Psi_{M,C}^{\psi}= \left( { 0\atop -i \sigma_2}{i \sigma_2 \atop 0}\right) \left( {\psi^{\ast}} \atop -i \sigma_2 \psi \right)= \left( \psi \atop -i \psi^{\ast} \right)= \Psi_M^{\psi} \; .
\end{equation}
Therefore they can be represented in a unique way by two component complex spinors $\psi$ or $\chi$ as well as by a four dimensional spinor
\begin{equation}
\Psi_M^{\psi}=\left(\psi \atop -i \sigma_2 \psi^{\ast} \right) \qquad \text{or equivalently} \qquad \Psi_M^{\chi}=\left(i \sigma \chi^{\ast} \atop \chi \right) \; .
\end{equation}
The spinors $\psi$ or $\chi$ are defined by using the chirality projection operators as
\begin{equation}
P_R \; \Psi=\left(\psi \atop 0 \right) \qquad \text{and} \qquad P_L \; \Psi=\left( 0 \atop \chi \right) \; ,
\end{equation}
with 
\begin{equation}
P_R=\left(\frac{1+ \gamma_5}{2}\right)  \qquad \text{and} \qquad P_L=\left(\frac{1- \gamma_5}{2}\right ).
\end{equation}

In~\cite{1992NuPhB.387..467D} Feynman rules for Majorana fermions are derived and it is proved that it is possible to treat them nearly equal to Dirac fermions. Therefore one chooses an arbitrary direction for the fermion flow and writes down the Feynman rules for the Majorana fermions in the same way as one does for Dirac fermions, going against the fixed flux. For the external Majorana fermions one uses the same spinors $u(\textbf{p},s), v(\textbf{p},s)$ and the same spin sums 
\begin{equation}
\sum \limits_s u^s(p) \bar{u}^s(p)= \slashed{p}+m \qquad \text{and} \qquad \sum \limits_s v^s(p) \bar{v}^s(p)= \slashed{p}-m
\end{equation}
as in the case of Dirac fermions. For the correct choice of the external spinor, depending on momentum and direction of the fermion flow, we use the Feynman rules given in~\cite{1992NuPhB.387..467D}.

Of course Majorana fermions have no arrow on their lines indicating if they are particle or anti-particle since they are both. These arrows, belonging to Dirac fermions, indicate the so-called fermion \textit{number} flow.

Since one fixes the fermion flow in the Feynman diagrams there are two different Feynman rules for a vertex containing Dirac and Majorana fermions, one with the fermion number flow parallel and one anti-parallel to the direction of the fermion flow. These two rules are, as in the case of the gluino-quark-squark vertex, associated with the helicity of the squark:
\begin{equation}
\mathcal{L}_{q \tilde{g} \tilde{q}}=- \sqrt{2} g_s t^{a}_{ij} \sum_{f=u,d} (\bar{\tilde{g}}_a \; P_L \; q_f^j \; \tilde{q}_{Lf}^{\ast i} + \bar{q}_f^i \; P_R \; \tilde{g}_a \; \tilde{q}_{Lf}^j-\bar{\tilde{g}}_a \; P_R \; q_f^j \; \tilde{q}_{Rf}^{\ast i} - \bar{q}_f^i \; P_L \; \tilde{g}_a \; \tilde{q}_{Rf}^j). \nonumber
\end{equation}
Here $\tilde{g}$ denotes the Majorana gluino, $q$ a usual Dirac quark and $\tilde{q}$ a scalar squark. Obviously so-called left-/right-handed squarks only couple to left-/right-handed quarks. The Feynman rules for SUSY-QCD, including quark-gluino-squark interactions, can be found in the appendix.


\section{Extra Dimensions}
Though Supersymmetry is the most frequently studied extension of the SM, there are also other theories solving problems like the dark matter and the hierachy problem. One of them is similar to the Standard Model extended by additional space-like dimensions, called Extra Dimensions. There are different kinds of extra dimensional theories discussed in the literature, beginning in 1919 when Kaluza and Klein unified gravity and electromagnetism using a hidden fifth dimension, compactified on a circle~\cite{Kaluza:1921tu},~\cite{Klein}.
Later it was realized that string theories can not be formulated in four dimensional space-time. In the following years, different ideas like flat, large extra dimensions on a millimeter-scale were proposed in~\cite{arkanihamed-1998-429} in 1998 (ADD). In this model only gravity, i.e.\ the graviton, can propagate into the extra dimension, while all SM particles are confined to the four dimensional space-time.
In contrast to this, a theory was proposed in 2001, where all SM particles can propagate into the extra dimension~\cite{appelquist-2001-64}. This kind of extra dimensional theory is therefore called Universal Extra Dimension (UED). In the literature one can additionally find calculations with fermions confined to the usual 4D world, while gauge bosons can also propagate into the extra dimension~\cite{PhysRevD.65.076007},~\cite{macesanu-2002-66}.

Various problems can be solved by UED models. Including boundary conditions the symmetry breaking in the electroweak sector can be performed higgsless, being an interesting option if Higgs searches at the LHC fail~\cite{csaki-2005}. There is also a solution for the dark matter problem of cosmology. When Kaluza-Klein (KK) parity, a quantum number being quite similar to the R-parity in SUSY, is conserved, a lightest electrically uncharged and colorless KK-particle is present. This particle is an interesting dark matter candidate~\cite{Bringmann}. 

It is known that higher dimensional operators violate global symmetries like baryon number and lepton number and can create flavor changing neutral currents. The SM, by chance, has no problems with baryon-number violating interaction terms mediating the proton decay, since all possible terms are of higher dimension and therefore non-renormalizable. In contrast UED in principle can include these higher dimensional operators. But since the proton decay is limited to very low decay rates by experiment, there should be a mechanism suppressing these higher dimensional terms in UED. Therefore propositions like \text{split fermions}~\cite{csaki-2004} or simply a suppression by the cut-off scale are made.

In a model with universal extra dimensions, i.e.\ all particles of the theory can propagate into the extra dimension, the size of the extra dimension is of the order of $R \sim \text{TeV}^{-1}$. Otherwise their existence would contradict observations since they have not yet been found. This can only be solved by compactifying the extra dimension on a scale where large energies are needed to see their effects. In our calculations we use this UED model and only one additional space-like extra dimension. 

Here we concentrate on the collider implications of UED. We do not want to give an overview of the different extra dimensional theories but instead explain in detail how to derive the Feynman rules and the UED Lagrangian in four dimensions. A collection of UED Feynman rules can be found in the appendix. For simplicity we concentrate on the QCD part of the Lagrangian and derive the couplings of the neutral KK-gauge bosons.
In this thesis we do not intend to show that UED is the theory realized by nature but contrast it to SUSY to investigate the differences coming from the different spin assignments. While the SUSY particles have different spin compared to their SM partners, particles in UED have the same spin statistic behavior as their SM partners. For example, gluons have bosonic partners in UED (heavy KK-gluons) while they have fermionic partners in SUSY (gluinos). One can find numerous papers on how UED could mimic the signature of a supersymmetric theory, e.g.~\cite{cheng-2002-66}, though SUSY and UED have fundamental differences. While usually the mass spectrum in UED is quite degenerate and thereby different from SUSY, radiative corrections and boundary conditions can make the spectrum more SUSY-like. Reviews on extra dimensional theories can be found in~\cite{csaki-2004},~\cite{kribs-2006} and in~\cite{Hooper:hep-ph0701197}. The latter one focuses on collider phenomenology and dark matter.

\subsection{Deriving a UED-QCD Lagrangian in 4D}
In this chapter we derive the 4D effective Lagrangian and the UED-QCD Feynman rules. The basic idea is to generalize the 4D Lagrangian to its analog in 5D, then to Fourier expand the Lagrangian in the coordinate $y$ of the extra dimension. Orbifold compactification of the fifth dimension is performed and the usual SM terms plus additional interactions of KK-excitations are obtained, i.e.\ one obtains the 4D Lagrangian by integrating out the extra dimension. From the Lagrangian we get the masses and couplings of the new particles. Since our discussion is parallel to those in~\cite{appelquist-2001-64},~\cite{PhysRevD.65.076007} and~\cite{macesanu-2002-66}, we denote the multiplets of the Standard Model by
\begin{equation}
Q_L^{SM}(x), U_R^{SM}(x) \;\text{and}\; D_R^{SM}(x) \; .
\end{equation}
As we will see later, each of the SM particles has infinitely many KK-partners. These are equidistant in mass and form a KK-tower. The different excitation levels of a tower are denoted by an index $n$. At the LHC, we can only expect the first KK-partner of this tower to be produced if $R$ is sufficiently small. Therefore in our calculations we do not take into account the heavier KK-partners. 

Since SM quarks are assumed massless in our calculations they can be denoted as two component Weyl-Spinors in 4D. In the 5D theory quark multiplets consist of massless four component quark fields $Q(x,y), U(x,y)$ and $D(x,y)$. 
There are no chiral fermions in a 5D theory, since it is not possible to construct a $\gamma$-matrix in an even number of spatial dimensions, having the same anti-commutativity property as $\gamma_5$ in 4D. Bilinears containing $\gamma_5$-matrices are not Lorentz-invariant in a 5D theory and can therefore not be included in the Lagrangian.

The additionally needed $\gamma$-matrix is chosen to be 
\begin{equation}
\gamma^{D} = (-i)^{D/2+s+1} \; \gamma^0 \gamma^1 \dots  \gamma^{D-1}\; ,
\end{equation}
where $\gamma^0, \dots ,\gamma^{D-1}$ are the $\gamma$-matrices in even dimensions $D$ and $s$ is the number of space-like dimensions. Since this $\gamma$-matrix anticommutes with $\gamma^0, \dots ,\gamma^{D-1}$, it can be used as $\gamma^{4}$ in five dimensions.
This choice for the additional $\gamma$-matrix is the usual choice in a higher dimensional supergravity theory, as explained e.g.\ in~\cite{vanproeyen-1999}. In the case of a 5D theory this yields
\begin{equation}
\gamma^4 = i^{2} \gamma^0 \gamma^1 \gamma^{2} \gamma^{3} = i \gamma_5 \;.
\end{equation}
For $\gamma^4= i\gamma_5$ the Clifford Algebra 
\begin{equation}
\gamma_{\mu} \gamma_{\nu}+\gamma_{\nu} \gamma_{\mu}=2 \; g_{\mu \nu} = 2\; \text{diag}(+1,-1,-1,-1,-1)
\end{equation}
still holds but it does not anticommute with $\gamma_5$ as the other $\gamma$-matrices $\gamma^0, \dots ,\gamma^3$.

We are forced to introduce two 5D fermionic fields with the quantum numbers of a left- and a right-chiral spinor. Decomposing these 5D fields into 4D one obtains a left- and a right-handed zero-mode. Of course, since right-handed doublets and left-handed singlets are not observed, at least their zero-modes have to vanish. Therefore one chooses a $S_1/Z_2$ compactification. Higher modes $n\geqslant1$ can, in general, contain both, left- and right-chiral particles without a contradiction to experiments.
In order to identify $Q_L^{SM}(x), U_R^{SM}(x) \; \text{and} \; D_R^{SM}(x)$ with the SM particles, the zero-modes of the Fourier expansion of $Q(x,y), U(x,y)$ and $D(x,y)$ must be even under $y \rightarrow -y$. This yields the following Fourier expansions for the doublet $Q$ and the singlets $U$ and $D$:
\vspace{-0pt} \begin{eqnarray*} Q (x,y) &\!\!\! = &\!\!\!
\frac{1}{\sqrt{\pi R}} \left[ \left( \! \begin{array}{c} u (x)
\\ d (x) \end{array} \! \right)_{\!L} + \sqrt{2} \sum_{n=1}^{\infty}
\left[ Q_L^n (x) \cos \left(\frac{n y}{R} \right) + Q_R^n (x) \sin
\left(\frac{n y}{R} \right) \right] \right] \label{eq:Qdecomp}\;, \\
U (x,y) &\!\!\! = &\!\!\! \frac{1}{\sqrt{\pi R}} \left[ u_R (x) +
\sqrt{2} \sum_{n=1}^{\infty} \left[ U_R^n (x) \cos \left(\frac{n
y}{R} \right) + U_L^n (x) \sin \left(\frac{n y}{R} \right) \right]
\right] \label{eq:Udecomp}\;, \\
D (x,y) &\!\!\! = &\!\!\! \frac{1}{\sqrt{\pi R}} \left[ d_R (x) +
\sqrt{2} \sum_{n=1}^{\infty} \left[ D_R^n (x) \cos \left(\frac{n
y}{R} \right) + D_L^n (x) \sin \left(\frac{n y}{R} \right) \right]
\right] \label{eq:Ddecomp} \, .
\end{eqnarray*}
Since the zero-mode of a gauge field, polarized along a direction in 4D, must be even under $y \rightarrow -y$ it must be odd when it is polarized along the $y$ axis. Otherwise we would find unphysical massless zero-modes in the 4D effective Lagrangian. Therefore its Fourier expansion is given by
\vspace{-0pt} \begin{eqnarray} \label{eq:expansionA} A_{\mu}^a (x,y) =
\frac{1}{\sqrt{\pi R}}\left[ A_{\mu 0}^a (x) +
\sqrt{2} \sum_{n=1}^{\infty}A_{\mu ,n}^{a} (x) \cos \left(\frac{n y}{R} \right) \right]\;,\\
A_{4}^a (x,y) = \frac{\sqrt{2}}{\sqrt{\pi R}}
\sum_{n=1}^{\infty}A_{4,n}^{a} (x) \sin\left(\frac{n y}{R} \right) \, .
\end{eqnarray}
In unitary gauge the last term disappears, due to the gauge choice $A_{4,n}^a(x)=0$. Normalization of the zero modes is different from the higher modes in order to obtain canonically normalized terms in the kinetic part of the Lagrangian and because of the limits for the integration of $y$.
The Lagrangian in 5D is given by
\vspace{-0pt} \begin{equation} \label{eq:LkinQ} \mathcal{L}_5  = i
\bar{Q} (x,y)\;  \gamma^m \big[ \partial_m + i g_5 t^a \; A_m^a
(x,y) \big] Q (x,y) \, 
\end{equation}
with $g = \frac{g_5}{\sqrt{\pi R}}$, where $g$ equals the usual SM coupling constant $g_s$ and $R$ is the radius of compactification. The index $m$ is equivalent to the usual 4D index $\mu$, here also including the extra dimension, i.e.\ $m \in \{\mu,4 \}$. 
The KK-mass eigenstates are given by
\vspace{-3pt} \begin{equation*} 
\label{eq:id} Q_{L,R}^n (x) \equiv
P_{L,R} \left( \! \begin{array}{c} u_{n,1} (x)
\\ d_{n,1} (x) \end{array} \! \right) \, , \;\;
U_{R,L}^n (x) \equiv P_{R,L} \; u_{n,2}(x) \, , \;\; D_{R,L}^n (x)
\equiv P_{R,L} \; d_{n,2}(x) \;,
\end{equation*}
with $P_{L,R}=\frac{1}{2}(1\mp \gamma_5)$. The fields $q_{n,1}$ and $q_{n,2}$ denote two towers of KK-partners, arising for each usual SM quark $q$. Here the first index denotes the excitation level of the KK-tower while the second index denotes the tower itself. Each of the fields $Q$,$U$ and $D$ gets a contribution to its mass from integrating the kinetic term. This is different from the SM, where fermionic masses only come from the Yukawa interaction terms and the electroweak symmetry breaking. In principle the fields $q_{n,1}$ and $q_{n,2}$ could also get contributions to their mass from symmetry breaking.
Integration of the kinetic terms, after insertion of the Fourier expansions of $Q$,$U$ and $D$, yields the effective 4D kinetic and mass terms
\vspace{-1pt} \begin{eqnarray} i \!\!
\int_{\mbox{\raisebox{-1.3ex}{\scriptsize{$\!\!\!\!
0$}}}}^{\mbox{\raisebox{.9ex}{\scriptsize{$\!\!\!\! \pi R$}}}} &
\!\!\!\!\!\!\!\! \bar{Q} (x,y) & \!\!\!\! \gamma^m
\partial_m Q (x,y) dy
= i \left[ (\bar{u}(x) \bar{d}(x))_L \gamma^{\mu}
\partial_{\mu} \left( \!
\begin{array}{c} u (x)
\\ d (x) \end{array} \! \right)_{\!L} \right. \nonumber \\
&\!\!\! + &\!\!\!\!\!\!\! \sum_{n=1}^{\infty} \left.
\bar{Q}_L^n(x) \gamma^{\mu}
\partial_{\mu} Q_L^n(x) + \bar{Q}_R^n(x) \gamma^{\mu}
\partial_{\mu} Q_R^n(x) \right. \\
&\!\!\! + &\!\!\!\!\!\!\! \left. i\frac{n}{R}\bar{Q}_L^n (x)
Q_R^n(x) + i\frac{n}{R}\bar{Q}_R^n (x) Q_L^n(x) \right] \nonumber
\, ,
\end{eqnarray}
\vspace{-3pt} \begin{eqnarray} i \!\!
\int_{\mbox{\raisebox{-1.3ex}{\scriptsize{$\!\!\!\!
0$}}}}^{\mbox{\raisebox{.9ex}{\scriptsize{$\!\!\!\! \pi R$}}}} &
\!\!\!\!\!\!\!\! \bar{U} (x,y) & \!\!\!\! \gamma^m
\partial_m U (x,y) dy
= i \left[ (\bar{u}(x))_R \gamma^{\mu}
\partial_{\mu} \left(
u (x) \right)_{\!R} \right. \nonumber \\
&\!\!\! + &\!\!\!\!\!\!\! \sum_{n=1}^{\infty} \left.
\bar{U}_R^n(x) \gamma^{\mu}
\partial_{\mu} U_R^n(x) + \bar{U}_L^n(x) \gamma^{\mu}
\partial_{\mu} U_L^n(x) \right. \\
&\!\!\! + &\!\!\!\!\!\!\! \left. i\frac{n}{R}\bar{U}_R^n (x)
U_L^n(x) + i\frac{n}{R}\bar{U}_L^n (x) U_R^n(x) \right] \nonumber
\, 
\end{eqnarray}
and similarly for the second singlet $D(x,y)$. The last term, i.e.\ the mass term mixing left- and right-handed fields, derives from integrating the extra dimensional term with $m=4$.
The $\gamma_4$-matrix disappears from the Lagrangian since 
\begin{equation}
 \gamma_5 (1 \mp \gamma_5)= \mp (1 \mp \gamma_5) \;.
\end{equation}
One also receives the mass of the heavy KK-partners $M_n =\frac{n}{R}$, depending on the size of the extra dimension. Contributions to the KK-masses from the Higgs part of the Lagrangian are neglected here.
The interaction term is the second term in our Lagrangian in eq.~(\ref{eq:LkinQ}), given by the usual covariant derivative, and yields
\vspace{-3pt} \begin{eqnarray}
\label{eq:example_Lag}
 -&\!\!\!\!\!g_{{}_5} &\!\!\!\!\!\!
\int_{\mbox{\raisebox{-1.3ex}{\scriptsize{$\!\!\!\!
0$}}}}^{\mbox{\raisebox{.9ex}{\scriptsize{$\!\!\!\! \pi R$}}}}
\bar{Q} (x,y) \gamma^m t^a A_m^a (x,y) Q (x,y) d y \nonumber \\
= -& \!\!\!\!\!\!\! g &\!\!\!\!\! \left.
\mbox{\raisebox{-.5ex}{\huge$[$}} \bar{q}_L (x) \gamma^{\mu} t^a
q_L (x) A_{\mu ,0}^{a} (x) + \sum_{n=1}^{\infty} \left[\bar{Q}_L^n
(x) \gamma^{\mu} t^a Q_L^n(x)
+ \bar{Q}_R^n \gamma^{\mu} t^a Q_R^n (x) \right] A_{\mu ,0}^{a}(x) \right. \nonumber \\
& \!\!\! + & \!\!\! \sum_{n=1}^{\infty} \left[ \bar{q}_L(x)
\gamma^{\mu} t^a Q_L^n(x) + \bar{Q}_L^n(x) \gamma^{\mu} t^a q_L(x)
\right]
A_{\mu ,n}^{a} (x) \\
& \!\!\! + & \!\!\! \frac{1}{\sqrt{2}} \sum_{n,m,\ell =1}^{\infty}
\left. \left[ \bar{Q}_L^n (x) \gamma^{\mu} t^a Q_L^m(x)
(\delta_{\ell,\mid m - n \mid}+\delta_{\ell,m+n}) \right. \right. \nonumber \\
& \!\!\! + & \!\!\! \left. \left. \bar{Q}_R^n (x) \gamma^{\mu} t^a
Q_R^m (x) (\delta_{\ell,\mid m - n \mid}-\delta_{\ell,m+n})
\right] A_{\mu ,\ell}^{a} \mbox{\raisebox{-.5ex}{\huge$]$}}
\right. \nonumber \, .
\end{eqnarray}
For $U$ and $D$ fields we again obtain a similar expression given by
\vspace{-3pt} \begin{eqnarray} -&\!\!\!\!\!g_{{}_5} &\!\!\!\!\!\!
\int_{\mbox{\raisebox{-1.3ex}{\scriptsize{$\!\!\!\!
0$}}}}^{\mbox{\raisebox{.9ex}{\scriptsize{$\!\!\!\! \pi R$}}}}
\bar{U} (x,y) \gamma^m t^a A_m^a (x,y) U (x,y) d y \nonumber \\
= -& \!\!\!\!\!\!\! g &\!\!\!\!\! \left.
\mbox{\raisebox{-.5ex}{\huge$[$}} \bar{u}_R (x) \gamma^{\mu} t^a
u_R (x) A_{\mu ,0}^{a} (x) + \sum_{n=1}^{\infty} \left[\bar{U}_R^n
(x) \gamma^{\mu} t^a U_R^n(x)
+ \bar{U}_L^n \gamma^{\mu} t^a U_L^n (x) \right] A_{\mu ,0}^{a}(x) \right. \nonumber \\
& \!\!\! + & \!\!\! \sum_{n=1}^{\infty} \left[ \bar{u}_R(x)
\gamma^{\mu} t^a U_R^n(x) + \bar{U}_R^n(x) \gamma^{\mu} t^a u_R(x)
\right]
A_{\mu ,n}^{a} (x) \\
& \!\!\! + & \!\!\! \frac{1}{\sqrt{2}} \sum_{n,m,\ell =1}^{\infty}
\left. \left[ \bar{U}_R^n (x) \gamma^{\mu} t^a U_R^m(x)
(\delta_{\ell,\mid m - n \mid}+\delta_{\ell,m+n}) \right. \right. \nonumber \\
& \!\!\! + & \!\!\! \left. \left. \bar{U}_L^n (x) \gamma^{\mu} t^a
U_L^m (x) (\delta_{\ell,\mid m - n \mid}-\delta_{\ell,m+n})
\right] A_{\mu ,\ell}^{a} \mbox{\raisebox{-.5ex}{\huge$]$}}
\right. \nonumber \, .
\end{eqnarray}
Expressing everything in terms of the mass eigenstates $q_{n,1}$ and $q_{n,2}$, and summing up the contributions of both heavy quark towers, one obtains
\vspace{-3pt} \begin{eqnarray} \mathcal{L}_{\text{int}} = &
\!\!\! -&\!\!\!\!\! g \mbox{\raisebox{-.5ex}{\huge$[$}} \!\!
\left. \bar{q} (x) \gamma^{\mu} t^a q (x) A_{\mu ,0}^{a} (x) \right. \nonumber\\ 
& \!\!\! + & \!\!\!
\sum_{n=1}^{\infty} \left[ \bar{q}_{n,1} (x) \gamma^{\mu} t^a q_{n,1} (x) + \bar{q}_{n,2}
(x) \gamma^{\mu} t^a q_{n,2} (x) \right]
 A_{\mu ,0}^{a}(x)  \nonumber \\
& \!\!\! + & \!\!\! \sum_{n=1}^{\infty} \left[ \bar{q}_L(x)
\gamma^{\mu} t^a q_{n,1} (x) + \bar{q}_{n,1} (x) \gamma^{\mu} t^a q_L(x)
\right] A{}_{\nu ,n}^{a} (x) \nonumber \\
& \!\!\! + & \!\!\! \sum_{n=1}^{\infty} \left[ \bar{q}_R(x)
\gamma^{\mu} t^a q_{n,2} (x) + \bar{q}_{n,2} (x) \gamma^{\mu} t^a q_R(x)\right]
A{}_{\nu ,n}^{a} (x) \\
& \!\!\! + & \!\!\! \frac{1}{\sqrt{2}} \sum_{n,m,\ell =1}^{\infty}
\left[ - \bar{q}_{n,1} (x) \gamma^{\mu} \gamma_5 t^a q_{m,1} (x) + \bar{q}_{n,2} (x)
\gamma^{\mu} \gamma_5 t^a q_{m,2} (x) \right] A_{\mu ,\ell}^{a} \,
\delta_{\ell, m + n} \nonumber \\
& \!\!\! + & \!\!\! \frac{1}{\sqrt{2}}\sum_{n,m,\ell =1}^{\infty}
\left. \left[ \bar{q}_{n,1} (x) \gamma^{\mu} t^a q_{m,1} (x) + \bar{q}_{n,2} (x)
\gamma^{\mu} t^a q_{m,2} (x) \right] A_{\mu ,\ell}^{a} \,
\delta_{\ell, \mid m - n \mid} \right.
\!\! \mbox{\raisebox{-.5ex}{\huge$]$}}  \nonumber \, .
\end{eqnarray}
Due to the field expansions, integrations include sine and cosine integrals, resulting in the Kronecker-$\delta s$. Interactions between different towers, $q_{n,1}$ and $q_{n,2}$, which are a mixture of the singlet and doublet particles, are automatically excluded for symmetry reasons and KK-number is automatically conserved.
The relation
\vspace{-3pt} \begin{equation} \label{eq:modes} \mid \! n_{{}_{1}}
\, \pm \, n_{{}_{2}} \, \pm \, \cdots \, \pm \, n_{{}_{N-1}} \!
\mid \,= n_{{}_N} \, ,
\end{equation}
deriving from the Kronecker-$\delta s$, implies KK-number conservation. But it is only valid at tree and broken at loop level. 

To show in an example that KK-number is conserved at each vertex we consider the vertex of two left handed fermions coupling to one gauge boson. This is one part of the Lagrangian already derived in eq.~(\ref{eq:example_Lag}). Therefore one has to integrate out the 5D Lagrangian
\vspace{-0pt}\begin{eqnarray}
\label{eq:dummy}
  -&\!\!\!\!\!g_{{}_5} &\!\!\!\!\!\!
\int_{\mbox{\raisebox{-1.3ex}{\scriptsize{$\!\!\!\!
0$}}}}^{\mbox{\raisebox{.9ex}{\scriptsize{$\!\!\!\! \pi R$}}}}
\bar{Q}_L (x,y) \gamma^m t^a A_m^a (x,y) Q_L (x,y) d y \; .
\end{eqnarray}
Inserting the Fourier expansion of the fields, the term including only the zero-modes of the fermions reads
\vspace{-0pt}\begin{eqnarray}
 -&\!\!\!\!\!g_{{}_5} &\!\!\!\!\!\!
\int_{\mbox{\raisebox{-1.3ex}{\scriptsize{$\!\!\!\!
0$}}}}^{\mbox{\raisebox{.9ex}{\scriptsize{$\!\!\!\! \pi R$}}}}
\bar{q}_L (x,y) \gamma^m t^a \frac{1}{\sqrt{\pi R}}\left[ A_{\mu 0}^a (x) +
\sqrt{2} \sum_{n=1}^{\infty}A_{\mu ,n}^{a} (x) \cos \left(\frac{n y}{R} \right) \right]  q_L (x,y) d y \nonumber \; .
\end{eqnarray}
Since integration of the cosine gives
\vspace{-0pt}\begin{eqnarray}
 &\!\!\!\!\! &\!\!\!\!\!\!
\int_{\mbox{\raisebox{-1.3ex}{\scriptsize{$\!\!\!\!
0$}}}}^{\mbox{\raisebox{.9ex}{\scriptsize{$\!\!\!\! \pi R$}}}}
\cos \left(\frac{n y}{R} \right) dy = \pi R \; \delta_{n,0} \; ,
\end{eqnarray}
the zero-modes of the fermions only interact with the zero-modes of the gauge bosons. 
Considering the interaction between one zero-mode fermion, excitation level $l$ of a second fermion and the gauge boson, one finds
\vspace{-0pt}\begin{eqnarray}
-&\!\!\!\!\!g_{{}_5} &\!\!\!\!\!\!
\int_{\mbox{\raisebox{-1.3ex}{\scriptsize{$\!\!\!\!
0$}}}}^{\mbox{\raisebox{.9ex}{\scriptsize{$\!\!\!\! \pi R$}}}}
\bar{q}_L (x,y) \gamma^m t^a \frac{1}{\sqrt{\pi R}}\left[ A_{\mu 0}^a (x) +
\sqrt{2} \sum_{l=1}^{\infty}A_{\mu ,l}^{a} (x) \cos \left(\frac{l y}{R} \right) \right] \\
 & \times & \frac{1}{\sqrt{\pi R}} \sqrt{2} \sum_{n=1}^{\infty}Q_{L}^{n} (x) \cos \left(\frac{n y}{R} \right) d y \nonumber \; .
\end{eqnarray}
Now integration of the cosines yields
\vspace{-0pt}\begin{eqnarray}
 &\!\!\!\!\! &\!\!\!\!\!\!
\int_{\mbox{\raisebox{-1.3ex}{\scriptsize{$\!\!\!\!
0$}}}}^{\mbox{\raisebox{.9ex}{\scriptsize{$\!\!\!\! \pi R$}}}}
\cos \left(\frac{n y}{R} \right) \cos \left(\frac{l y}{R} \right) dy = \pi R \; \delta_{n,l} \; .
\end{eqnarray}
Here it becomes obvious that no mixing between different KK-excitation levels occurs since there is only interaction if $n=l$.
The last interaction term derives from two KK-fermions and one KK-gauge boson. Therefore we integrate eq.~(\ref{eq:dummy}) after inserting all Fourier expansions and find the integral
\vspace{-0pt}\begin{eqnarray}
 &\!\!\!\!\! &\!\!\!\!\!\!
\int_{\mbox{\raisebox{-1.3ex}{\scriptsize{$\!\!\!\!
0$}}}}^{\mbox{\raisebox{.9ex}{\scriptsize{$\!\!\!\! \pi R$}}}}
\cos \left(\frac{l y}{R} \right) \cos \left(\frac{m y}{R} \right) \cos \left(\frac{n y}{R} \right) dy = \pi R \; \delta_{n,l} = \delta_{l,\vert m-n \vert}+ \delta_{l,m+n} \;, \hphantom{anbc}
\end{eqnarray}
where $l$ is the KK-level index of the gauge boson and $n$ and $m$ are the indices of the excitations of the fermions. We again find that KK-number is conserved also for this vertex. Of course, we can easily violate KK-number at loop level by choosing one vertex with a zero-mode scattering into two first level excitations and the other vertex with the two first level excitations creating one second level excitation. But nevertheless KK-parity, given by \vspace{-3pt}\begin{equation}
P_{KK}=(-1)^n\; ,
\end{equation}
is still conserved at the loop level. If one assumes that the SM fermions can not propagate into the extra dimension and therefore multiplies the Lagrangian in eq.~(\ref{eq:dummy}) by $\delta(y)$, the SM fermions couple to all gauge boson excitations with the same strength. Since the absence of the $\delta$-distribution results in KK-parity conservation, this can be understood as a consequence of translation invariance in the fifth dimension. This conservation is the reason for the lowest lying KK-modes to be stable and thereby an interesting candidate for dark matter. It also implies that at colliders, KK-particles can only be produced in pairs.

We also want to derive the purely gluonic part of UED-QCD which we have not taken into account yet. 
In equivalence to the Standard Model, the purely gluonic part is given by
\vspace{-0pt} \begin{eqnarray} \label{eq:L5}
\mathcal{L}_5  =  & \!\!\! - & \!\!\!\!\! \frac{1}{4} F_{\mathit{mn}}^a F^{\mathit{mna}}  \nonumber \\
= & \!\!\! - & \!\!\!\!\! \frac{1}{4}\mbox{\raisebox{-.6ex}{\huge $($}}F_{\mu \nu}^a F^{\mu \nu a}  +  2 F_{\mu 4}^a F^{\mu 4 a}\mbox{\raisebox{-.6ex}{\huge $)$}} \
\end{eqnarray}
with the field strength tensor in 5D given by
\vspace{-3pt}
\begin{eqnarray}
F_{\mathit{mn}}^a  =  \partial_m A_n^a - \partial_n A_m^a
- g_{{}_5} f^{\mathit{abc}} A_{\mathit{m}}^b A_{\mathit{n}}^c
\end{eqnarray}
and $m,n \in \lbrace\mu,4 \rbrace$ are the 5D space time indices. The first term in brackets in eq.~(\ref{eq:L5}) contains interactions and kinetic terms of gluons $g$ and their heavy partners $g_n^*$, given by
\vspace{-3pt} \begin{eqnarray}-\frac{1}{4} F_{\mu \nu}^a F^{\mu \nu a} = &
\!\!\! & \!\!\!\!\!\!\!\!\!\!
-\frac{1}{4} \Big[\partial_{\mu} A_{\nu}^a
\partial^{\mu} A^{\nu a} \,
- \, \partial_{\nu} A_{\mu}^a \partial^{\mu} A^{\nu a} \nonumber\\
&\!\!\! - & \!\!\! \partial_{\mu} A_{\nu}^a \partial^{\nu} A^{\mu a}  +  \partial_{\nu} A_{\mu}^a \partial^{\nu} A^{\mu a} \nonumber \\
&\!\!\! - & \!\!\! 2 g_{{}_5} f^{\mathit{abc}} A_{\mu}^b A_{\nu}^c \mbox{\raisebox{-.6ex}{\huge $($}}\partial^{\mu} A^{\nu a}  -  \partial^{\nu} A^{\mu a}\mbox{\raisebox{-.6ex}{\huge $)$}}  \\
&  \!\!\! + & \!\!\! g_{{}_5}^2 f^{\mathit{abc}} f^{\mathit{ade}}
A_{\mu}^b A_{\nu}^c A^{\mu d} A^{\nu e} \Big] \nonumber \, ,
\end{eqnarray}
while the second term in brackets in eq.~(\ref{eq:L5}) becomes the heavy gluons mass term
\vspace{-3pt} \begin{equation} \label{eq:KKgmass} 
-\frac{1}{4} F_{\mu 4}^a F^{\mu 4 a} = -\frac{1}{2}
\partial_4 A_{\mu}^a
\partial^4 A^{\mu a} \, ,
\end{equation}
where we again apply the gauge $A_4^a=0$. 

Now integration over the Lagrangian has to be performed after inserting the expansion~(\ref{eq:expansionA}). For the three-gluon interaction term one obtains
\vspace{-3pt} \begin{eqnarray} 
\frac{1}{2}&\!\!\!\!\!\!\!g_{{}_5} &\!\!\!\!\!\! f^{\mathit{abc}}
\int_{\mbox{\raisebox{-1.3ex}{\scriptsize{$\!\!\!\!
0$}}}}^{\mbox{\raisebox{.9ex}{\scriptsize{$\!\!\!\! \pi R$}}}} A_{\mu}^b (x,y) A_{\nu}^c (x,y) \mbox{\raisebox{-.6ex}{\huge $[$}}\partial^{\mu}A^{\nu a} (x,y)  -  \partial^{\nu} A^{\mu a} (x,y)\mbox{\raisebox{-.6ex}{\huge $]$}} d y \nonumber \\
= & \!\!\! +&\!\!\!\!\!\frac{1}{2} g f^{\mathit{abc}} \mbox{\raisebox{-.6ex}{\huge $[$}} A{}^{  }{}_{\mu 0}^b (x) A{}^{  }{}_{\nu 0}^c (x) \mbox{\raisebox{-.6ex}{\huge $[$}}\partial^{\mu}A{}^{  }{}_0^{\nu a} (x)  -  \partial^{\nu} A{}^{  }{}_0^{\mu a} (x)\mbox{\raisebox{-.6ex}{\huge $]$}} \nonumber \\
& \!\!\! + & \!\!\! 3 A{}^{  }{}_{\mu 0}^b (x) \sum_{n=1}^{\infty} A{}^{  }{}_{\nu n}^c (x) \mbox{\raisebox{-.6ex}{\huge $[$}}\partial^{\mu}A{}^{  }{}_n^{\nu a} (x)  -  \partial^{\nu} A{}^{  }{}_n^{\mu a} (x)\mbox{\raisebox{-.6ex}{\huge $]$}} \\
& \!\!\! + & \!\!\! \frac{1}{\sqrt{2}} \sum_{n,m,\ell =1}^{\infty}
A{}^{  }{}_{\mu n}^b (x) A{}^{  }{}_{\nu m}^c
\mbox{\raisebox{-.6ex}{\huge $[$}}
\partial^{\mu}A{}^{  }{}_{\ell}^{\nu a} (x)  -  \partial^{\nu}
A{}^{  }{}_{\ell}^{\mu a} (x)\mbox{\raisebox{-.6ex}{\huge $]$}}
\delta_{\ell, \pm m \pm n } \mbox{\raisebox{-.6ex}{\huge $]$}}\nonumber \
.
\end{eqnarray}
The four gluon interaction in 4D is given by 
\vspace{-3pt} \begin{eqnarray} - \frac{1}{4}&\!\!\!\!\!\!\!
g_{{}_5}^2 &\!\!\!\!\!\! f^{\mathit{abc}} f^{\mathit{ade}}
\int_{\mbox{\raisebox{-1.3ex}{\scriptsize{$\!\!\!\!
0$}}}}^{\mbox{\raisebox{.9ex}{\scriptsize{$\!\!\!\! \pi R$}}}} A_{\mu}^b (x,y) A_{\nu}^c (x,y) A^{\mu d} (x,y) A^{\nu e} (x,y) d y \nonumber \\
= &\!\!\! - &\!\!\!\!\! \frac{1}{4} g^2 f^{\mathit{abc}} f^{\mathit{ade}} \mbox{\raisebox{-.6ex}{\huge $[$}}A{}^{  }{}_{\mu 0}^b (x) A{}^{  }{}_{\nu 0}^c (x) A{}^{  }{}_0^{\mu d} (x) A{}^{  }{}_0^{\nu e} (x) \nonumber \\
& \!\!\! + & \!\!\! 6 A{}^{  }{}_{\mu 0}^b (x) A{}^{  }{}_{\nu 0}^c (x) \sum_{n=1}^{\infty} A{}^{  }{}_n^{\mu d} (x) A{}^{  }{}_n^{\nu e} (x)  \\
& \!\!\! + & \!\!\! \frac{4}{\sqrt{2}} A{}^{  }{}_{\mu 0}^b (x) \sum_{n,m,\ell=1}^{\infty} A{}^{  }{}_{\nu n}^c (x) A{}^{  }{}_m^{\mu d} (x) A{}^{  }{}_{\ell}^{\nu e} (x) \delta_{\ell, \pm m \pm n  } \nonumber \\
& \!\!\! + & \!\!\! \frac{1}{2} \sum_{n,m,\ell,k=1}^{\infty} A{}^{
 }{}_{\mu n}^b (x) A{}^{  }{}_{\nu m}^c (x) A{}^{ 
}{}_{\ell}^{\mu d} (x) A{}^{  }{}_k^{\nu e} (x) \delta_{k, \pm m
\pm n \pm \ell } \mbox{\raisebox{-.6ex}{\huge $]$}} \nonumber \, .
\end{eqnarray}
The Kronecker-$\delta s$ again derive from integrations over sine and cosine functions and have to be understood in the following way:
\begin{eqnarray}
\delta_{\ell, \pm k \pm m}
= \delta_{\ell, k+m}  +  \delta_{\ell, k-m}  + \delta_{\ell, m-k}
 +  \delta_{\ell, -m-k} \; .
\end{eqnarray}
The last Kronecker-$\delta$ can never be non-zero, since $n,m$ and $l$ are defined as positive integers.

The heavy gluons mass term from eq.~(\ref{eq:KKgmass}) gives
\vspace{-3pt} \begin{equation}
-\frac{1}{2}\int_{\mbox{\raisebox{-1.3ex}{\scriptsize{$\!\!\!\!
0$}}}}^{\mbox{\raisebox{.9ex}{\scriptsize{$\!\!\!\! \pi R$}}}}
\partial_{4} A_{\mu}^a (x,y)
\partial^{4} A^{\mu a} (x,y) dy  =
-\frac{1}{2}\frac{n^2}{R^2}\sum_{n=1}^{\infty}A{}^{  }{}_{\mu
n}^a (x) A{}^{  }{}_n^{\mu a} (x) \, .
\end{equation}
For the KK-gauge bosons we again find $M_n=\frac{n}{R}$. This leads to the naive conclusion that the mass spectra within one mode in UED are highly degenerate, neglecting other symmetry-consistent boundary terms and the Higgs mechanism. This means that the mass spectra, which are typically expected in UED, are quite different from a typical SUSY spectrum, e.g.\ the SPS 1a benchmark point. 

All Feynman-rules and Lagrangians were re-derived for UED-QCD. Feynman-rules can be found in the appendix. They agree with the rules given in~\cite{PhysRevD.65.076007} and~\cite{macesanu-2002-66}.

As will be explained in section~\ref{sec:HELAS}, we include a width in the propagator of the massive gluons, denoted by $g_n^*$, and the massive quarks, denoted by $q_{n,1}$, $q_{n,2}$. This Breit-Wigner propagator will be used for the decay chains in order to regularize divergences at the pole.

Instead of the usual polarization sum for massless gluons $g$, for the massive external gauge bosons $g_n^*$ with $n\geq1$, we use a polarization sum given in the unitary gauge by
\vspace{-0pt} \begin{equation} \sum_{\sigma} \epsilon_{\mu
n}^{a\ast}(k,\sigma)\epsilon_{\nu n}^b(k,\sigma)  =
\mbox{\raisebox{-.6ex}{\huge $($}} \!\! -g_{\mu \nu}
 +  \frac{k_{\mu}k_{\nu}}{M_n^{2}}\mbox{\raisebox{-.6ex}{\huge
$)$}}\delta^{ab} \; .
\label{eq:massive_pol}
\end{equation}
This is equal to the polarization sum for usual massive gauge bosons. It contains three physical degrees of freedom.

\subsection{Electroweak sector in UED}

In general it has to be taken into account that all particles with equal quantum numbers can mix. This is important for the couplings of the neutral heavy gauge bosons from the electroweak Lagrangian. In this thesis we neglected the Higgs part in the electroweak Lagrangian. One can in principle add it, though this is not necessary since also higgsless models in UED exist~\cite{csaki-2005}. Taking the Higgs terms into account, one obtains a second mass term for the quarks. These two mass terms mix singlets and doublets of KK-particles with the mass matrix
\vspace{-0pt} \begin{equation} (\bar{Q}'^{\;n}(x) , \bar{U}'^{\;n}(x) )
\left(
\begin{array}{cr} \frac{n}{R} & M_{SM} \\ M_{SM} & -\frac{n}{R}
\end{array}\right) \left( \! \begin{array}{c} Q'^{\;n} (x)
\\ U'^{\;n} (x) \end{array} \! \right) \;.  \end{equation}
After diagonalization one obtains mass eigenstates with
\begin{equation}
 M_n=\sqrt{\frac{n^2}{R^2}+(M_{SM})^2} \; .
\end{equation}
The mixing of the weak eigenstates $Q'^{\;n}_u (x)$ and $U'^{\;n} (x)$, forming the mass eigenstates $u_{n,1}$ and $u_{n,2}$, is then denoted by
\begin{eqnarray}
\label{eq:mixings}
u_{n,1} \;&=& \text{cos} \,\alpha^{(n)} \; Q'^{\;n}_u (x) + \text{sin} \,\alpha^{(n)} \; U'^{\;n} (x)\; ,\nonumber\\
u_{n,2} \; &=& \text{sin} \,\alpha^{(n)} \gamma_5 \;Q'^{\; n}_u (x) - \text{cos} \, \alpha^{(n)} \gamma_5 \; U'^{\;n} (x)\; .
\end{eqnarray}
The mixing angle $\alpha^{(n)}$ between singlets and doublets is given, without NLO corrections, by 
\begin{eqnarray}
\text{tan}2\alpha^{(n)}= \frac{m_f}{\frac{n}{R}} \; .
\end{eqnarray}
Here $Q'^{\;n}_u (x)$ denotes only the upper component of the weak doublet eigenstate and $U'^{\;n}(x)$ denotes the $u$-type singlets. The mixing structure for the $d$-type component of $Q'^{\;n}_d (x)$ and singlet $D'^{\;n}(x)$ is equal. This results in the mass eigenstates $d_{n,1}$ and $d_{n,2}$.
Since it is suppressed by the KK-excitation mass, the mixing between singlets and doublets will, except for the top quark, be quite small. 
The mass term from the kinematic part of the Lagrangian is not present for the zero-modes. That is consistent with the non-mixing singlets and doublets in the SM. 

The treatment of the QCD gauge part was exemplary for the SM gauge group $SU(3)_c \times SU(2)_L \times U(1)$. In the same way we calculate the Feynman rules from the electroweak sector of UED, using the following Lagrangian
\vspace{-0pt} \begin{eqnarray} \label{eq:EW_part}
\mathcal{L}_5^{EW}
&=&  \Big(\bar{u}'(x,y), \bar{d}'(x,y)\Big)_L \; \gamma^{m} \left( g \frac{\sigma^r}{2} \; A_m^r+ Y_d\,g_Y \ B_m\right)
 \left( \!
\begin{array}{c} u' (x,y)
\\ d' (x,y) \end{array} \! \right)_{\!L} \nonumber \\
& &+\; \bar{U}' (x,y) \,\gamma^m\, (Y_{s,U}\, g_Y \; B_m)\; U'(x,y)\nonumber\\
& &+\; \bar{D}' (x,y) \,\gamma^m \,(Y_{s,D}\, g_Y \; B_m)\;  D'(x,y) \; .
\end{eqnarray}
Here $U'$ and $D'$ denote the eigenstates of the weak interaction. Integration of the fifth dimension is performed as usual. But since for our later calculations we only need the uncharged gauge boson fields, we only compute the neutral part of the effective Lagrangian in 4D.

As a short example we want to give the Lagrangian for the interaction of the first level KK-partner of the SM photon with a SM $u$-type quark and a first level excitation of the KK-quark. As given in eq.~(\ref{eq:EW_part}), the interaction reads
\vspace{-0pt} \begin{eqnarray} 
\bar{Q}_u'^{(0)}(x,y) Y_{d,U} g_Y  \gamma^{m}B_{m} Q_u'^{(0)}(x,y)+\bar{U}'^{(0)} (x,y) Y_{s,U} g_Y  \gamma^{m}B_{m}	 U'(x,y)\, ,
\end{eqnarray}
where the first term derives from a doublet and the second terms derives from a singlet interaction. After insertion of the Fourier expansion for the KK-quark, compactification and insertion of the mixing matrix of the weak eigenstates one finds
\vspace{-0pt} \begin{eqnarray}
& &\bar{u} \,\gamma^{\mu} \,g_Y\, B_{1\,\mu}\, Y_{d,U} \,P_L \,\text{cos} \,\alpha^{(1)} \,q_{1,1}-\bar{u} \,\gamma^{\mu}\, g_Y\, B_{1\,\mu} \,Y_{d,U} \,P_L \,\text{sin}\, \alpha^{(1)}\, q_{1,2} \nonumber\\
& +& \bar{u} \,\gamma^{\mu}\, g_Y \,B_{1\,\mu} \,Y_{s,U} \,P_R \,\text{sin} \,\alpha^{(1)} \,q_{1,1}-\bar{u} \,\gamma^{\mu}\, g_Y \,B_{1\,\mu}\, Y_{s,U}\, P_R \,\text{cos} \,\alpha^{(1)} \,q_{1,2} \nonumber \; .
\end{eqnarray}
This corresponds to the Feynman rules given in the appendix.

In~\cite{Hooper:hep-ph0701197} the mixing of the electroweak gauge bosons is investigated and, including corrections from NLO, found to be very small for the first excitation mode, given by the corresponding mass matrix
\vspace{-1pt}
\begin{equation}
\left( \begin{array}{cc} 
\frac{n^2}{R^2} + \delta m^2_{B^{(n)}} + \frac{1}{4} g'^2 v^2 
  & \frac{1}{4} g'g v^2 \\
\frac{1}{4} g'g v^2 
  & \frac{n^2}{R^2} + \delta m^2_{W^{(n)}} + \frac{1}{4} g^2 v^2 
\end{array} \right).
\end{equation}
In our calculations we assume, that the first level KK-excitations $A_{3,1}$ and $B_1$ already are the mass eigenstates since the mixing angle $\theta_W^{(n)} $, defined by
\begin{eqnarray}
\gamma_n &=& \text{sin} \theta_W^{(n)} \; A_{3,\, n}+\text{cos}\theta_W^{(n)}\;B_n\\
Z_n &=& \text{cos} \theta_W^{(n)}\; A_{3,\, n}-\text{sin}\theta_W^{(n)}\;B_n
\end{eqnarray}
approaches zero because of the degenerate masses for higher level excitations in $\text{cos} \theta_W= \frac{m_W}{m_Z}$.

We explicitly calculated only four vertices for the neutral gauge bosons. A complete collection of electroweak UED Feynman rules is given in the appendix of~\cite{Bringmann}.
  \clearpage{\pagestyle{empty}\cleardoublepage}

  \chapter{SUSY- and UED-QCD Processes at the LHC} 
\label{ch:4}
\chaptermark{SUSY- and UED-QCD Processes}
In this chapter we want to calculate the partonic and hadronic cross sections and present the contributing Feynman diagrams for squark and gluino production. Since the Large Hadron Collider (LHC) is a proton-proton collider it is necessary to calculate the hadronic cross sections, according to the parton model, using the PDFs for the inner structure of the proton. Since the presented processes are all SUSY-QCD processes, they are very important for hadronic colliders. 
\section{Partonic Cross Sections}
On the partonic level the following processes contribute to gluino and squark production at hadron colliders at leading order:\\
\bigskip 
%
%
%
%
%
%
\begin{alignat*}{4}
\sq\sqb \,\text{ production:}\quad  &q_i &+&\qb_j &\,\longrightarrow\,& 
  \sq_k &+&\sqb_l  \nonumber \\[0.2cm]
  &g   &+&g     &\,\longrightarrow\,& 
  \sq_i &+&\sqb_i \\[0.2cm]
\sq\sq \,\text{ production:}\quad  &q_i &+&q_j   &\,\longrightarrow\,& 
  \sq_i &+&\sq_j \,\text{ and } c.c.\\[0.2cm]
\gl\gl \,\text{ production:}\quad  &q_i &+&\qb_i &\,\longrightarrow\,& 
  \gl   &+&\gl\\[0.2cm]
  &g   &+&g     &\,\longrightarrow\,& 
  \gl   &+&\gl\\[0.2cm]
\sq\gl \,\text{ production:}\quad  &q_i &+&g     &\,\longrightarrow\,& 
  \sq_i &+&\gl \,\text{ and } c.c.\; .
\end{alignat*}
\newline
The incoming momenta are denoted $k_{1}$ and $k_{2}$ while the outgoing momenta are called $p_{1}$ and $p_{2}$. The given processes exist for both chiralities of the squarks denoted by $\tilde q_{l}$ and $\tilde q_{r}$ which are not given explicitly above. In the case of $\tilde q \tilde q $ and $\tilde q \tilde g $ production charge conjugated processes also have to be considered. 
When presenting the results for squark-antisquark final states, the processes of quark-antiquark and gluon-gluon scattering have to be taken into account. Similarly for gluino-gluino production, there are contributions coming from quark-antiquark and gluon-gluon initial states. The only initial states contributing to squark-squark pair production are two incoming quarks. In the case of squark-gluino pairs in the final state, only an incoming quark-gluon pair gives a contribution. 
\begin{figure}[t!]
\centering
\begin{tabular}{p{5mm} c@{}c@{}c@{}c}
\raisebox{5.3ex}{a)} &
  \includegraphics[width=0.30\textwidth]{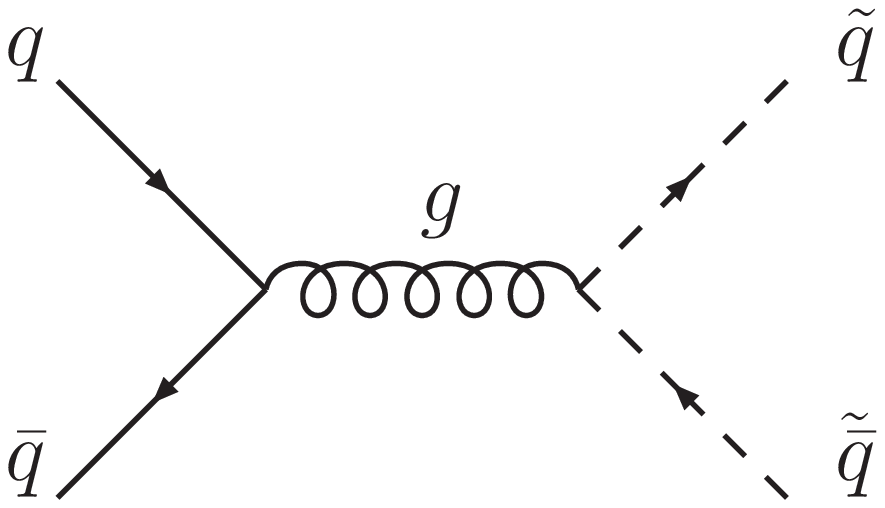}  &
  \includegraphics[width=0.23\textwidth]{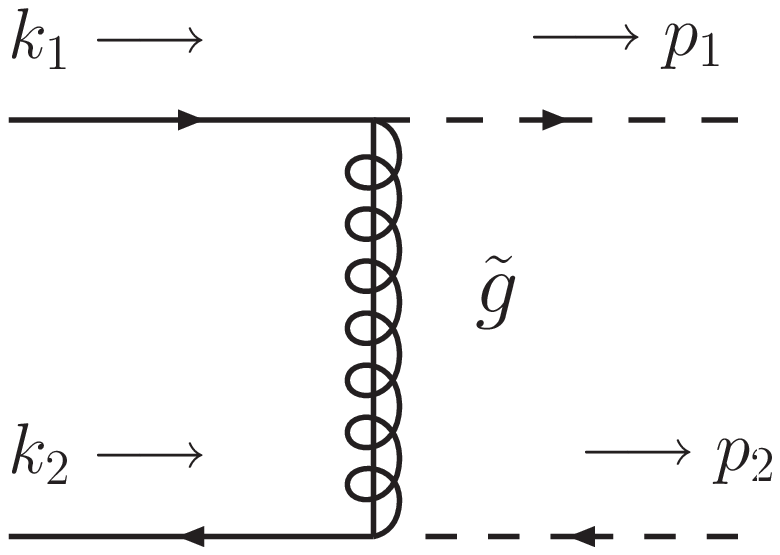} & \\[4mm]
\raisebox{5.0ex}{b)} &
  \includegraphics[width=0.23\textwidth]{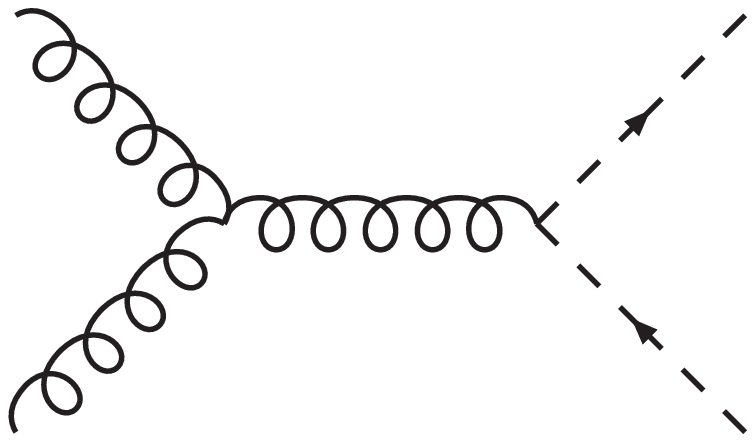}  &
  \includegraphics[width=0.24\textwidth]{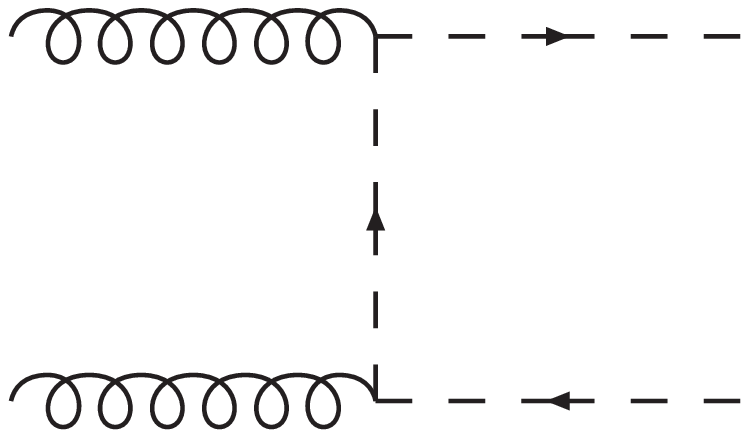}  &
  \includegraphics[width=0.23\textwidth]{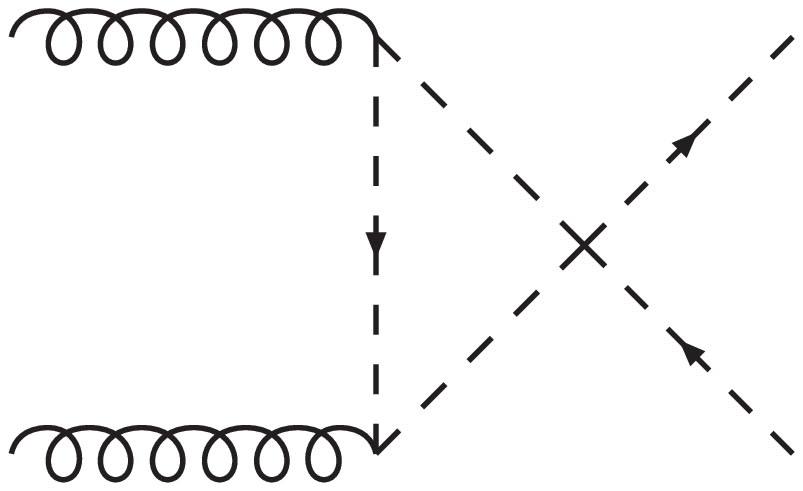} \\ [4mm]
 & \includegraphics[width=0.21\textwidth]{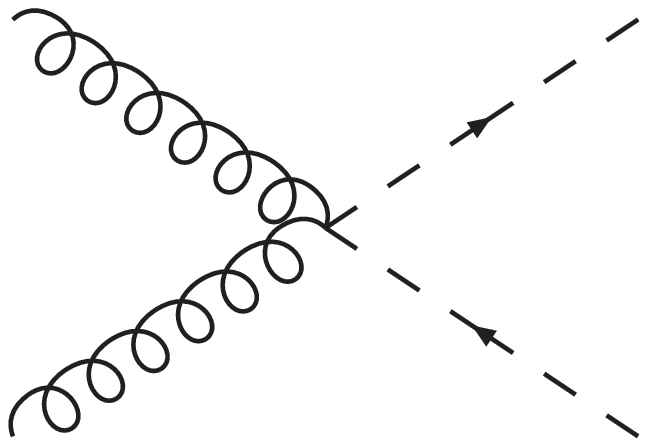} \\[4mm]
 \raisebox{5.0ex}{c)}&
  \includegraphics[width=0.23\textwidth]{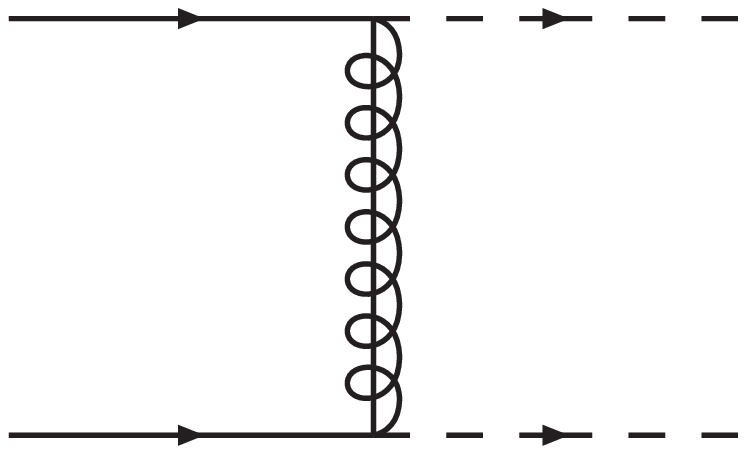}  &
  \includegraphics[width=0.23\textwidth]{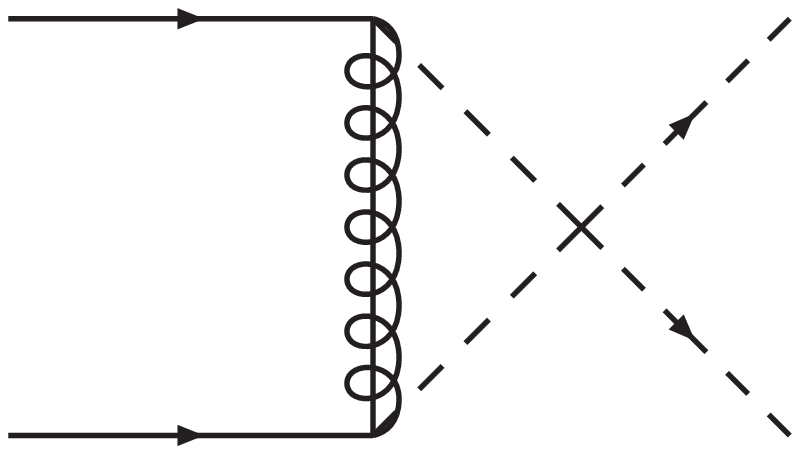} \\[4mm]
\raisebox{5.0ex}{d)} &
  \includegraphics[width=0.23\textwidth]{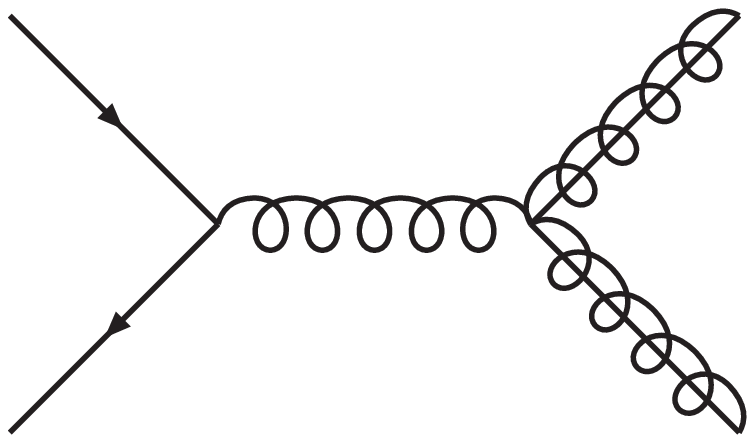}  &
  \includegraphics[width=0.23\textwidth]{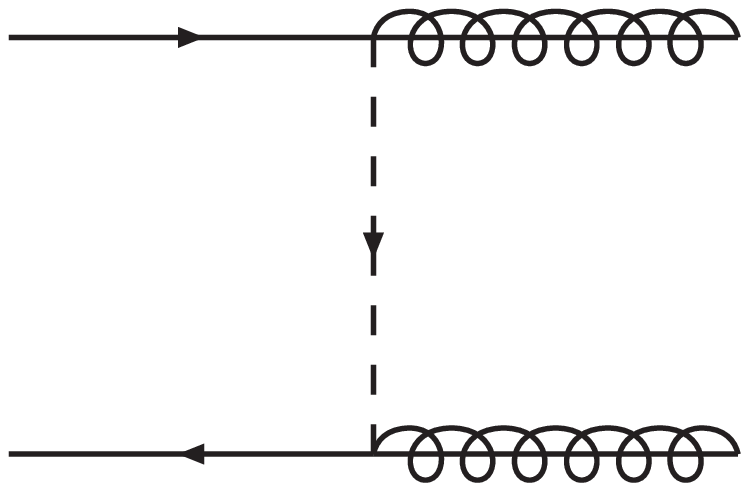} &
  \includegraphics[width=0.23\textwidth]{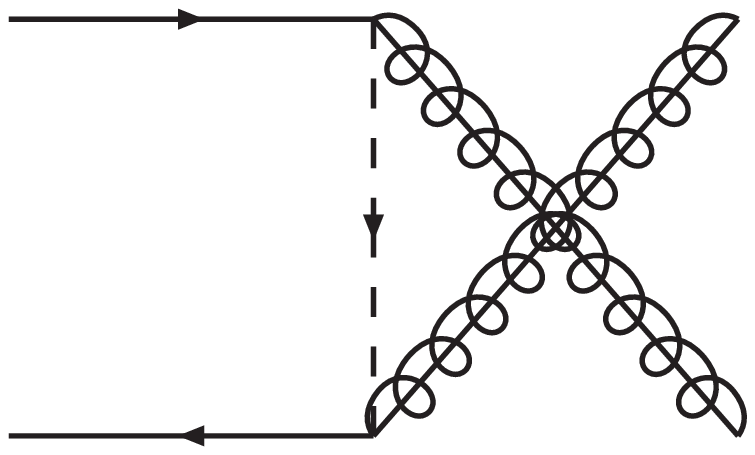} \\[4mm]
\raisebox{5.0ex}{e)} &
  \includegraphics[width=0.26\textwidth]{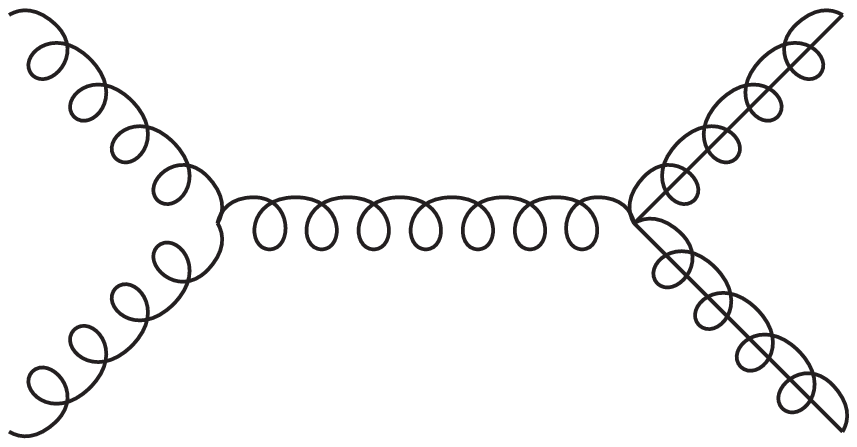}  &
  \includegraphics[width=0.23\textwidth]{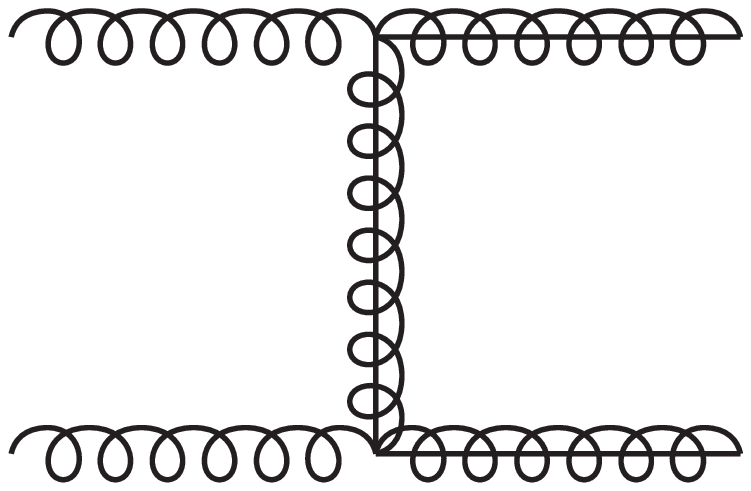} &
  \includegraphics[width=0.25\textwidth]{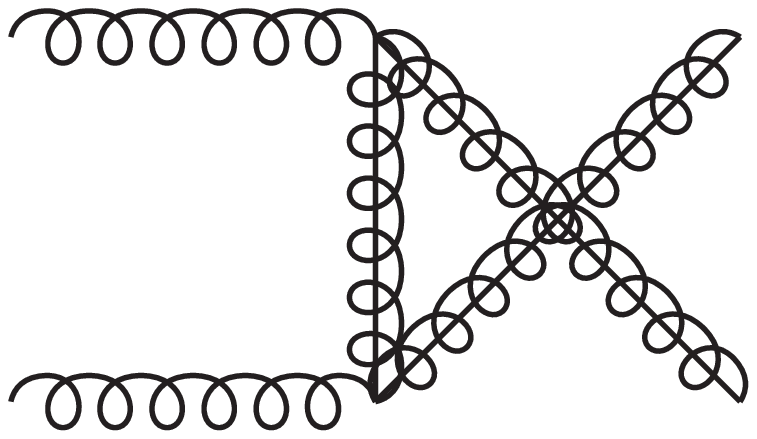}  \\[4mm]
\raisebox{5.0ex}{f)} &
  \includegraphics[width=0.25\textwidth]{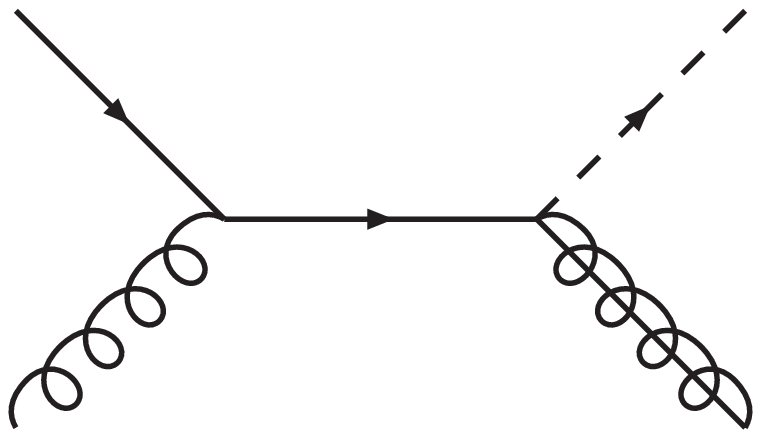} &
  \includegraphics[width=0.23\textwidth]{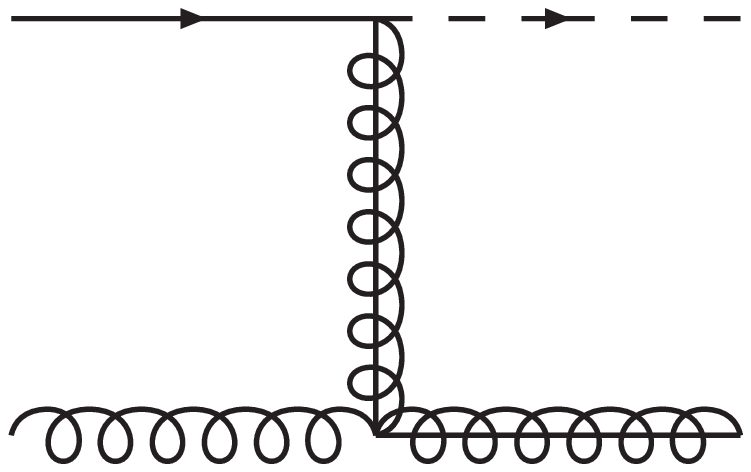} &
  \includegraphics[width=0.25\textwidth]{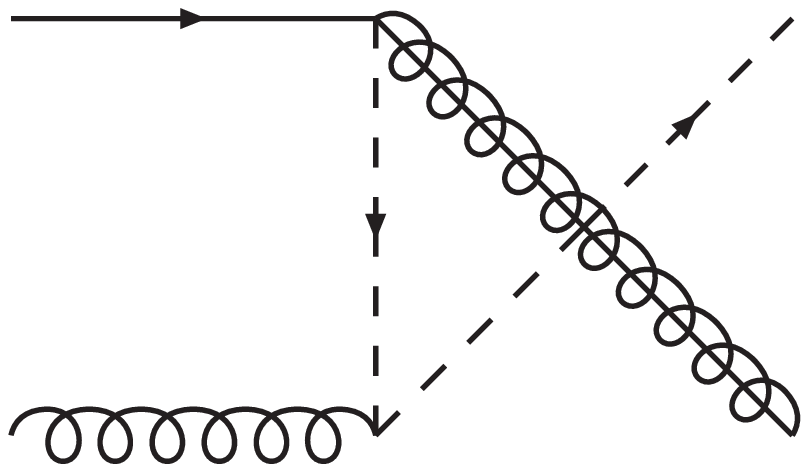} \\[4mm]
\end{tabular}
\caption{LO contributions to the production of squarks and gluons at hadron colliders in SUSY.
  \label{fig:SUSY_production}}
\end{figure}
In our calculations all outgoing squark flavor and chirality states are considered to have the same mass while the top-squark is excluded from the final state. Incoming top-quarks will always be neglected, since in the PDFs the top-quark content is expected to be approximately zero, due to the high top mass. All the other quark flavors, $n_{f}=5$, are treated as massless particles since the energy at the LHC ($\sqrt{s}=14 \; \text{TeV}$) is much higher than the mass of the bottom-quark. Therefore for on-shell particles one uses $k_i^2=0$ and  $p_{i}=m_{i}^2$.

Our notation for the matrix elements follows the notation of~\cite{Beenakker}. We use the Mandelstam variables which are kinematical invariants and are defined as given in eq.~(\ref{eq:Mandelstams}).
For the Mandelstam variables one finds
\begin{equation} 
 s+t+u=k_1^2+k_2^2+p_1^2+p_2^2=m_1^2+m_2^2 \; .
\end{equation}
We also introduce the following abbreviations:
\begin{eqnarray}
t_1 = (k_2-p_2)^2-m_{\tilde q}^2 \;,\qquad	 \hspace{20pt} u_1 = (k_1-p_2)^2-m_{\tilde q}^2 \; ,\\
t_g = (k_2-p_2)^2-m_{\tilde g}^2 \;,\qquad  \hspace{20pt} u_g = (k_1-p_2)^2-m_{\tilde g}^2 \; .
\end{eqnarray}
When calculating the squared matrix elements of a process with external gluons, one needs to sum over all polarization states. In the axial gauge the polarization sum for external gluons is given by\\
\begin{equation}
\label{eq:pol_sum}
P_i^{\mu \nu}= \sum \limits_{T} \epsilon_{T}^{\mu \ast}(k_i) \epsilon_{T}^{\nu}(k_i)=-g^{\mu \nu}+\frac{n_i^{\mu} k_i^{\nu}+k_i^{\mu} n_i^{\nu}}{(n_i k_i)}-\frac{n_i^2 k_i^{\mu} k_i^{\nu}}{(n_i k_i)^2} \; ,
\end{equation}
with $n^{\mu} \epsilon_{T \mu}=0$, where $n^{\mu}$ is an arbitrary four-vector. This polarization sum as well as the polarization vectors $\epsilon_T^{\mu}$ obey the transversality relations\\
\begin{equation}
 k_{i \mu} P_i^{\mu \nu}=n_{i \mu} P_i^{\mu \nu}=0
\end{equation}
and 
\begin{equation}
 k_{i \mu} \epsilon_T^{\mu}(k_i)=\epsilon_{T \mu}(k_i) k_i^{\mu}=0
\label{eq:transmom}
\end{equation}
for all transverse polarization vectors. This is also true for the polarization sum of a massive gauge boson, given in eq.~(\ref{eq:massive_pol}).

\section{Squark and Gluino Production Cross Sections in SUSY}
\label{sec:WQ}
For the computation of the partonic cross sections of all processes given in fig.~\ref{fig:SUSY_production} we used the Feynman rules from appendix~\ref{SUSY-QCD Feynman-Rules}. A way for a simplified treatment of external gluons, called ghost subtraction, is treated in appendix~\ref{Explicit Calculation with External Gluons}. The resulting squared matrix elements agree with~\cite{Beenakker} and are given by:
\begin{small}
\begin{eqnarray*}
  \sum |{\cal M}^B|^2 (q_i\qb_j\to\sq\sqb) & = & 
  \delta_{ij}\left[8 n_f g_s^4 \,N C_F \,\frac{t_1 u_1 -\ms^2 s}{s^2} 
  + 4 \ghat^4 \,N C_F \,\frac{t_1 u_1 -(\ms^2 -\mg^2) s}{t_g^2}
\right. \label{born1} \\ 
  & &\hphantom{\delta_{ij}a} \left. {} \hspace*{-1.0cm}
  - 8 g_s^2 \ghat^2 \,C_F \,\frac{t_1 u_1 -\ms^2 s}{s t_g}
\right] 
  + (1-\delta_{ij}) 
  \left[ 4 \ghat^4 \,N C_F \,\frac{t_1 u_1 -(\ms^2 -\mg^2)s}{t_g^2}\right]
    \nonumber \; ,\\[0.2cm] 
    \sum |{\cal M}^B|^2 (gg\to\sq\sqb) & = & \!\!4 n_f g_s^4
    \left[C_O\!\left(1 -2\,\frac{t_1 u_1}{s^2}\right) -\!C_K\right]\! \left[ 1
      -\!2\,\frac{s\ms^2}{t_1 u_1}\left(1 -\!\frac{s\ms^2}{t_1
        u_1}\right)\right]\; , \quad\qquad\\[0.2cm] 
\end{eqnarray*}
\end{small}
\begin{small}
\begin{eqnarray*}
  \sum |{\cal M}^B|^2 (q_i q_j\to\sq\sq) & = & 
  \delta_{ij}\Bigg[ 2\ghat^4 \,N C_F \left(t_1 u_1 -\ms^2 s\right)
  \left(\frac{1}{t_g^2} +\frac{1}{u_g^2}\right)  \\ 
  & &{} \hphantom{\delta_{ij}a} + 4\ghat^4 \,\mg^2  s \left( N C_F 
  \left(\frac{1}{t_g^2} +\frac{1}{u_g^2}\right)
  - 2 C_F \,\frac{1}{t_g u_g}\right) \Bigg]\nonumber\\
  & &{} + (1-\delta_{ij}) 
  \left[ 4 \ghat^4 \,N C_F \,\frac{t_1 u_1 -(\ms^2-\mg^2) s}{t_g^2}\right]
    \nonumber \; ,\\[0.2cm] 
  \sum |{\cal M}^B|^2 (q\qb\to\gl\gl) & = & 
  4 g_s^4 \,C_O \left[\frac{2 \mg^2 s +t_g^2 +u_g^2}{s^2} \right]
+ 4 g_s^2 \ghat^2 \,C_O \left[\frac{\mg^2 s +t_g^2}{s t_1}
  +\frac{\mg^2 s +u_g^2}{s u_1} \right]\nonumber\\ 
  & &{} + 2\ghat^4 \left[ C_O \left(\frac{t_g^2}{t_1^2} +
  \frac{u_g^2}{u_1^2}\right) +C_K \left(2\,\frac{\mg^2 s}{t_1\,u_1} -
  \frac{t_g^2}{t_1^2} - \frac{u_g^2}{u_1^2} \right)\right]\nonumber\; ,\\[0.2cm] 
  \sum |{\cal M}^B|^2 (gg\to\gl\gl) & = & 8g_s^4 \,N C_O \left(1
  -\frac{t_g u_g}{s^2}\right) \left[\frac{s^2}{t_g\,u_g}
   -2  +4\,\frac{\mg^2 s}{t_g u_g} \left (1-\frac{\mg^2 s}{t_g u_g}\right)
  \right]\; ,\\[0.2cm]
%
  \sum |{\cal M}^B|^2 (qg\to\sq\gl) & = & 
  2  g_s^2 \ghat^2 \left[ C_O \left( 1 -2\,\frac{su_1}{t_g^2}
  \right) - C_K \right] \\
  & &{} \times \Bigg[- \,\frac{t_g}{s} \label{born6} 
+ \frac{2(\mg^2-\ms^2)\,t_g}{su_1} \left(1 +\frac{\ms^2}{u_1}
  +\frac{\mg^2}{t_g} \right) \Bigg] \nonumber\; ,
\end{eqnarray*}
\end{small}
%
%
%
%
%
%
%
\hspace*{-3.4mm} with $N=3$, $C_0=N(N^2-1)=24$, $C_K=(N^2-1)/N=8/3$ and $C_F=(N^2-1)/(2N)$. The QCD gauge coupling $g_s$ is identical to the Yukawa coupling $\hat {g}_s$. One has to be very careful with the formula given for $\sum |{\cal M}^B|^2 (q_i q_j\to\sq\sq)$. Since the first term derives from the squark production of particles with different chiralities the second one comes from the production of equal chiralities. Therefore the second term was multiplied by a factor of $\frac{1}{2}$, relative to the first term. As usual this symmetry factor has to be taken into account when the partonic cross sections are calculated and integration over two identical outgoing particles is performed.

All matrix elements where calculated completely by hand or with FeynCalc~\cite{FeynCalc}. All results agree with the results of~\cite{Beenakker} and~\cite{gehrmann-2004-703}. In the latter results are also given for a hypothetical ``polarized LHC``. This seems, at least in principle, to be an interesting possibility for distinguishing between different models of new physics by using spin asymmetries.

In order to find the leading order cross section at the partonic level we integrate the differential cross section
\begin{eqnarray}
\begin{aligned}
\frac{d^2\sigma^B}{dt du} &= F_{ij} \, \frac{1}{16 \pi s^2} \, \theta ([t-p_2^2][u-p_2^2]-p_2^2 s) \, \theta
(s-4m^2) \times \\ & \hphantom{K_{ij}} \delta (s+t+u-p_1^2-p_2^2) \sum \vert M^B \vert ^2 \; ,
\label{eq:partcross}
\end{aligned}
\end{eqnarray}
where $m=(\sqrt{p_1^2}+\sqrt{p_2^2})/2$. Here $F_{ij}$ is introduced for averaging over the initial-states colors and spins:
\begin{eqnarray}
\vspace*{-10mm}
F_{q q}=\frac{1}{4 N^2}, \quad F_{g g}=\frac{1}{4 (N^2-1)^2}, \quad F_{qg}=\frac{1}{4 N (N^2-1)} \; ,
\end{eqnarray}
with $N=3$ for the SUSY-QCD group $\text{SU(N)}$.
Integration over the two Mandelstam variables t and u yields the partonic cross section at leading-order. The integration over one of the two variables, e.g.\ u without loss of generality, is trivial since integration is performed over the $\delta$-distribution in the double differential cross section.

Integration over t can be performed analytically. Limits arise from the first $\theta$-function and are given by
\begin{equation}
t_g^{\pm}=-\frac{s+m_{\tilde g}^2-m_{\tilde q}^2}{2} \pm \frac{1}{2} \sqrt{(s-m_{\tilde g}^2-m_{\tilde q}^2)^2-4 m_{\tilde q}^2 m_{\tilde g}^2}
\end{equation}
for the case of one outgoing squark and one outgoing gluino in fig.~\ref{fig:SUSY_production} f). The second $\theta$-function is the constraint from the production threshold. It is implicit in the following cross section formulae. 

A factor of $\frac{1}{2}$ for identical particles in the final state is included. The resulting partonic cross sections agree with~\cite{Beenakker} and are given by:\\
\begin{small}
\begin{eqnarray*}
  \sigma^B(q_i\qb_j\to\sq\sqb) & = &
  \delta_{ij} \,\frac{n_f\pi\as^2}{s}\,\bs\left[\frac{4}{27}
  -\frac{16\ms^2}{27s} \right] \\
  &&{} +\delta_{ij}\,\frac{\pi\as\hat{\alpha}_s}{s}\left[\bs\left(\frac{4}{27}
  +\frac{8\md^2}{27s}\right) +\left(\frac{8\mg^2}{27s}
  +\frac{8\md^4}{27s^2} \right) L_1\right] \nonumber \\ 
  &&{}  +\frac{\pi\hat{\alpha}_s^2}{s}\left[\bs\left(
  -\frac{4}{9}-\frac{4\md^4}{9(\mg^2s+\md^4)}\right)
  +\left(-\frac{4}{9} -\frac{8\md^2}{9s}\right) L_1 \right]\nonumber \; , \\[0.2cm]
  \sigma^B(gg\to\sq\sqb) & = &
  \frac{n_f\pi\as^2}{s}\left[
  \bs \left(\frac{5}{24} +\frac{31\ms^2}{12s} \right)
  +\left(\frac{4\ms^2}{3s}+ \frac{\ms^4}{3s^2}\right)
  \log\left(\frac{1-\bs}{1+\bs}\right) \right] \; , \\[0.2cm]
  \sigma^B(q_i q_j\to\sq\sq) & = &
  \frac{\pi\hat{\alpha}_s^2}{s}\left[\bs\left(
  -\frac{4}{9}-\frac{4\md^4}{9(\mg^2s+\md^4)}\right)
  +\left(-\frac{4}{9} -\frac{8\md^2}{9s}\right) L_1 \right] \\ 
  & &{} + \delta_{ij}\,\frac{\pi\hat{\alpha}_s^2}{s}
  \left[ \frac{8\mg^2}{27(s+2\md^2)} L_1 \right] \nonumber \; ,\\[0.2cm]
  \sigma^B(q\qb\to\gl\gl) & = &
  \frac{\pi\as^2}{s}\, \bg\left(\frac{8}{9} +\frac{16\mg^2}{9s}\right)
  \\
  & &{} +\frac{\pi\as\hat{\alpha}_s}{s}\left[
  \bg\left(-\frac{4}{3}-\frac{8\md^2}{3s} \right)
  +\left(\frac{8\mg^2}{3s} +\frac{8\md^4}{3s^2}\right)L_2\right] \nonumber\\
  & &{} +\frac{\pi\hat{\alpha}_s^2}{s}\left[
  \bg\left(\frac{32}{27}+\frac{32\md^4}{27(\ms^2s+\md^4)}\right)
  +\left(-\frac{64\md^2}{27s} -\frac{8\mg^2}{27(s-2\md^2)}\right)
  L_2\right] \nonumber\; ,\\[0.2cm]
  \sigma^B(gg\to\gl\gl) & = &
  \frac{\pi\as^2}{s}\left[
  \bg\left(-3-\frac{51\mg^2}{4s}\right)
  + \left(-\frac{9}{4}
  -\frac{9\mg^2}{s} +\frac{9\mg^4}{s^2} \right)
  \log\left(\frac{1-\bg}{1+\bg} \right) \right] \; ,\\[0.2cm]
  \sigma^B(qg\to\sq\gl) & = &
  \frac{\pi\as\hat{\alpha}_s}{s}\,\left[
  \frac{\kappa}{s}\left(-\frac{7}{9} -\frac{32\md^2}{9s}\right)
  + \left(-\frac{8\md^2}{9s}+\frac{2\ms^2\md^2}{s^2}
  +\frac{8\md^4}{9s^2} \right) L_3 \right. \\
  & &\hphantom{\frac{\pi\as\hat{\alpha}_s}{s}a} \left. {}
  + \left(-1-\frac{2\md^2}{s}+\frac{2\ms^2\md^2}{s^2}
  \right) L_4 \right] \nonumber \; ,
\end{eqnarray*}
with
\vspace*{10pt}
\begin{eqnarray*}
\begin{aligned}{}
  L_1 & = \text{ln}\frac{s+2\md^2-s\bs}{s+2\md^2+s\bs} \; ,&\qquad 
  L_2 & = \text{ln}\frac{s-2\md^2-s\bg}{s-2\md^2+s\bg} \; ,\\
\end{aligned}
\end{eqnarray*}
\begin{eqnarray*}
\begin{aligned}{}
  L_3 & = \text{ln}\frac{s-\md^2-\kappa}{s-\md^2+\kappa} \; ,&\qquad
  L_4 & = \text{ln}\frac{s+\md^2-\kappa}{s+\md^2+\kappa} \; ,\\
  \bs & = \sqrt{1-\frac{4\ms^2}{s}} \; ,&\qquad 
  \bg & = \sqrt{1-\frac{4\mg^2}{s}} \; ,\\
  \md^2 & = \mg^2 -\ms^2 \; ,&\qquad
  \kappa & = \sqrt{(s-\mg^2-\ms^2)^2-4\mg^2\ms^2} \; ,\\
  \as & = g_s^2/4\pi \; ,&\qquad
  \hat{\alpha}_s & = \ghat^2/4\pi \;. 
\end{aligned}
\end{eqnarray*}
\end{small}

\section{Hadronic Transverse Momentum and Rapidity Distributions} 
\label{sec:hadronic_param}
Up to now we have only investigated the collisions of quarks and gluons. Since the LHC is a hadron collider, one has to take into account the inner structure of the proton. The necessary PDFs were already introduced in section~\ref{the_parton_model}. For our analysis of the resulting hadronic cross sections, we choose the PDFs from the CTEQ 6 collaboration~\cite{Pumplin:hep-ph0512167}.

When hadronic cross sections are calculated, all processes from different initial states and with equal final states have to be summed up.

Hadronic cross sections are calculated numerically since the PDFs can not be derived from first principles within perturbation theory and have to be extracted from experimental data. Therefore we use Vegas~\cite{2006NIMPA.559..273H}, a Monte Carlo integration routine.

In this chapter we will explicitly calculate the total cross section and the differential cross sections in transverse momentum and rapidity for the process of quark-gluon collision producing a squark-gluino pair. Kinematics are very important for the comparison of different models of beyond-standard-model physics. Since the transverse momentum is invariant under Lorentz boosts along the beam axis and can be calculated and measured easily it is a very convenient quantity, especially for hadronic processes where the center of momentum is not fixed.

In analogy to~\cite{Beenakker}, we derived a special parametrization for the hadronic integration of transverse momentum and rapidity. As an example let us investigate the process
\begin{equation}
  h_1(K_1) + h_2(K_2) \longrightarrow \sq(p_1) +\gl(p_2) \; .
\end{equation}
$K_1$ and $K_2$ are the momenta of the incoming protons while $p_1$ and $p_2$ are the momenta of the outgoing squark and gluino. Of course we only have to cover the case of massless incoming particles. Both outgoing particles are massive. Analogously to the partonic Mandelstam variables the hadronic ones are defined by
\begin{alignat}{2}
 \nonumber \\[-0.7cm]
S &= (K_1 + K_2)^2 \; , & \nonumber\\
T_{g} &= (K_2 -p_2)^2 -\mg^2 \; ,&\qquad  T_1 &= (K_2 -p_2)^2 -\ms^2 \; , \\
U_{g} &= (K_1 -p_2)^2 -\mg^2 \; ,&\qquad  U_1 &= (K_1 -p_2)^2 -\ms^2  \; .\nonumber
\end{alignat}
So obviously one can write
\begin{equation}
  s = x_1 x_2 S \; ,\qquad t_g = x_2 T_g \; ,\qquad u_g = x_1 U_g \; .
\end{equation}
The definitions for $p_t$ and $y$ of the gluino are given by\\
\begin{eqnarray}
  p_t^2 = \frac{T_g\,U_g}{S} -\mg^2=\frac{t_g\,u_g}{s}
  -\mg^2\; , \qquad
  y = \frac{1}{2} \text{ln} \left (\frac{T_g}{U_g}\right ) \; .
\label{eq:rap1}
\end{eqnarray}
The first relation can be easily motivated by geometrical considerations. The formula for $y$ can be shown to be equivalent to 
\begin{equation}
y_{\text{hadr.}}=\frac{1}{2} \left(\text{ln} \left (\frac{x_1}{x_2} \right )+\text{ln} \left (\frac{p^0_{2}+\vert \vec{p_2} \vert \text{cos} \theta}{p^0_{2}-\vert \vec{p_2} \vert \text{cos} \theta } \right ) \right )
\end{equation}
for massless incoming particles. The second term corresponds to the usual definition of rapidity in the partonic frame
\begin{equation}
y_{\text{part.}}=\frac{1}{2} \text{ln} \left (\frac {p^0_{2}+{p_{2,L}}}{p^0_{2}-{p_{2,L}}} \right) \; ,
\label{eq:rap2}
\end{equation}
while the first term takes into account that the rapidity in eq.~(\ref{eq:rap1}) is given in the hadronic frame. Depending on the fractions of momenta $x_1$ and $x_2$, one has to boost the rapidity for calculating it in the lab frame, since the center of momentum is different for every pair of incoming momenta.

The double differential hadronic cross section in the hadronic Mandelstam variables is given by the convolution of the PDFs with the partonic double differential cross section
\begin{small}
\begin{multline}
  \frac{d^2\sigma}{dT_g dU_g}(S,T_g,U_g,Q^2)
  = \\
  = \sum_{i,j=g,q,\bar{q}}
  \int_{x_1^-}^1 dx_1 \int_{x_2^-}^1 dx_2 \, x_1 f_{i}^{h_1} (x_1,Q^2)\,
  x_2 f_{j}^{h_2} (x_2,Q^2)
  \,\frac{d^2\hat{\sigma}_{ij}(s,t_g,u_g,Q^2)}{dt_g\, du_g}  \; ,
\end{multline}
\end{small}
\hspace*{-2.6mm} where the partonic cross section is given by eq.~(\ref{eq:partcross}) and multiplication of $x_1$ and $x_2$ is due to the Jacobi determinant. Distributions in transverse momentum $p_t$ or rapidity $y$ are gained by using
\begin{equation}
    \frac{d^2\sigma}{dp_t \, dy} = 2 p_t S \frac{d^2\sigma}{dT_g \, dU_g} 
\end{equation}
and then integrating out $p_t$ or $y$ 
\begin{small}
\begin{equation}
  \sigma(S,Q^2) = \int_0^{p_t^{\text{max}}(0)} dp_t 
  \int_{-y^{\text{max}}(p_t)}^{y^{\text{max}}(p_t)} dy\, \frac{d^2\sigma}{dp_t dy}
  = \int_{-y^{\text{max}}(0)}^{y^{\text{max}}(0)} dy \int_0^{p_t^{\text{max}}(y)} dp_t\,
  \frac{d^2\sigma}{dp_t dy} \; .
\end{equation}
\end{small}

The limits for the integrations for the three variables, $x_1, x_2, y$ or $x_1, x_2, p_t$, in order to obtain a differential distribution in $p_t$ or $y$, are simply calculated by using the $\theta$ functions and the momentum conserving $\delta$-distribution in eq.~(\ref{eq:partcross}). One finds
\begin{equation}
  x_2^- = \frac{-x_1 U_g -\mg^2 +\ms^2}{x_1 S + T_g},
\qquad x_1^- = \frac{-T_g-\mg^2 +\ms^2}{S+U_g} \; ,
\end{equation}
where $x_2^-$ comes from integrating over the $\delta$ -distribution while the lower limit for $x_1$ is obtained by setting $x_2=1$.
The limits for $p_t$ and $y$ integration are easily derived to be
\begin{eqnarray}
   p_t^{\text{max}}(y) & = &
  \frac{1}{2\sqrt{S}\,\text{cosh}y}\sqrt{\left(S+\mg^2-\ms^2\right)^2 
  -4\mg^2 S \,\text{cosh}^2y} \; , \\[5mm]
  y^{\text{max}}(p_t) & = & 
  \text{arccosh}\left(\frac{S+\mg^2-\ms^2}{2\sqrt{S(p_t^2+\mg^2)}}\right) \; .
\end{eqnarray}

Choosing another parametrization to calculate the cross section is of course possible and was numerically done as a check. Since integrations with Vegas are performed automatically in an interval from $0$ to $1$, one has to make a substitution for the limits of integration and then multiply the integrand with the appropriate Jacobi determinant. In cases where integrations are more involved the so called \textit{histogram method} is used, since then one does not have to derive a special parametrization. In the later calculation of $2\rightarrow5$ processes we calculate the differential cross sections for every observable from the four vectors of the particles, using any parametrization and within only one run of the program.

\begin{figure}
\centering
\includegraphics[width=0.65\textwidth]{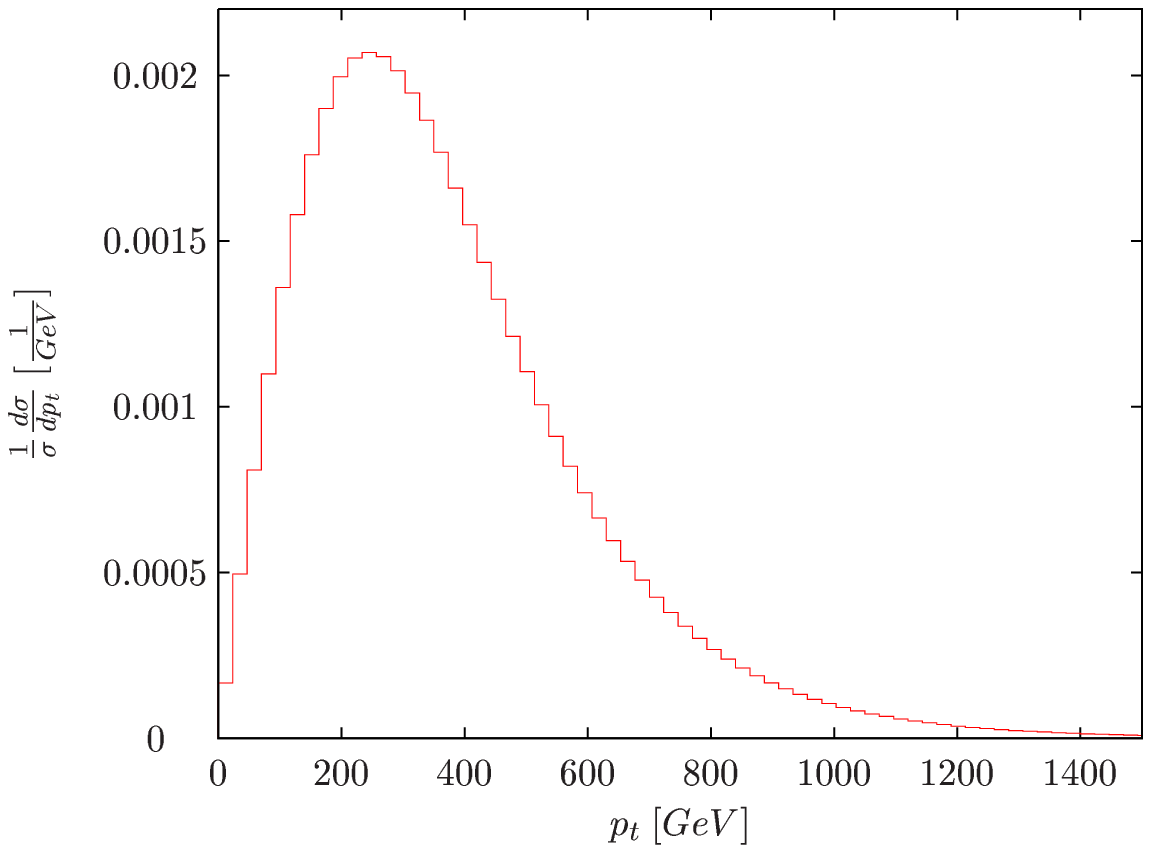}
\vspace*{-5mm}
\caption{Transverse momentum distribution of gluino and squark in the \mbox{SPS 1a} scenario.
  \label{fig:ULGO_Trans_mom}}
\vspace*{+5mm}
\includegraphics[width=0.65\textwidth]{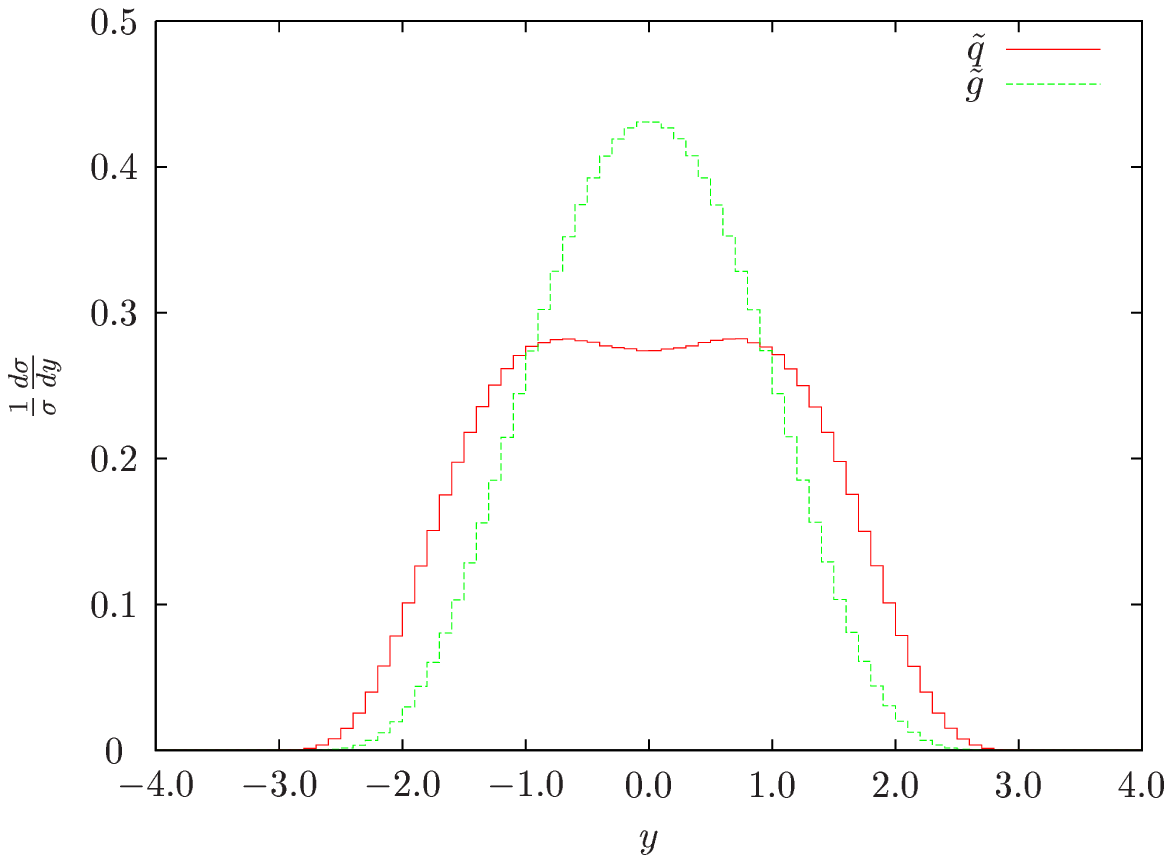}
\vspace*{-5mm}
\caption{Rapidity distribution of squark and gluino in the SPS1a scenario.
  \label{fig:ULGO_rap}}
\vspace*{+5mm}
\includegraphics[width=0.65\textwidth]{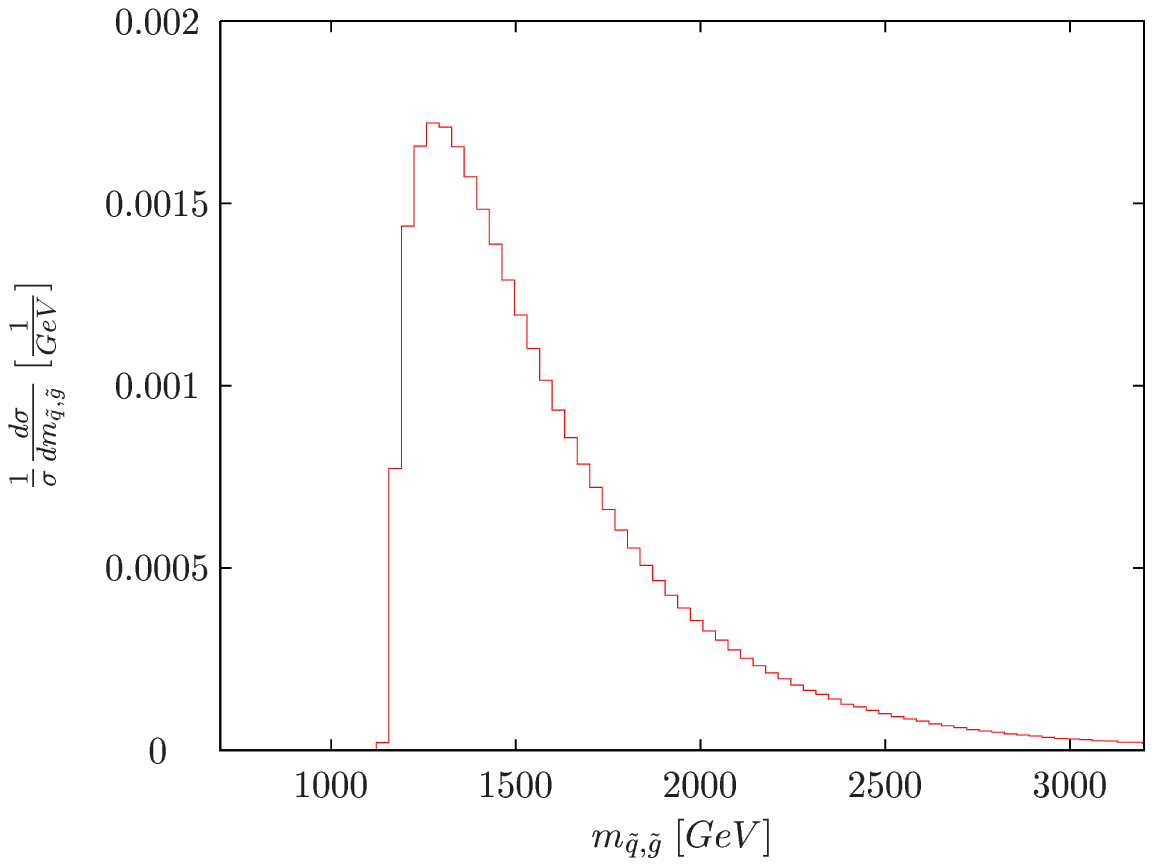}
\vspace*{-5mm}
\caption{The invariant mass distribution of the squark-gluino pair in the SPS1a scenario.
  \label{fig:ULGO_InvMasse}}
\end{figure}

For the hadronic differential cross sections of gluino-squark pair production by quark-gluon collision one obtains the distributions shown in figs.~\ref{fig:ULGO_Trans_mom} to~\ref{fig:ULGO_InvMasse}. Calculations were performed using the partonic cross sections from section~\ref{sec:WQ} and using the helicity amplitude generators SMadgraph~\cite{cho-2006-73},~\cite{Alwall:hep-ph0706.2334} and HELAS~\cite{Murayama:1992gi}. These results agree within the numerical accuracy of our integration. The calculation of the hadronic cross sections was also performed for the other SUSY-QCD processes given in fig.~\ref{fig:SUSY_production} a)-e). But since they are not primarily important for us, and show no significant new results compared to the squark-gluino production, we omit their presentation. For all plots we summed over all quark flavors except for the top-quark. 

As given in the SPS1a scenario, we assumed masses to be $m_{\tilde{q}_L}=562.260 \; \text{GeV}$, $m_{\tilde{q}_R}=545.890 \; \text{GeV}$ and $m_{\tilde{g}}=606.105 \; \text{GeV}$. For the strong coupling we chose a value of $\alpha_s(580 \; \text{GeV})=0.100375$. Masses of all squark flavors are assumed to be equal.

Of course the transverse momentum of squark and gluino are identical for this 2$\rightarrow$2 process. As one can see from~\cite{Beenakker}, where calculations for the squark and gluino production processes are also performed in next-to-leading-order (NLO), the NLO-corrections to $p_t$ and $y$ distributions are very small. The normalized NLO-distribution of transverse momentum and rapidity are quite well described by the lowest-order approximation.

Therefore is seems legitimate to only use the LO distributions to compare the kinematics of SUSY to other models.

\begin{figure}[h!]
\centering
\includegraphics[width=0.65\textwidth]{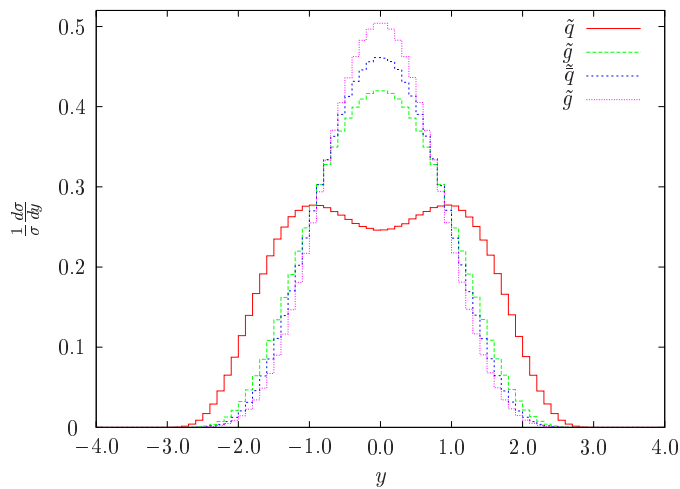}
\vspace*{-5mm}
\caption{Rapidity of squark and gluino from quark-gluon collision and antisquark and gluino from antiquark-gluon collision in the SPS1a scenario.
  \label{fig:ULGO_rap_both_1}}
\end{figure}

If the PDFs are set equal to one, the shapes of both rapidity distributions of squark and gluino are equal. This implies, that the shape of the rapidity is governed by the PDFs. In fig.~\ref{fig:ULGO_rap_both_1} we show that significant differences for the distributions of particles and antiparticles exist. The red and the green line derive from quark-gluon collisions while the blue and the pink line derive from antiquark-gluon collisions. The differences are due to the fact that the parton distribution function of the quarks is very different from the gluon PDF. This is different for the antiquark PDF, which is more similar to the gluon PDF, cf.\ fig.~\ref{fig:pdfs}. This results in a large boost for the $u$- or $d$-quark and gluon system, because $u$- and $d$-quarks are more often included in the proton with larger values of $x$ than their antiparticles are. This can be seen immediately from the dent in their PDFs. Therefore the rapidity distribution for the squarks and gluinos in the case of incoming quarks is broadened, compared to the case of incoming antiquarks. The shape of the rapidity distributions for squark, antisquark and gluino are differently influenced by the PDFs, since in t- and u-channel diagrams the outgoing squark directly couples to the incoming quark and the outgoing gluino couples to the incoming gluon. As a consequence the rapidity distribution of the antisquark in fig.~\ref{fig:ULGO_rap_both_1} obtains a minimum at $y=0$. The rapidity distributions of a LHC process always have to be symmetric, since the colliding partons can come from both of the protons. There is no favored direction at the LHC even for the more complicated $2 \rightarrow 5$ process.

The numerical results for the cross sections are given in table~\ref{tab:WQ_SUSY} separately for all quark flavors. After summing over all quark flavors, we obtain a total cross section of $21.64 \; \text{pb}$.
\bigskip
\clearpage
\section{The Production of a Kaluza-Klein Gluon-Quark Pair in UED}
\begin{figure}[t!]
\centering
\begin{tabular}{c p{1mm} c@{      } p{1mm} c@{      }}

  \includegraphics[width=0.27\textwidth]{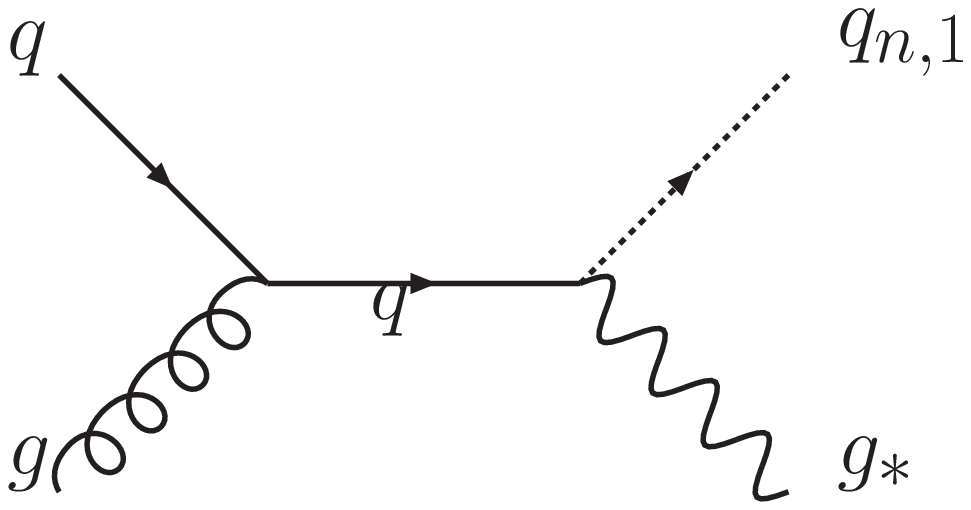} &
&
  \includegraphics[width=0.27\textwidth]{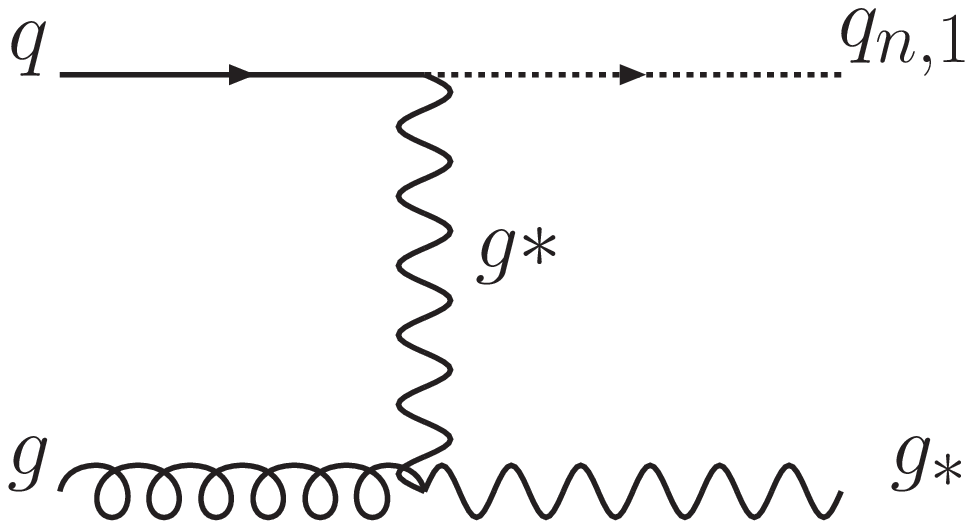} &
&
  \includegraphics[width=0.27\textwidth]{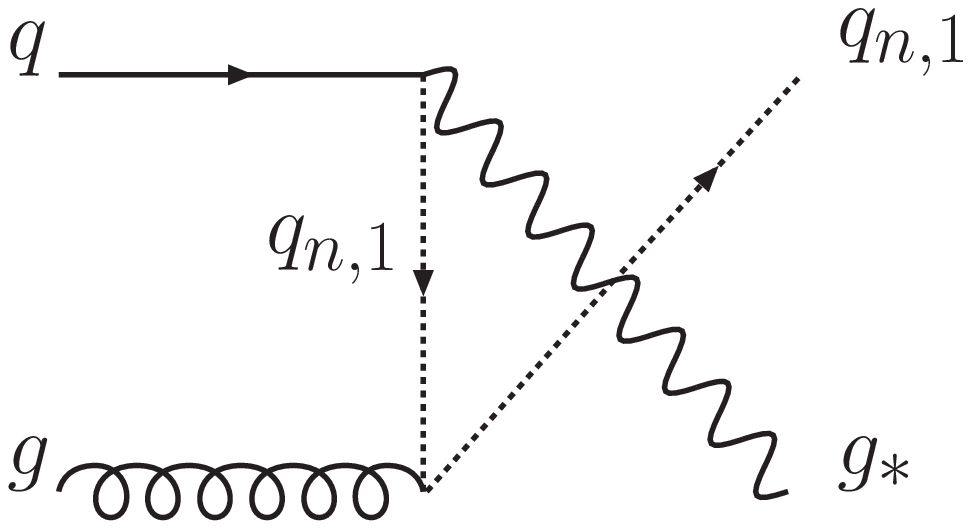} \\[4mm]

\end{tabular}
\caption{LO contributions to the production of heavy KK-quarks and heavy KK-gluons in UED.
  \label{fig:UED_production}}
\end{figure}

\begin{figure}[p!]
\centering
\includegraphics[width=0.65\textwidth]{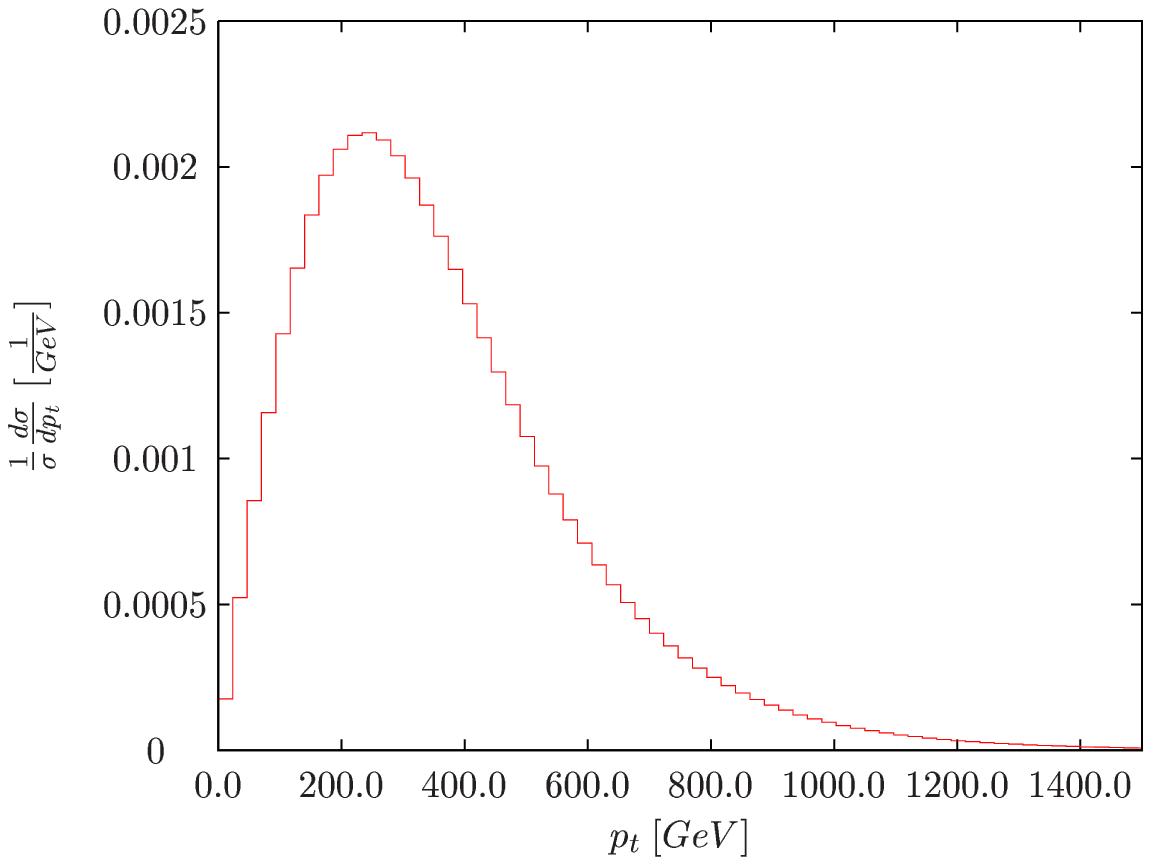}
\vspace*{-5mm}
\caption{Transverse momentum distribution of KK-gluon and KK-quark with masses of the SPS1a scenario.
  \label{fig:UED_Trans_mom}}
\vspace*{+5mm}
\includegraphics[width=0.65\textwidth]{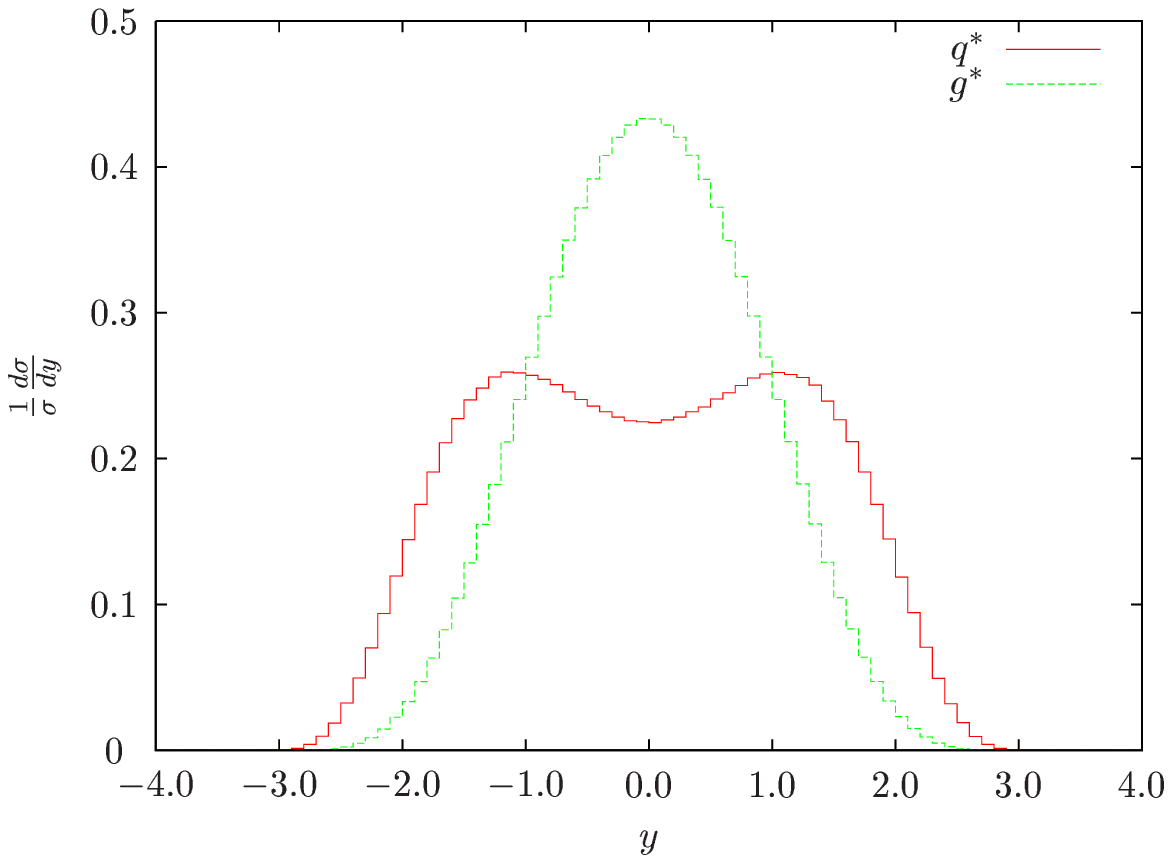}
\vspace*{-5mm}
\caption{Rapidity distribution of the KK-quark and KK-gluon with masses of the SPS1a scenario.
  \label{fig:UED_rap}}
\vspace*{+5mm}
\includegraphics[width=0.65\textwidth]{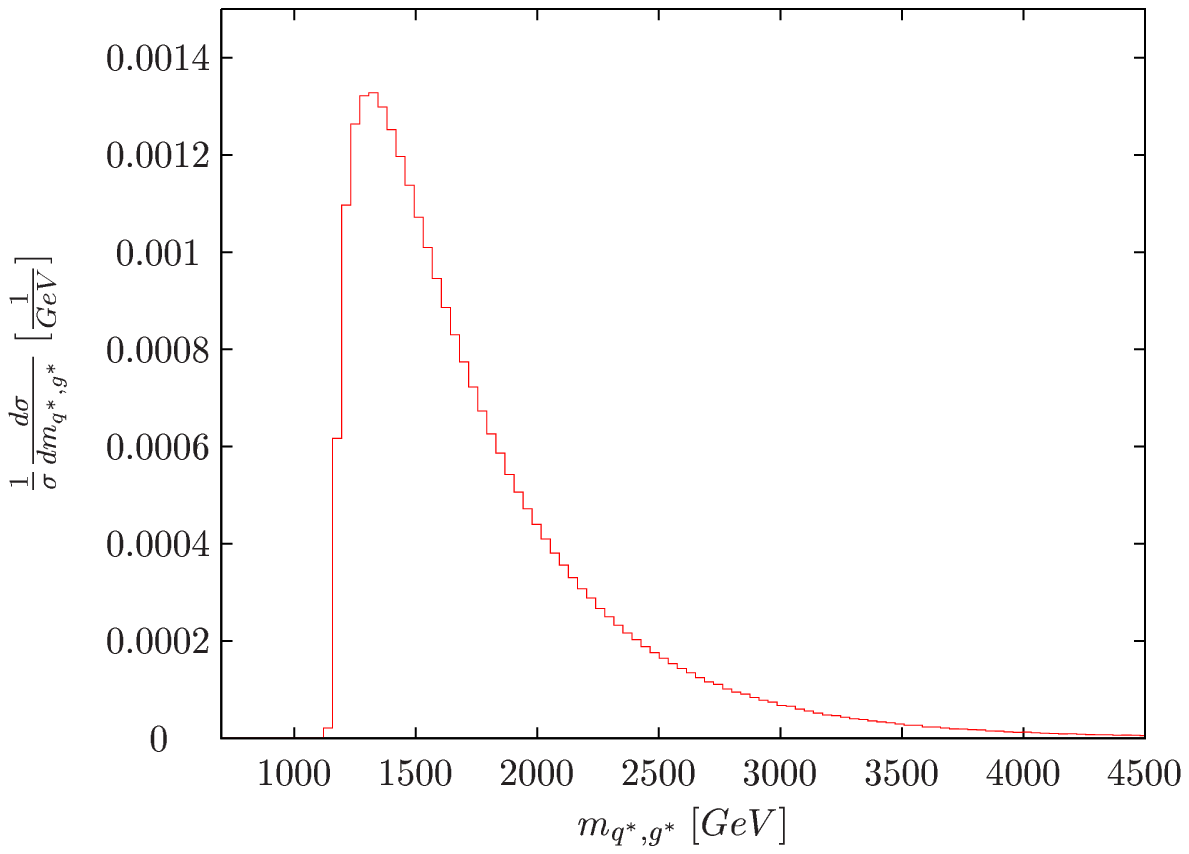}
\vspace*{-5mm}
\caption{The invariant mass distribution of the KK-quark-KK-gluon pair.
  \label{fig:UED_InvMasse}}
\end{figure}

In this section we compute the $2 \rightarrow 2$ cross section for KK-gluon and KK-quark ($g^*,q_{n,i}$) production. The diagrams taken into account are presented in fig.~\ref{fig:UED_production}. First calculations in the literature were performed in~\cite{PhysRevD.65.076007} and~\cite{macesanu-2002-66}. Their results were corrected in~\cite{smillie-2005-0510}. Both calculations were done for equal masses of KK-gluon and KK-quark. This is reasonable for a UED calculation since masses are nearly degenerate in a UED spectrum. But since we want to compare SUSY and UED for the SPS 1a mass spectrum, we derive a more general cross section with different masses of the final state particles. As a check we could show that in the limit of equal masses for the KK-quark and KK-gluon, the result of~\cite{smillie-2005-0510} agrees with our calculation.
Using the Feynman rules from the appendix and the polarization sum $-g_{\mu \nu}$ for gluons and eq.~(\ref{eq:massive_pol}) for KK-gluons, we obtain the squared matrix element given by
\vspace{-8pt} \begin{eqnarray}
\label{eq:neues_matrixel}
\hspace*{-15mm}& &\hspace*{-10mm}\mbox{\raisebox{-.5ex}{\Large$\bar{\Sigma}$}}\! \mid \!
\mathcal{M}(q g \rightarrow q_{n,1} g_n^{\ast})\!\mid ^2 \, = -\frac{g_s^4(Q)}{1152\, m_{g^{\ast}}^2 \,s\, t_{g^{\ast}}^2\, u_{q^{\ast}}^2} \Big [ 32 \lbrace 4 t_{g^{\ast}}^2 + 9 u_{q^{\ast}} t_{g^{\ast}} + 9 u_{q^{\ast}}^2 \rbrace \nonumber \\[2mm]
& & \times \Big ( 4 u_{q^{\ast}} m_{g^{\ast}}^6 +\lbrace m_{q^{\ast}}^2 \lbrace 4 t_{g^{\ast}} - 2 u_{q^{\ast}} \rbrace + 4 t_{g^{\ast}} u_{q^{\ast}} \rbrace m_{g^{\ast}}^4+2 \lbrace s m_{q^{\ast}}^4-t_{g^{\ast}} u_{q^{\ast}} m_{q^{\ast}}^2 \nonumber \\[2mm]
& & +u^{3}_{q^{\ast}}+s^2 u^{2}_{q^{\ast}} \rbrace m_{g^{\ast}}^2+ m_{q^{\ast}}^2 \,t_{g^{\ast}} \lbrace -2 m_{q^{\ast}}^4- 2 u_{q^{\ast}} m_{q^{\ast}}^2+t_{g^{\ast}} u_{q^{\ast}} \rbrace \Big )\Big ]
\;,
\end{eqnarray}
with $u_q=u-m_{q^{\ast}}^2$ and $t_g=t-m_{g^{\ast}}^2$. The masses $m_{q^{\ast}}$ and $m_{g^{\ast}}$ here denote the masses of the KK-quark and KK-gluon.
The squared matrix element in the limit of equal masses of the outgoing KK-gluon and KK-quark, as it is the case in a typical UED mass spectrum, is given by
\vspace{-1pt} \begin{eqnarray*}
\mbox{\raisebox{-.5ex}{\Large$\bar{\Sigma}$}}\! \mid \!
\mathcal{M}(q g  \rightarrow q_{n,1} g_n^{\ast})\!\mid ^2 \, =& &
\!\!\!\!\!\!\!\!\! \frac{-1}{3}\ g_s^4(Q)\left( 
\frac{5 {s}^2}{12 t'^2} +  \frac{{s}^3}{t'^2 u'} +
\frac{11 {s} u'}{6 t'^2} + \frac{5 u'^2}{12 t'^2} +
\frac{u'^3}{{s} t'^2}\right) \,,
\end{eqnarray*}
with $u'=u-m_n^2$ and $t'=t-m_n^2$.
We calculate the total cross section from eq.~(\ref{eq:neues_matrixel}), according to eq.~(\ref{eq:partcross}), and obtain
\begin{align}
\label{eq:smillie}
\sigma^B(qg &\rightarrow q_{n,1} g_n^{\ast}) = \frac{\alpha^2_s(Q) \, \pi}{36 \,m^2_{{g^{\ast}}}\, s^3} \Big[16 \mu m^6_{{g^{\ast}}}+ 4 \,\lbrace3\,(\mu-3 \nu)m^2_{{q^{\ast}}}+\eta \rbrace m^4_{{g^{\ast}}} \nonumber \\[2mm]
&+ 2\, \lbrace-(9\,(\mu-\nu)m^2_{{q^{\ast}}} + \eta ) m^2_{{q^{\ast}}} + 11 s \,\xi + 2 s^2 \,(2 \mu + 9 \nu )\rbrace m^2_{{g^{\ast}}}\nonumber \\[2mm]
 &+36 s^2 \xi -2 m^4_{{q^{\ast}}} \eta + m^2_{{q^{\ast}}} s\, \lbrace 3 \xi + 4 s \mu \rbrace - 2 m^6_{{q^{\ast}}}\, \lbrace 5 \mu -9 \nu \rbrace \Big]\; ,
\end{align}
with
\begin{align}
\xi&=\sqrt{m_g^4+ \lbrace m_q^2 -s \rbrace^2 -2 m_g^2 \lbrace m_{q^{\ast}}^2+s\rbrace}\; , \qquad
\mu=\frac{\text{ln}\left( m_{q^{\ast}}^2-m_{g^{\ast}}^2+s+\xi \right )}{\text{ln}\left( m_{q^{\ast}}^2-m_{g^{\ast}}^2+s-\xi \right )}\;\nonumber ,\\[2mm]
\nu&=\frac{\text{ln}\left( m_{g^{\ast}}^2-m_{q^{\ast}}^2+s-\xi\right )}{\text{ln}\left( m_{g^{\ast}}^2-m_{q^{\ast}}^2+s+\xi\right )}\; , \qquad
\hspace*{21.5mm} \eta=16 \xi -4 s \mu + 9 s \nu \; .
\end{align}
\newpage
Calculating the kinematic distributions in UED for the parameters of the SUSY SPS 1a benchmark point using Madgraph, one obtains the graphs given in figs.~\ref{fig:UED_Trans_mom} to~\ref{fig:UED_InvMasse}. We checked the agreement with the given formulae numerically. The transverse momentum of KK-quark and KK-gluon are equal due to momentum conservation, as it is also the case in our SUSY calculation. The invariant masses of the outgoing particles in SUSY and UED show a different behavior. This is shortly discussed in section~\ref{sec:A SUSY-UED Comparison at the Threshold}. The difference between the squark and KK-quark rapidity distributions are discussed in section~\ref{sec:A SUSY-UED Comparison of Angular Distributions} in more detail.

Later a comparison of the hadronic SUSY and UED distributions will be performed in chapter~\ref{ch:6}. The numerical results for the cross sections are given  separately for all quark flavors in table~\ref{tab:WQ_SUSY}. From these values one can see that UED cross sections are much larger than SUSY cross sections. Accordingly, the total cross section itself could be used to discriminate UED and SUSY at the LHC, provided that the masses of all participating particles are known. After summing over all quark flavors, we obtain a total cross section of $170.08 \; \text{pb}$.

\begin{table}
\begin{center}
\begin{tabular}{|c|cc||cc|} 
  \hline 
  \rule[-5pt]{0pt}{20pt} Process & $\sigma^{SUSY}$ in pb & $\text{Error}^{SUSY}$ in pb & $\sigma^{UED}$ in pb & $\text{Error}^{UED}$ in pb\\[2pt]
\hline
  \rule{0pt}{13pt}$u\;g \rightarrow \bar{\tilde{u}}_L \; \tilde{g}$         &  $6.0563 $ & $ \pm 3 \cdot 10^{-3}$  &  $ 48.870 $ & $ \pm 3 \cdot 10^{-2}$     \\[4pt]
\hline
 \rule{0pt}{13pt}$u\;g \rightarrow \bar{\tilde{u}}_R \; \tilde{g}$         &  $ 6.4399$ & $ \pm 3 \cdot10^{-3} $  &  $ 51.182$ & $ \pm 3 \cdot 10^{-2}$     \\[4pt]
\hline
  \rule{0pt}{13pt}$d\;g \rightarrow \bar{\tilde{d}}_L \; \tilde{g}$        &  $ 2.5726 $ & $ \pm 1 \cdot10^{-3}$  &  $ 20.331$ & $ \pm 1 \cdot 10^{-2}$     \\[4pt]
\hline
\rule{0pt}{13pt}$d\;g \rightarrow \bar{\tilde{d}}_R \; \tilde{g}$         &  $ 2.7435 $ & $ \pm 1 \cdot 10^{-3}$  &  $ 21.368$ & $ \pm 1 \cdot 10^{-2}$     \\[4pt]
\hline
  \rule{0pt}{13pt}$c\;g \rightarrow \bar{\tilde{c}}_L \; \tilde{g}$       &  $ 0.13976$ & $ \pm 8 \cdot 10^{-5}$  &  $ 1.0270$ & $ \pm 7 \cdot 10^{-4}$     \\[4pt]
\hline
  \rule{0pt}{13pt}$c\;g \rightarrow \bar{\tilde{c}}_R \; \tilde{g}$      &  $ 0.15097 $ & $ \pm 8 \cdot 10^{-5}$  &  $ 1.0886$ & $ \pm 7 \cdot 10^{-4}$     \\[4pt]
\hline
  \rule{0pt}{13pt}$s\;g \rightarrow \bar{\tilde{s}}_L \; \tilde{g}$      &  $ 0.25875 $ & $ \pm 1 \cdot 10^{-4}$  &  $ 1.9298$ & $ \pm 1 \cdot 10^{-3}$     \\[4pt]
\hline
\rule{0pt}{13pt}$s\;g \rightarrow \bar{\tilde{s}}_R \; \tilde{g}$        &  $ 0.27868 $ & $ \pm 1 \cdot 10^{-4}$  &  $ 2.0396$ & $ \pm 1 \cdot 10^{-3}$     \\[4pt]
\hline
  \rule{0pt}{13pt}$b\;g \rightarrow \bar{\tilde{b}}_L \; \tilde{g}$      &  $ 0.087215$ & $ \pm 5 \cdot 10^{-5}$  &  $ 0.63831$ & $ \pm 4 \cdot 10^{-4}$     \\[4pt]
\hline
 \rule{0pt}{13pt}$b\;g \rightarrow \bar{\tilde{b}}_R \; \tilde{g}$       &  $ 0.094239$ & $ \pm 5 \cdot 10^{-5}$  &  $ 0.67685$ & $ \pm 4 \cdot 10^{-4}$     \\[4pt]
\hline 
 \rule{0pt}{13pt}$\bar{u}\;g \rightarrow \bar{\tilde{u}}_L \; \tilde{g}$ &  $ 0.37626$ & $ \pm 2 \cdot 10^{-4}$  &  $ 2.8260$ & $ \pm 2 \cdot 10^{-3}$     \\[4pt]
\hline
\rule{0pt}{13pt}$\bar{u}\;g \rightarrow \bar{\tilde{u}}_R \; \tilde{g}$  &  $ 0.40481$ & $ \pm 2 \cdot 10^{-4}$ &  $ 2.9833 $& $ \pm 2 \cdot 10^{-3}$     \\[4pt]
\hline
  \rule{0pt}{13pt}$\bar{d}\;g \rightarrow \bar{\tilde{d}}_L \; \tilde{g}$&  $ 0.49811$ & $ \pm 3 \cdot 10^{-4}$  &  $ 3.7413$ & $ \pm 2 \cdot 10^{-3}$     \\[4pt]
\hline
  \rule{0pt}{13pt}$\bar{d}\;g \rightarrow \bar{\tilde{d}}_R \; \tilde{g}$&  $ 0.53561 $ & $ \pm 3 \cdot 10^{-4}$  &  $ 3.9480$ & $ \pm 3 \cdot 10^{-3}$     \\[4pt]
\hline
  \rule{0pt}{13pt}$\bar{c}\;g \rightarrow \bar{\tilde{c}}_L \; \tilde{g}$&  $ 0.13976$ & $ \pm 8 \cdot 10^{-5}$  &  $ 1.0270$ & $ \pm 7 \cdot 10^{-4}$     \\[4pt]
\hline
 \rule{0pt}{13pt}$\bar{c}\;g \rightarrow \bar{\tilde{c}}_R \; \tilde{g}$ &  $ 0.15097 $ & $ \pm 8 \cdot 10^{-5}$  &  $1.0886 $ & $ \pm 7 \cdot 10^{-4}$     \\[4pt]
\hline  
\rule{0pt}{13pt}$\bar{s}\;g \rightarrow \bar{\tilde{s}}_L \; \tilde{g}$  &  $ 0.25875 $ & $ \pm 1 \cdot 10^{-4}$  &  $ 1.9298$ & $ \pm 1 \cdot 10^{-3}$     \\[4pt]
\hline
\rule{0pt}{13pt}$\bar{s}\;g \rightarrow \bar{\tilde{s}}_R \; \tilde{g}$  &  $ 0.27868$ & $ \pm 1 \cdot 10^{-4}$  &  $ 2.0396$ & $ \pm 1 \cdot 10^{-3}$     \\[4pt]
\hline
  \rule{0pt}{13pt}$\bar{b}\;g \rightarrow \bar{\tilde{b}}_L \; \tilde{g}$&  $ 0.087215 $ & $ \pm 5 \cdot 10^{-5}$  &  $ 0.63831$ & $ \pm 4 \cdot 10^{-4}$     \\[4pt]
\hline
  \rule{0pt}{13pt}$\bar{b}\;g \rightarrow \bar{\tilde{b}}_R \; \tilde{g}$&  $ 0.094239 $ & $ \pm 5 \cdot 10^{-5}$  &  $ 0.67685$ & $ \pm 4 \cdot 10^{-4}$     \\[4pt]
\hline
\end{tabular}
\end{center}
\caption{Cross sections for all quark flavors for quark-gluon collision in SUSY and UED}
\label{tab:WQ_SUSY}
\end{table}
\newpage
\begin{figure}
\centering
\includegraphics[width=0.65\textwidth]{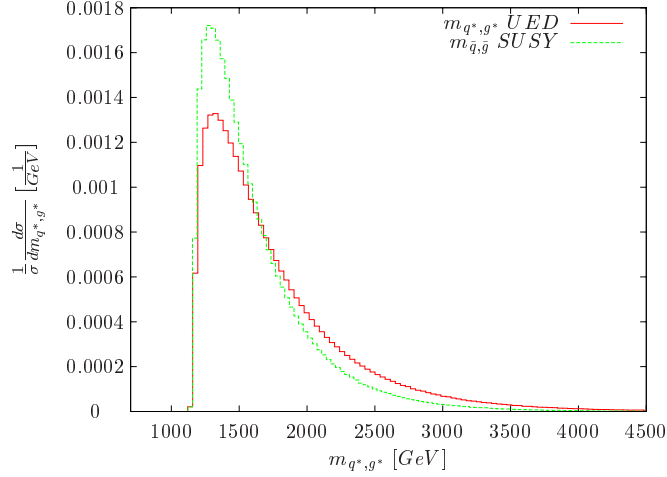}
\vspace*{-5mm}
\caption{Comparison of the hadronic invariant mass distributions of the given particles in SUSY and UED.
  \label{fig:gnu_plots/Threshold/InvMasse_SUSY_und_UED.eps}}
\end{figure}
\section{A SUSY-UED Comparison at the Threshold}
\label{sec:A SUSY-UED Comparison at the Threshold}
Investigating the behavior of the cross section close to the threshold is interesting due to its strong energy dependence. It can be expected that SUSY and UED models differ significantly in this region. As we can see in fig.~\ref{fig:gnu_plots/Threshold/InvMasse_SUSY_und_UED.eps}, where the invariant masses of a squark-gluino and a KK-quark-KK-gluon pair are shown, differences right behind the production threshold are obvious. This could be due to a different threshold behavior of the partonic cross section in both scenarios. Therefore it should already be visible at the partonic level.\\
As in~\cite{Beenakker}, the so called phase-space suppression factor $\beta$, is defined as
\begin{eqnarray}
\vspace{-12pt}
\beta=\sqrt{1-\frac{4 m_{\tilde{q}}m_{\tilde{g}}}{s-(m_{\tilde{q}}-m_{\tilde{g}})^2}} \,.
\end{eqnarray}
Close to the threshold, i.e.\ when the produced particles have a small velocity in their center of mass system, $\beta \ll 1$, one can give analytical expressions for the expansion of the cross section in $\beta$.
The expansion of the SUSY cross section for quark-gluon collisions, as given in section~\ref{sec:WQ}, yields
\begin{eqnarray}
\sigma_{\text{SUSY}}^{\text{B}\; \text{approx}}(qg \rightarrow \tilde{q} \tilde{g})=\alpha_s^2(Q) \frac{4 \pi \beta}{(m_{\tilde{q}}+m_{\tilde{g}})^5} \, \left[ \frac{2}{9} m_{\tilde{q}} m_{\tilde{g}}^2 + \frac{1}{2} m_{\tilde{q}}^2 m_{\tilde{g}}+\frac{1}{2} m_{\tilde{q}}^3 \right] \, .
\end{eqnarray}
This agrees with the result given in~\cite{Beenakker}. In the UED scenario one obtains for the equivalent process, expanding eq.~(\ref{eq:smillie})
\begin{align}
\sigma_{\text{UED}}^{\text{B}\; \text{approx}}&(qg \rightarrow q_{n,1} g_n^{\ast})=\\[2mm]
&\hspace*{-3mm}=\alpha_s^4(Q) \pi \beta \frac{(8 m_{\tilde{g}}^4+34 m_{\tilde{q}} m_{\tilde{g}}^3+74 m_{\tilde{q}}^2 m_{\tilde{g}}^2 +81 m_{\tilde{q}}^3 m_{\tilde{g}} +45 m_{\tilde{q}} ^4)}{9 m_{\tilde{g}} (m_{\tilde{g}}+m_{\tilde{g}})^5} \nonumber \; .
\end{align}

\begin{figure}[h!]
\centering
\includegraphics[width=0.65\textwidth]{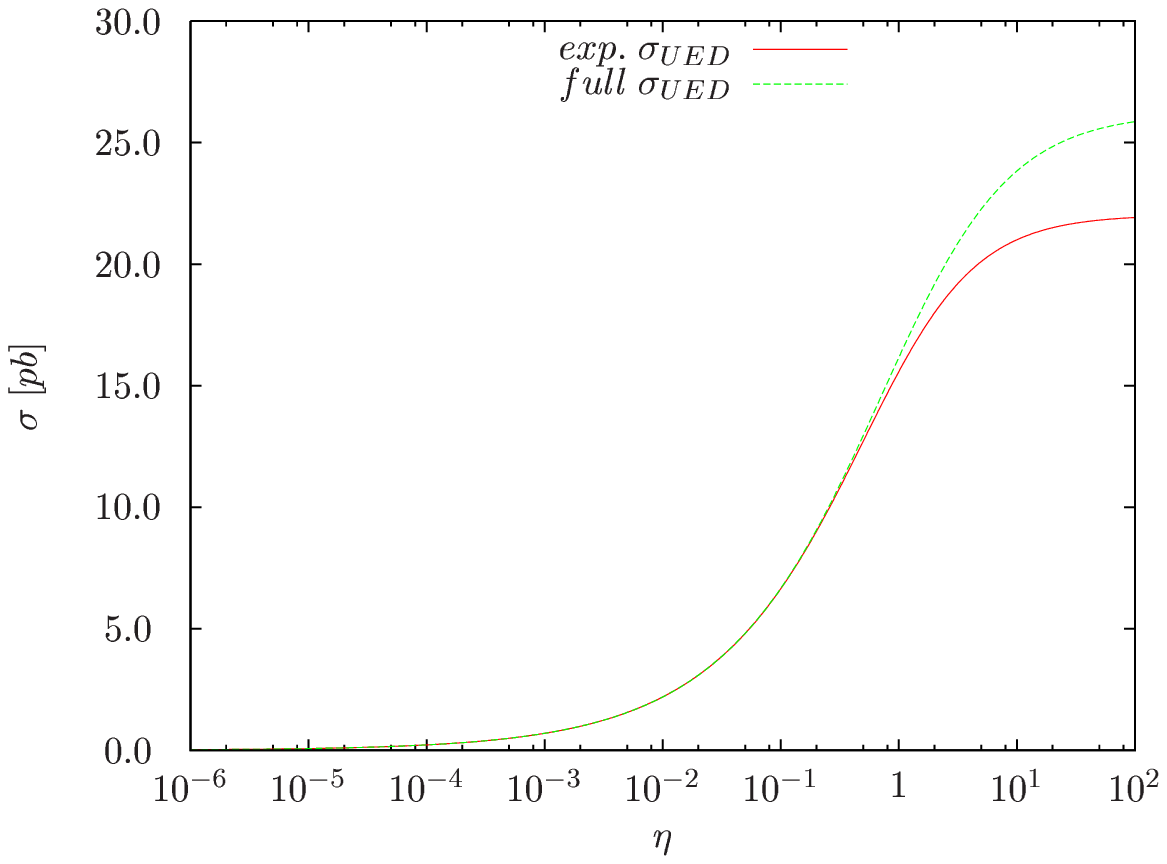}
\vspace*{-5mm}
\caption{Partonic leading-order cross section in UED for production of a KK-quark-KK-gluon pair.
  \label{fig:UED_partonic_2_2}}
\vspace*{+5mm}
\includegraphics[width=0.65\textwidth]{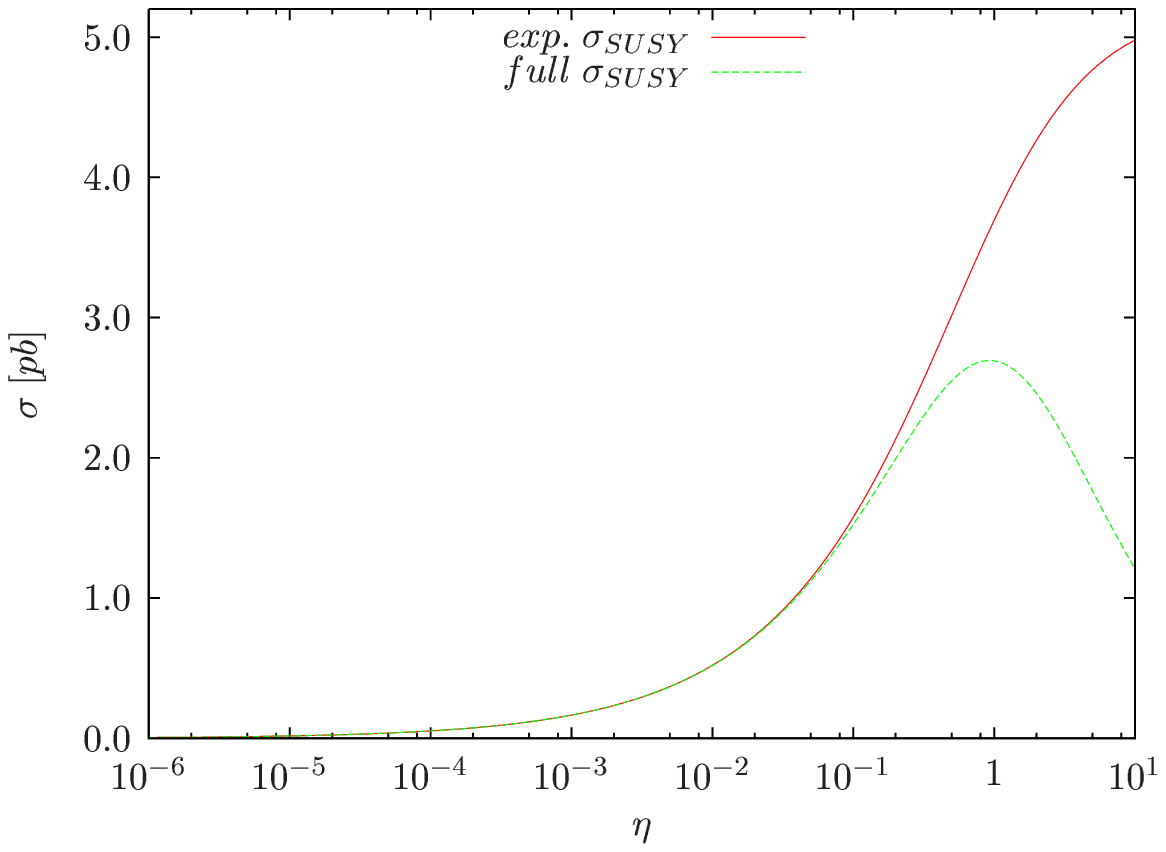}
\vspace*{-5mm}
\caption{Partonic leading-order cross section in SUSY for production of a squark-gluino pair.
  \label{fig:SUSY_partonic_2_2}}
\vspace*{+5mm}
\includegraphics[width=0.65\textwidth]{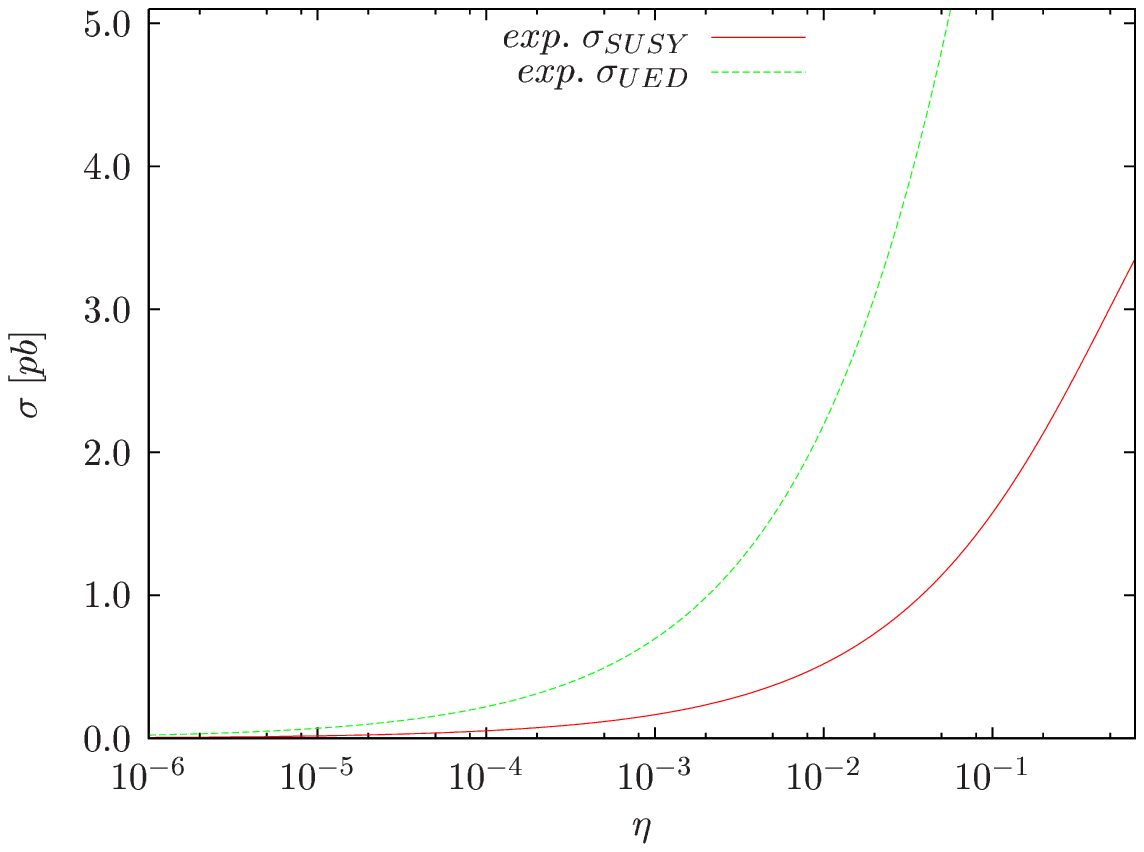}
\vspace*{-5mm}
\caption{Comparison of the UED and SUSY cross section threshold expansions.
  \label{fig:comparison_approx_2_2}}
\end{figure}
\clearpage
In fig.~\ref{fig:UED_partonic_2_2} and fig.~\ref{fig:SUSY_partonic_2_2} we compare the partonic SUSY and UED threshold expansions to the full partonic cross sections. In fig.~\ref{fig:comparison_approx_2_2} we compare both threshold expansions. The masses of the outgoing particles are given by $m_{\tilde{g}}=606.105 \; \text{GeV}$ and $m_{\tilde{q}}=562.260 \; \text{GeV}$. 
The factorization scale is set to $Q=580 \, \text{GeV}$. 
The quantity $\eta$ is defined by
\begin{equation}
\eta=\frac{s}{(m_{\tilde{q}}+m_{\tilde{g}})^2}-1 \; .
\end{equation}

There are obvious differences between the partonic SUSY and UED cross sections. While the SUSY cross section peaks behind the threshold and decreases for $s \rightarrow \infty$, the UED cross section does not decrease. This is in contradiction to unitarity. But since we did not take into account higher KK-excitations, due to suppression by their higher mass, the result is only meaningful below the second KK-excitation, i.e.\ not too far above the threshold. Therefore it seems to be reasonable to compare UED and SUSY at the threshold. Directly finding the second excitation level at the LHC would, of course, be a clear sign for the theory of Universal Extra Dimensions being realized in nature. As one can see easily, the UED cross section increases much faster than the SUSY cross section. Of course this partonic quantity can not be measured at the LHC. But it shows that typical UED cross sections are usually much higher than typical SUSY cross sections, assuming the same mass spectrum.

\section{A SUSY-UED Comparison of Angular Distributions}
\label{sec:A SUSY-UED Comparison of Angular Distributions}

\begin{figure}[b!]
\centering
\includegraphics[width=0.65\textwidth]{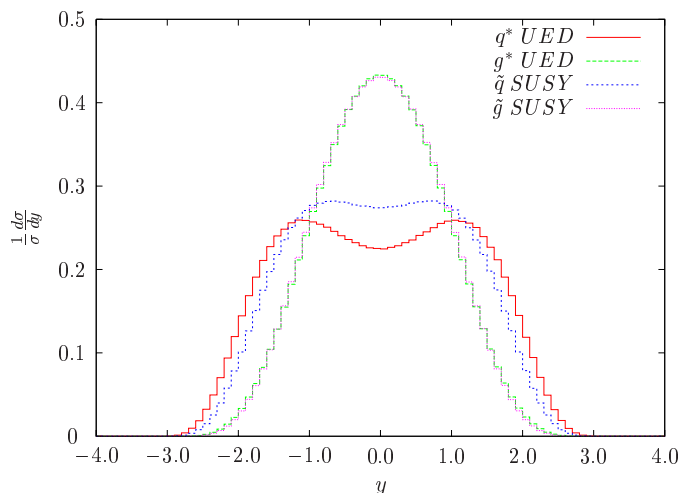}
\vspace*{-5mm}
\caption{Comparison of the hadronic rapidity distributions of the given particles in SUSY and UED.
  \label{fig:Rapiditaet_both_SUSY_ANS_UED.eps}}
\end{figure}

As we find in fig.~\ref{fig:ULGO_rap} and fig.~\ref{fig:UED_rap}, the rapidity distributions in SUSY and UED are quite different. These hadronic rapidity distributions are all shown again in fig.~\ref{fig:Rapiditaet_both_SUSY_ANS_UED.eps}. One finds that the gluino and KK-gluon rapidities are quite similar, while there are differences for the outgoing squark and KK-quark. For the hadronic cross section the sum is taken over all quark flavors $u$, $d$, $c$, $s$ and $b$.
\begin{figure}[b!]
\centering
\includegraphics[width=0.65\textwidth]{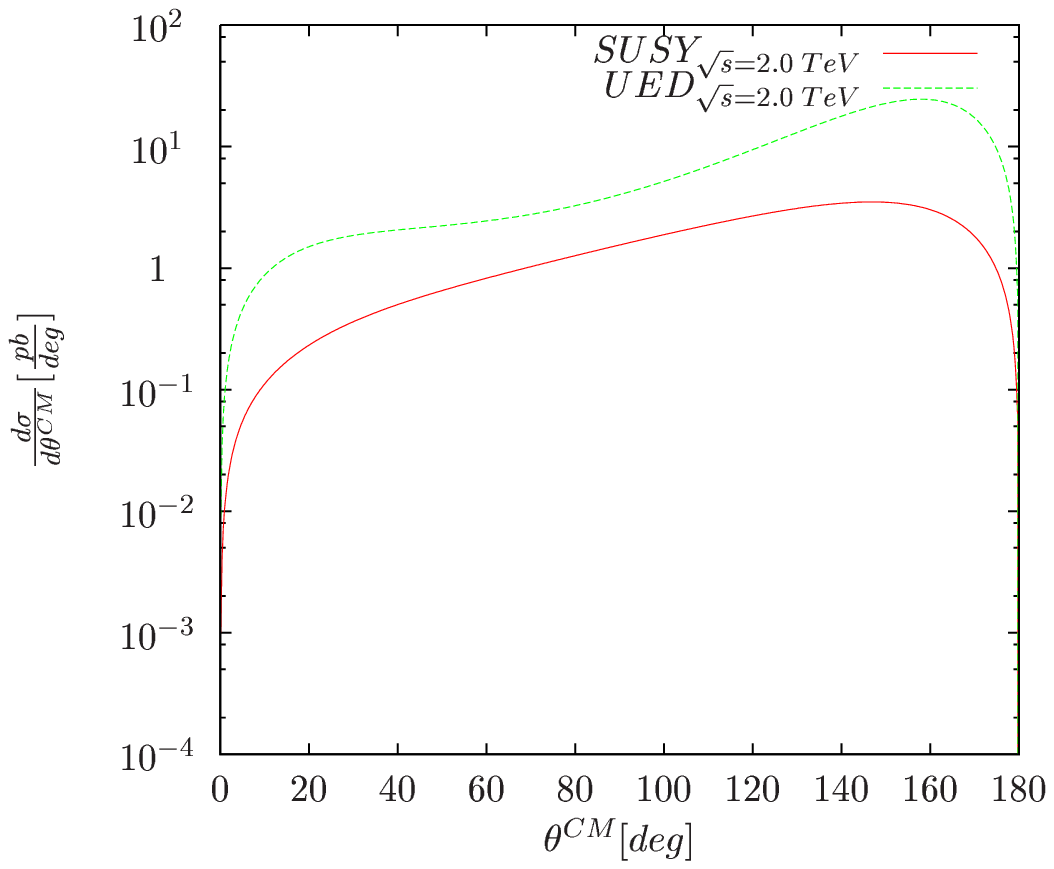}
\vspace*{-5mm}
\caption{Distribution of the angle between the squark/ KK-quark and the beam axis in the CMS for $\sqrt{s}=2.0 \;\text{TeV}$.
  \label{fig:Rap_plot_UED_3.eps}}
\vspace*{+20mm}
\includegraphics[width=0.65\textwidth]{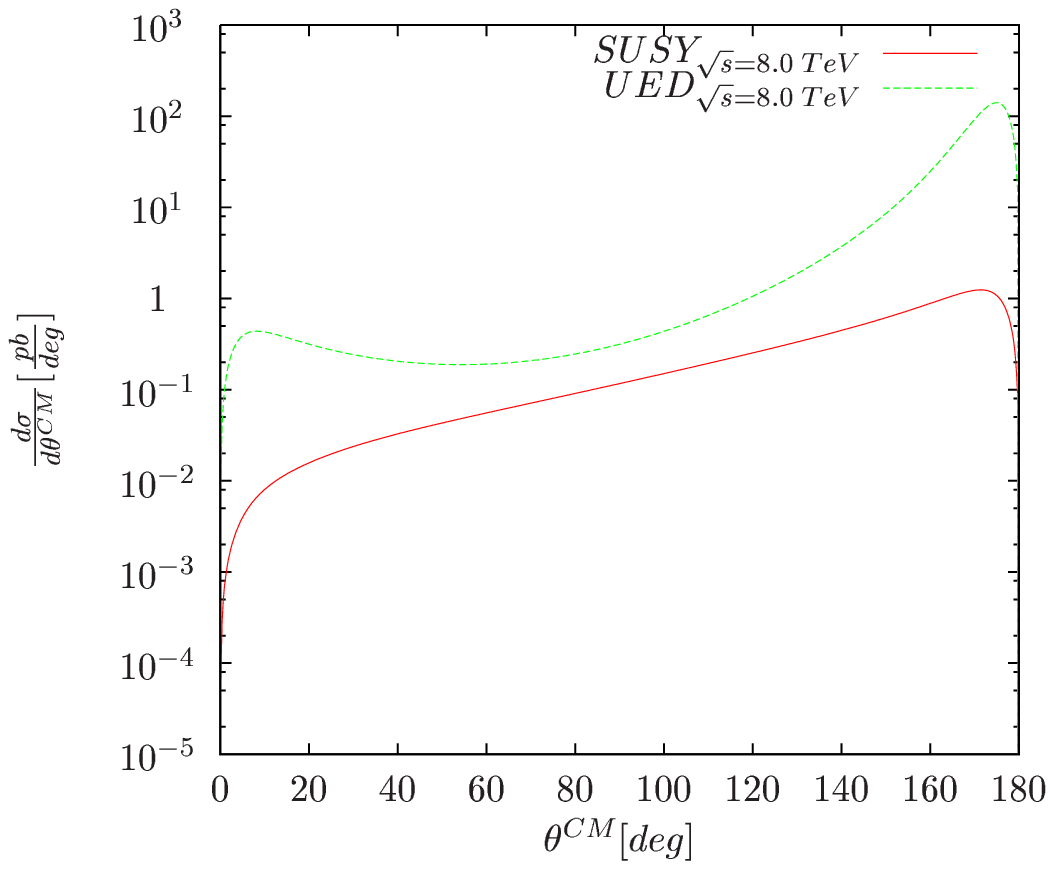}
\vspace*{-5mm}
\caption{Distribution of the angle between the squark/ KK-quark and the beam axis in the CMS for $\sqrt{s}=8.0 \;\text{TeV}$.
  \label{fig:Rap_plot_UED_64.eps}}
\end{figure}

A plausible explanation is found by analyzing the angular distributions of the partonic cross sections in the CMS, given in fig.~\ref{fig:Rap_plot_UED_3.eps} and fig.~\ref{fig:Rap_plot_UED_64.eps}. For the case of a moderate center of mass energy, e.g.\ $\sqrt{s}=2.0\;  \text{TeV}$, one finds that the angular distributions of the squark and KK-quark in SUSY and UED have a very similar shape. For a high center of mass energy, e.g.\ $\sqrt{s}=8.0 \; \text{TeV}$, one finds increasing differences between SUSY and UED. These differences could be due to the momentum dependent three gluon vertex in UED, while the corresponding gluon-gluino-gluino-vertex in SUSY is independent of the incoming momenta. Instead the squark-squark-gluon vertex in SUSY is dependent on the squark momenta. Therefore it would be very surprising if the angular distributions in SUSY and UED had an equal shape for all center-of-momentum energies.

The distributions are computed using the relations for the partonic cross sections in chapter~\ref{ch:2}. The angle $\theta_{CM,1}$ between the beam axis and the outgoing squark, respectively the KK-quark, is given in the center-of-momentum frame as defined in eq.~(\ref{eq:def_theta}). In the CMS, the outgoing gluino, respectively the KK-gluon, has an angle of $\theta_{CM,2}=180^{\circ}-\theta_{CM,1}$. Since the hadronic rapidity distributions include this information from the angular distributions on parton level, it is plausible that the rapidity distributions can differ on the hadronic level. The $u$- and $d$-quark are comparatively often included in the proton with a high momentum fraction. This yields a high center of momentum energy $s$, where angular distributions in SUSY and UED are somewhat different, which can lead to differences in the hadronic rapidity distributions.

As we will see in chapter~\ref{ch:6}, the difference in the hadronic differential cross sections of the rapidity for SUSY and UED yields a difference for the kinematics of the decay products of the squark, compared to the decay products of the KK-quark. 
  \clearpage{\pagestyle{empty}\cleardoublepage}

  \chapter{Decay Chains}
At the LHC one will not observe supersymmetric particles directly. Instead people are searching for their decay products. It is important to choose a signature that is as clear as possible, considering the background from Standard Model processes. Therefore decay chains play an important role in the detection of supersymmetric particles. To be certain that one observes SUSY at the LHC one has to measure the spins of the decaying particles. In a comparison of the kinematics of an extra-dimensional and a supersymmetric theory, differences in kinematic distributions should show up, due to different spin correlations in the applied models. In a theory with extra dimensions all ``extra-dimensional'' partners, the Kaluza-Klein particle towers, have opposite spin statistics compared to the supersymmetric partner particles. In our calculations we assume that all masses are already measured. Masses can be found by using kinematical distributions like two- and three-particle invariant mass distributions~\cite{lester},~\cite{hinchliffe-1999-60}.
In this chapter we discuss the kinematics of decay chains and explain the setup of our program.

\section{Decay Kinematics and Phase Space Generation}
If a particle with momentum $p$ decays into $n$ particles with momenta $p_1,\dots,p_{n}$, the phase-space element, described by $(3n-4)$ independent variables, is given by
\begin{equation}
d \Phi_{1 \rightarrow n}=\left[  \prod \limits ^{n}_{i=1}  \frac{d^{3}p_i}{2 E_{i}} \right] \delta^{(4)}(p-\sum\limits _{i=1}^{n}p_i) \; .
\end{equation}
Hence, the phase space for one particle decaying into two particles reads
\begin{eqnarray}
\int  d \Phi (p^2,m_1^2,m_2^2) &=& \int \frac{d^{3}p_1}{2E_{1}} \frac{d^{3}p_2}{2E_{2}} \delta ^{(4)}(p-p_1-p_2) \nonumber\\
&=&\frac{\lambda^{\frac{1}{2}}(p^2,m_1^2,m_2^2)}{8 p^2} \int \limits _{0}^{2 \pi} d\phi \int \limits_{-1}^{1} dcos{\theta} \;,
\label{eq:two-phase}
\end{eqnarray}
where $\lambda$ is given by eq.~(\ref{eq:lambda_kinematic}).
\begin{figure}[t!]
 \begin{center}
  \includegraphics[width=0.25\textwidth]{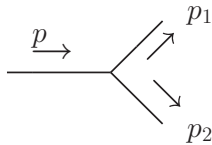}
 \end{center}
\caption{Our decay chains are built of $1 \rightarrow 2$ particle decays.}
\label{fig:1_2_Zerfall}
\end{figure}
The momenta of the outgoing particles in fig.~\ref{fig:1_2_Zerfall} are given in the rest frame of the decaying particle by 
\begin{equation}
\vert \vec{p}_1 \vert =\vert \vec{p}_2 \vert= \sqrt{\vec{p}_1^{\;2}}=\frac{\lambda^{\frac{1}{2}}(p^2,p_1^2,p_2^2)}{2 \sqrt{p^2}} \; .
\end{equation}
In order to find the momentum of the particles in the lab frame one has to boost them to the frame where the decaying particle has the momentum $p$. The usual Lorentz boost along the z axis, from the rest frame into a frame moving with $\beta_z$, is given by
\begin{equation}
p'_{1}= L(\gamma,\beta_z) \; p_{1} \; ,
\end{equation}
with $L(\gamma,\beta_z)$ given by
\begin{equation}
L(\gamma,\beta_z)=\begin{pmatrix} \gamma & 0 & 0 & -\gamma\beta_z \\ 0 & 1 & 0 & 0 \\ 0 & 0 & 1 & 0 \\ -\gamma \beta_z & 0 & 0 & \gamma \end{pmatrix}
\end{equation}
and 
\begin{equation}
\gamma=\frac{p_0}{m}\; \; , \qquad \gamma \beta_z=\frac{\vert \vec {p} \vert }{m}\; .
\end{equation}
For a chain of multiple decays we need a more general boost matrix into a frame moving with $\vec{\beta}$ in the direction $\frac{\vec{\beta}}{\vert \beta \vert}$, given in~\cite{Jackson} by
\begin{equation}
\label{eq:trafo_boost}
L(\gamma,\vec{\beta})=\begin{pmatrix} \gamma & -\gamma \beta_x & -\gamma \beta_y & -\gamma \beta_z \\[1.4mm]
 -\gamma \beta_x & 1+\frac{(\gamma -1)\beta_x^2}{\beta ^2} & \frac{(\gamma -1)\beta_x \beta_y}{\beta ^2} & \frac{(\gamma -1)\beta_x \beta_z}{\beta ^2} \\[1.4mm]
 -\gamma \beta_y & \frac{(\gamma -1)\beta_x \beta_y}{\beta ^2} & 1+\frac{(\gamma -1)\beta_y^2}{\beta ^2} & \frac{(\gamma -1)\beta_y \beta_z}{\beta ^2} \\[1.4mm]
 -\gamma \beta_z & \frac{(\gamma -1)\beta_x \beta_z}{\beta ^2} & \frac{(\gamma -1)\beta_y \beta_z}{\beta ^2} & 1+\frac{(\gamma -1)\beta_z^2}{\beta ^2} \end{pmatrix} \;.
\end{equation}
Our phase space is generated by calculating four vectors from transverse momentum and rapidity. Four vectors simply transform by eq.~(\ref{eq:trafo_boost}) under arbitrary boosts. Transverse momentum and rapidity also have a simple transformation behavior under Lorentz boosts, while a parametrization in angles and modulus of the momenta of the particles has a very complicated behavior under general Lorentz boosts.

Since a real cascade decay involves more than one decay, we need to extend the phase space element towards a particle decaying into $n$ particles, by joining numerous two-particle decays. A general discussion on that can be found in a chapter on multiparticle production in~\cite{Byckling}.

\begin{figure}[t!]
\begin{center}
	\includegraphics[width=0.7\textwidth]{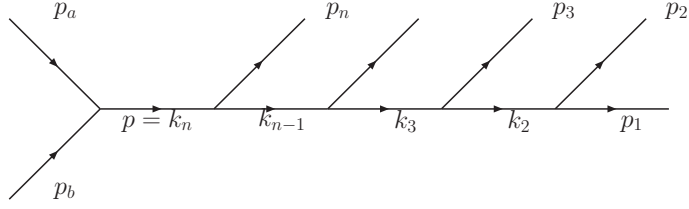}\\
\end{center}
\caption{The process $p_a+ p_b \rightarrow p_1+\dots+p_n$ as a sequence of two-particle decays. 
  \label{fig:Byckling6.2.3}}
\end{figure}

For the phase space element of a chain as shown in figure~\ref{fig:Byckling6.2.3}, one obtains the recursive relation
\begin{eqnarray}
\int d \Phi_{1 \rightarrow n} &=&  \int \int \frac{d^3p_n}{2 E_n} \prod \limits _{i=1}^{n-1} \frac{d^3p_i}{2 E_i} \delta^{(4)} \left\{ (p-p_n)- \sum \limits_{i=1}^{n-1} p_i \right\} \nonumber \\
&=& \int \frac{d^3p_n}{2 E_n} R_{n-1}(p-p_n) \, .
\label{eq:recurs}
\end{eqnarray}
When one inserts
\begin{eqnarray}
1&=& \int dM_{n-1}^2 \delta(M^2_{n-1}-k^2_{n-1})\\
1&=& \int d^4 k_{n-1} \delta^{(4)}(p-p_n-k_{n-1})
\end{eqnarray}
into eq.~(\ref{eq:recurs}) and uses the phase space for the two particle decay eq.~(\ref{eq:two-phase}), one obtains
\begin{eqnarray}
\int d \Phi_{1 \rightarrow n} &=&\int \limits _{\mu_{n-1}^2}^{(M_n-m_n)^2} dM_{n-1}^2 R_2(k_n^2,k_{n-1}^2,p_n^2) R_{n-1}(M_{n-1}^2)\\
&=& \int \limits _{\mu_{n-1}^2}^{(M_n-m_n)^2} dM_{n-1}^2 \int d\Omega_{n-1} \frac{\lambda^{\frac{1}{2}}(M_n^2,M_{n-1}^2,m_n^2)}{8 M_n^2}R_{n-1}(M_{n-1}^2) \; ,\nonumber
\end{eqnarray}
with the trivial limits
\begin{equation}
\mu_i =m_1+...+m_i \;\; ,   \quad M^2_{n-1}=(p-p_n)^2=(p_1+p_2+...+p_{n-1})^2 \equiv k_{n-1}^2 \; .\nonumber\\
\end{equation}
The decay is split into a two particle decay with the momenta $p_n$ and $k_{n-1}$ and the phase space for the decays of $k_{n-1}$ into the particles with momenta $p_1,\dots,p_{n-1}$. Integration over all invariant masses in the intermediate states has to be performed, since the particles do not necessarily need to be on the mass shell. The integration limits are trivial since the decay only takes place if $M_n>M_{n-1}+m_n$ and $M_{n-1}>\mu_{n-1}$.
For the case of $n=3$, the corresponding phase space element explicitly reads
\begin{equation}
\begin{aligned}
\int d \Phi_{1 \rightarrow 3} =&\int d\Phi(p^2,p_3^2,k_2^2) d\Phi(k_2^2,p_1^2,p_2^2) d k_2^2 \\
=&\int \limits _{\mu_2}^{(M_2-m_3)^2} \frac{\lambda^{\frac{1}{2}}(p^2,p_3^2,k_2^2)}{8 M_3^2} \frac{\lambda^{\frac{1}{2}}(k_2^2,p_1^2,p_2^2)}{8 M_2^2} d k^2_2 \\
 & \hphantom{abc} \int \limits _{0}^{2 \pi} d \phi_1 \int \limits _{-1}^{1} d cos \Theta_1 \int \limits _{0}^{2 \pi} d\phi_2 \int \limits _{-1}^{1} d cos \Theta_2 .
\end{aligned}
\end{equation}
In our application the decay chain follows the production of the decaying particle in a $2 \rightarrow 2$ scattering process. Therefore the decaying particle's momentum itself is not on-shell and also has to be integrated over. 

\section{Matrix Element}
\label{sec:HELAS}
The matrix element for a decay chain, in principle, is nothing special. But of course it is hard to square, since formulae are quite long even for a leading order calculation. Therefore we explain in this chapter how the matrix element is calculated numerically. We also describe what is done to avoid divergences from propagating particles.

In order to keep the perturbative error to a tolerable level, one takes some effects of higher orders of perturbation theory into account. One example is the running of the coupling $\alpha_s$ which was already mentioned. Since usual propagators of massive particles are divergent at the pole, one has to take into account the width of the particles propagating in the graph. Otherwise divergences in the s-channel diagrams occur.
Due to the perturbative expansion \\
  \begin{minipage}{0.85\textwidth}
      \vspace{3.0mm}
	\hspace{10mm}
      \includegraphics[width=\textwidth]{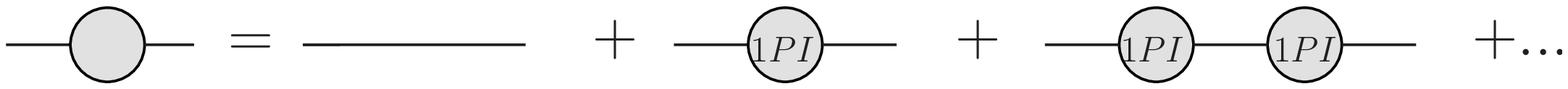}
  \end{minipage}
\vspace*{2mm}
\newline
one finds for the propagator
\begin{align}
\begin{minipage}{0.25\textwidth}
      \vspace{-0.6mm}
	\hspace{-3mm}
      \includegraphics[width=0.65\textwidth]{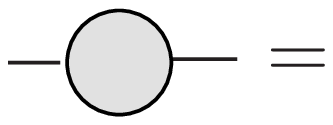}
  \end{minipage}
\hspace{-20pt}&\frac{i}{p^2-m_0^2}+\frac{i}{p^2-m_0^2}(-iM^2)\frac{i}{p^2-m_0^2}+ \dots \nonumber\\
=\hphantom{abcfh} &\frac{i}{p^2-m_0^2-M^2(p^2)} \; .
\end{align}
When the propagating particle is unstable, $M^2(p^2)$ acquires an imaginary part and can be written as 
\begin{align}
\begin{minipage}{0.25\textwidth}
      \vspace{-0.6mm}
	\hspace{-4mm}
      \includegraphics[width=0.45\textwidth]{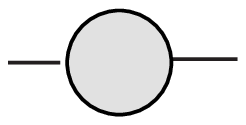}
  \end{minipage}
\hspace{-60pt}\approx \hspace{25pt} &\frac{i\,Z}{p^2-m_0^2-Re\,M^2(p^2)-i\,Z\,Im\,M^2(p^2)} \; ,
\end{align}
where Z is called the field renormalization constant. Therefore, close to the pole, i.e.\ when the particle is nearly on-shell, one finds the following dependence of the cross section
\begin{equation}
\sigma \varpropto \left \vert \frac{1}{p^2-m^2+i\,m\,\Gamma}\right \vert^2 \; ,
\label{eq:BW_prop}
\end{equation}
with $m^2=m_0^2 + Re \,M^2(m^2)$ and 
\begin{equation}
\Gamma= -\frac{Z}{m}\, Im \, M^2(m^2) \; .
\end{equation}
The propagator including the physical mass $m$ and the particle width $\Gamma$ is called Breit-Wigner propagator.
A more detailed discussion on how to treat divergences by regularization can be found in~\cite{Peskin}.

If we plot the value of eq.~(\ref{eq:BW_prop}) against the squared momentum of the particle, where the ratio of the particles mass $m$ divided by its width $\Gamma$ is small, one can see that the propagator strongly peaks at the pole. Therefore factorizing the matrix element at the Breit-Wigner propagator and forcing the propagating particle on-shell seems practicable to make computations easier. This approximation is also called ``Narrow Width Approximation`` (NWA). It corresponds to the multiplication of the cross section for on-shell production of the decaying particle, with the branching ratio of the decay. Integrating over eq.~(\ref{eq:BW_prop}), using the residue theorem, yields
\begin{equation}
\int_{\infty}^{-\infty} dk^2 \frac{f(k^2)}{\vert k^2 -m^2 +i \Gamma m \vert^2} \approx \frac{\pi}{m \Gamma} f(m^2)	
\label{eq:BW_approx}
\end{equation}
if $\Gamma \ll m$.
Since $\Gamma$ for the decay of a particle p into the final state f is defined as
\begin{equation}
\Gamma=\frac{1}{(2 \pi)^{3n-4}} \frac{1}{2m_1} \sum_f \int d \Phi_{1 \rightarrow n} \vert \mathcal{M}(p \rightarrow f) \vert^2 \; ,
\end{equation}
one finds for the example of a $1 \rightarrow 3$ process
\begin{eqnarray}
\label{eq:br}
\Gamma(k_3 \rightarrow p_1,p_2,p_3) = &\frac{1}{(2 \pi)^{2}} \frac{1}{2m_{k_3}} \sum_f \int d \Phi_{k_1 \rightarrow p_1,k_2} \vert \mathcal{M}(k_1 \rightarrow p_1,k_2) \vert^2 \nonumber \\
 \times & \hspace{-10pt} \frac{1}{(2 \pi)^{2}} \frac{1}{2m_{k_2}} \sum_f \int d \Phi_{k_2 \rightarrow p_2,p_3} \vert \mathcal{M}(k_2 \rightarrow p_2,p_3) \vert^2\\
= & \hspace{-30pt} \Gamma(k_3 \rightarrow p_1,k_2) \cdot Br(k_2 \rightarrow p_2,p_3)\; , \nonumber 
\end{eqnarray}
with momenta defined as in fig.~(\ref{fig:Byckling6.2.3}). 
Applying this NWA is problematic if the mass of the decay products is close to the mass of the decaying particle. This is due to threshold effects and was investigated in~\cite{kauer-2007-649} and~\cite{berdine-2007}.

In our calculations we do not assume that particles are totally on-shell. But making use of the fact that the matrix element far from the pole is small, we only have to integrate over a certain width around the pole. We will later give a numerical justification for this approximation.

This Breit-Wigner propagator only has to be taken into account in s-channels, where particles can be on-shell. Therefore the s-channel is effectively lowered in its order in the coupling constant $\alpha_s$ by division by $\Gamma$ from eq.~(\ref{eq:BW_approx}), since it is proportional to the squared coupling constant.
In decay cascades involving t- and u-channels, the squared momentum is far from the pole. Therefore the contribution of such diagrams is not dominated by the width of the Breit-Wigner propagator. In this sense these contributions to our decay can be regarded as a NLO contribution and can therefore be neglected in our LO calculation. This is discussed in detail in section~\ref{sec:Neglected Topologies in the Final Decay Chain}.

Since we do not want to square the matrix element of a cascade decay process by hand, we use the helicity amplitude formalism~\cite{1986NuPhB.2741H}. For our calculations we use the tool SMadgraph~\cite{cho-2006-73},~\cite{Alwall:hep-ph0706.2334}, which generates a Fortran code calling HELAS~\cite{Murayama:1992gi}, a helicity amplitude generator. Madgraph evaluates all possible helicity combinations for all topologies for the given external particles and produces a Fortran file with HELAS calls. HELAS calculates and squares the matrix element numerically,
using Dirac four-spinors given by
\begin{eqnarray}
u^{\lambda}(p)&=& \begin{pmatrix} \omega_{- \lambda}(p) \; \chi_{\lambda}(\vec{p})\\ \omega_{\lambda}(p) \; \chi_{\lambda}(\vec{p}) \end{pmatrix} \; , \nonumber\\
v^{\lambda}(p)&=& \begin{pmatrix} -\lambda \omega_{\lambda}(p) \; \chi_{-\lambda}(\vec{p}) \\ \lambda \omega_{- \lambda}(p) \; \chi_{-\lambda}(\vec{p}) \end{pmatrix}
\label{eq:hel_eigen}
\end{eqnarray}
with 
\begin{equation*}
\omega_{\mp}=\sqrt{E \mp \vert \vec{p} \vert} \quad.
\end{equation*}
Here the two component helicity eigenstates are given by
\begin{eqnarray}
\chi_+(\vec{p})&=& \frac{1}{\sqrt{2\vert \vec{p}\vert (\vert \vec{p} \vert +p_z)}} \begin{pmatrix} \vert \vec{p} \vert +p_z \\ p_x + i p_y   \end{pmatrix} \; ,\nonumber\\
\chi_-(\vec{p})&=& \frac{1}{\sqrt{2\vert \vec{p}\vert (\vert \vec{p} \vert +p_z)}} \begin{pmatrix} -p_x + i p_y   \\ \vert \vec{p} \vert +p_z\end{pmatrix}
\end{eqnarray}
for $\vert \vec{p} \vert \neq -p_z$. For $\vert \vec{p} \vert = -p_z$ one uses 
\begin{eqnarray}
\chi_+(\vec{p})&=& \begin{pmatrix} 0 \\ 1   \end{pmatrix} \; ,\nonumber\\
\chi_-(\vec{p})&=& \begin{pmatrix} -1   \\ 0 \end{pmatrix} \; .
\end{eqnarray}
These helicity eigenstates satisfy
\begin{equation}
\frac{\vec{\sigma} \cdot \vec{p}}{\vert \vec{p} \vert} \chi_{\lambda}(\vec{p})=\lambda \; \chi_{\lambda}(\vec{p})
\end{equation}
with $\lambda =\pm 1$.

\begin{figure}
\centering
      \vspace{0.0mm}
	\hspace{10mm}
      \includegraphics[width=0.55\textwidth]{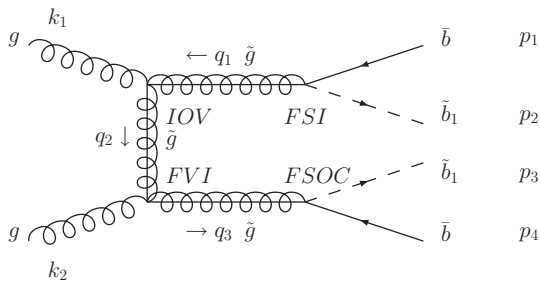}
\caption{One of the diagrams contributing to the process $ g g \rightarrow \bar{b} \tilde{b}_1 \bar{b} \tilde{b}_1 $}
\label{fig:2-4-bsp}
\end{figure}

Checking the calculations of Madgraph and HELAS is easy since one can have a direct look at all vertices. HELAS performs calculations in a model independent way. Since the structure of a vertex is given by the spin and polarization properties of the particles coupling to it, HELAS provides general vertices for scalars, vectors and fermions. We want to give a short example of a helicity amplitude of a $2 \rightarrow 4$ matrix element. In fig.~\ref{fig:2-4-bsp} one of the diagrams for the gluon-gluon to sbottom-anti-bottom, sbottom-anti-bottom process is presented, using the HELAS abbreviations for the vertices, which can be found in~\cite{Murayama:1992gi}. The helicity matrix element is given by
\begin{align}
\mathcal{M^{\lambda,\lambda'}}= \hphantom{a}&\bar{v}^{\lambda'}_{\bar{b}}(p_4) \left(i g_1 \frac{1+\gamma_5}{2}+i g_2 \frac{1-\gamma_5}{2} \right) \frac{i (\slashed{q}_3+m)}{q_3^2-m_{\tilde{g}}^2+i m_{\tilde{g}} \Gamma_{\tilde{g}}} \slashed{\epsilon}(k_2) \nonumber \\
 &\times\left(i g_1 \frac{1-\gamma_5}{2}+i g_2 \frac{1+\gamma_5}{2} \right) \frac{i (\slashed{q}_2+m)}{q_2^2-m_{\tilde{g}}^2+i m_{\tilde{g}} \Gamma_{\tilde{g}}} \left(i g_1 \frac{1-\gamma_5}{2}+i g_2 \frac{1+\gamma_5}{2} \right) \nonumber\\
 & \times \frac{i (\slashed{q}_1+m)}{q_1^2-m_{\tilde{g}}^2+i m_{\tilde{g}} \Gamma_{\tilde{g}}} \slashed{\epsilon}(k_1) \left(i g_1 \frac{1-\gamma_5}{2}+i g_2 \frac{1+\gamma_5}{2} \right) v^{\lambda}_{b}(p_1)  \; .
\end{align}
Here $\lambda$ and $\lambda '$ denote helicity eigenstates, according to eq.~(\ref{eq:hel_eigen}). The matrix element is evaluated numerically for every possible combination of helicities and polarizations of the external particles.

In order to calculate the cross sections we need all couplings and masses. In the formula above, the coupling $g_1$ is the coupling to the chirality left fermions while the coupling $g_2$ belongs to the chirality right fermions. They are simply derived from the strong coupling constants as given by the Feynman rules.

As we mentioned earlier, we assume the masses of the SUSY particles to be already known. In our calculations we use SDECAY~\cite{muhlleitner-2005-168} which calculates branching ratios for the supersymmetric particles, making use of the spectrum generator SuSpect~\cite{djouadi-2007-176}. SDECAY puts masses, mixing angles and branching ratios in a SLHA format file~\cite{skands-2004-0407}, which is read into our program and used by our matrix element and phase space generator. In SuSpect a phenomenological MSSM with 22 free parameters is implemented. When parameters are set by the user, renormalization group running is performed, going iteratively up to the GUT scale and down to the electroweak scale, taking into account threshold effects of all particles. All masses and couplings are then derived for the electroweak scale, also including higher order effects.

Our whole program is build up step by step on the basis of the first version of the simplest $2 \rightarrow 2$ scattering cross section program. But since the number of degrees of freedom $(3n-4)$ increases with the number of the outgoing particles $n$, the integration gets more complicated.
Therefore it is mandatory to introduce mappings, forcing the integration routine to evaluate more phase space points in the region where the value of the integrand is large.
The integration variable $x$ is mapped to a set of random numbers $y$ by 
\begin{equation}
x=h(y), \qquad 0 \leqslant y \leqslant1 \; .
\end{equation}
The integral can then be calculated as
\begin{equation}
\label{eq:mapping}
I=\int f(x) dx=\int_0^1 f(h(y)) \frac{\partial h(y)}{\partial y} dy = \int_0^1 \frac{f(h(y))}{g(h(y))}dy \; ,
\end{equation}
where $g(h(y))$ is called the density.
The integral is then calculated by sampling the integral $N$ times and averaging as
\begin{equation}
\bar{I}=\frac{1}{N} \sum_{i=1}^N \frac{f(h(y_i))}{g(h(y_i))}  \; .
\end{equation}
By mapping the integration variables $x$ in a way that $\frac{f(h(y))}{g(h(y))}$ is smoother than $f(x)$, convergence of the integral is improved and more points are evaluated in those regions where $f(x)$ is steep.

The most important mapping in our program is the one for the Breit-Wigner propagator $\frac{1}{x-m^2+im\Gamma}$ , as given in~\cite{Byckling} by
\begin{equation}
h(y,m^2-i m \Gamma,x_{\text{min}},x_{\text{max}})=m \Gamma \; \text{tan}\left[z_1+(z_2-z_1)y\right]+m^2 \; .
\end{equation}
This results in a density given by
\begin{equation}
g(y,m^2-i m \Gamma,x_{\text{min}},x_{\text{max}})=\frac{m \Gamma}{(z_2-z_1)\left[(x-m^2)^2+m^2 \Gamma^2 \right]}
\end{equation}
with 
\begin{equation}
z_{1/2}=\text{arctan} \left(\frac{x_{\text{min}/\text{max}}-m^2}{m \Gamma} \right) \; .
\end{equation}
Leaving out this mapping results in a very inefficient and slowly converging integration.
The function we integrate over, the squared Breit-Wigner propagator, is given by 
\begin{equation}
f(p)=\frac{1}{(p^2-m^2)^2+m^2 \Gamma^2} \; .
\end{equation}
As we will see in section~\ref{sec:Testing_chains}, our limits for the momentum of the propagating particle are determined by its width
\begin{equation}
\label{eq:on_shell_quasi}
 -n \cdot \Gamma + m \; \leqslant \;  p \; \leqslant m+n \cdot \Gamma \; ,
\end{equation}
which yields 
\begin{equation}
z_{1/2}=\text{arctan} \left(\frac{(m\pm n \Gamma)^2-m^2}{m \Gamma} \right) \; .
\end{equation}
Calculating the integrand in eq.~(\ref{eq:mapping}) one effectively finds a constant function
\begin{equation}
\frac{f(h(y))}{g(h(y))}=\frac{z_2-z_1}{m \Gamma}\; ,
\end{equation}
which has to be integrated from zero to one. This constant function can be integrated easily while the squared Breit-Wigner propagator is strongly peaked at $p^2=m^2$.

Other mappings, e.g.\ a mapping for the integration over $p_t$ using $p_t=\frac{1}{y}$, are also included. Angles are mapped as
\begin{equation}
\text{cos} \theta =2y-1  \qquad \text{and} \qquad \phi=2 \pi y \; .
\end{equation}
Since our Monte Carlo routine always integrates from 0 to 1, Jacobians have to be included for different integration limits.

\section{Testing of a Decay Chain Program}
\label{sec:Testing_chains}

In the following we explain the testing of our phase space generator. Beginning with a usual parametrization of a $2 \rightarrow 2 $ process in terms of $y$ and $p_t$ and integrations over $x_1$ and $x_2$, we extend the phase space by attaching a $1 \rightarrow 2$ decay to one of the final state particles. Then the squared momentum of this decaying particle has to be integrated over. We make use of the fact that all intermediate state particles have a small width compared to their mass, when we discuss the emerging topologies.
For a three particle final state there are three different topologies:\\
\begin{align*}
  & a)\begin{minipage}{0.35\textwidth}
      \vspace{0.0mm}
	\hspace{10mm}
      \includegraphics[width=1.00\textwidth]{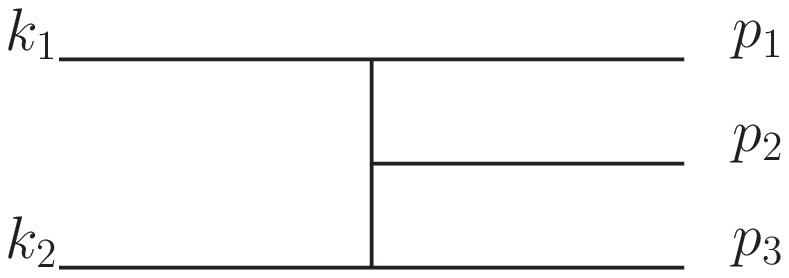}
  \end{minipage} \qquad
b) \begin{minipage}{0.35\textwidth}
      \vspace{0.0mm}
	\hspace{+10mm}
     \includegraphics[width=1.00\textwidth]{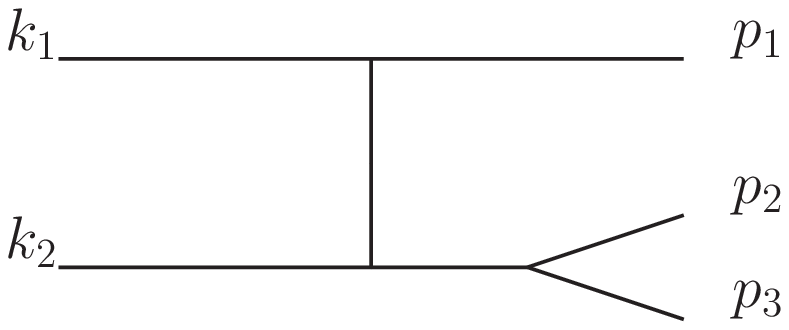}
  \end{minipage}
\\
&c)\begin{minipage}{0.35\textwidth}
      \vspace{0.0mm}
	\hspace{10mm}
      \includegraphics[width=1.00\textwidth]{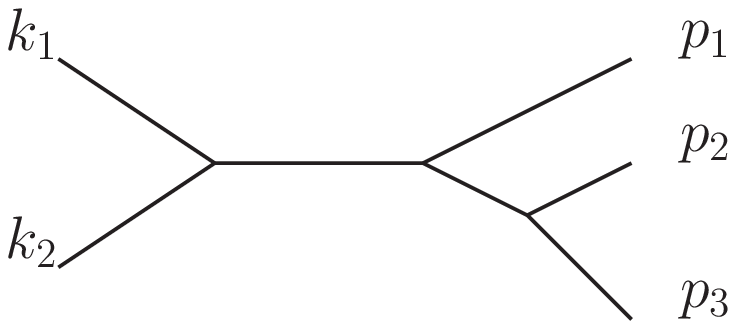}
  \end{minipage}
\end{align*}
In our calculations we take into account the topologies of b) and c). Since the momentum in the t-channel is far from the pole, the topology a) is not enhanced by the Breit-Wigner and therefore suppressed, compared to the other topologies.
The following topologies can be found for four particle final states:\\
\begin{align*}
&a) \begin{minipage}{0.35\textwidth}
      \vspace{0.0mm}
	\hspace{10mm}
      \includegraphics[width=1.00\textwidth]{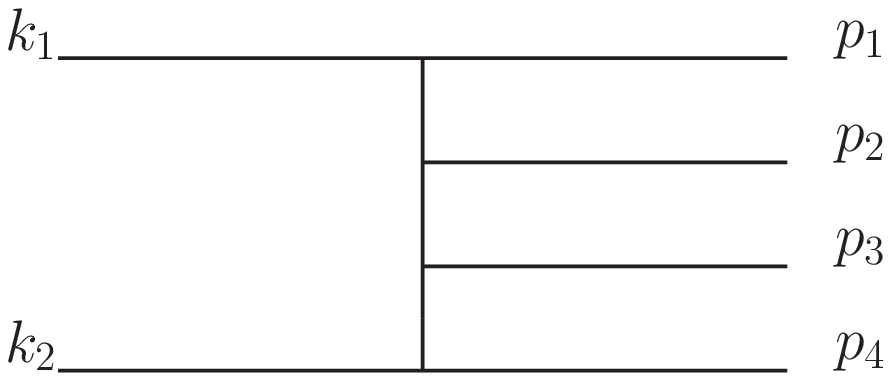}
  \end{minipage}\qquad
b) \begin{minipage}{0.35\textwidth}
      \vspace{0.0mm}
	\hspace{10mm}
      \includegraphics[width=1.00\textwidth]{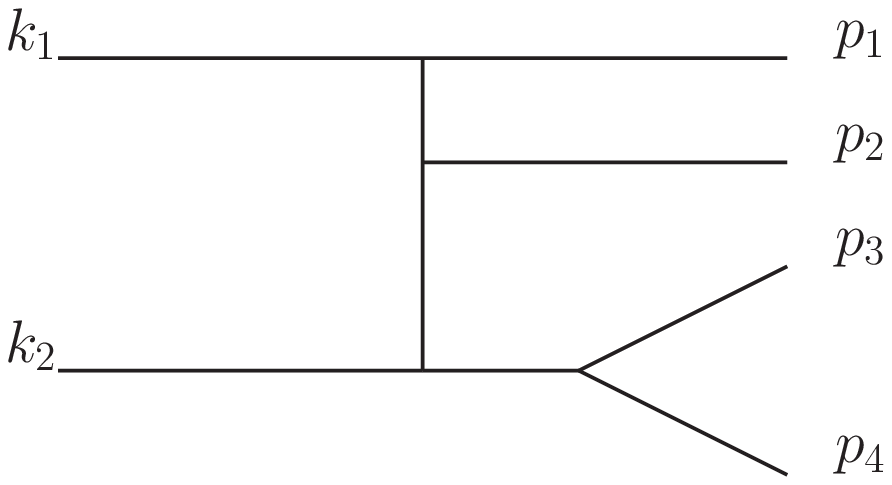}
  \end{minipage}
\\
&c) \begin{minipage}{0.35\textwidth}
      \vspace{0.0mm}
	\hspace{10mm}
      \includegraphics[width=1.00\textwidth]{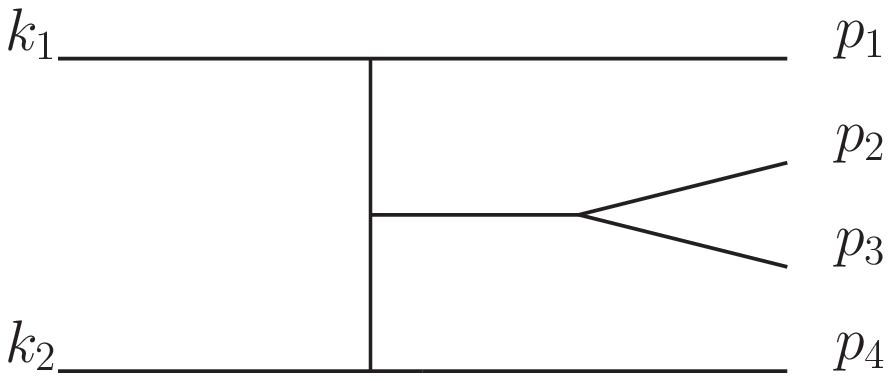}
  \end{minipage}\qquad
d) \begin{minipage}{0.35\textwidth}
      \vspace{0.0mm}
	\hspace{10mm}
      \includegraphics[width=1.00\textwidth]{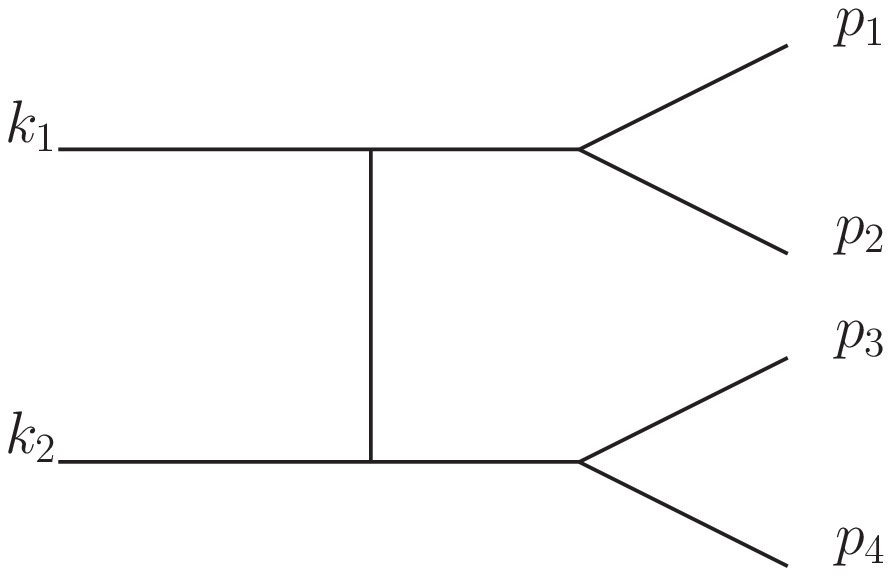}
  \end{minipage}
\\
& e) \begin{minipage}{0.35\textwidth}
      \vspace{0.0mm}
	\hspace{10mm}
      \includegraphics[width=1.00\textwidth]{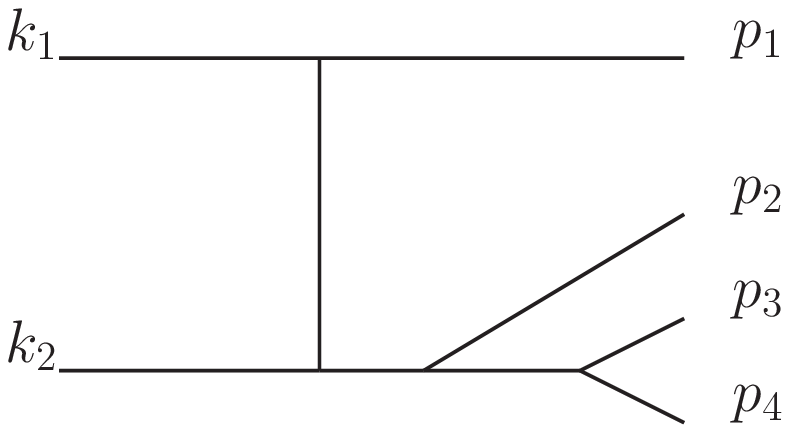}
  \end{minipage}\qquad
f) \begin{minipage}{0.35\textwidth}
      \vspace{0.0mm}
	\hspace{10mm}
      \includegraphics[width=1.00\textwidth]{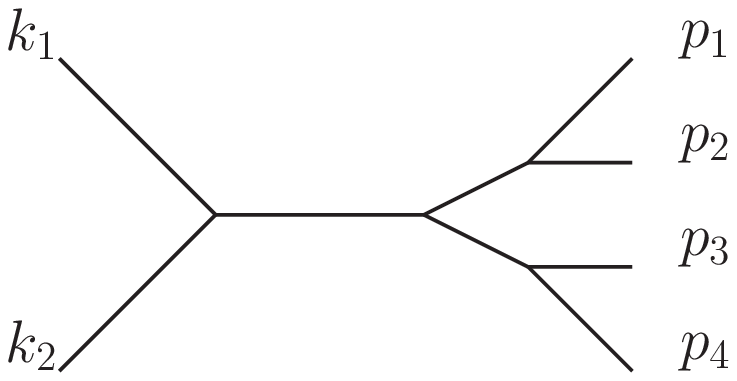}
  \end{minipage}\\
&g) \begin{minipage}{0.35\textwidth}
      \vspace{0.0mm}
	\hspace{10mm}
      \includegraphics[width=1.00\textwidth]{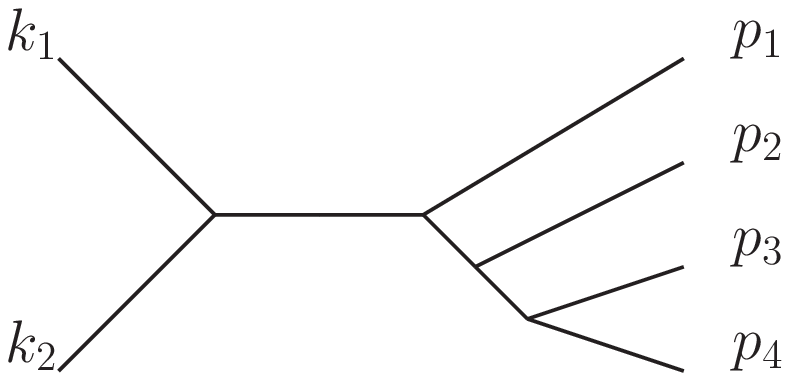}
  \end{minipage}
\end{align*}
The topologies a), b) and c) are again neglected since their momenta are far from the poles. In general, all other topologies are taken into account. Depending on the mass hierachy of the intermediate particles, it is possible that topology g) is also suppressed. This happens when an intermediate quasi-on-shell particle, i.e.\ a particle with a momentum within the limits of eq.~(\ref{eq:on_shell_quasi}), is followed in the decay cascade by a heavier quasi-on-shell particle. It is not possible for both particles to be close to their mass shell at the same time. 

The topologies of our special $2 \rightarrow 5$ process will be investigated in chapter~\ref{ch:6}.
Leaving out all effective NLO topologies, we are left with s-channel decays and can compare, e.g.\ the $2 \rightarrow 5$ process to the $2 \rightarrow 4$ process multiplied by the branching ratio of the last decaying particle, because the decaying particle is nearly on-shell.
To give a feeling for the contribution of the on-shell momenta, we give some total cross sections in table~\ref{tab:cross_check}.
Calculations were performed for the $2 \rightarrow 2$ processes 
\begin{equation}
u \quad g \rightarrow \tilde{u} \quad \tilde{g} \qquad \text{and} \qquad g \quad g \rightarrow \tilde{g} \quad \tilde{g} \; .
\label{eq:proc}
\end{equation}
To obtain a $2 \rightarrow 5$ process, we attach step by step the decays
\begin{equation}
(1)\; \tilde{g} \rightarrow b, \tilde{b_i} \qquad (2)\;  \tilde{b} \rightarrow b, N_2 \qquad (3)\;  \tilde{u} \rightarrow u, N_1
\end{equation}
to the first process in eq.~(\ref{eq:proc}).
This was done for an intermediate $\tilde{u}_L$ and $\tilde{u}_R$. We also repeated the procedure for the second process, given in eq.~(\ref{eq:proc}). Since the masses in this process are different, it gives us a second check for the generated phase space. 

\begin{figure}
\begin{align*}
  \begin{minipage}{0.45\textwidth}
      \vspace{-0.6mm}
    \hspace*{-20pt} \includegraphics[width=\textwidth]{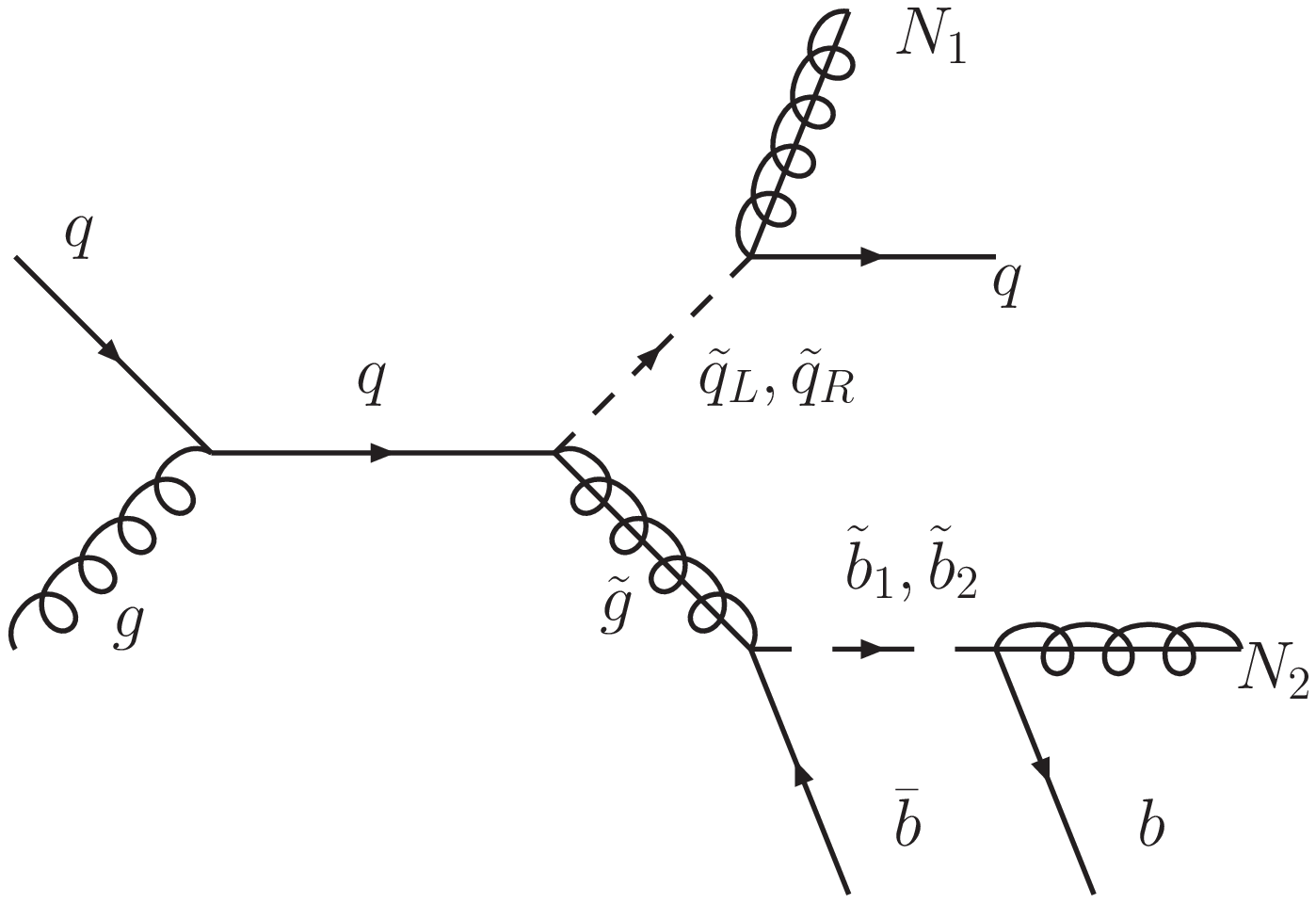}
  \end{minipage}
 &\begin{minipage}{0.45\textwidth}
      \vspace{-0.6mm}
  \hspace*{20pt}  \includegraphics[width=\textwidth]{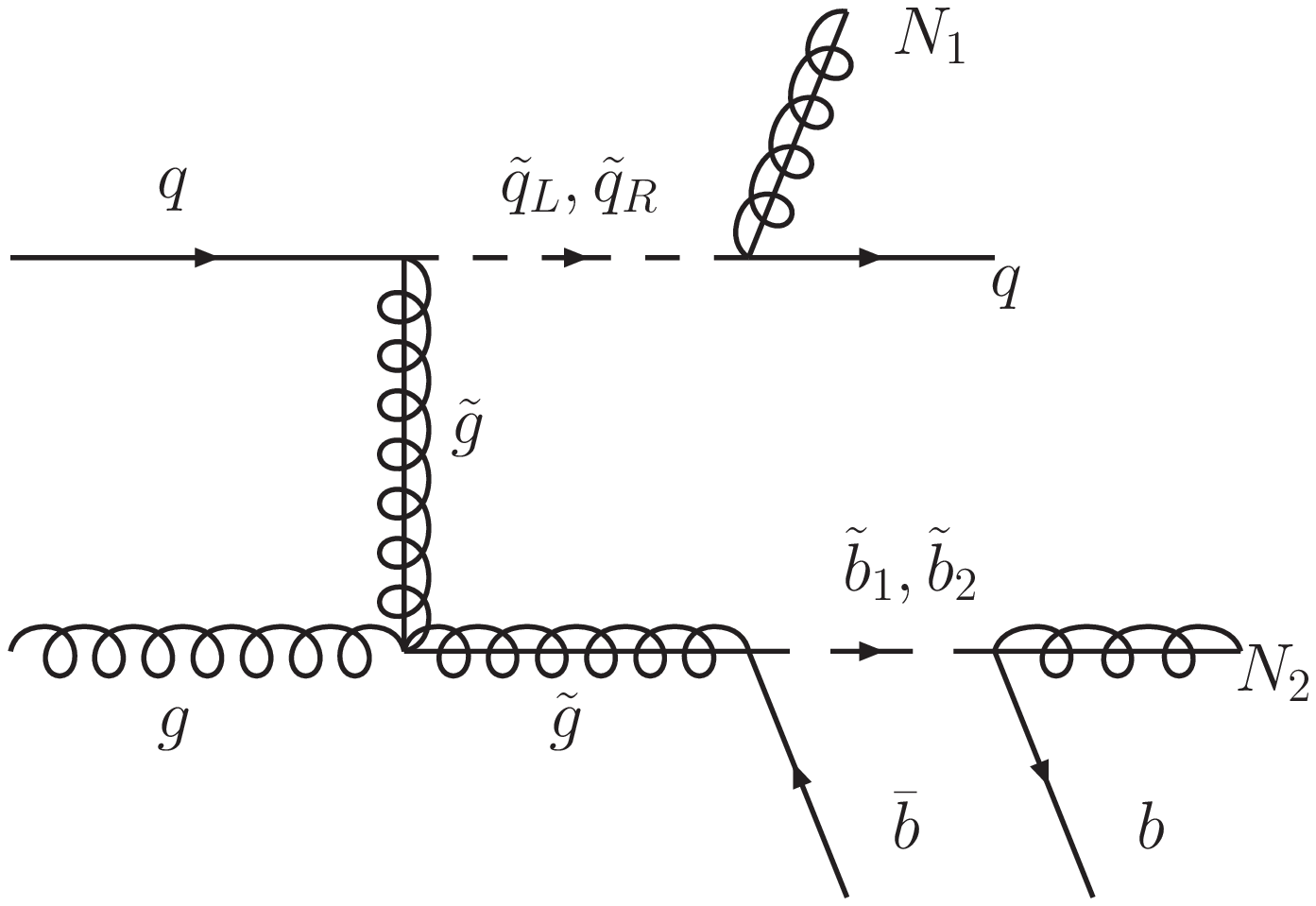} 
\end{minipage}\\
\vspace*{25pt}
  \begin{minipage}{0.45\textwidth}
	\vspace*{20pt}
      \vspace{-0.6mm}
    \hspace*{-20pt} \includegraphics[width=\textwidth]{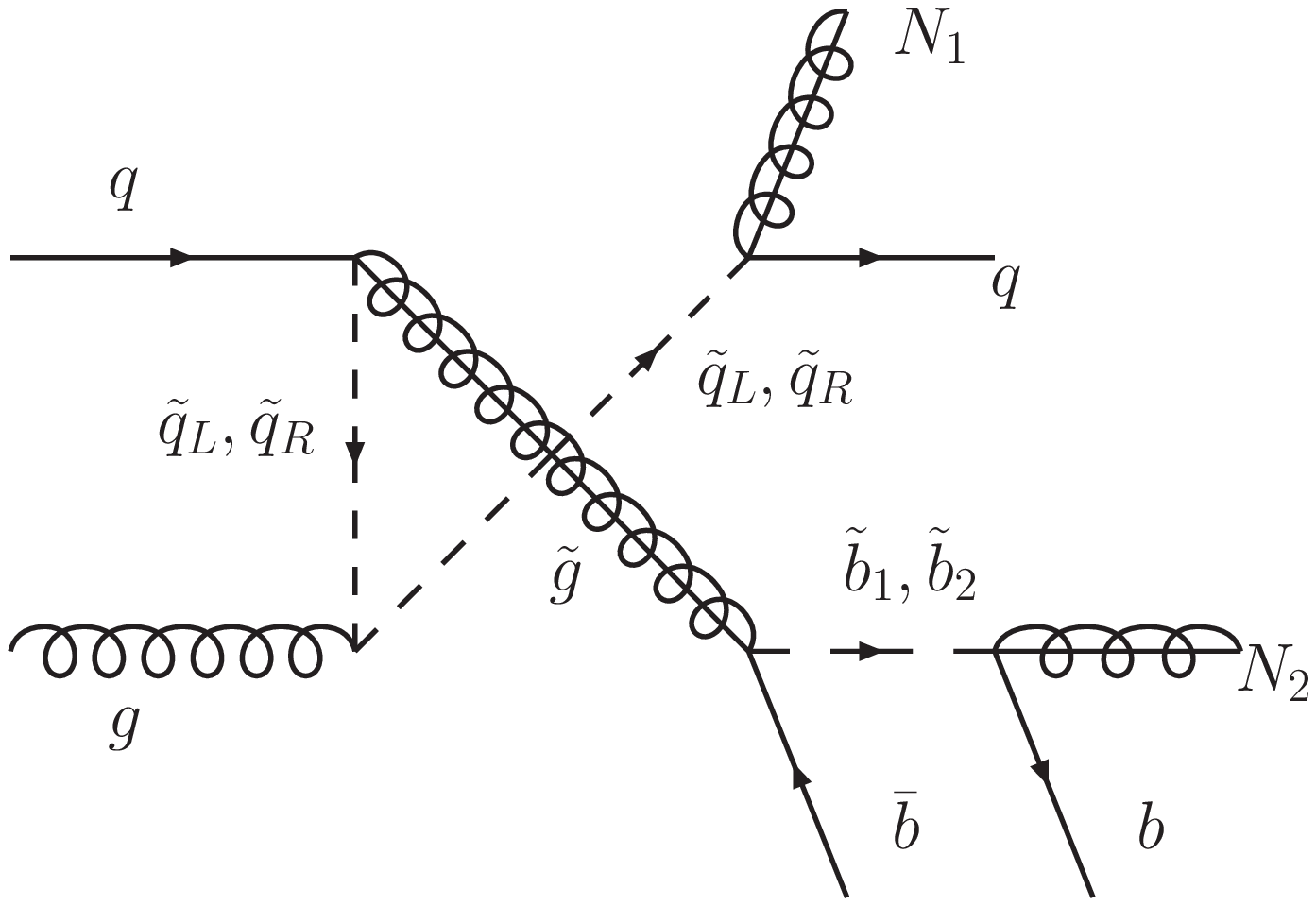}
  \end{minipage}
\end{align*}
\caption{Diagrams contributing in LO to the $2 \rightarrow 5$ process.}
\label{fig:cascade_diagrams}
\end{figure}

The columns in table~\ref{tab:cross_check} represent three different integration limits for the intermediate momenta. Due to the Breit-Wigner propagator, contributions are only large at the pole and therefore we integrate an integer multiple of widths, here denoted $n \cdot \Gamma$, around the pole, i.e.\ our explicit integration limits are given by
\begin{equation}
\label{n_parameter}
 (-n \cdot \Gamma + m)^2 \; \leqslant \;  p^2 \; \leqslant (m+n \cdot \Gamma)^2 \; .
\end{equation}
The diagrams contributing to the cross sections are shown in fig.~\ref{fig:cascade_diagrams}. Though other diagrams and more complicated topologies in principle exist, their couplings were commented out in Madgraph in order to make a comparison with the branching ratio possible. Row one to three in table~\ref{tab:cross_check} correspond to intermediate $\tilde{b}_1$ and $\tilde{u}_L$ while row four to six correspond to intermediate $\tilde{b}_1$ and $\tilde{u}_R$. If there are more constellations for the final state, e.g.\ the commutation of the $b$ and $\bar{b}$ jets, we left them out since we just want to understand the behavior of our phase space here. We only included the $\tilde{b}_1$ squark and left out $\tilde{b}_2$ in the intermediate state. Therefore these calculations are by no means measurable results, though we could easily include all particles.

The masses for this arbitrarily chosen scenario are given by $m_{\tilde{g}}=746.01\; \text{GeV}$, $m_{\tilde{u}_L}=692.72\; \text{GeV}$, $m_{\tilde{b}_1}=625.12 \;\text{GeV} $, $m_{N_1}=93.83 \;\text{GeV}$, $m_{N_2}=191.15 \;\text{GeV}$. For the strong coupling we used $\alpha_S=0.1003750$. The used particle widths are given by $\Gamma_{\tilde{g}}=8.9834\; \text{GeV}$, $\Gamma_{\tilde{u}_L}=7.7452\; \text{GeV}$ and  $\Gamma_{\tilde{b}_1}=6.3995 \;\text{GeV}$. The branching ratios corresponding to this spectrum are calculated with SDECAY and are given below.
\begin{table}
\begin{center}
	\includegraphics[width=0.9\textwidth]{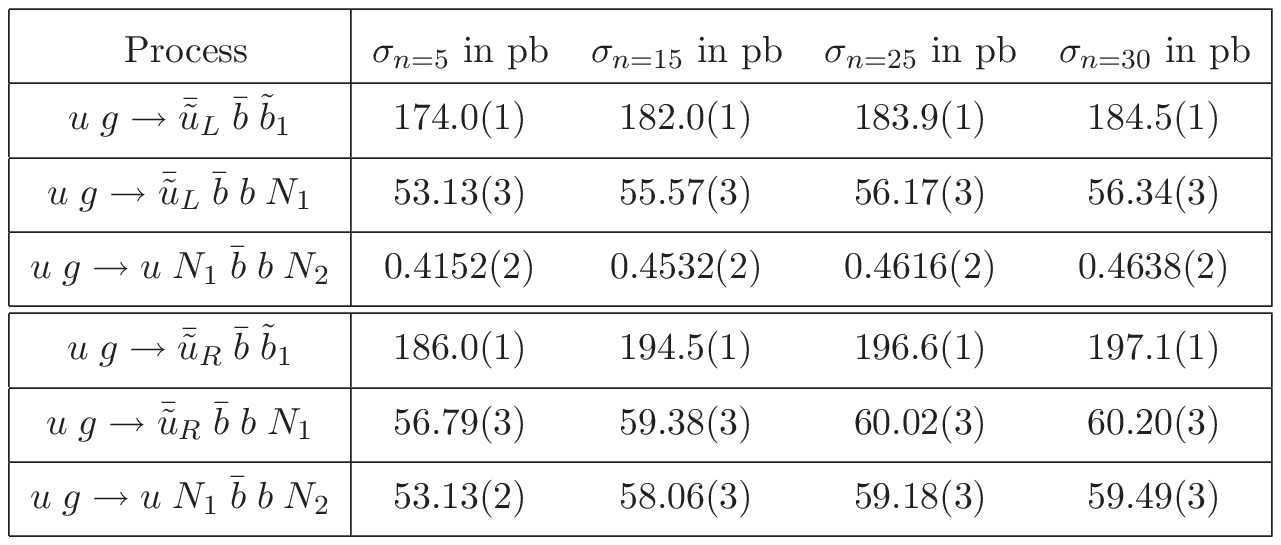}\\
\end{center}
\vspace*{0mm}
\caption{Cross sections for an exemplary calculation.}
\label{tab:cross_check}
\end{table}
\clearpage
We again stress that our result does not take into account off-shell contributions in general, since beyond the $2 \rightarrow 2$ process we only consider s-channel decay topologies, while we still allow some deviation of the squared momentum from the pole. In a region not too far from the pole, as given by eq.~(\ref{n_parameter}), contributions to the total cross section can be quite significant. From table~\ref{tab:cross_check} we find, that the main contributions to the process come from intermediate on-shell momenta. In our later calculations we integrate out a region with $n=25$, since then only small further changes in the cross sections appear. We also find that our integration is equally stable for all cases. Integration errors are given in brackets for the last digit.
%

For checking the phase space, we use the narrow-width-approximation. For the case of $n=5$ and $\tilde{u}_L$, taking the branching ratios from SDECAY, we multiply the $2 \rightarrow 3$ result by 
\begin{equation}
BR(\tilde{b}_1 \rightarrow N_2\; b)=0.305532
 \end{equation}
to obtain $ 53.17 \;\text{pb}$, which is close to the calculated value of $ 53.13\;\text{pb}$. For the decay of the $\tilde{u}_L$ squark we multiply by
\begin{equation}
 BR(\tilde{u}_L \rightarrow N_2\; b)=0.0082336 
\end{equation}
and obtain $0.4375 \; \text{pb}$, compared to the calculated cross section of $ 0.4152\;\text{pb}$. 

For the case of $n=25$ and $\tilde{u}_R$, we multiply the $2 \rightarrow 3$ result by $BR(\tilde{b}_1 \rightarrow N_2\; b)$ to obtain $60.06 \;\text{pb}$, which is close to the calculated value of $ 60.02\;\text{pb}$. For the decay of the $\tilde{u}_R$ squark we multiply by  \begin{equation}
BR(\tilde{u}_R \rightarrow N_2\; b)=0.995188
\end{equation}
and obtain $59.73 \; \text{pb}$, compared to the calculated cross section of $ 59.18\;\text{pb}$. This approximation is quite good for larger and smaller $n$. Obviously $\tilde{u}_R$ couples much stronger to the neutralino $N_2$, which is due to the mixing angles of neutralinos and squarks in the given scenario.

As a last useful check for the extensions of our program, we used various plots to qualitatively understand the behavior of the program and check for consistency. The shape of transverse momentum and rapidity distributions as well as invariant mass distributions have to stay nearly unchanged for the intermediate particles when the phase space is enlarged by an additional $1 \rightarrow 2$ process, since only small off-shell influences are included. This was tested and various distributions will be presented in the next chapter. All these agreements confirm that our phase space generator works correctly. Since for the case of UED calculations we only have to exchange the matrix element, phase space checks were only performed for SUSY calculations.
%
%
  \clearpage{\pagestyle{empty}\cleardoublepage}

  \chapter{A SUSY-UED Decay Chain Comparison}
\label{ch:6}
Finding like-sign dilepton signatures is a promising strategy to confirm new physics, as explained in~\cite{barnett}. Like-sign signatures are expected to emerge from processes involving Majorana fermions, e.g.\ the gluino in SUSY. Processes like $q \bar{q}/ gg \rightarrow \tilde{g} \tilde{g}$ and their decays are candidates for this discovery. The Majorana gluinos decay to $q \bar{\tilde{q}}$ or $\bar{q} \tilde{q}$, leaving like-sign leptons in the final state. But the idea of using like-sign dileptons to claim the existence of a heavy Majorana fermion is incomplete. If the particle decaying in the gluino-like cascade is a boson in the adjoint representation, e.g.\ a KK-gluon in UED, like-sign signatures also arise.

To solve this problem and show that SUSY is the theory realized by nature, one has to show that the decaying particle is really a fermion. This can be done by comparing two scenarios of different spin assignments for the gluino-like particle. UED is a popular candidate to compare to SUSY. It has exactly the same spin assignments as the Standard Model for all partner particles. Since the UED mass spectrum is \textit{typically} very different from a SUSY spectrum, there are many ways to distinguish between UED and SUSY. The mass spectrum itself, the discovery of higher KK-excitations, ratios of branching fractions, threshold behavior and cross sections can be expected to be very different for both theories. 
Since we want to use UED as a toy model with different spin assignments for verifying SUSY, we are not interested in the total cross section of a UED process. As usual we extract spin information from angular correlations. Therefore we divide all distributions by the total cross section and exclude influences from different masses and coupling constants. An entanglement of the spin information with the couplings of the left and right handed sfermions in our decay chain can not be excluded.

We show in this chapter that the spin information can be extracted from kinematic distributions. Differences in the boost distributions and angular correlations are studied. We also investigate the origin of the differences in angular correlations.

Various recent publications deal with the issue of comparing new-physics scenarios to find the spin of new particles emerging at the LHC. Knowing spin and masses is crucial for knowing the Lagrangian. While masses can be extracted from various edges and thresholds of invariant mass distributions~\cite{lester},~\cite{hinchliffe-1999-60}, the measurement of the spin is non-trivial. 

The idea first came up in~\cite{barr-2004-596}, where it is demonstrated that spin information can be extracted from angular distributions and invariant masses. This idea is used in~\cite{smillie-2005-0510} to perform a comparison of UED and SUSY decay chains to determine the spin of the squark. Analytical results for the invariant mass in the decay of a squark are given. The leading order UED-QCD cross sections are also presented in this paper.

Due to the like-sign dilepton argument the spin measurement of the gluino is especially interesting. Therefore LHC decay chains involving a gluino and KK-gluon are analyzed in~\cite{alves-2006-74}, in addition to our decay chain also including the further decay of the NLSP into the LSP and two leptons. For the processes of gluon-gluon and gluon-quark collisions various asymmetries of outgoing leptons and bottom jets are studied. It is shown that by using these asymmetries the spin of the gluino can be determined. Significant differences in the angle between two outgoing bottom jets are found. It is claimed that this difference is mainly due to the different boost of the gluino and the KK-gluon.

In a recent publication~\cite{csaki-2007} the question arises if the boost of the gluino-like particle or the different helicity structure of the couplings is responsible for the kinematics in the final state of a decay chain. This is studied for the three-particle decay $\tilde{g} \rightarrow \bar{b} \; b \; N_1$. It is argued that none of both influences can be excluded from the calculations. Even in the case that masses and spins are treated as independent parameters which are used to fit one model to the other, it is shown that SUSY and UED can always be differentiated by invariant mass distributions of the decay products. It is explained that the longitudinal modes of the neutralino, if highly boosted, are mainly responsible for the characteristic differences. Calculations are performed assuming $m_{\tilde{g}}>m_{\tilde{q}}$ for all squarks.
\begin{figure}
\begin{align*}
  \begin{minipage}{0.5\textwidth}
      \vspace{-0.6mm}
    \hspace*{-20pt} \includegraphics[width=\textwidth]{figures_chains/chain_5_1.eps}
  \end{minipage}
 &\begin{minipage}{0.5\textwidth}
      \vspace{-0.6mm}
  \hspace*{20pt}  \includegraphics[width=\textwidth]{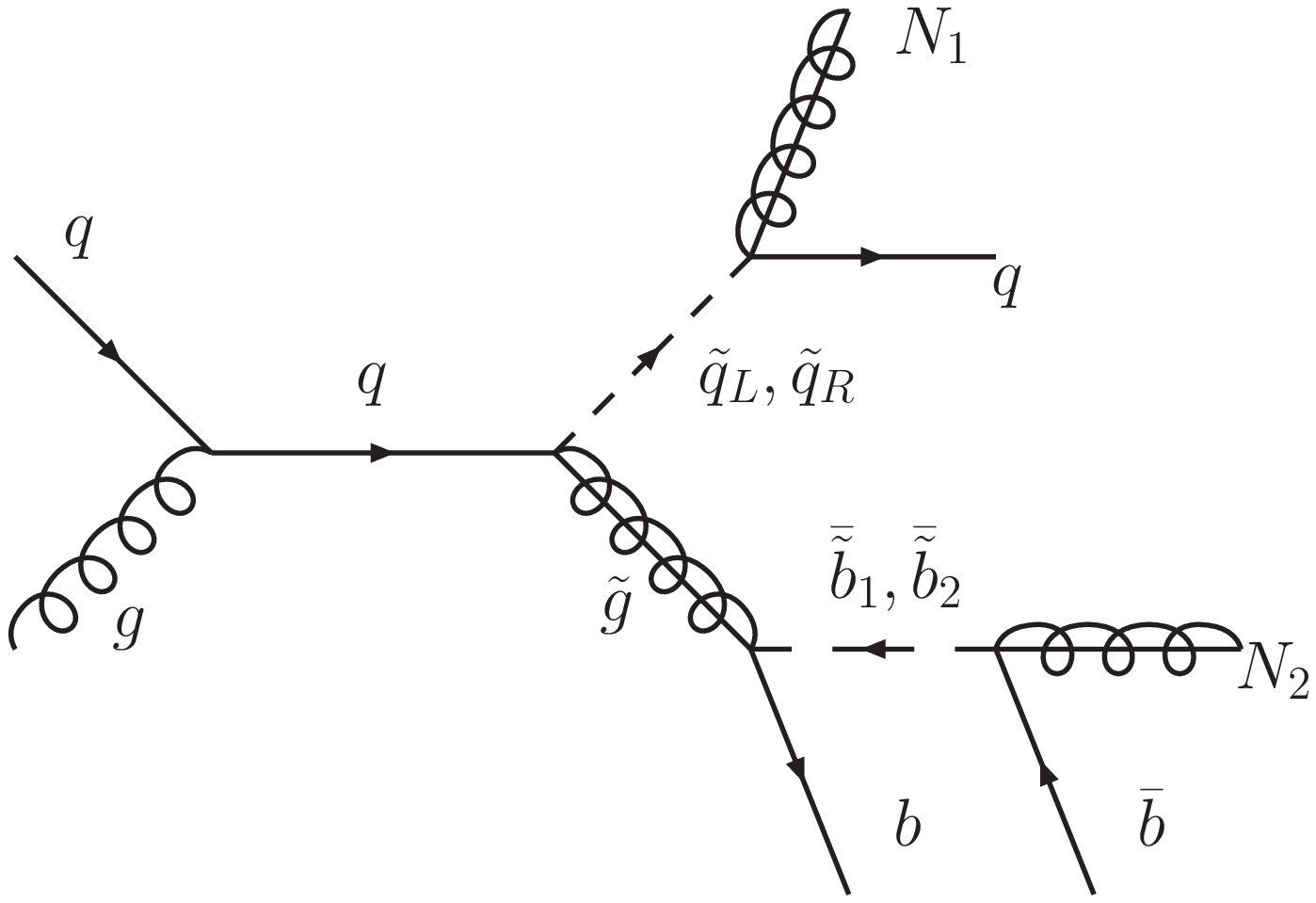} 
\end{minipage}\\
  \begin{minipage}{0.5\textwidth}
	\vspace*{20pt}
      \vspace{-0.6mm}
    \hspace*{-20pt} \includegraphics[width=\textwidth]{figures_chains/chain_5_2.eps}
  \end{minipage}
 &\begin{minipage}{0.5\textwidth}
      \vspace{2.5mm}
  \vspace{2.5mm} \hspace*{20pt}  \includegraphics[width=\textwidth]{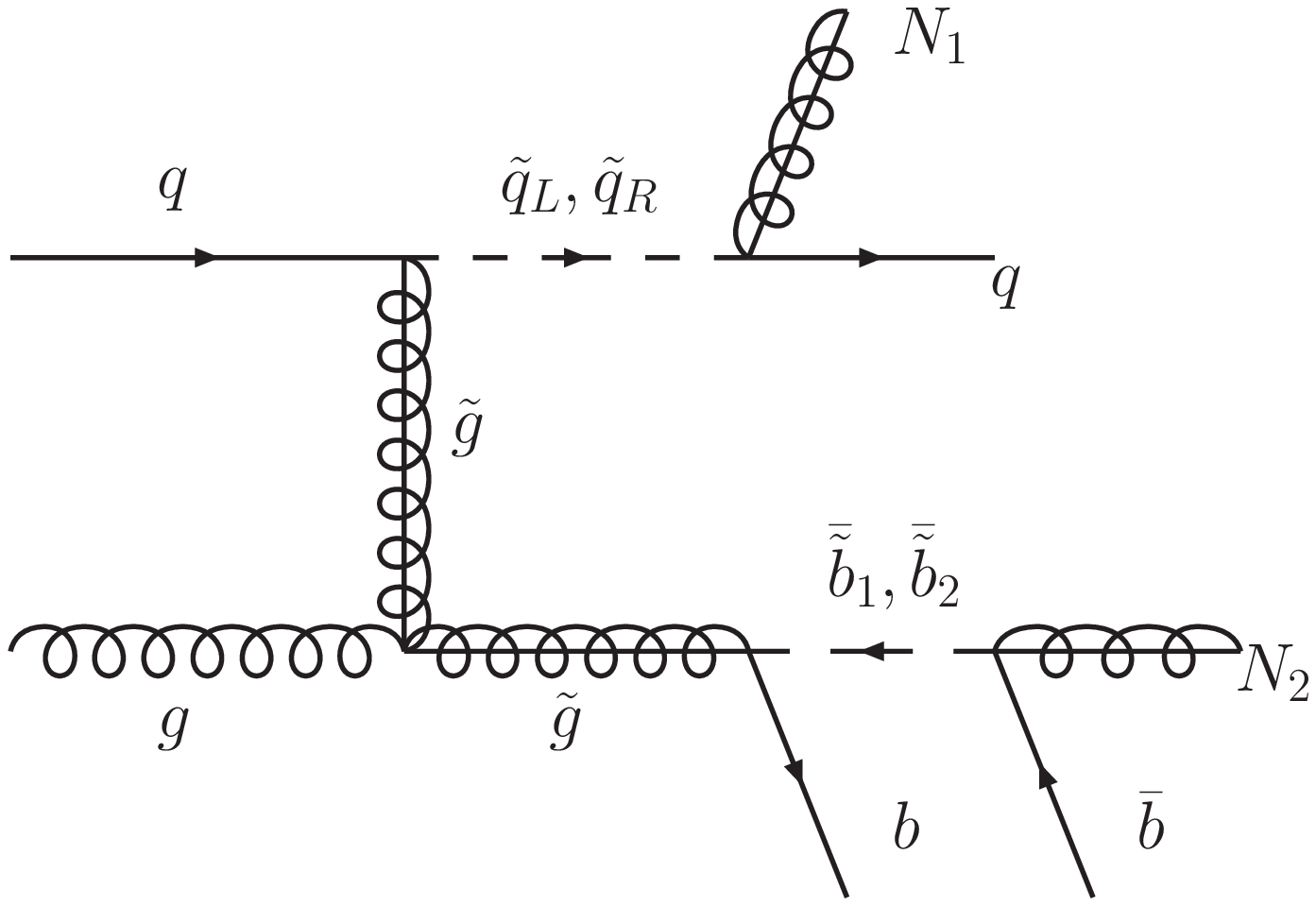} 
\end{minipage}\\
  \begin{minipage}{0.5\textwidth}
	\vspace*{20pt}
      \vspace{-0.6mm}
    \hspace*{-20pt} \includegraphics[width=\textwidth]{figures_chains/chain_5_3.eps}
  \end{minipage}
 &\begin{minipage}{0.5\textwidth}
      \vspace{2.5mm}
  \vspace{2.5mm} \hspace*{20pt}  \includegraphics[width=\textwidth]{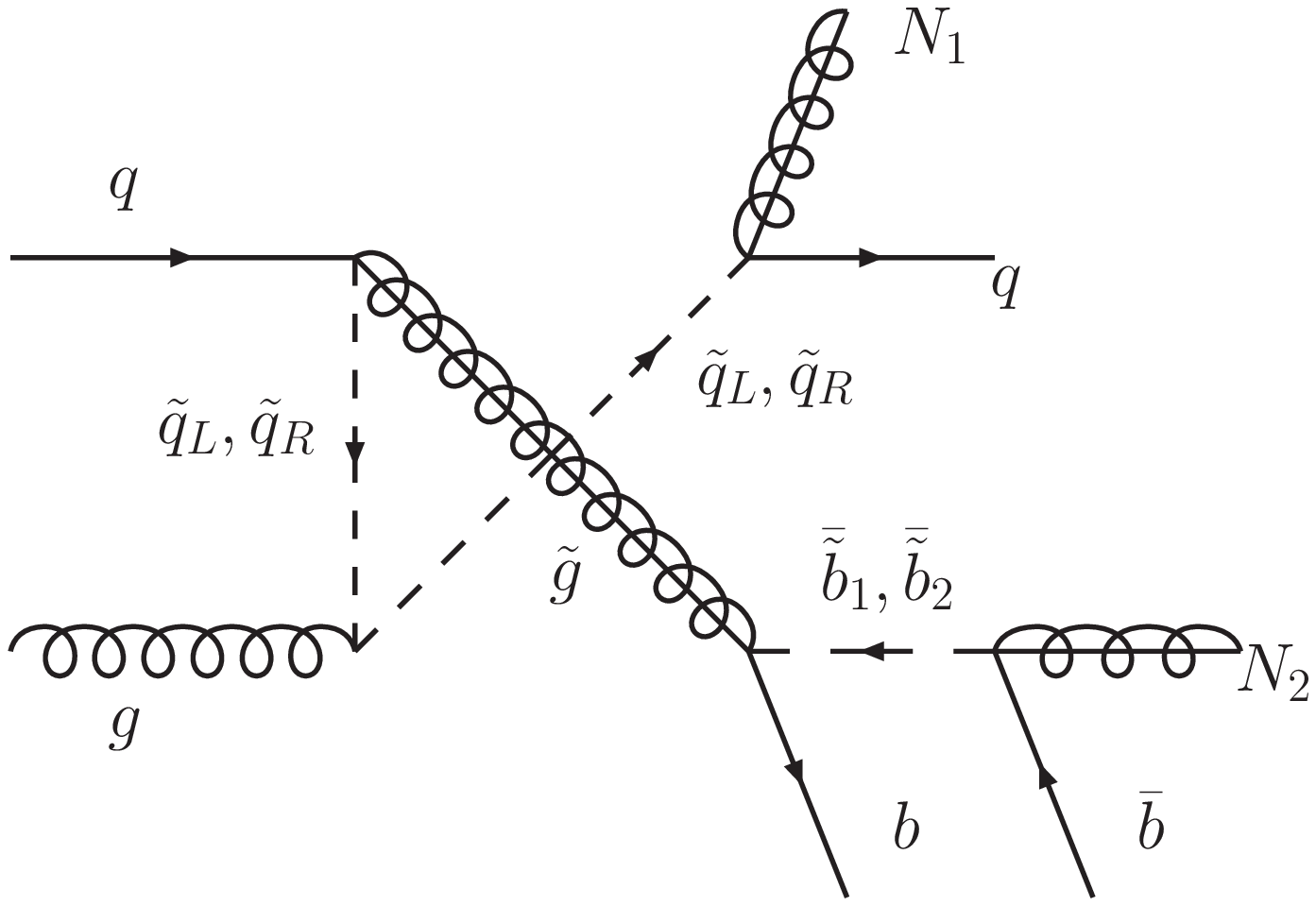} 
\end{minipage}\\
\vspace*{0pt}
\end{align*}
\caption{Diagrams contributing in LO to the $2 \rightarrow 5$ process in SUSY. All light incoming quark flavors can not be distinguished and have to be summed over.}
\label{fig:all_chains}
\end{figure}

In this chapter we calculate the gluon-quark collision and analyze the differences in the gluino and KK-gluon boost and the angle between the outgoing jets. We leave out the decay of the NLSP. In the SPS 1a scenario, the first neutralino $N_1$ is the LSP and the second neutralino $N_2$ is the NLSP. Together with these neutralinos, two $b$-jets and a light quark jet are outgoing. In comparison to~\cite{alves-2006-74} and~\cite{csaki-2007}, we additionally try to find out numerically if the impact of the boost on the angular correlations between the $b$-jets is significant. This is done by mapping the boost distributions of the SUSY gluino and the UED KK-gluon onto each other and thereby taking out the effect of the different boosts.

\section{Decay Chains in SUSY and UED}

The decay chain we want to consider is presented in fig.~\ref{fig:all_chains}. Since jets of quarks and antiquarks are not distinguishable one has to sum over all flavors of quarks and antiquarks. Using SDECAY we obtain different masses for up- and down-like flavors and left- and right-chirality of squarks as given in table~\ref{tab:SPS1_masses}. We therefore include all possible combinations of $u_L$, $u_R$, and respectively $d_L$, $d_R$, and $b_1$, $b_2$ in our calculations. Since the down-like coupling to the LSP is different from the up-like one, it is essential to evaluate the matrix elements separately for these two kinds of flavors due to its dependence on the weak isospin. 
This dependence is not present in the case of the $2 \rightarrow 2 $ process in chapter~4. There one can simply sum over all PDFs and multiply them by the matrix element of the $2 \rightarrow 2 $ process if masses of all flavors are assumed to be equal.

\begin{table}
\centering
 \includegraphics[width=0.70\textwidth]{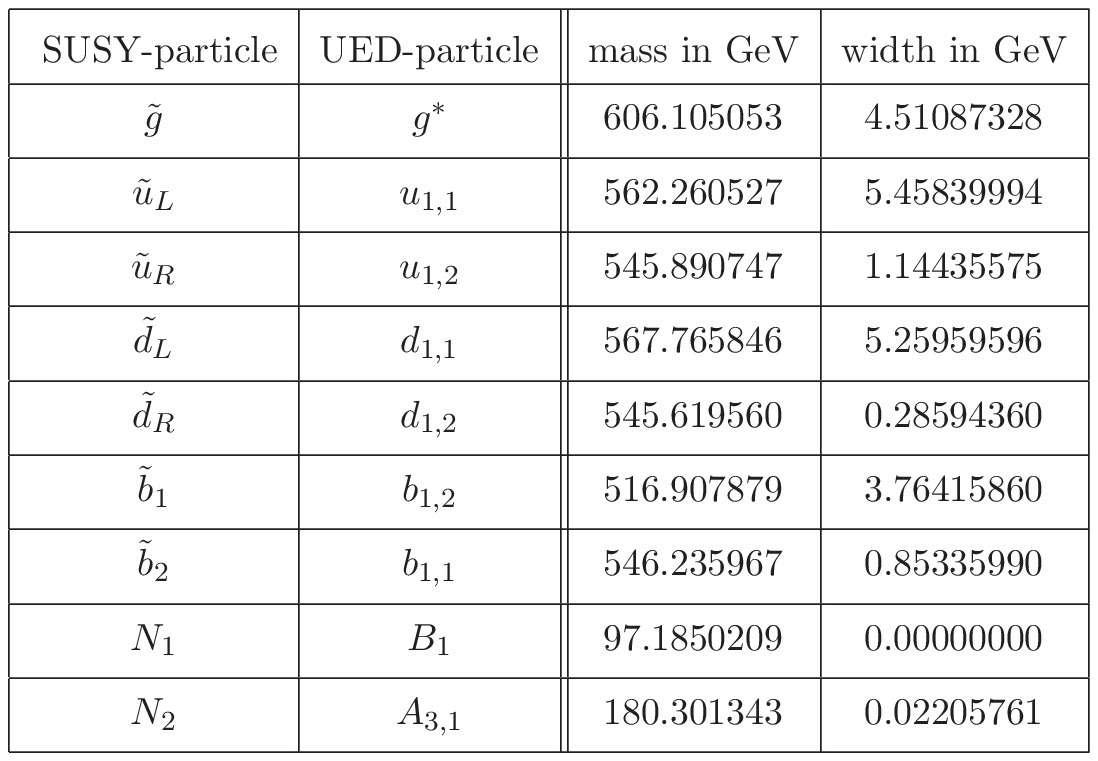}
\caption{Masses and widths for the SUSY particles used in our calculations.}
\label{tab:SPS1_masses}
\end{table}
Exchanging the $b$- and $\bar{b}$-jets in the final state of the $2 \rightarrow 5$ process is also necessary. The sbottoms are then substituted by anti-sbottoms. We do not consider incoming bottom-quarks since there would be an additional bottom-quark in the final state, leading to additional permutations with the other bottoms in the final state. This leads to a highly increased number of diagrams that have to be evaluated. Moreover this is not necessary from an experimental point of view since it is practically possible to differentiate between a bottom jet and a light jet, i.e.\ $u$,$d$,$c$ and $s$-jet.

Since the same final state can also be reached by exchanging $N_1$ and $N_2$, one has to exchange them as well. Both particles can in principle also couple to the other vertex. But since we do not claim to perform a calculation with a directly measurable result, we leave out the exchange of $N_1$ and $N_2$ and respectively the KK-particles $B^1$ and $A_{3,1}$. This does not harm gauge invariance explicitly since there are no gauge parameters involved. Therefore the calculation is a reasonable but not directly observable one. As already stated, we are interested in the origin of the angular correlations of the bottom jets. This is a theoretical question and exchanging these particles would result in doubled run time, not providing a much deeper insight into the angular correlations. In the SPS 1a scenario the branching fraction of the $u_R$ into $u$ and $N_1$ is approximately one. Therefore this gives the leading contribution to the process we are interested in. In principle, our calculation can be seen as a part of the longer decay chain, where the $N_2$ decays into two leptons and the LSP. This larger chain could be needed anyway, since outgoing muons are easy to detect which could reduce background. Here we do not consider the background from SM or other SUSY processes and the smearing of the signal from limited detector resolution. 

Of course there is no mixing between the processes with different incoming quark flavors. All contributions are simply added. But for each flavor interference terms in principle occur, after exchanging the $b$- and $\bar{b}$-jets. They are simply evaluated by Madgraph, by calculating the matrix element numerically and then squaring it. We tried to compute the diagrams, including interference terms in this way. Unfortunately this integration is quite unstable and converges very slowly. 
\begin{figure}
\begin{align*}
  \begin{minipage}{0.5\textwidth}
      \vspace{-0.6mm}
    \hspace*{-20pt} \includegraphics[width=\textwidth]{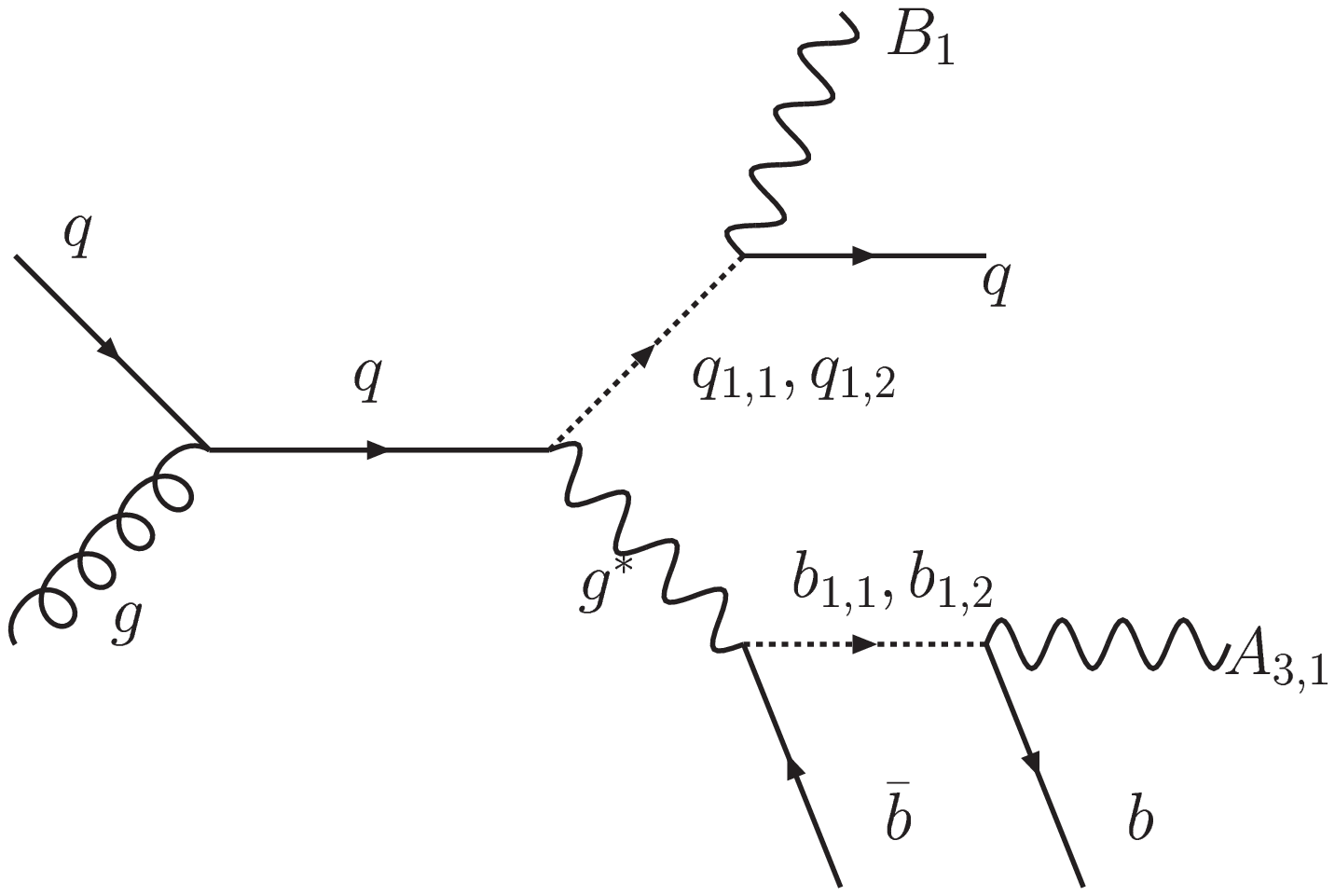}
  \end{minipage}
 &\begin{minipage}{0.5\textwidth}
      \vspace{-0.6mm}
  \hspace*{20pt}  \includegraphics[width=\textwidth]{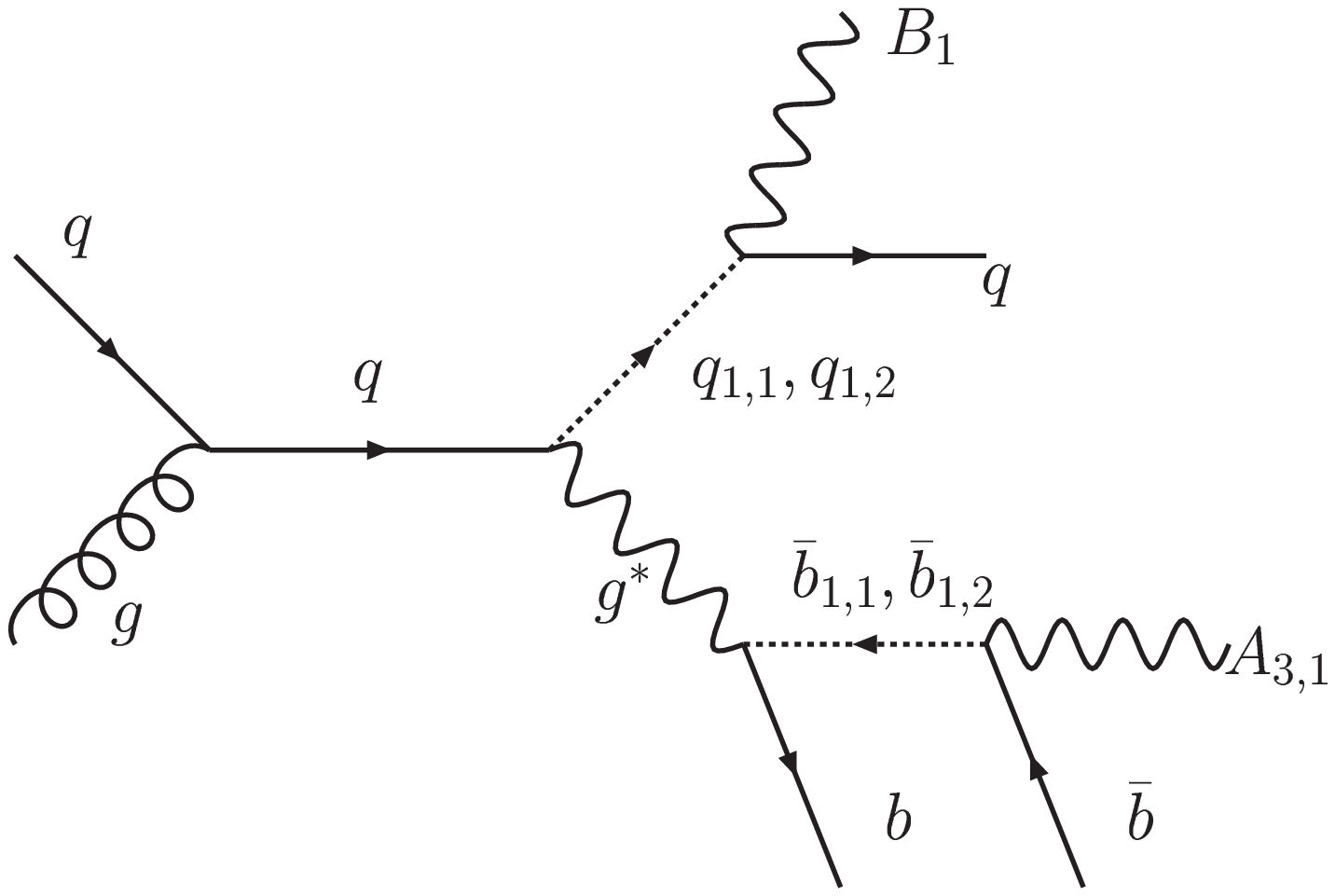} 
\end{minipage}\\
  \begin{minipage}{0.5\textwidth}
	\vspace*{20pt}
      \vspace{-0.6mm}
    \hspace*{-20pt} \includegraphics[width=\textwidth]{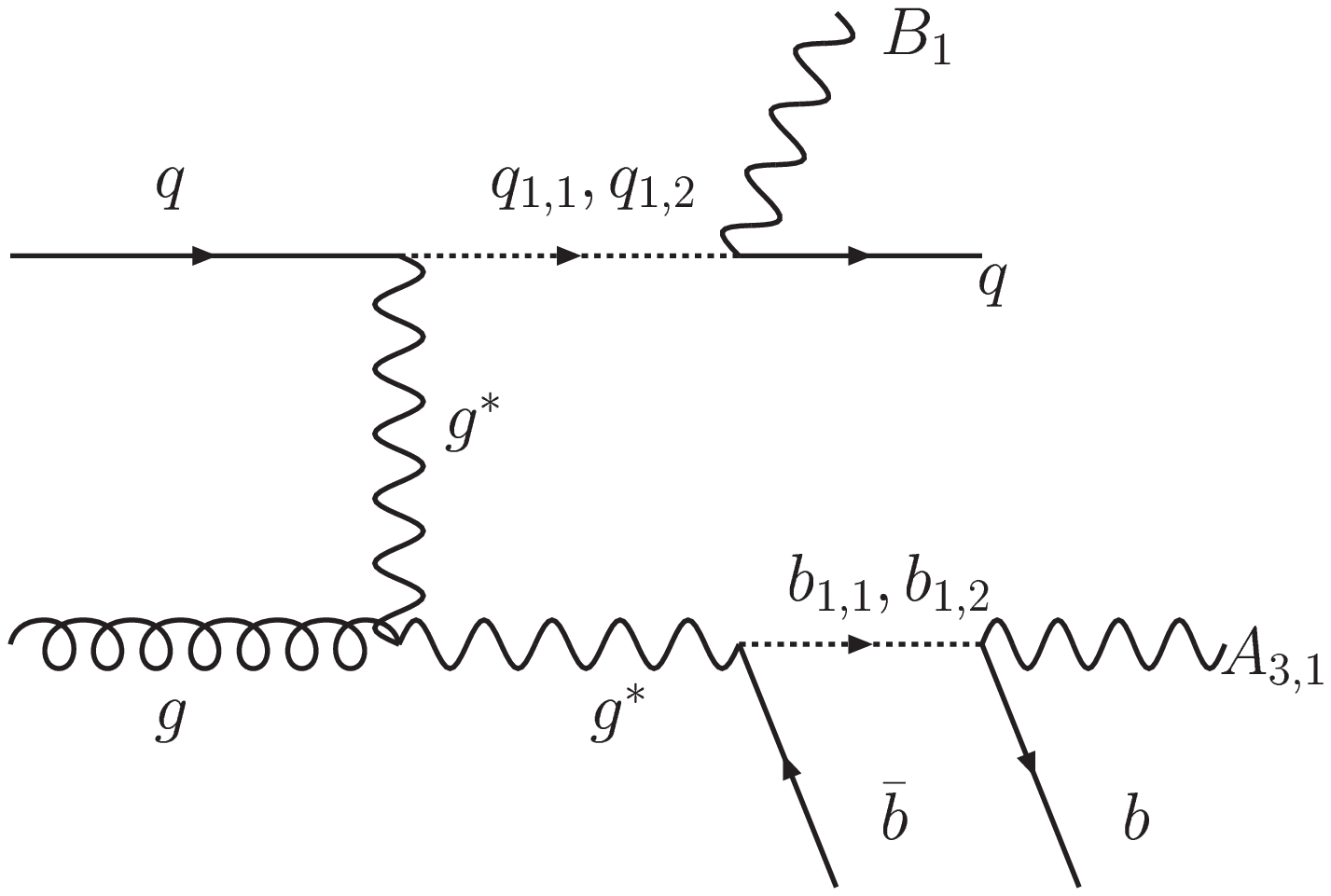}
  \end{minipage}
 &\begin{minipage}{0.5\textwidth}
      \vspace{2.5mm}
  \vspace{2.5mm} \hspace*{20pt}  \includegraphics[width=\textwidth]{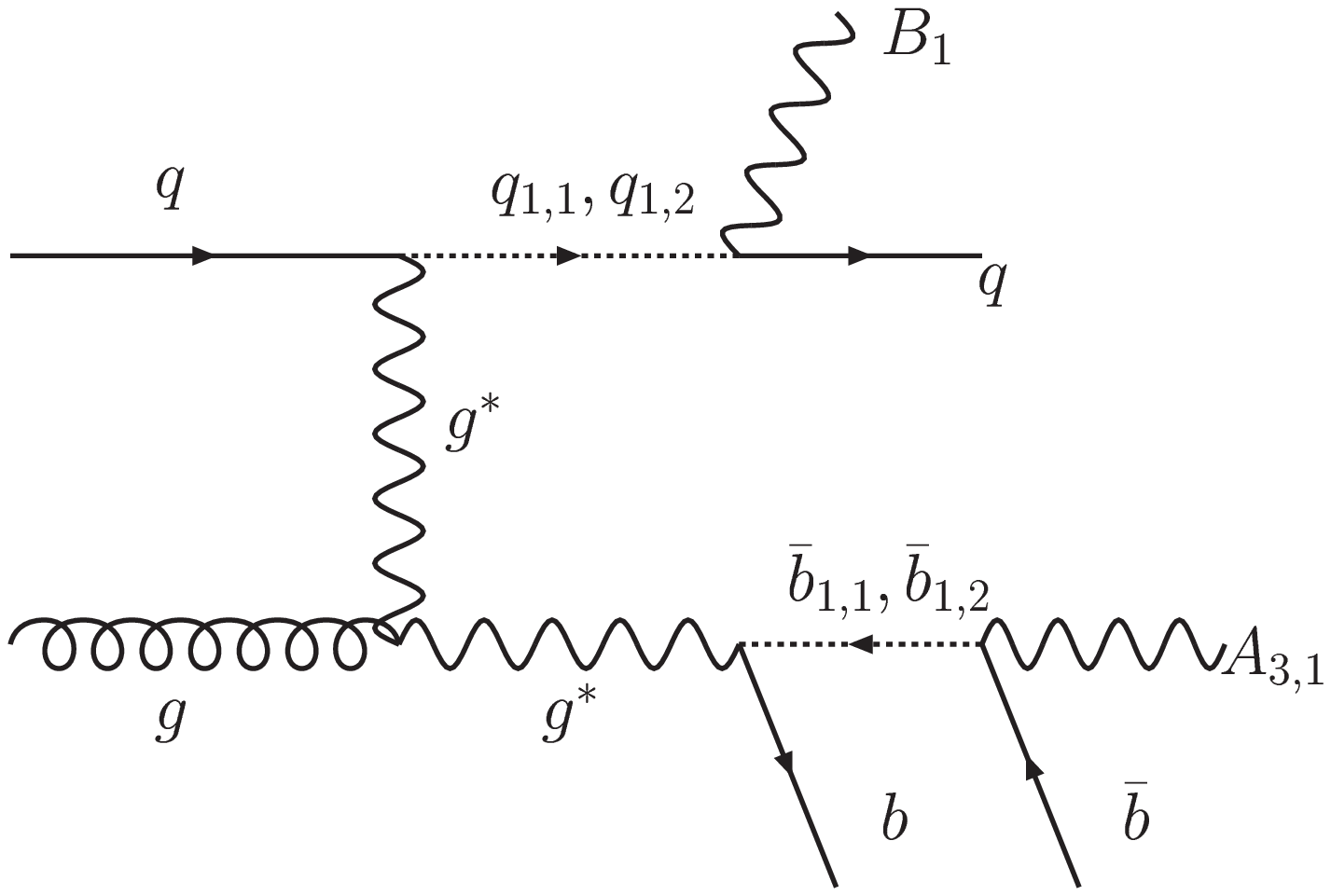} 
\end{minipage}\\
  \begin{minipage}{0.5\textwidth}
	\vspace*{20pt}
      \vspace{-0.6mm}
    \hspace*{-20pt} \includegraphics[width=\textwidth]{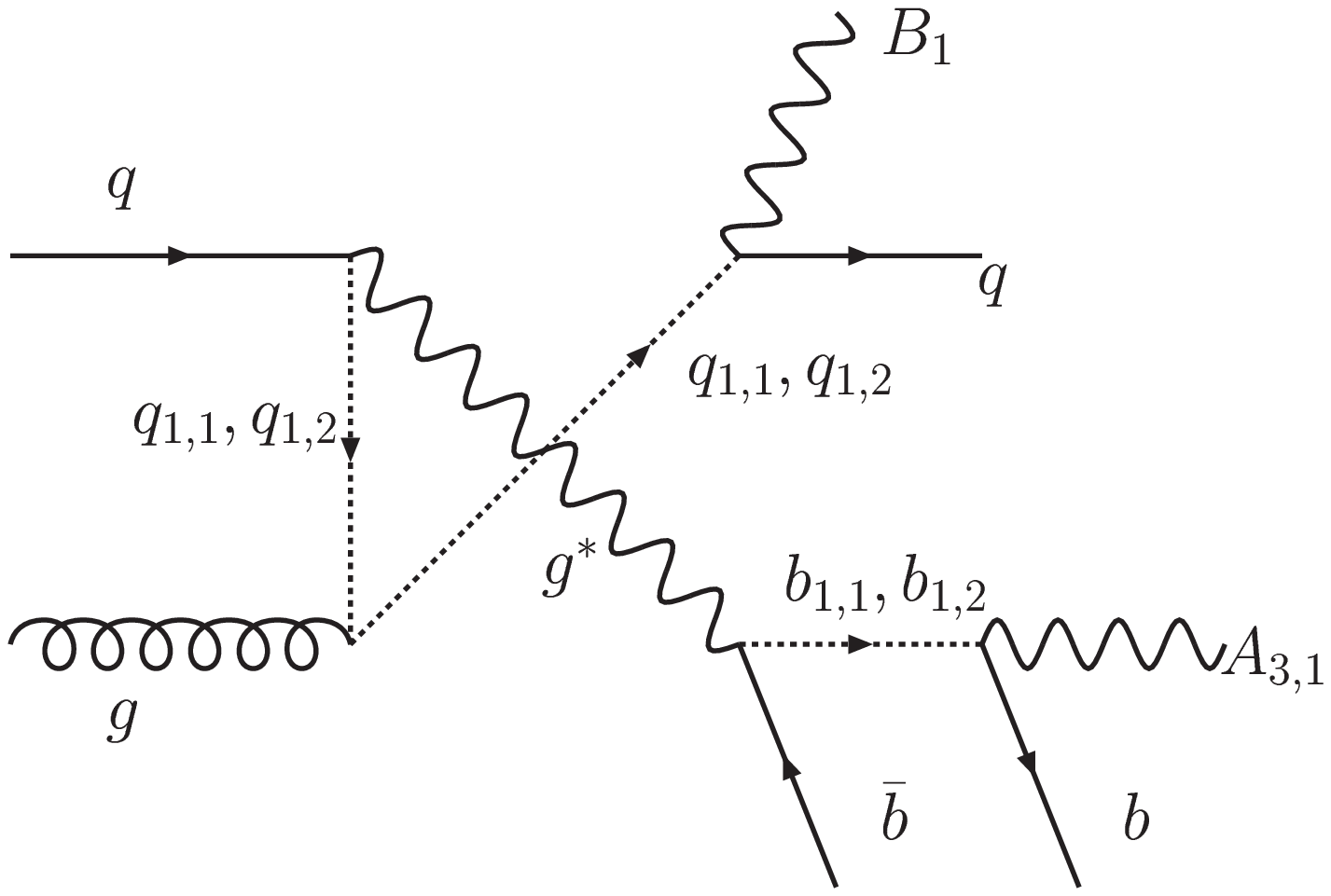}
  \end{minipage}
 &\begin{minipage}{0.5\textwidth}
      \vspace{2.5mm}
  \vspace{2.5mm} \hspace*{20pt}  \includegraphics[width=\textwidth]{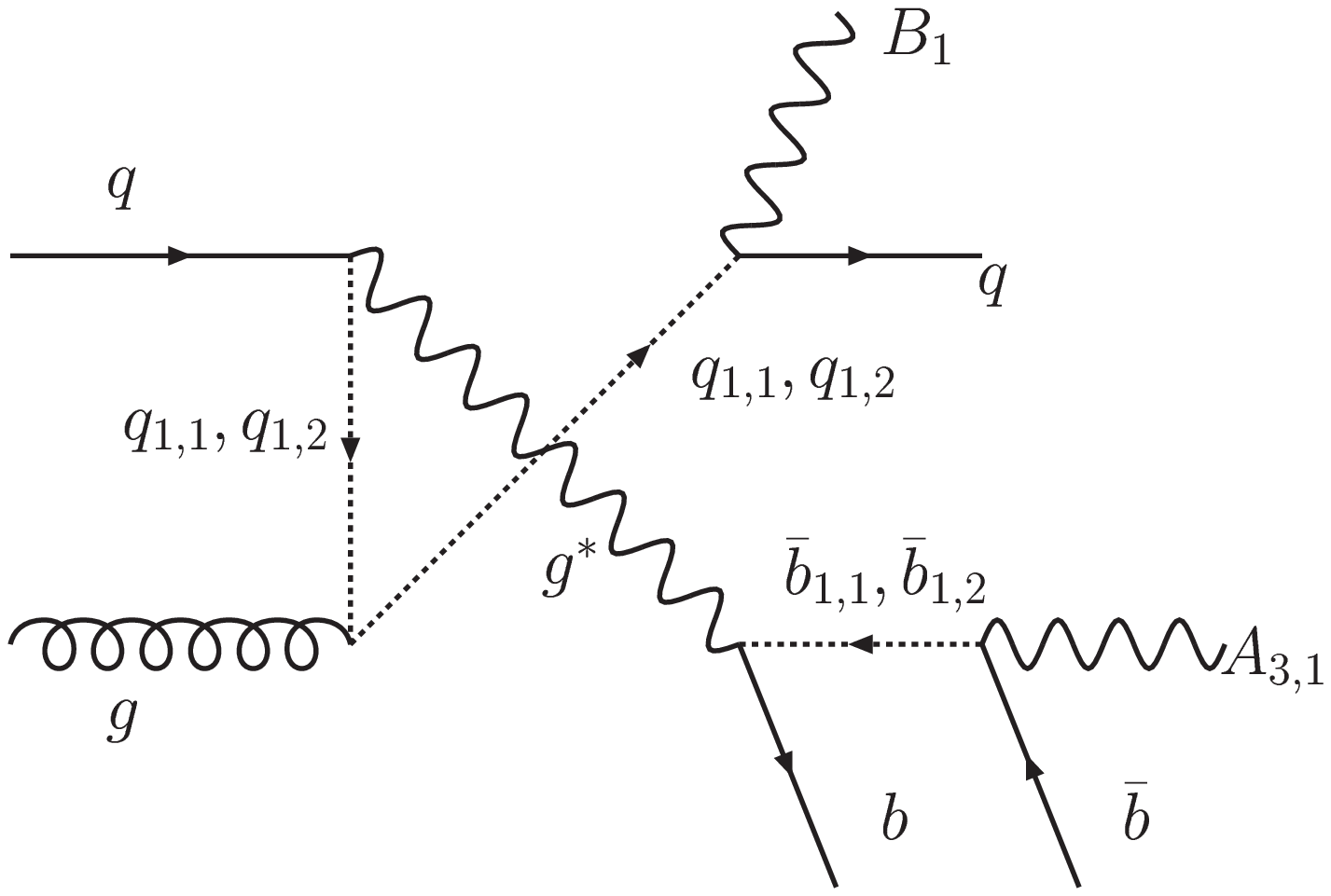} 
\end{minipage}\\
\vspace*{0pt}
\end{align*}
\caption{Diagrams contributing in LO to the $2 \rightarrow 5$ process in UED. All light incoming quark flavors can not be distinguished and have to be summed over.}
\label{fig:all_UED_chains}
\end{figure}
This can be understood by having a look at fig.~\ref{fig:all_chains}. When the momentum of the $\bar{b}$-jet from the decay of the quasi-on-shell gluino to the sbottom is calculated, it is found to be much softer than the $b$-jet from the decay of the sbottom to the NLSP, as later shown in fig.~\ref{fig:both_b_jets}. If these momenta are now exchanged for the calculation of the interference term, the gluino and sbottom can not both be on-shell. This makes our integration very inefficient because the Breit-Wigner mapping effectively only evaluates points at the pole. Since the interference terms are not increased by the Breit-Wigner propagator, they are suppressed in comparison to the squared matrix elements without the exchange of the $b$- and $\bar{b}$-jet. This Breit-Wigner suppressed contribution to the angular distribution is expected to have the same shape as the pure squared matrix elements without interference terms and can therefore be left out as a NLO effect. As a consequence we calculate the diagrams with exchanged $b$- and $\bar{b}$-jets separately. We then find that then the integration is stable and converges much faster.

Another problem occurs from the contributions deriving from different intermediate squarks. Of course, there is interference between the diagrams with the same final and initial states and including $\tilde{q}_L$ or $\tilde{q}_R$ in the intermediate states. But it is not trivial to calculate interference of such diagrams, since the phase space has poles from the Breit-Wigner propagators at different energies. This problem can not be solved in our calculation since the pole either is set to $m_{\tilde{q}_L}$ or $m_{\tilde{q}_R}$. The phase space can only be generated for one of these poles at a time, e.g.\ $m_{\tilde{q}_L}$. Then the resulting four momenta do not generate on-shell contributions to the other diagram with the pole at $m_{\tilde{q}_R}$. The region of the phase space where many points are evaluated for a pole at $m_{\tilde{q}_L}$ is Breit-Wigner suppressed for the diagrams with the pole at $m_{\tilde{q}_R}$. Therefore interference terms between these diagrams are again of NLO and therefore left out. Of course it is crucial for this approximation that the difference between $m_{\tilde{q}_L}$ and $m_{\tilde{q}_R}$ is much larger than the width of the particles, as one can see in table~\ref{tab:SPS1_masses}. The same argument is used for the sbottoms in the intermediate state and equally in the case of the UED scenario.

If we want to exchange the $N_1$ and $N_2$ in the final state we would also have to neglect their interference term with the same argument since their masses are quite different.

All these different contributions to the final state are added and filled into histograms. Altogether sixteen integrations of different phase spaces have to be evaluated, each summing over particles, antiparticles and the four light flavors and including the three different topologies as given in fig.~\ref{fig:all_chains}. This results in a run time of several hours. For the case of UED we included the particle spectrum with quantum numbers and names of the particles into the file \verb|particles.dat| of the Madgraph code. All Feynman rules needed for our process are included into the file \verb|interactions.dat| to generate the correct couplings. It is necessary to use the correct couplings, even if one normalizes all cross sections to one, since the mixing angle between singlets and doublets will be varied later.
Since we want to exclude some interference terms as well as some topologies explicitly, as explained in the section~\ref{sec:Neglected Topologies in the Final Decay Chain}, we can comment out the unwanted couplings in the \verb|interactions.dat| file.

Using the quantum numbers of all particles as given in the Madgraph files, HELAS calculates the matrix element as the sum of combinations of helicity eigenstates of the external particles as we explained in section~\ref{sec:HELAS}. Each SUSY contribution is calculated as the sum of 128 helicity combinations. In the case of the UED scenario there are 288 different helicity combinations. The number of helicity combinations is different, since the UED final state contains the massive gauge bosons $A_{3,1}$ and $B_1$, both having three helicity eigenstates. In SUSY, the fermions $N_1$ and $N_2$ only have two helicity eigenstates. The boson $B_1$ in UED is the KK-partners of the photon, corresponding to the SUSY LSP $N_1$, while the $A_{3,1}$ corresponds to the NLSP $N_2$. Due to the larger number of helicity combinations the UED program has a much longer run-time because the matrix element routines are the slowest routines of the code.
\begin{figure}
\begin{align*}
  \begin{minipage}{0.5\textwidth}
      \vspace{-0.6mm}
    \hspace*{-7pt} \includegraphics[width=\textwidth]{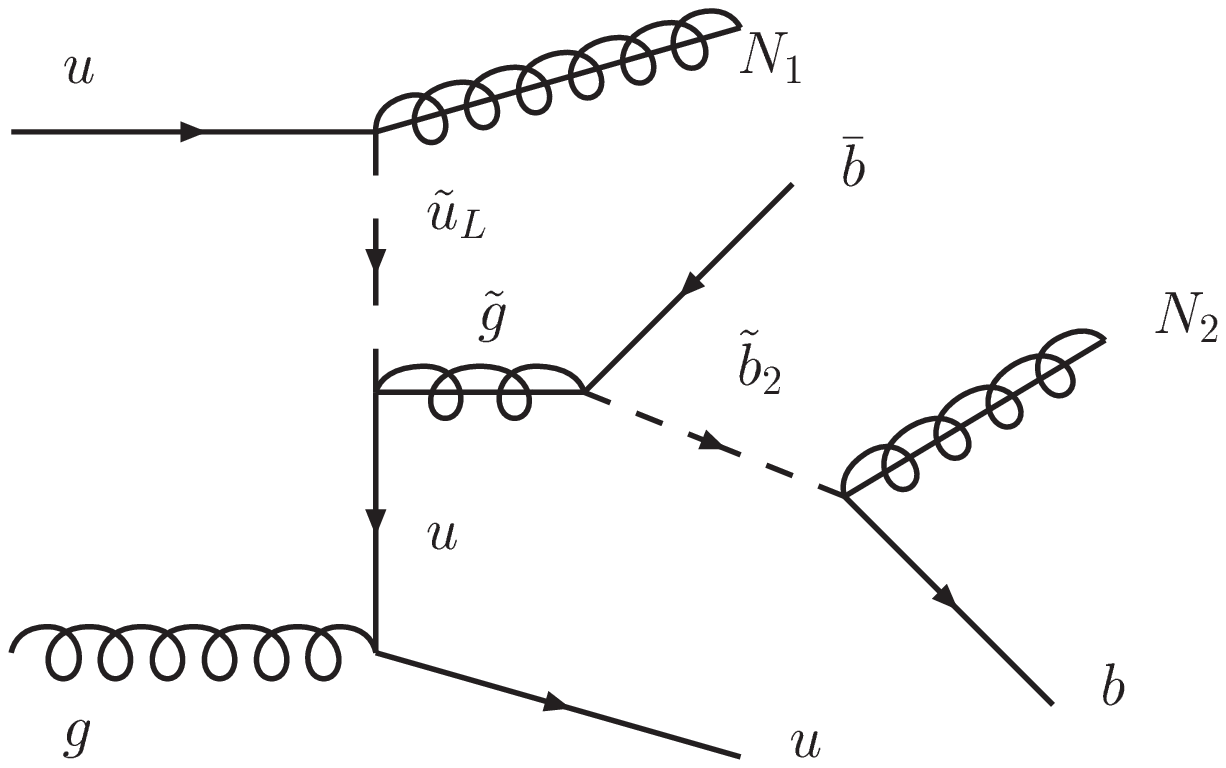}
  \end{minipage}
 &\begin{minipage}{0.5\textwidth}
      \vspace{-0.6mm}
  \hspace*{5pt}  \includegraphics[width=\textwidth]{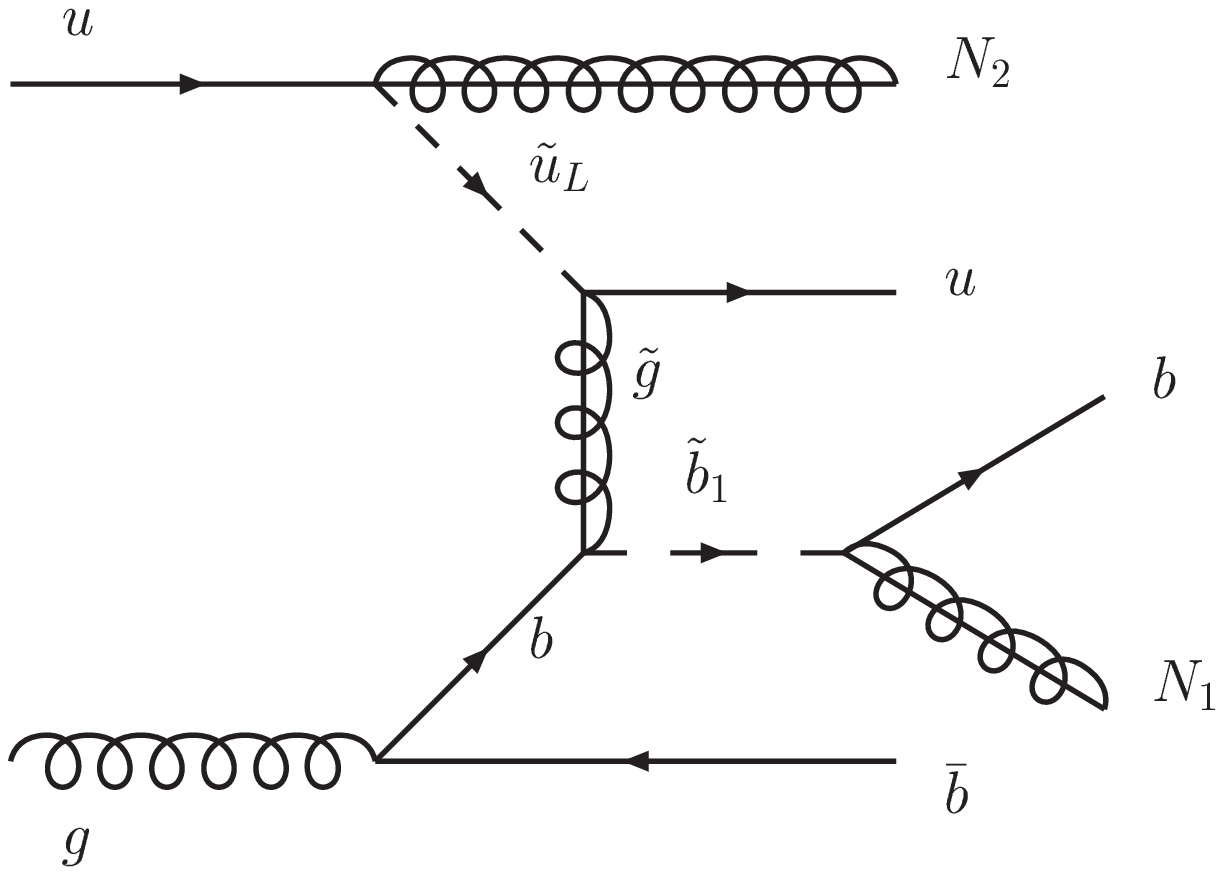} 
\end{minipage}\\
\end{align*}
\caption{Diagrams are suppressed since the particles in t- and u-channels can not be on-shell.}
\label{fig:u_and_t_channels}
\end{figure}

\section{Neglected Topologies in the Final Decay Chain}
\label{sec:Neglected Topologies in the Final Decay Chain}

Using SUSY-Madgraph for the construction of the topologies and taking into account all existing MSSM couplings, SUSY-Madgraph finds 460 different diagrams for the process denoted by the external particles
\begin{equation*}
u \; g \rightarrow N_1 \; u \; b \; \bar{b} \; N_2 \; .
\end{equation*}
The same number of diagrams comes from
\begin{equation*}
d \; g \rightarrow N_1 \; d \; b \; \bar{b} \; N_2 \; ,
\end{equation*}
where the quantum numbers of the quarks from the electroweak sector are different. Also antiquarks contribute to this process as we mentioned in the last section. In this section we shortly present the diagrams we neglect in our calculation. 

By using special examples from these Madgraph diagrams, we want to explain that diagrams different from those in fig.~\ref{fig:all_chains} can be neglected. Due to the Breit-Wigner propagators, s-channel-like decays give the dominant contribution to the resulting cross section. The diagrams presented in fig.~\ref{fig:all_chains} involve three Breit-Wigner propagators, all being on-shell in our calculation. 

As we already stated for the topologies of $2\rightarrow 3$ and $2\rightarrow 4$ processes, particles in t- and u-channels can not be on the mass shell. Therefore the topologies from fig.~\ref{fig:u_and_t_channels} are suppressed. There are a lot more combinations of particles which can be included into these topologies. Additional t-channel topologies also exist, including one or two $1 \rightarrow 2$ particle decays on the right side of the t-channel.

\begin{figure}
  \begin{minipage}{0.8\textwidth}
      \vspace{-0.6mm}
    \hspace*{35pt} \includegraphics[width=\textwidth]{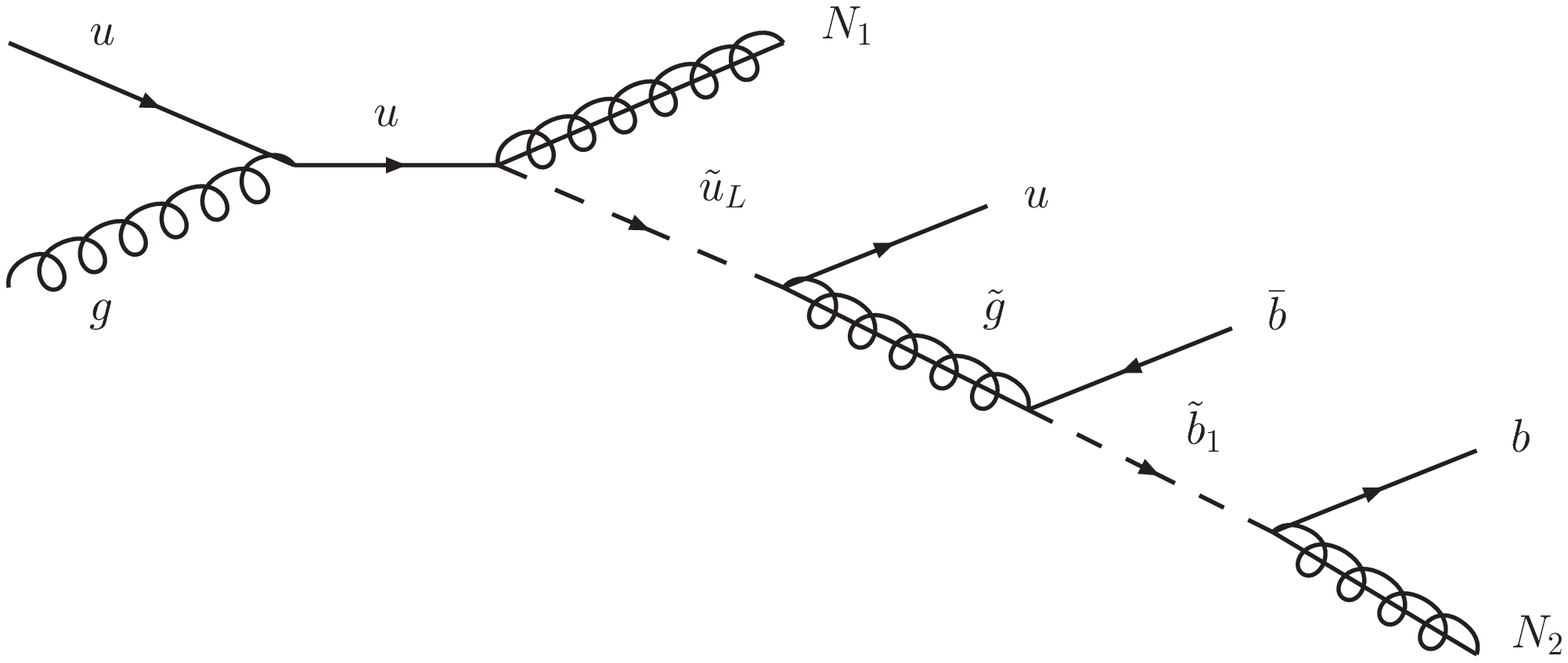}
  \end{minipage}
\caption{This diagram is suppressed since $u_L$ and $\tilde{g}$ can not be on-shell at the same time.}
\label{fig:long_chain}
\end{figure}

Another kind of topology that was neglected is given in fig.~\ref{fig:long_chain}. Considering the mass spectrum of the parameter point SPS 1a, one finds that the gluino is heavier than all squarks. Therefore either the $\tilde{u}_L$ or the $\tilde{g}$ can not be on-shell in fig.~\ref{fig:long_chain}. Although only s-channel decays are involved, this diagram is suppressed by a missing on-shell Breit-Wigner. This constrains the validity of our calculations to mass spectra having the same mass hierachy.

For a UED-like degenerate mass spectrum this topology could not be neglected. For the case of a nearly degenerate mass spectrum one also faces the problem that outgoing $b$-jets are very soft and therefore hard to see at LHC experiments.

\begin{figure}
  \begin{minipage}{0.5\textwidth}
      \vspace{-0.6mm}
    \hspace*{-7pt} \includegraphics[width=\textwidth]{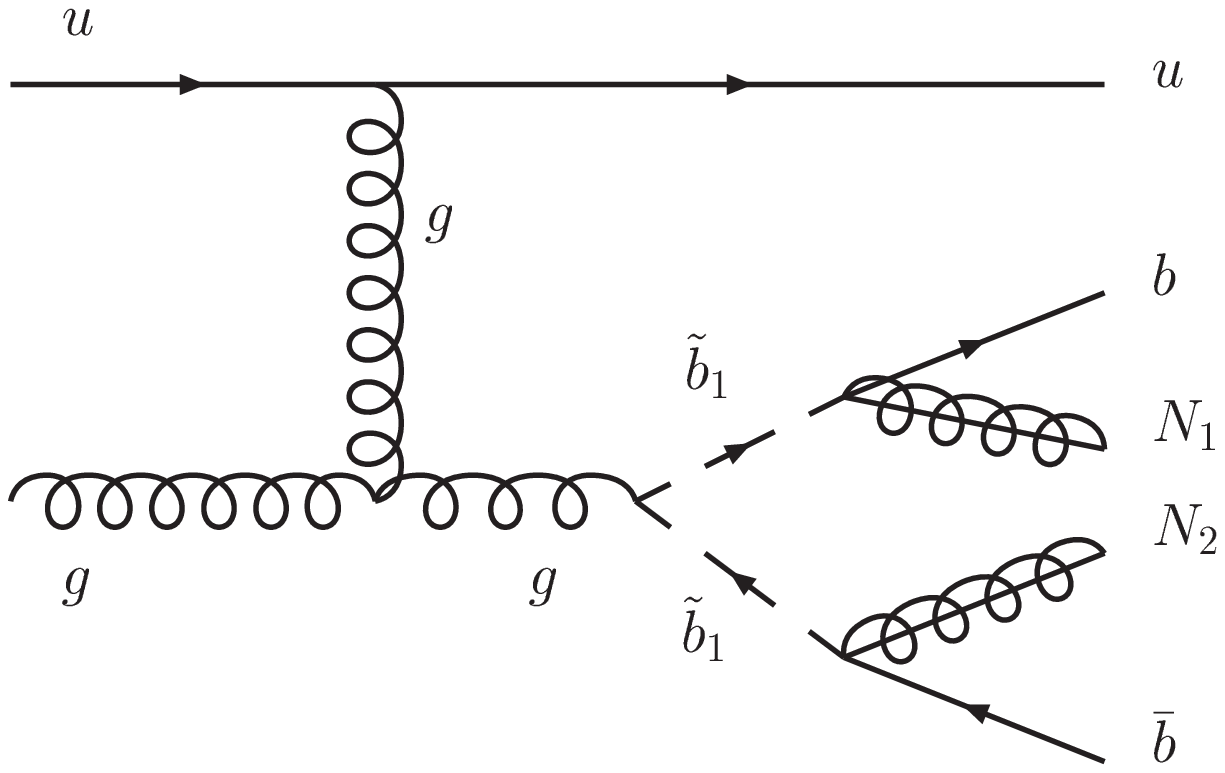}
  \end{minipage} 
\hspace{20pt}
  \begin{minipage}{0.42\textwidth}
      \vspace{-0.6mm}
    \hspace*{-7pt} \includegraphics[width=\textwidth]{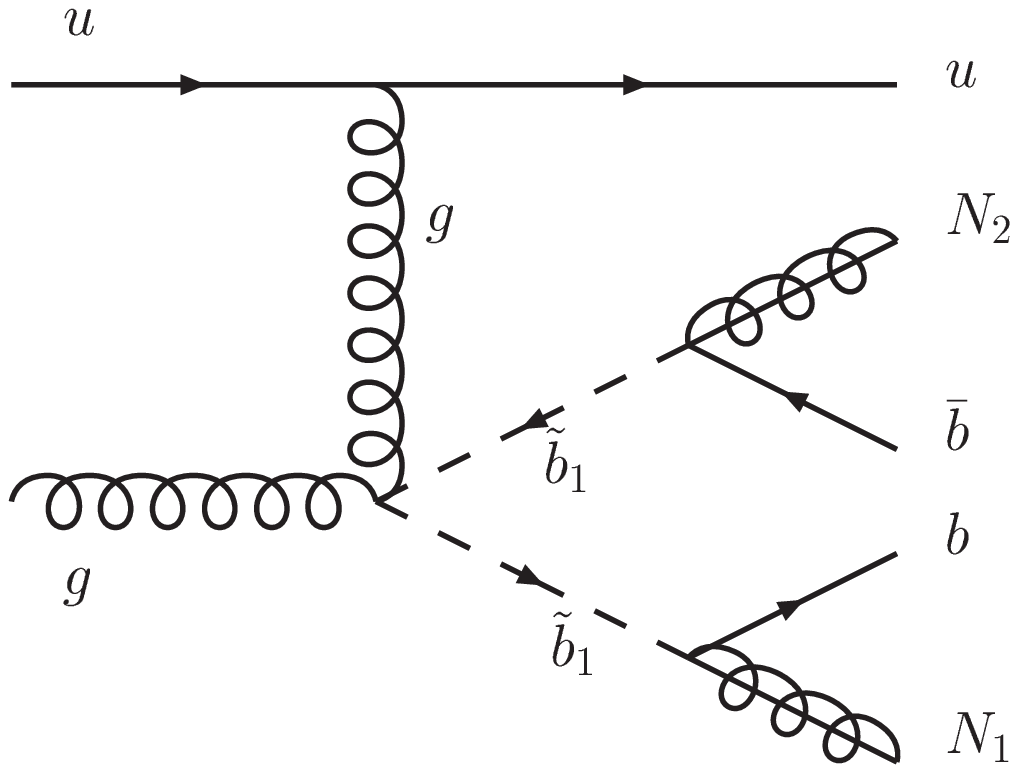}
  \end{minipage}
\caption{These diagrams are of NLO since there are only two on-shell Breit-Wigner propagators included.}
\label{fig:internal_gluon}
\end{figure}

Due to the gluons and quarks in the initial and final state there are different diagrams including gluon propagators. One of them is given on the left side in fig.~\ref{fig:internal_gluon}. Since the gluon has no finite width included in its propagator, the diagram is suppressed. This holds for a large number of diagrams with internal gluons and only two additional Breit-Wigner propagators. Another example for that is given on the right side in fig.~\ref{fig:internal_gluon}.  Here even a four boson vertex is involved. This diagram would not occur in UED in the same way, since the $\tilde{b}_1$ would be a fermionic partner of the bottom quark then. 

Of course, for our UED calculation we consider exactly the same topologies as in the case of SUSY. In principle, diagrams with higher excitations of KK-towers also exist. These are suppressed by the higher mass of the particles.

\newpage
\section{Results of the SUSY-UED Comparison}
\label{sec:The_SUSY_UED_COMPARISION}
To become more familiar with kinematics in decay chains we present some plots of invariant masses and momenta for UED and SUSY in this chapter. We also present the results for the angular distributions and the boost of the gluino/KK-gluon. Though it is in principle fixed by the masses, we use the mixing angle $\alpha^{(1)}$, introduced already in chapter~\ref{ch:3}, to test if the SUSY angular distributions can be reproduced by a UED decay chain with modified couplings. We therefore calculate the chain for $0^{\circ}$ and $45^{\circ}$ mixing angle of the KK-quark towers, beginning with the $0^{\circ}$ scenario. Both KK-towers in general can be expected not to mix due to the low mass of the SM $b$-quark. All distributions are normalized to one in order to make both theories comparable. Since the error of our integration of the total cross section is below $1\%$ there are no error bars presented in the histograms.

\subsection{A SUSY-UED Comparison for $\alpha^{(1)}=0^{\circ}$}
As we already stated in section~\ref{sec:Testing_chains}, the invariant masses and transverse momenta and rapidities of intermediate particles served as a good check when building up our decay chain program stepwise. Since contributions nearly completely derive from on-shell momenta, these curves do not change for all steps of our decay chain programs. For example, the transverse momentum of the gluino stays unchanged, no matter what particles it decays to as long as the decay-vertex itself is not dependent on momenta. Since we did not present these distributions yet, we now want to present them for the $2 \rightarrow 5$ process discussed in the last section. 
\begin{figure}[h!]
\centering
    \includegraphics[width=0.65\textwidth]{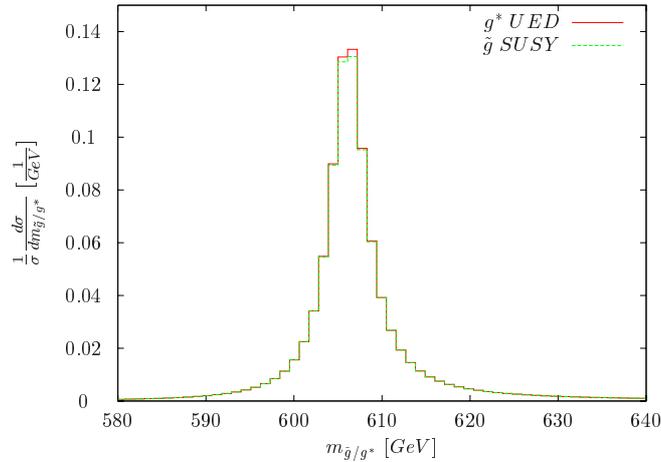}
\vspace*{-5mm}
\caption{Invariant mass distribution of the gluino and KK-gluon.}
\label{fig:plot_m_gl}
\end{figure}

In fig.~\ref{fig:plot_m_gl} we present the invariant mass distribution of the gluino and the KK-gluon for the SUSY and the UED decay chain. One finds two strongly peaked Breit-Wigner distribution exactly at the same value for the invariant mass. The invariant mass is given by
\begin{equation}
m_{ij}=\sqrt{(p_i \, + \, p_j)^2}\; ,
\end{equation}
where $p_i$ and $p_j$ can be four vectors of two different particles. For the invariant mass of one single particle with momentum $p_i$ we use $m_{i}=\sqrt{(p_i)^2}$.

\begin{figure}
 \centering
    \includegraphics[width=0.65\textwidth]{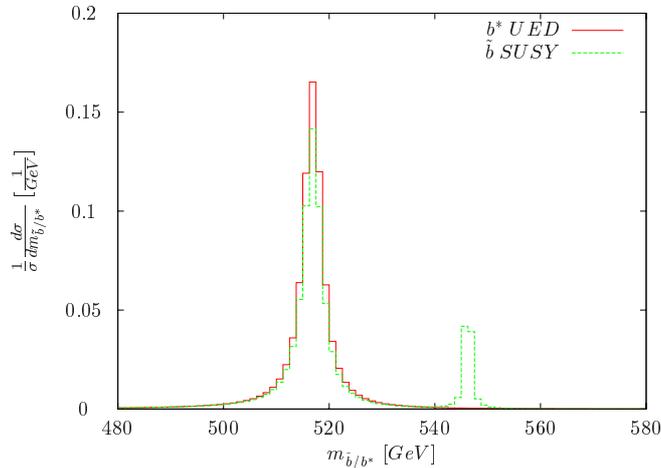}
\vspace*{-5mm}
\caption{Invariant mass distribution of the $b$-squark and KK-$b$-quark.}
\label{fig:plot_m_sb}
\end{figure}

For the invariant mass distribution of the sbottom $\tilde{b}$ and KK-$b$-quark one finds similar distributions, shown in fig.~\ref{fig:plot_m_sb}. Depending on the mixing angle of the two KK-towers in UED, either the first or second tower does not contribute to the $2 \rightarrow 5$ process when the mixing matrix is diagonal. Here we assume a mixing angle of $\alpha^{(1)}=0^{\circ}$ between the two KK-towers and obtain a peak for the KK-quark $b_{1,2}$ at $516.91 \; \text{GeV}$. This is also reflected by the Feynman rules in appendix~\ref{app:electroweak}. While the coupling strength of one tower is at the maximum, the coupling to $A_{3,1}$ for the second tower is equal to zero. In the case of SUSY both sbottoms contribute to the process and therefore a second peak for sbottom $\tilde{b}_2$ at $546.24 \; \text{GeV}$ is present in addition to the sbottom $\tilde{b}_1$ peak at $516.91 \; \text{GeV}$. 

\begin{figure}
 \centering
    \includegraphics[width=0.65\textwidth]{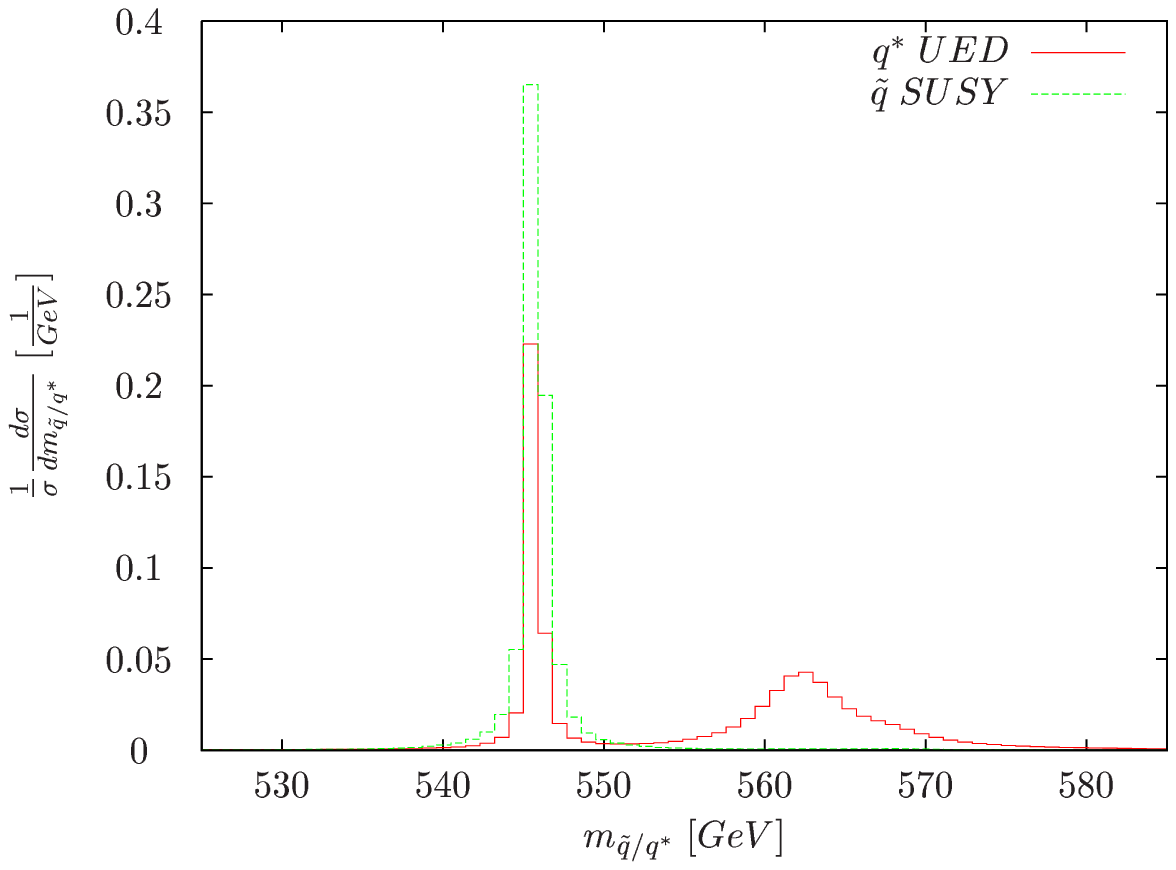}
\vspace*{-5mm}
\caption{Invariant mass distribution of the $u$,$d$,$c$ and $s$-squark and KK-quark.}
\label{fig:plot_m_usq}
\end{figure}

The invariant mass of the $u$,$d$,$c$ and $s$-squark/KK-quark is given in fig.~\ref{fig:plot_m_usq}, showing that $\tilde{u}_L$ does not decay to $N_1$ with a large branching ratio. The two masses of $q_{1,1}$ and $q_{1,2}$ in UED are chosen such, that the larger peak for UED is to be found at the same place as it is the case for SUSY. Also in the case of $\alpha^{(0)}=0^{\circ}$ mixing angle, there are two peaks for the UED scenario, since none of the couplings becomes zero.


\begin{figure}[p!]
\centering
\includegraphics[width=0.65\textwidth]{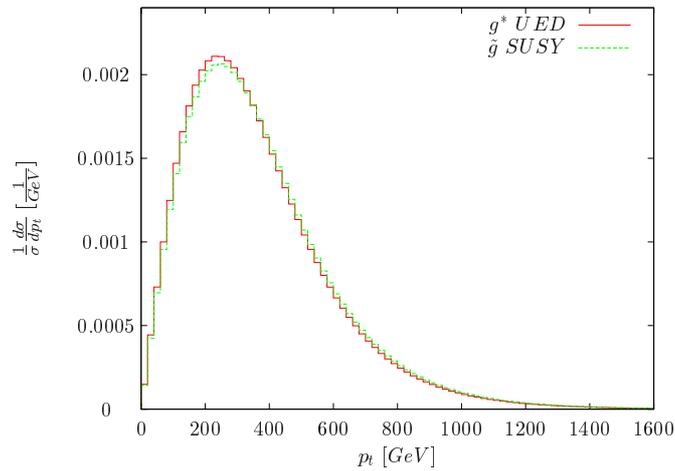}
\vspace*{-5mm}
\caption{Transverse momentum distribution of the gluino and the KK-gluon.
  \label{fig:plot_trans_mom_g.eps}}
\end{figure}
\begin{figure}
\centering
\vspace*{+5mm}
\includegraphics[width=0.65\textwidth]{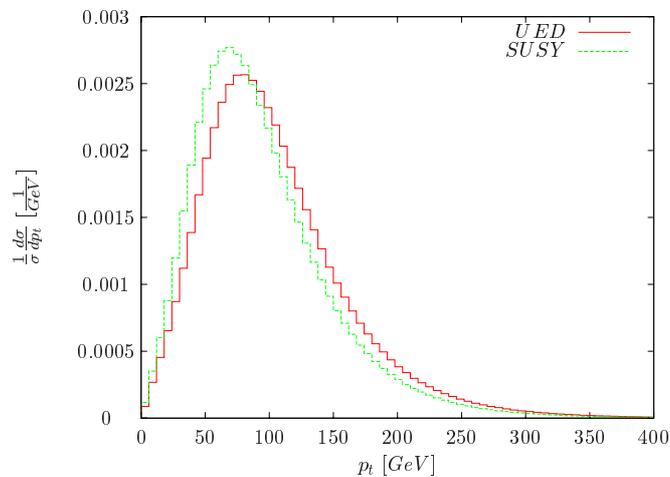}
\vspace*{-5mm}
\caption{Transverse momentum distribution of the light quark jet.
  \label{fig:light_squark_pt}}
\end{figure}
\begin{figure}
\centering
\vspace*{+5mm}
\includegraphics[width=0.65\textwidth]{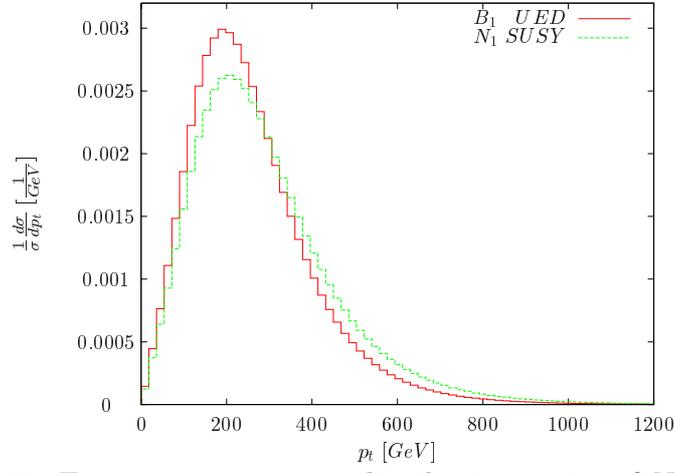}
\vspace*{-5mm}
\caption{Transverse momentum distribution section of $N_1$ and $B_1$.
  \label{fig:light_jet_pt}}
\end{figure}
\begin{figure}
\centering
\includegraphics[width=0.65\textwidth]{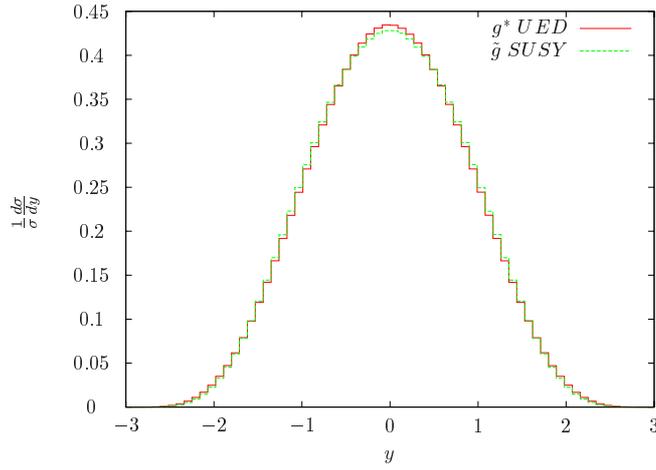}
\vspace*{-5mm}
\caption{Rapidity distribution of the gluino and the KK-gluon.
  \label{fig:plot_rapidity_g.eps}}
\end{figure}
\begin{figure}
\centering
\vspace*{+5mm}
\includegraphics[width=0.65\textwidth]{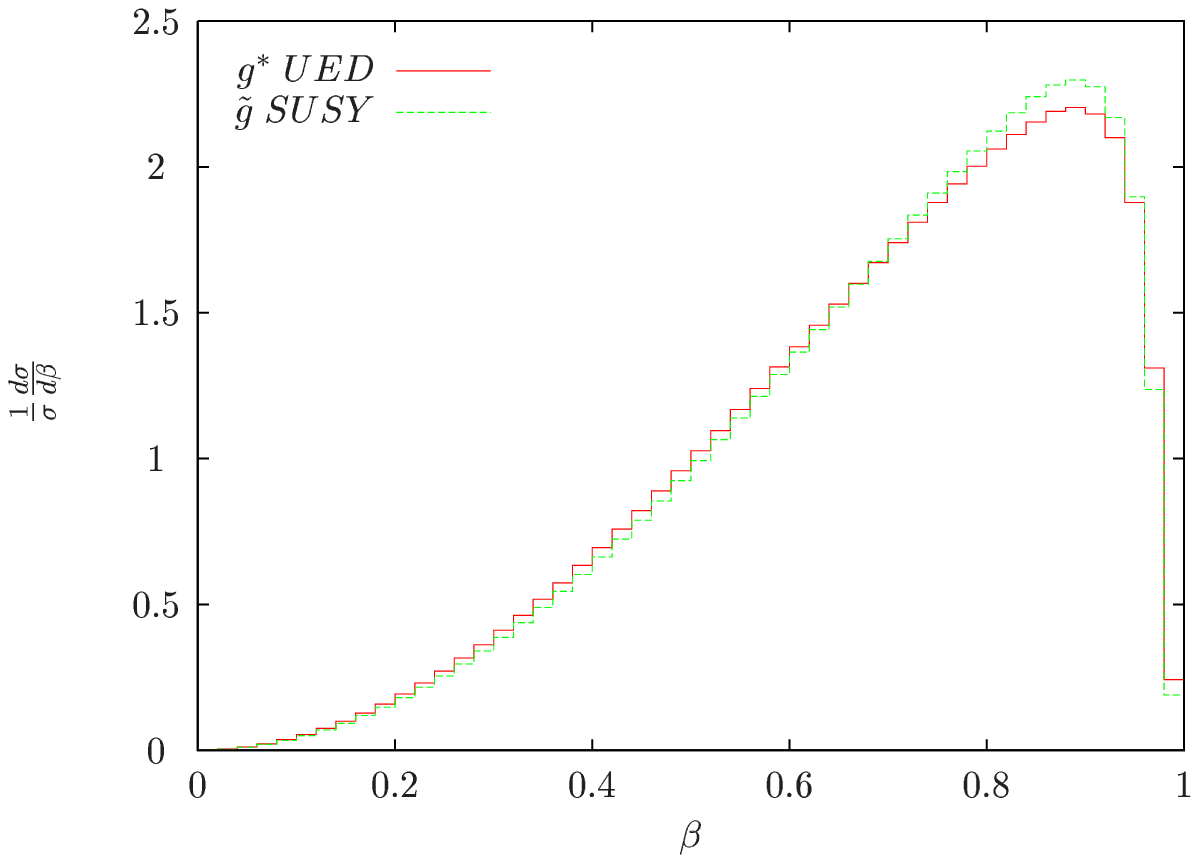}
\vspace*{-5mm}
\caption{Distribution of the boost parameter $\beta$ of the gluino and KK-gluon.
  \label{fig:plot_beta_g.eps}}
\end{figure}
In fig.~\ref{fig:plot_trans_mom_g.eps} we show the differential cross section for the transverse momentum of the gluino, of the light jet in fig.~\ref{fig:light_squark_pt} and of the non-measurable neutralino in fig.~\ref{fig:light_jet_pt}. Of course, neither the transverse momentum nor the invariant mass distributions of the intermediate gluinos/KK-gluons and squarks/KK-quarks can be measured directly in experiment. While the transverse momentum of gluino and KK-gluon are quite equal, there are differences between the transverse momentum distributions of the light quarks in both scenarios and between the transverse momentum distributions of $N_1$ and $B_1$. These differences are due to the different rapidity distributions of the $u$,$d$,$c$ and $s$-squark in SUSY and UED. This can already be observed for our $2 \rightarrow 2$ process in chapter~\ref{ch:4}. The shape of the rapidity distribution close to $y=0$ differs for SUSY and UED and stays unchanged also for the $2 \rightarrow 5$ process. The $N_1$, which is the LSP and the $B_1$, the lightest Kaluza-Klein particle (LKP), can not be observed directly.

From the rapidity distribution in fig.~\ref{fig:plot_rapidity_g.eps} we also do not see significant differences between gluino and KK-gluon. As we expect, these distributions are nearly equal to those of the $2 \rightarrow 2$ process.
%
%
%

As a result the distributions of the boost parameter $\beta$ for the gluino and the KK-gluon, given in fig.~\ref{fig:plot_beta_g.eps}, are quite similar. Here $\beta$ is given by
\begin{equation}
\beta=\sqrt{1-\left(\frac{m}{E}\right)^2} \; .
\end{equation}
One finds that the boost distribution for large values of $\beta$ is slightly higher for the gluinos than for the KK gluons. Our plot of the $\beta$ distribution qualitatively agrees with the one given in~\cite{alves-2006}. There gluon-gluon collisions are taken into account additionally, while we only considered the process of quark-gluon collision. In~\cite{alves-2006} the difference of the boosts seems to be more significant which could be due to the acceptance cuts assumed there. In our analysis we do not take into account any cuts or smearing of the signal due to finite resolution of the detector. In principle, cuts on the $b$-jet momenta are necessary since the detector can not distinguish between both jets if they are very close to each other. One usually uses transverse momentum cuts which can have an impact on the shape of the curves.

\begin{figure}
\centering
\includegraphics[width=0.65\textwidth]{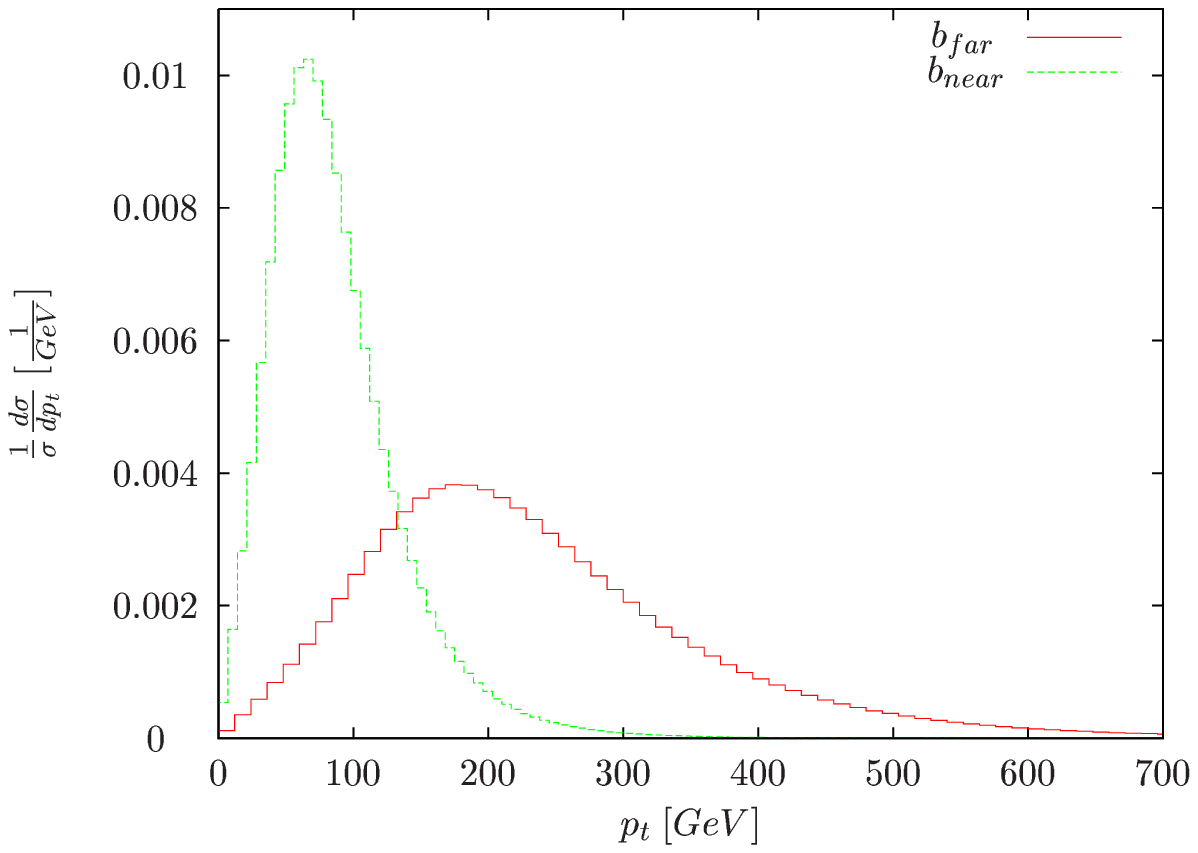}
\vspace*{-5mm}
\caption{Transverse momentum distribution of the near and far $b$-jets in UED.
  \label{fig:both_b_jets}}
\vspace*{+5mm}
\includegraphics[width=0.65\textwidth]{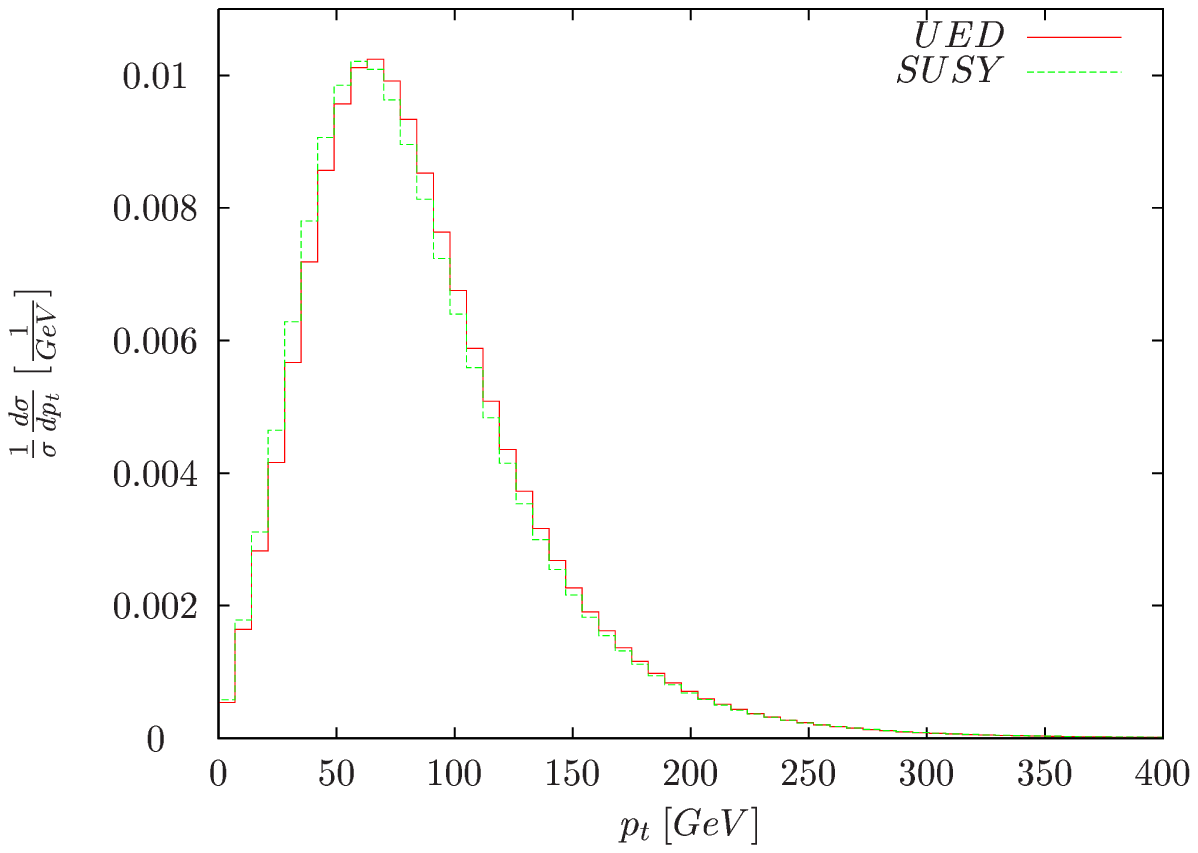}
\vspace*{-5mm}
\caption{Transverse momentum distribution of the outgoing near $b$-jet.
  \label{fig:plot_trans_mom_b_near.eps}}
\end{figure}

\begin{figure}
\centering
\includegraphics[width=0.65\textwidth]{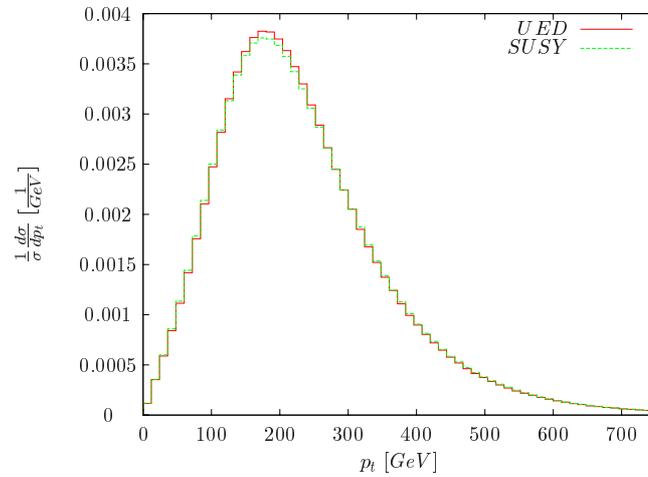}
\vspace*{-5mm}
\caption{Transverse momentum distribution of the outgoing far $b$-jet.
  \label{fig:plot_trans_mom_b_far.eps}}
\end{figure}

The $b$-jet coupling directly to the gluino/KK-gluon, is usually called the \textit{near} $b$-jet while the second $b$-jet, emerging from the decay of the $b$-squark/KK-$b$-quark is called the \textit{far} $b$-jet. They have, due to the mass hierachy in the decay chain, very different transverse momenta. Because of the invariant mass differences between gluino and sbottom and between sbottom and the NLSP, the second $b$-jet is much harder than the first one. This results in very different transverse momentum distributions for the near and the far jet, as shown in fig.~\ref{fig:both_b_jets} for the UED scenario. A comparison of the near and far jet transverse momentum for SUSY and UED is shown in fig.~\ref{fig:plot_trans_mom_b_near.eps} and fig.~\ref{fig:plot_trans_mom_b_far.eps}. One finds that the peaks are only slightly shifted.


\begin{figure}
\centering
\includegraphics[width=0.65\textwidth]{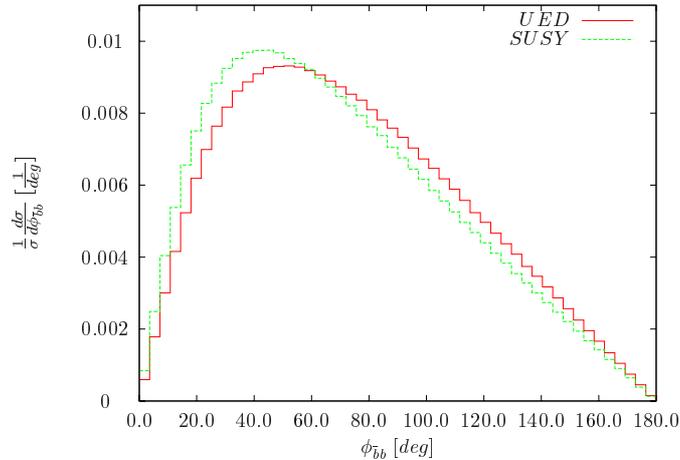}
\vspace*{-5mm}
\caption{Distribution of the angle between the two bottom jets in the lab frame.
  \label{fig:plot_thetabb.eps}}
\vspace*{+5mm}
\end{figure}

\begin{figure}
\centering
\vspace*{0mm}
\includegraphics[width=0.65\textwidth]{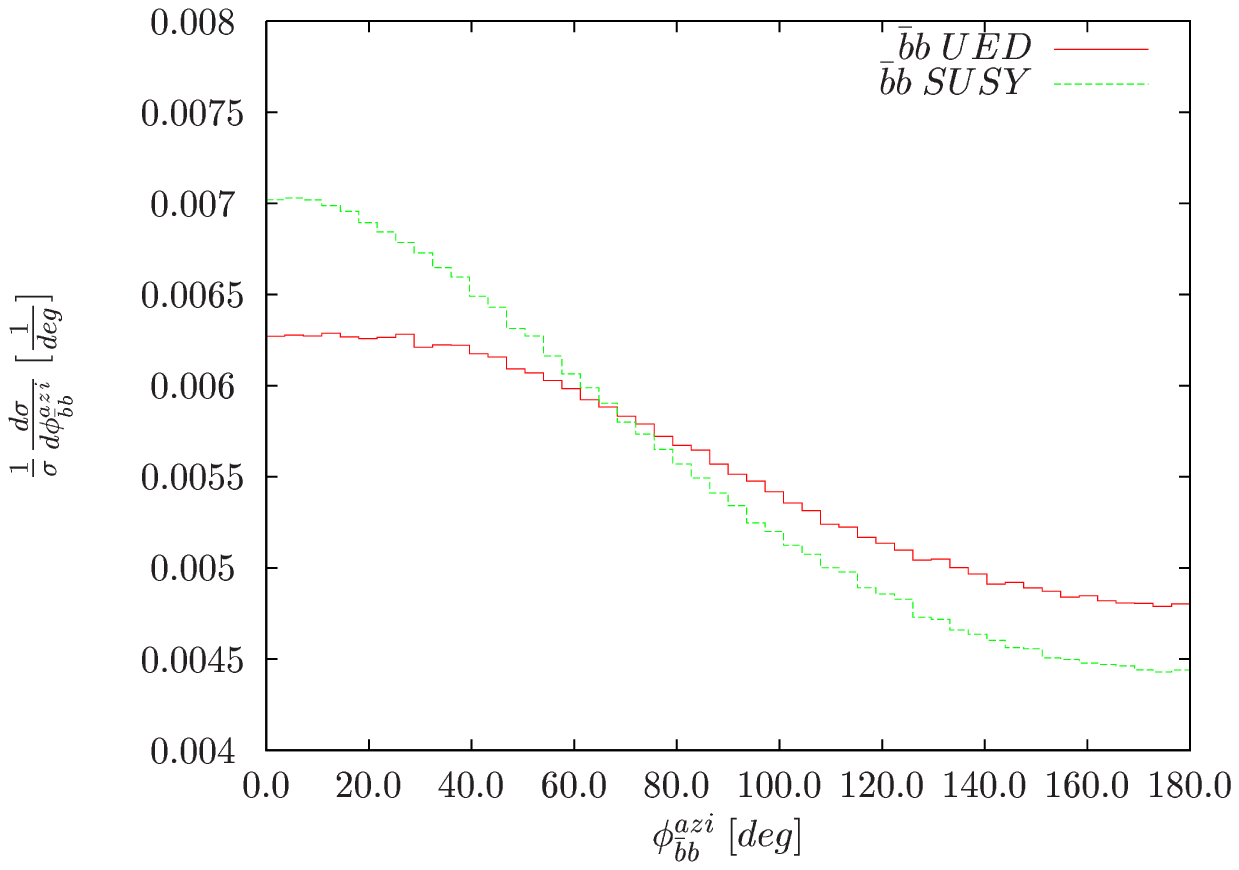}
\vspace*{-5mm}
\caption{Distribution of the azimuth angle between the two bottom jets.
  \label{fig:plot_azimuth.eps}}
\vspace*{+10mm}
\includegraphics[width=0.65\textwidth]{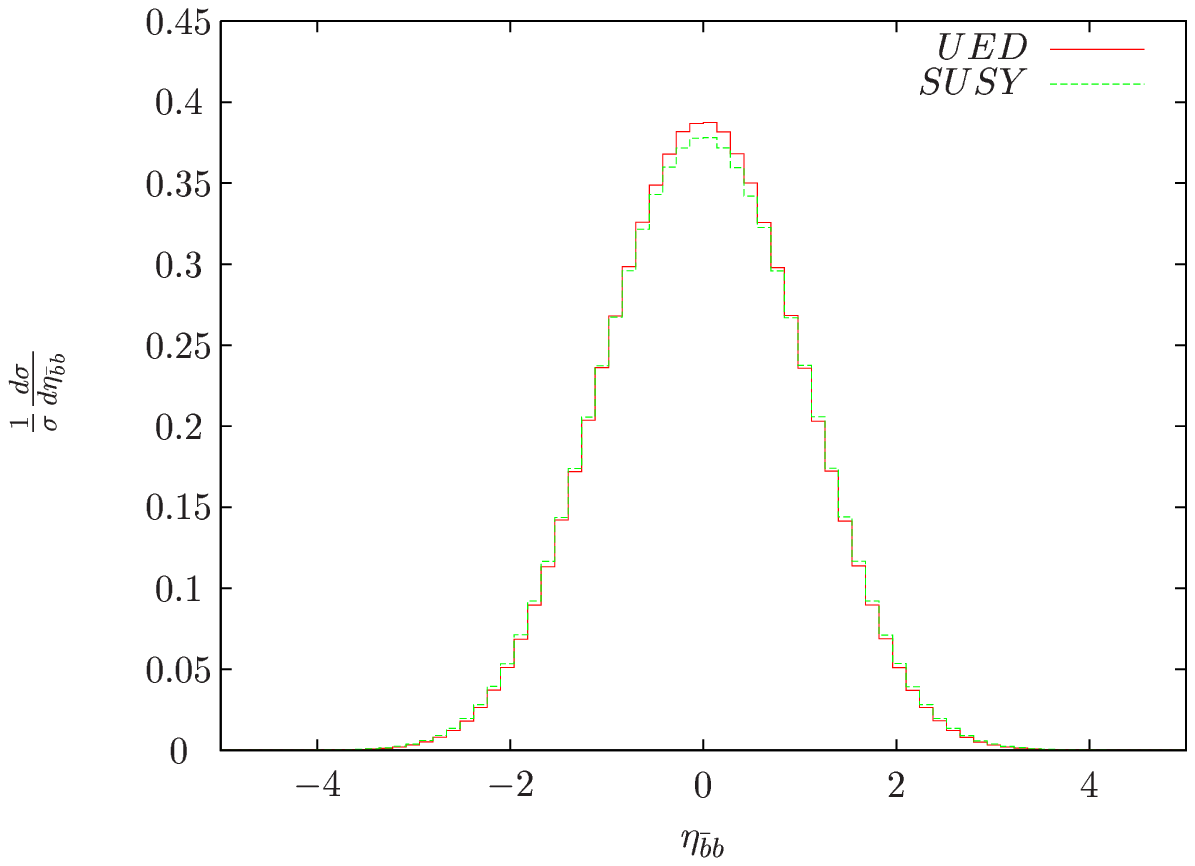}
\vspace*{-5mm}
\caption{Distribution of the average rapidity of the outgoing $b/ \bar{b}$-jets.
  \label{fig:plot_medium_rap.eps}}
\end{figure}

Due to different spins of the gluino and KK-gluon and the $b$-squarks and KK-$b$-quark, we expect to find significant differences in the angle $\theta_{b\bar{b}}$ between these two $b$-jets. Since in our program we boost all momenta to the lab frame, we show the angle between the two $b$-jets in the lab frame in fig.~\ref{fig:plot_thetabb.eps}. This angle $\theta_{b\bar{b}}$ is not Lorentz invariant under boosts along the beam axis. But this is true for the azimuth angle $\theta_{b\bar{b}}^{azi}$ between the two jets, i.e.\ the angle in the plane orthogonal to the beam axis. The azimuth angle is plotted in fig.~\ref{fig:plot_azimuth.eps}. 
For the calculation of $\theta_{b\bar{b}}$ we used
\begin{equation}
\theta_{b\bar{b}} \;  = \; \text{arccos} \left(\; \frac{\vec{p}_b \cdot \vec{p}_{\bar{b}}}{\vert \vec{p}_b \vert \vert \vec{p}_{\bar{b}} \vert} \; \right) \quad \text{with} \quad 0^{\circ}<\theta_{b\bar{b}}<180^{\circ} \; .
\end{equation}
The same relation is used for $\theta_{b\bar{b}}^{azi}$ with the third component of both vectors equal to zero, which is the direction of our beam axis.

In both plots we find significant differences between the two models. It seems that a large part of this effect is not due to the different boosts $\beta$, which are quite close for both models. The origin of this difference in the angular distribution is discussed in more detail in the following sections. The angular distribution given in~\cite{alves-2006} can be distinguished equally well, again containing quark-gluon and gluon-gluon collisions. The plot for the azimuth angle looks different in~\cite{alves-2006} which could again be due to the cuts on the $b$-jet momenta. To quantify the difference between the angular distributions in SUSY and UED the following asymmetry seems to be a reasonable measure:
\begin{equation}
\mathcal{A}=\frac{\sigma(\Delta \phi_{b\bar{b}}<90^{\circ})-\sigma(\Delta \phi_{b\bar{b}}>90^{\circ})}{\sigma(\Delta \phi_{b\bar{b}}<90^{\circ})+\sigma(\Delta \phi_{b\bar{b}}>90^{\circ})} \; .
\end{equation}
For SUSY we obtain $\mathcal{A}=0.1443$ while for UED we find $\mathcal{A}=0.0903$ in the case of mixing angle $\alpha^{(1)}=0^{\circ}$. 

As in~\cite{alves-2006} we also present the distribution of the averaged rapidity of the outgoing $b$- and $\bar{b}$-jets, given by
\begin{equation}
\eta=\frac{y_{b}+y_{\bar{b}}}{2} \; .
\end{equation}
While they are different in~\cite{alves-2006}, we do not find a significant difference between them in fig.~\ref{fig:plot_medium_rap.eps}.

\begin{figure}
\centering
\vspace*{+5mm}
\includegraphics[width=0.65\textwidth]{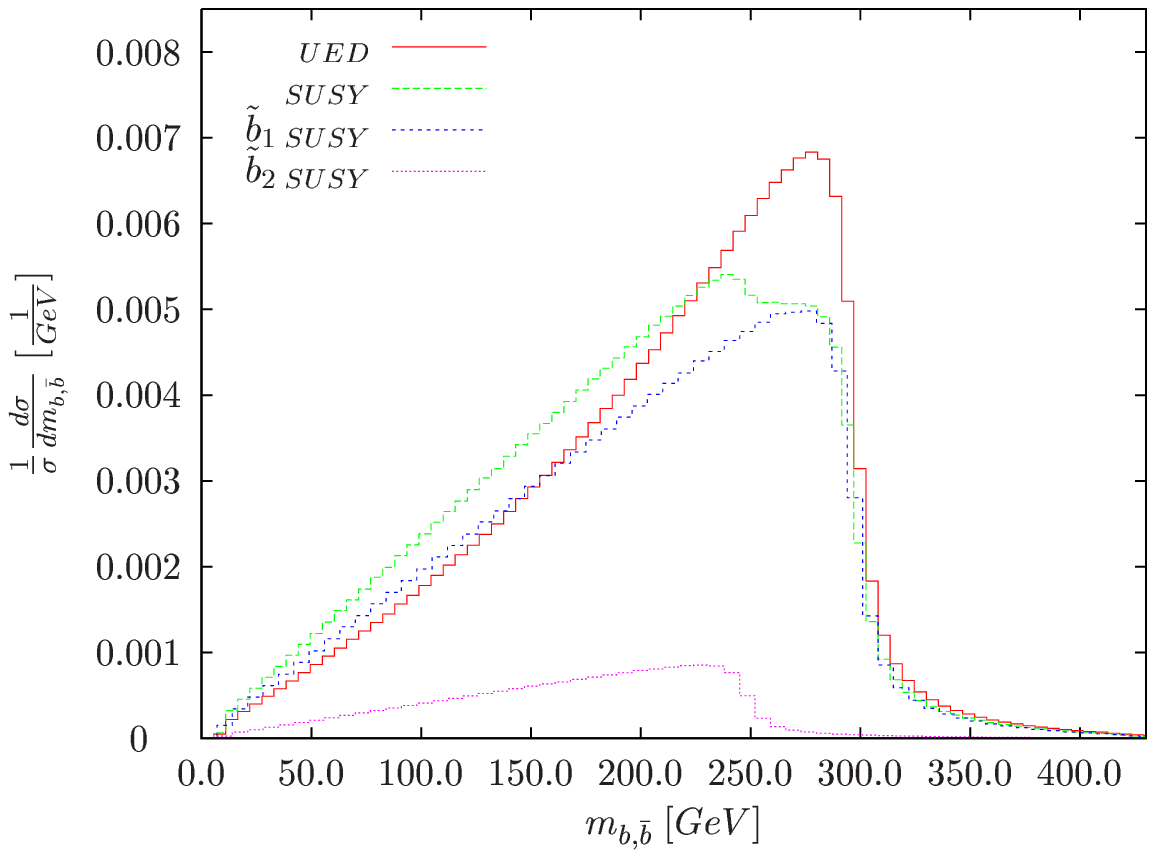}
\vspace*{-5mm}
\caption{Invariant mass distribution of the $b \bar{b}$-jet pair.
  \label{fig:plot_inv_Masse_b_jets.eps}}
\end{figure}

As a last plot we present the differential cross section for the invariant mass of the two bottom jets in fig.~\ref{fig:plot_inv_Masse_b_jets.eps}. We already mentioned that we assume all masses to be already measured in this thesis.  At the LHC, masses are extracted from invariant masses of outgoing particles. As discussed in references~\cite{lester} and~\cite{hinchliffe-1999-60}, the edges of invariant mass distributions will be used to measure the masses of the particles. In the case of the gluino decaying into two $b$-jets and the NSLP, the edge is given by
\begin{equation}
m^{max}_{b \bar{b}}=m_{\tilde{g}} \;  \sqrt{1-\frac{m^2_{\tilde{b}}}{m^2_{\tilde{g}}}} \; \sqrt{1-\frac{m^2_{N_2}}{m^2_{\tilde{b}}}}
\end{equation}
and equally in the case of UED. Using the SPS 1a mass spectrum, we find a value of $m_{b \bar{b}}=296.62 \; \text{GeV}$ for the right edge with an intermediate $\tilde{b}_1$. For the left edge with an intermediate $\tilde{b}_2$ one finds $m_{b \bar{b}}=247.93 \; \text{GeV}$. This agrees quite well with the edges in fig.~\ref{fig:plot_inv_Masse_b_jets.eps}.


The small amount of events lying on the right side of the edges is due to off shell effects, i.e.\ when the gluino/KK-quark or the sbottom/KK-$b$-quark is off-shell the $b$-jet momentum can be increased. We checked that the number of points on the right side of the edge lowers if the integration parameter $n$, the number of widths integrated over as given in eq.~(\ref{n_parameter}), is reduced. Of course, there is only one KK-$b$-quark contributing to the invariant mass distribution in the case of $\alpha^{(0)}=0^{\circ}$.


\subsection{A SUSY-UED Comparison for $\alpha^{(1)}=45^{\circ}$}
Until now we always assumed that there is no mixing between the two KK-quark towers. In the following we present the results for the maximal mixing of \mbox{$\alpha^{(1)}=45^{\circ}$}. 

\begin{figure}
\centering
\vspace*{+5mm}
\includegraphics[width=0.65\textwidth]{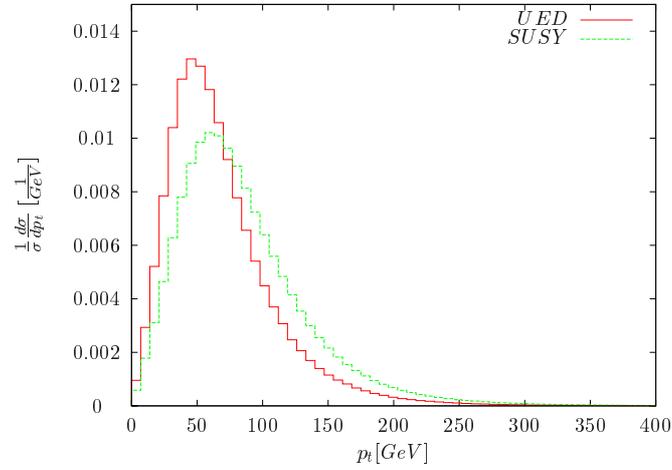}
\vspace*{-5mm}
\caption{Transverse momentum distribution of the outgoing near $b$-jet.
  \label{fig:gnu_plots/Chapter_6_last/45_Grad_schmaler/plot_trans_mom_b_near.eps}}
\end{figure}

\begin{figure}
\centering
\vspace*{+5mm}
\includegraphics[width=0.65\textwidth]{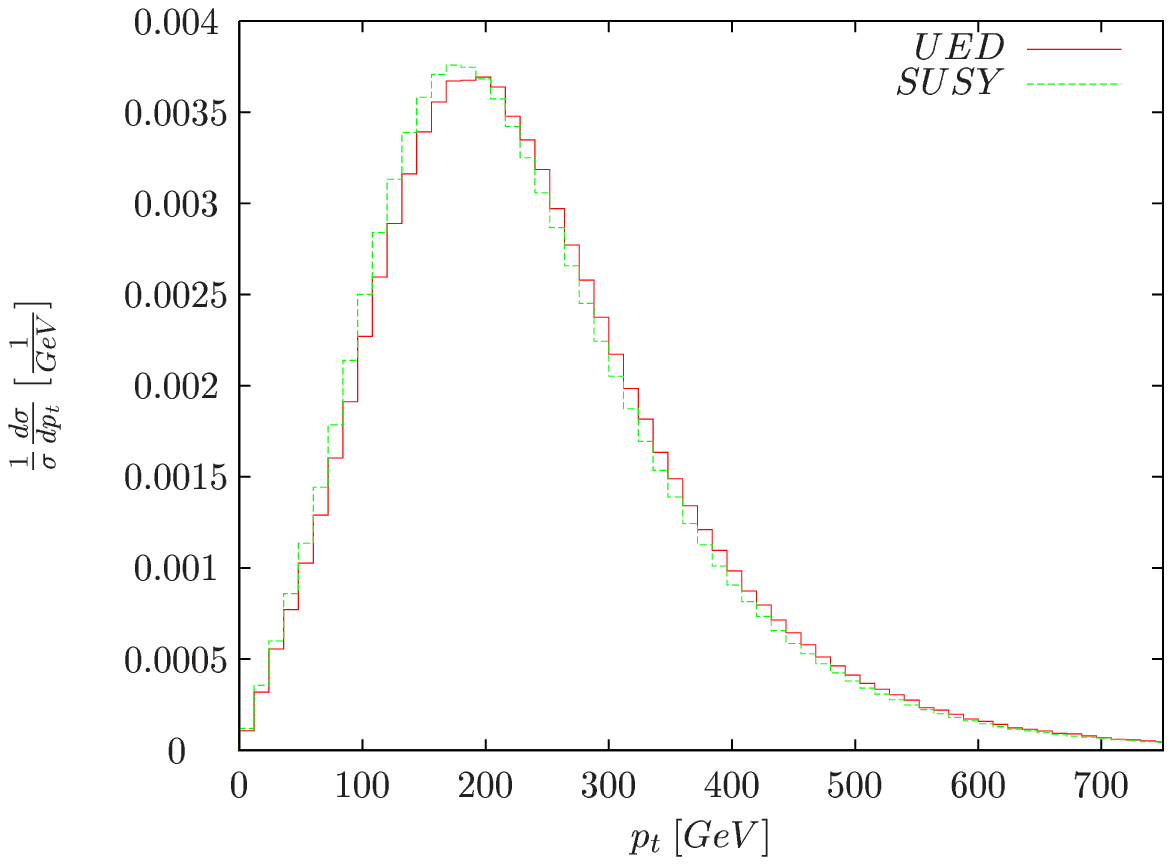}
\vspace*{-5mm}
\caption{Transverse momentum distribution of the outgoing far $b$-jet.
  \label{fig:gnu_plots/Chapter_6_last/45_Grad_schmaler/plot_trans_mom_b_far.eps}}
\end{figure}

\begin{figure}
\centering
\vspace*{+5mm}
\includegraphics[width=0.65\textwidth]{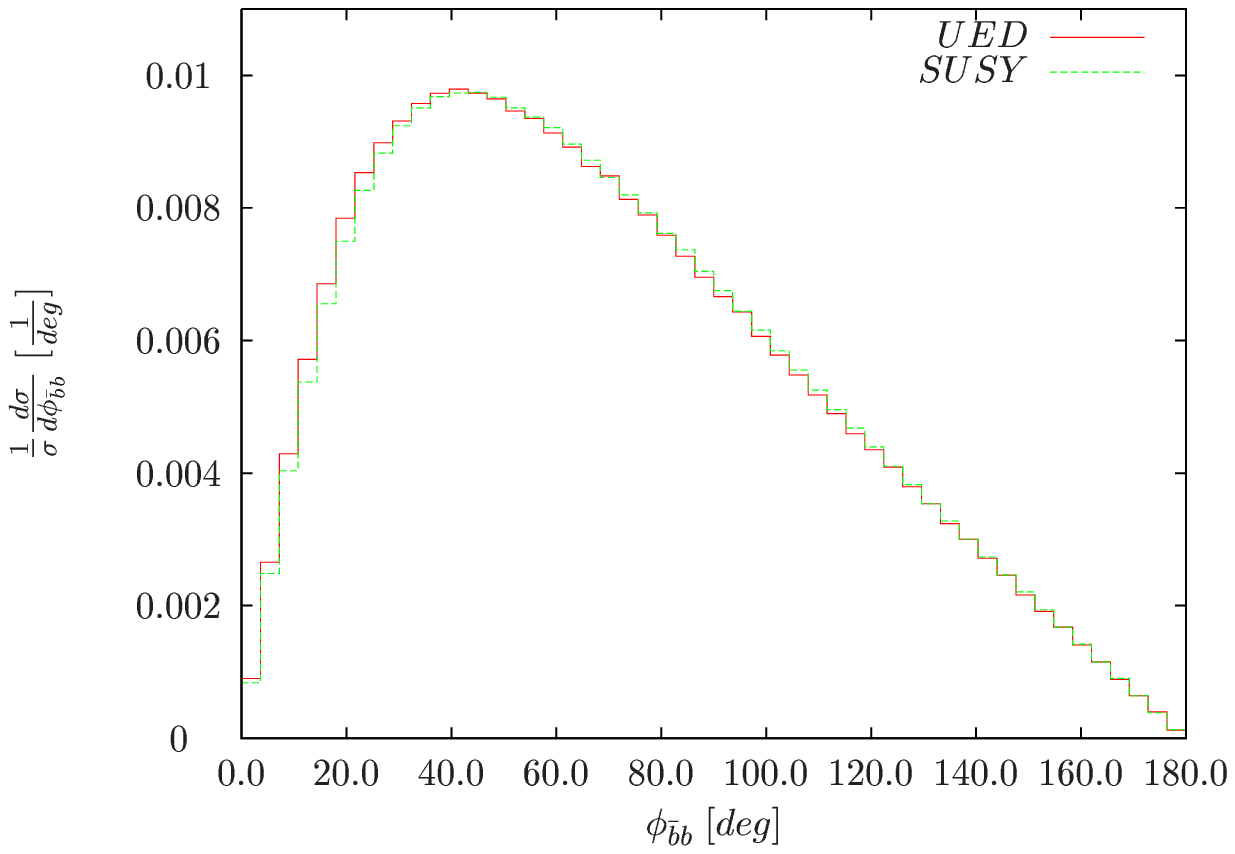}
\vspace*{-5mm}
\caption{Distribution of the angle between the two bottom jets in the lab frame.
  \label{fig:gnu_plots/Chapter_6_last/45_Grad_schmaler/plot_thetabb.eps}}
\end{figure}

\begin{figure}
\centering
\vspace*{+5mm}
\includegraphics[width=0.65\textwidth]{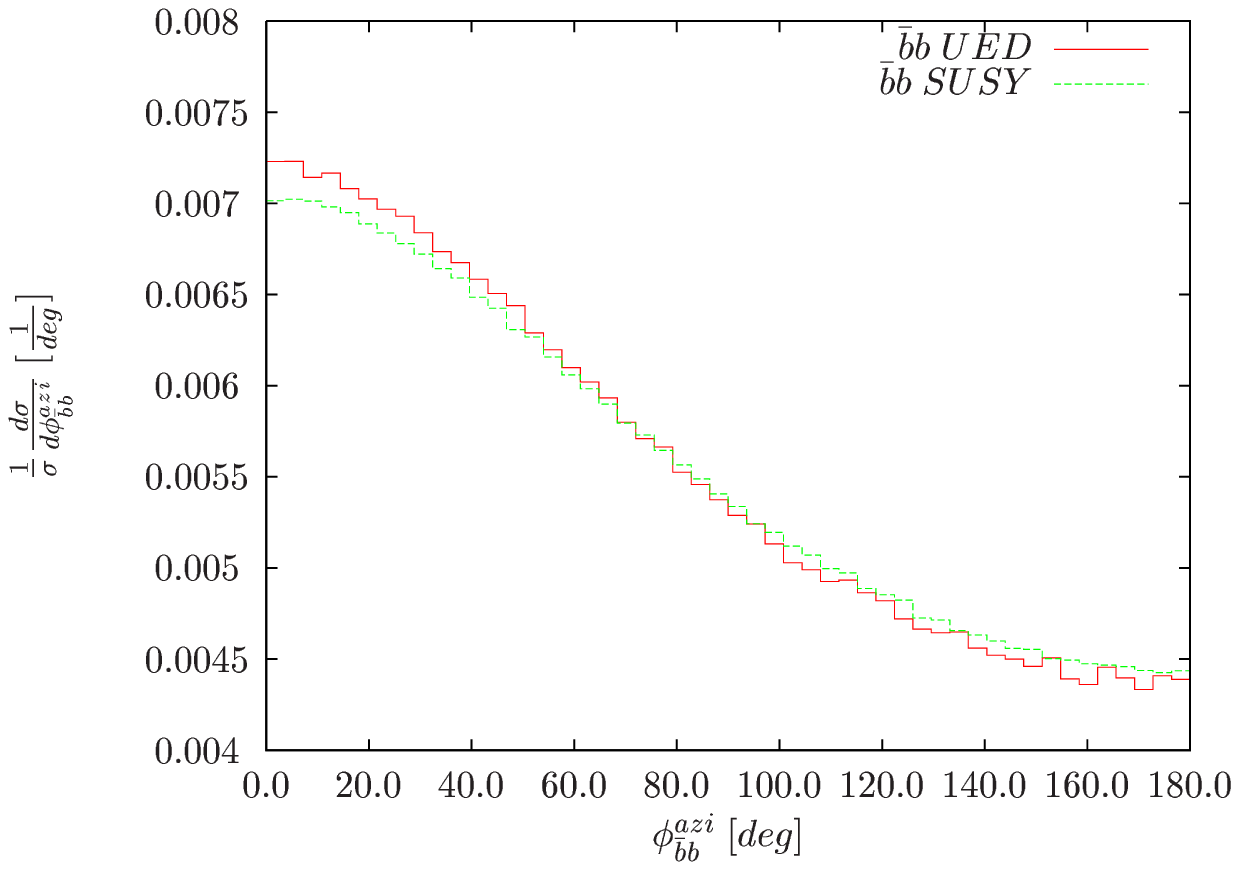}
\vspace*{-5mm}
\caption{Distribution of the azimuth angle between the two bottom jets.
  \label{fig:gnu_plots/Chapter_6_last/45_Grad_schmaler/plot_azimuth.eps}}
\end{figure}

One finds significantly different angular distributions. Since the couplings of all squarks change, the distributions of $p_t$ and $y$ of the $b$-jets also change. Their transverse momentum distributions are given in fig.~\ref{fig:gnu_plots/Chapter_6_last/45_Grad_schmaler/plot_trans_mom_b_near.eps} and fig.~\ref{fig:gnu_plots/Chapter_6_last/45_Grad_schmaler/plot_trans_mom_b_far.eps}. Especially the transverse momentum distribution of the near $b$-jet exhibit a different shape. 
In fig.~\ref{fig:gnu_plots/Chapter_6_last/45_Grad_schmaler/plot_thetabb.eps} and fig.~\ref{fig:gnu_plots/Chapter_6_last/45_Grad_schmaler/plot_azimuth.eps} we find the differential cross sections for the angle between both $b$-jets in the lab frame and the azimuth angle as defined before. Both curves can not be discriminated as in the case of $0^{\circ}$ mixing angle. Obviously the coupling structure seems to have a strong impact on the angular distributions.

\begin{figure}
\centering
\vspace*{+5mm}
\includegraphics[width=0.65\textwidth]{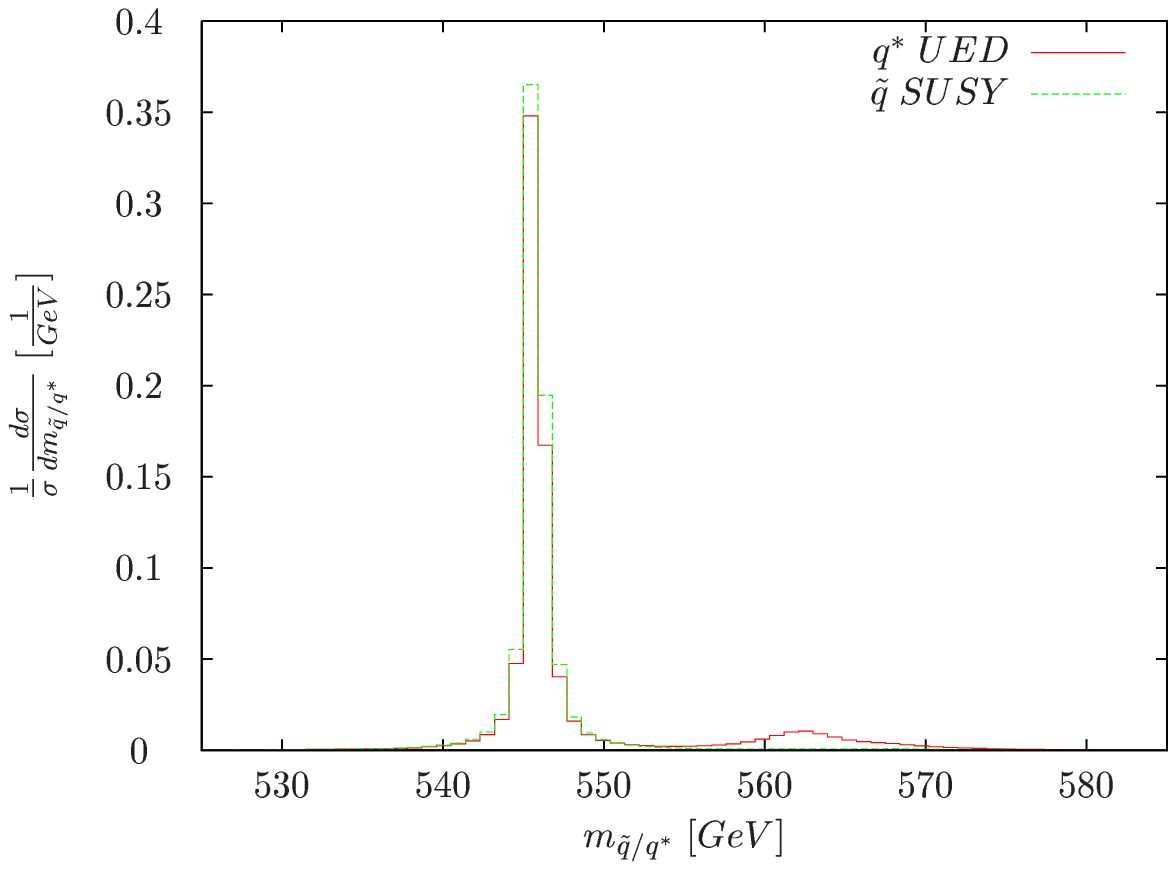}
\vspace*{-5mm}
\caption{Invariant mass distribution of the $u$,$d$,$c$ and $s$-squark/KK-quark.
  \label{fig:gnu_plots/Chapter_6_last/45_Grad_schmaler/plot_m_usq.eps}}
\end{figure}

The distribution of the boost parameter $\beta$ is found to be equal to the case of $\alpha^{(1)}=0^{\circ}$. Other distributions like invariant mass of the gluino do not change either. The invariant mass of the squarks is somewhat special since there are two masses that can be chosen in two different ways. For our calculations we choose the masses of $b_{1,1}$ and $b_{1,2}$ such that the invariant mass distribution in the case of $\alpha^{(1)}=0^{\circ}$ has its highest peak at the same place as it is the case in SUSY. This is obvious from fig.~\ref{fig:plot_m_sb}. The same choice applies to the invariant mass of the $u$,$d$,$c$ and $s$ squark/KK-quark. Its invariant mass for $\alpha^{(1)}=45^{\circ}$ is shown in fig.~\ref{fig:gnu_plots/Chapter_6_last/45_Grad_schmaler/plot_m_usq.eps}. 

\begin{figure}
\centering
\vspace*{5mm}
\includegraphics[width=0.65\textwidth]{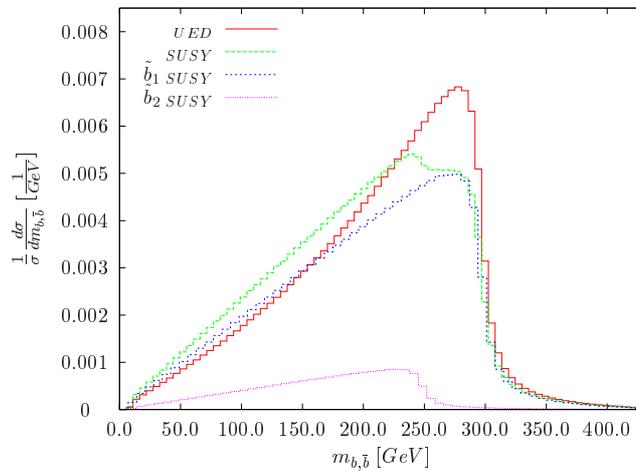}
\vspace*{-5mm}
\caption{Invariant mass distribution of the $b \bar{b}$-jet pair.
  \label{fig:gnu_plots/Chapter_6_last/45_Grad_schmaler/plot_inv_Masse_b_jets.eps}}
\end{figure}

The invariant mass distribution of the $b$-jets for $\alpha^{(1)}=45^{\circ}$ also looks significantly different. In fig.~\ref{fig:gnu_plots/Chapter_6_last/45_Grad_schmaler/plot_inv_Masse_b_jets.eps} one finds that the main contribution to the UED cross section derives from an intermediate $b_{1,2}$. But due to mixing there is also a contribution from an intermediate $b_{1,1}$ as opposed to the case of $\alpha^{(1)}=0^{\circ}$.

\clearpage

\section{Influence from the Boost of the Gluino/KK-Gluon}
As mentioned in~\cite{csaki-2007} the difference in the angular distributions in fig.~\ref{fig:plot_azimuth.eps} could be due to the boost and the helicity structure of the squark couplings. To find out how strong the influence from the boost really is, we map the UED boost onto the curve of the SUSY boost by multiplying the internal Vegas phase space weights with the ratio of the height of the histogram bins from SUSY and UED. The same procedure is applied to all distributions, multiplying them for each point in phase space with the appropriate factor belonging to the bin of the boost at that phase space point. This is done for each phase space point evaluated by Vegas. Thereby we obtain all other distributions, especially the angular distributions, assuming at the same time that the boosts for the gluino and the KK-gluon are equal. The boost distributions are then \textit{by construction} given by the SUSY curve in fig.~\ref{fig:plot_beta_g.eps}.

\begin{figure}[h!]
\label{fig:gnu_plots/Chapter_6_last/Helicity_plot/plot_azimuth.eps}
\centering
\vspace*{20mm}
\vspace*{0mm}\raisebox{+0.0ex}{ \includegraphics[width=0.65\textwidth]{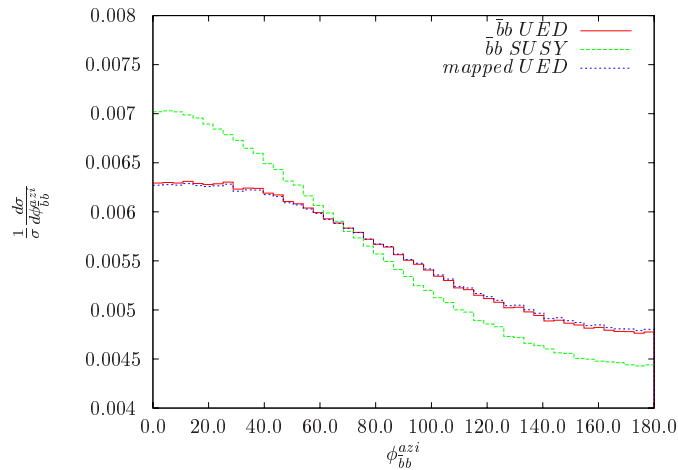}}
\vspace*{0mm}
\caption{Distribution of the azimuth angle between the two bottom jets.}
\end{figure}

The distributions for the azimuth angle between both jets are given in fig.~\ref{fig:gnu_plots/Chapter_6_last/Helicity_plot/plot_azimuth.eps}. We find that the influence from the boost is negligible because the boost of the gluino and the KK-gluon are quite similar. Therefore we expect the difference in the angular distributions to be mainly determined by the coupling structure of the KK-quarks and the different helicity eigenstates of the outgoing $N_1$ and $N_2$, respectively $B_1$ and $A_{3,1}$. Other kinematic observables do not change either, when the UED boost for $\alpha^{(1)}=0^{(\circ)}$ is mapped onto the SUSY boost.

The strong influence from the coupling structure can also be seen in section~\ref{sec:The_SUSY_UED_COMPARISION}, where a mixing angle of $\alpha^{(1)}=45^{\circ}$ is used. The angular distributions are much closer in this case, though then other kinematic distributions, like the transverse momentum of the near $b$-jets, significantly differ.
  \clearpage{\pagestyle{empty}\cleardoublepage}

  \chapter{Conclusions}
In order to distinguish between different scenarios of new physics at the LHC, it is important to measure as many properties as possible of all new particles. In this thesis we concentrated on the determination of the spin of an intermediate particle by using a decay chain. The gluino is especially interesting for SUSY searches at the LHC, since it is a Majorana fermion which can produce like-sign dileptons in the final state. In UED the like-sign dilepton signature could occur as well from the decay of a bosonic KK-gluon. Therefore it is crucial to measure the spin of the decaying particle to differentiate between different  beyond-Standard-Model scenarios.

In the case of SUSY, we calculated the partonic cross sections of all $2 \rightarrow 2$ SUSY-QCD processes. The quark- gluon collision was analyzed numerically on the hadronic level. After calculating the partonic cross section for the production of a KK-quark-gluon pair in UED, a comparison to the corresponding SUSY process was performed at the production threshold. It was found that the UED $2 \rightarrow 2$ cross section for quark-gluon collision increases much faster at the threshold, compared to the one in SUSY. In a comparison of the squark and KK-quark rapidities a slight difference between the curves could be identified.

After explaining general kinematics in decay chains, we demonstrated how to test a decay chain program by using invariant masses, transverse momenta and rapidity distributions of intermediate particles. The shape of the curves of an intermediate particle in a chain does not change if the chain is extended by an additional decay. This is due to our integration procedure, only considering momenta close to the Breit-Wigner pole. Within this argumentation the properties of the Breit-Wigner propagator are crucial. Since a Breit-Wigner propagator effectively lowers the order in the coupling constant, some topologies can be neglected as they are effectively of higher order. Our program was compared to the narrow width approximation, i.e. only considering the on-shell contributions by multiplication of the particle production by the branching fractions.

Hadronic cross sections were compared for the case of a well-known decay chain involving the SUSY gluino or a UED KK-gluon and assuming equal masses for the corresponding particles in both scenarios. In a comparison of this decay chain for SUSY and UED we could show that significant differences in the distributions of the azimuth angle between the outgoing $b$-jets occur. As a consequence, we showed that it is possible to distinguish a SUSY gluino and a UED KK-gluon by using angular correlations and assuming a naturally mass suppressed UED mixing angle of $\alpha^{(1)}=0^{\circ}$. 

By mapping the boosts of UED and SUSY onto each other, a comparison of the distributions of the azimuth angle between the $b$-jets and other kinematic distributions is performed while effects from the difference in the gluino and KK-gluon boost are effectively eliminated. We found that the difference in the boost distributions is too small and its effect is not large enough to cause significant differences in the azimuth angle between the outgoing $b$-jets. 

For the outgoing light jet and the lightest supersymmetric particle, respectively the lightest Kaluza-Klein particle, we found differences in the kinematic distributions. These derive from differences in the rapidity distribution of the squark and KK-quark in the final state of the $2 \rightarrow 2$ process. As we could show, this is most probably due to different angular distributions that can already be found on the partonic level in the $2 \rightarrow 2$ process.

It seems to be possible to make the azimuth angle distributions from SUSY and UED more similar by changing the mixing angle between the KK-quark towers for singlets and doublets. In the case of $\alpha^{(1)}=45^{\circ}$ the transverse momentum distribution of the near $b$-jet shows significant differences for SUSY and UED, while the azimuth angles between the two outgoing $b$-jets are nearly equal. Therefore it is most likely that the coupling structure and additional helicity states in the final state of the UED decay chain cause the differences in the angular distributions. The boost of the gluino and the KK-gluon remains nearly without any measurable effect.

Though we could show that distinguishing the chains in a SUSY and UED scenario by kinematic observables is indeed possible, a more complete treatment, including off-shell effects, background effects and simulation of the finite resolution of the detector, would be needed to obtain a measurable result. Including next-to-leading order effects would improve the accuracy of the calculation. Especially QCD corrections can have a considerable size.
  \clearpage{\pagestyle{empty}\cleardoublepage}

\appendix
\renewcommand{\myheaderpart}{\appendixname}
\renewcommand{\appendixname}{Chapter}
\fancyhead[C]{Appendix}
  \chapter{Appendix}
\section{Color Factors in QCD-couplings}
Quantum Chromodynamics is a non-abelian gauge theory, i.e.\ there are symmetry transformations 
\begin{equation}
\psi \rightarrow e^{i \alpha^k t^k} \psi
\end{equation}
which leave the theory unchanged. The $t^k$ are called symmetry generators. The symmetry in the color space of QCD has an underlying $\text{SU}(N)$ symmetry with $N=3$. Two of their representations, occurring in the QCD Feynman rules, are the fundamental and the adjoint representation. Quark fields transform under the former while gluon fields transform under the latter representation. When calculating QCD diagrams, sums over colors and flavors, i.e.\ traces over the symmetry generators, appear in the formulae. We give the formulae which were used in the calculations of the partonic cross sections in chapter~\ref{ch:4}.
The generators of the fundamental representation are denoted as $t^a$. The appropriate Lie-algebra is given by
\begin{equation}
\lbrack t^a,t^b \rbrack=t^a t^b-t^b t^a=i f^{abc} t^c \; ,
\end{equation}
where $f^{abc}$ are the structure constants of the Lie-algebra.
In the following formulae, the lower indices of $t_{ij}^k$ are always implicit.
\begin{align}
tr(t^a t^b t^a t^b)&=tr\left( \left(C_2(r)-\frac{1}{2} C_2(G) \right) t^a t^a \right)=-\frac{2}{3} \nonumber\\
tr(t^a t^b t^c)&= \left (\frac{d^{abc}}{4}+\frac{i}{4} f^{abc} \right )\nonumber\\
tr(t^a t^b t^b t^a)&=3 \left( C_2(r) \right )^2=\frac{16}{3}\\
f^{acd}f^{bcd}&=C_2(G) \delta^{ab}\nonumber\\
t^a_r t^a_r&=C_2(r) \cdot \textbf{1}\nonumber
\end{align}
with \textbf{1} being the $d(r) \times d(r)$ unit matrix and $C_2(r)$ and $C_2(G)$ being the quadratic Casimir operators of both representations. The structure constant $f^{abc}$ is totally antisymmetric while $d^{abc}$ is totally symmetric.
\newpage
\hspace*{-20pt} The Casimir operators for the fundamental representation are given by
\begin{equation}
C(N)=\frac{1}{2}\; \; , \qquad C_2(N)=\frac{N^2-1}{2 N} \; .
\end{equation}
The Casimir operators for the adjoint representation are given by
\begin{equation}
C_2(G)=C(G)=N \; .
\end{equation}
The so called Jacobi identity gives
\begin{equation}
\label{eq:jacobi}
f^{ade}f^{bcd}+f^{bde}f^{cad}+f^{cde}f^{abd}=0 \; .
\end{equation}

\section{An Explicit Calculation with External Gluons}
\label{Explicit Calculation with External Gluons}
When more than one external gluon is involved in a calculation, one has to properly take into account their non-physical longitudinal degrees of freedom. These longitudinal states come from the fact that Quantum Chromodynamics (QCD) is a non abelian gauge theory. They appear in the path integral quantization of QCD as non-physical fields usually called ghosts, spin zero fields with Fermi-Dirac statistics. In QED path integral quantization ghost fields also appear but decouple from the rest of the theory.

Since calculating squared matrix elements by hand is a tedious work if one has to use the transverse polarization sum given in eq.~(\ref{eq:pol_sum}), it would be nice if one could simply use a shorter polarization sum equal to the one in QED, $-g_{\mu \nu}$. In general the arbitrary vector $n_{\mu}$ has to drop out in the end since the result can not depend on this arbitrary, non-physical object. The result has to be gauge independent. Since the additional terms in the polarization sum given in eq.~(\ref{eq:pol_sum}) take out the longitudinal degrees of freedom, one has to do this by hand, if $-g_{\mu \nu}$ is used instead.

In leading order matrix elements this is possible by using a trick called ghost subtraction. Here the non-physical polarizations of the gluon are explicitly subtracted from the matrix element by crossing out all terms proportional to momenta which have the Dirac index of the polarization vector and the same momentum. These are exactly those terms in the matrix element that drop out when eq.~(\ref{eq:pol_sum}) is used. It corresponds to using eq.~(\ref{eq:transmom}) already at the amplitude level before squaring the matrix element, i.e.\ a projection on physical degrees of freedom. The only exception is the case that all external particles are gluons. Then one still has to use eq.~(\ref{eq:pol_sum}) for one of them.

Longitudinal degrees of freedom are taken out in order to make the (on-shell)-Ward-identities from QED also hold for QCD. If $M(k)=\epsilon_{\mu}(k) \mathcal{M}^{\mu}(k)$ is the amplitude for a given QED process with an external photon $\epsilon_{\mu}$, the amplitude vanishes if $\epsilon_{\mu}$ is replaced by $k_{\mu}$, i.e.\ it stays unchanged under the gauge transformation $\epsilon_{\mu}=\epsilon_{\mu}+k_{\mu}$. The so-called Ward identity is then denoted by 
\begin{equation}
k_{\mu} \mathcal{M}^{\mu}(k)=0 \; .
\end{equation}
In the case of QCD the equivalent relations are called Slavnov-Taylor identities. 

Following an argument in~\cite{Peskin}, we want to explain how ghost subtraction explicitly works. Therefore we check the Ward identity explicitly for the example of gluino-gluino production from gluon-gluon collision. In this case the physical incoming gluons must have transverse polarization since they are on-shell. The matrix elements on Born level for the three contributing diagrams in fig.~\ref{fig:SUSY_production} e), are given by
\begin{align*}
&i \hspace*{0mm}{\mathcal{M}}^{\mu \nu}_{\text{1.diagram}} \epsilon_{\mu}(k_1) \epsilon_{\nu}(k_2)=\\
 &  \hphantom{abcdesft}\hspace*{-20pt}=(-g_s) f^{abc} \big [g^{\mu \nu} (k_1-k_2)^{\rho'}+g^{\nu \rho'} (k_2+k_1+k_2)^{\mu}+g^{\rho' \mu} (-k_1-k_2-k_1)^{\nu} \big]\\ 
 & \hphantom{abjsdfg} \hspace*{-14pt}\times \frac{-i g_{\rho' \eta} \delta_{c k}}{(k_1+k_2)^2} \bar{u}(p_2) (-g_s)f^{kji} \gamma^{\eta} v(p_1) \epsilon_{\mu}(k_1) \epsilon_{\nu}(k_2) \; ,\\
\end{align*}
\begin{align*}
\vspace*{+35pt}
\hspace*{-45pt} & \hspace*{-20pt} i\hspace*{0mm} {\mathcal{M}}^{\mu \nu}_{\text{2.diagram}} \epsilon_{\mu}(k_1) \epsilon_{\nu}(k_2)=\\
&  \hphantom{abdfggdft}\hspace*{-20pt} =\bar{u}(p_2) \gamma^{\nu} f^{bjk} (-g_s) \frac{i(\slashed{p}+m_{\tilde g}) \delta_{c k}}{p^2-m_{\tilde g}^2+i \epsilon} \gamma^{\mu} (-g_s) f^{aci} v(p_1) \epsilon_{\nu}(k_2) \epsilon_{\mu}(k_1)\; ,\\
\end{align*}
\begin{align*}
\vspace*{+35pt}
\hspace*{-55pt}& \hspace*{-12pt} i \hspace*{0mm}{\mathcal{M}}^{\mu \nu}_{\text{3.diagram}} \epsilon_{\mu}(k_1) \epsilon_{\nu}(k_2)=\\
&  \hphantom{adfhgcdft}\hspace*{-20pt}= \bar{u}(p_2) \gamma^{\mu} f^{bki} (-g_s) \frac{i(-\slashed{p}^{\, '}+m_{\tilde g}) \delta_{c k}}{p'^2-m_{\tilde g}^2+i \epsilon} \gamma^{\nu} (-g_s) f^{ajc} v(p_1) \epsilon_{\nu}(k_2) \epsilon_{\mu}(k_1)\; ,
\end{align*}
\vspace{-20pt}
\bigskip
\\
with 
\begin{equation*}
p=k_1-p_1 \qquad \text{and} \qquad p'=k_1-p_2 \; .
\end{equation*}
\vspace{-25pt}
\bigskip
\\
The last two diagrams sum to 
\begin{equation*}
\begin{aligned}
i {\mathcal{M}}_{2,3}^{\mu \nu} \epsilon_{\mu}(k_1) \epsilon_{\nu}(k_2) &= \bar{u}(p_2) \Big[ \gamma^{\nu} (-g_s) f^{bjk} \frac{i (\slashed{p}+m_{\tilde g}) \delta_{ck}}{p^2-m_{\tilde g}^2} \gamma^{\mu} (-g_s) f^{aci}+ \\
& + \gamma_{\mu} f^{bki} (-g_s) \frac{(-\slashed{p}^{\; '}+m_{\tilde g}) \delta_{ck}}{p'^2-m_{\tilde g}^2}(-g_s) f^{ajc} \Big] v(p_1) \epsilon_{\mu}(k_1) \epsilon_{\nu}(k_2) \; .
\end{aligned}
\end{equation*}
Using eq.~(\ref{eq:transmom}) and for the gluinos
\begin{equation}
\label{eq:Dirac}
(\slashed{p}-m) u(p)=0 \qquad \text{and} \qquad \bar{v}(p)\, (-\slashed{p}-m)=0 \;,
\end{equation}
and replacing $\epsilon_{\nu}(k_2)$ by $k_{2 \nu}$ one obtains
\begin{eqnarray*}
i {{\mathcal{M}}_{2,3}^{\mu \nu}} \epsilon_{\mu}(k_1) k_{2 \nu} =-i g_s \bar{u}(p_2) \big[ f^{bjk} f^{aci}- f^{bki} f^{ajc} \big] \gamma^{\mu} \epsilon_{\mu}(k_1) \;.
\end{eqnarray*}
This term needs to be canceled by the contribution of the first term since the on-shell Ward identity has to hold. For the first term one finds
\begin{align}
\label{eq:brackets}
i {\mathcal{M}}_{1.\text{diagr.}}^{\mu \nu} \epsilon_{\mu}(k_1) k_{2 \nu}  &=-ig_s^2 f^{abc} f^{kji} \frac{1}{(k_1+k_2)^2} \big[g^{\rho \mu} k_1^2-g^{\rho \mu} k_3^2-k_1^{\rho} k_1^{\mu}+k_3^{\rho} k_3^{\mu} \big] \nonumber\\
& \hspace{0pt}\times \bar{u}(p_2) \gamma^{\rho} v(p_1) \epsilon_{\mu}(k_1)
\end{align}
with $k_3=(k_1+k_2)$.
For the third term in brackets we now assume transversality of the external gluons. Therefore it is equal to zero when it is contacted with the polarization vector.
The first term in brackets disappears for on-shell incoming gluons and the last term vanishes when it is contracted with fermionic currents, due to eq.~(\ref{eq:Dirac}).
Therefore one obtains the result
\begin{equation*}
i {\mathcal{M}}_{1.diagr.}^{\mu \nu} \epsilon_{\mu}(k_1) k_{2 \nu}  =i g_s^2 (-1) \bar{u}(p_2) \gamma^{\mu} v(p_1) \epsilon_{\mu}(k_1) f^{abc} f^{kji} \; ,
\end{equation*}
which with the Jacobi identity from eq.~(\ref{eq:jacobi}) exactly cancels the contribution from the last two diagrams. 

Therefore the on-shell Ward identity is fulfilled and shows that it is indeed possible to treat the incoming particles as exclusively transverse polarized gluons by using \textit{ghost subtraction}. Now $-g_{\mu \nu}$ may be used instead of the full polarization sum, since longitudinal degrees of freedom are no longer included in the matrix element. Using the Feynman rules for the ghosts given in~\cite{Peskin}, one finds that the third term in brackets in eq.~(\ref{eq:brackets}) exactly cancels against the ghost fields if they are taken into account explicitly and cancellation is not performed by hand.

When we used this method in section~\ref{sec:WQ}, we also checked gauge invariance of the result by using the full polarization sum from eq.~(\ref{eq:pol_sum}) and calculating the result in a general gauge without making use of the transversality condition, given in eq.~(\ref{eq:transmom}).

\newpage
\section{SUSY-QCD Feynman-Rules}
\label{SUSY-QCD Feynman-Rules}
\begin{align*}
  \begin{minipage}{0.24\textwidth}
      \vspace{-0.6mm}
      \includegraphics[width=\textwidth]{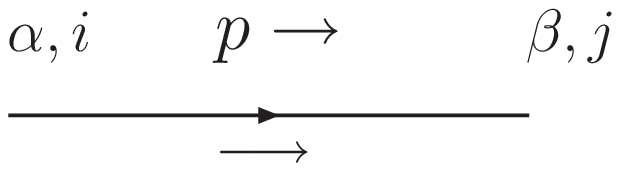}
  \end{minipage}
  &=\,\frac{i(\slashed{p}+m_q)_{\beta \alpha} \delta_{ij}}{p^{2}-m_{q}^2+i\epsilon}\, ,\quad
 \begin{minipage}{0.24\textwidth}
      \vspace{-0.6mm}
     \qquad  \includegraphics[width=\textwidth]{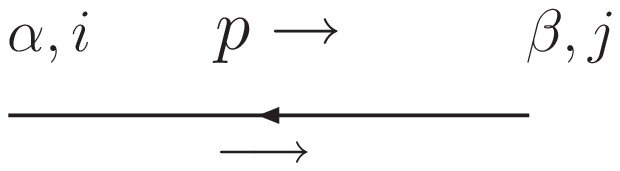}
  \end{minipage}
  =\, \frac{i(\slashed{p}+m_q)_{\beta \alpha} \delta_{ij}}{p^{2}-m_{q}^2+i\epsilon}\,,\\
  \begin{minipage}{0.24\textwidth}
      \vspace{-0.6mm}
      \includegraphics[width=\textwidth]{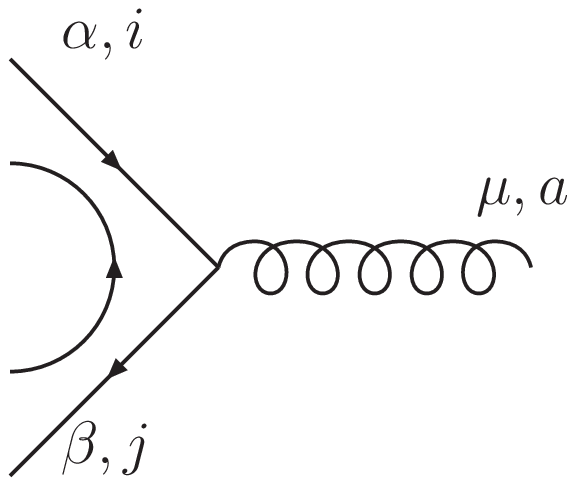}
  \end{minipage}
  &=+ig_s(t^a)_{ji} (\gamma^{\mu})_{\alpha \beta} ,\quad
 \begin{minipage}{0.24\textwidth}
      \vspace{-0.6mm}
      \includegraphics[width=\textwidth]{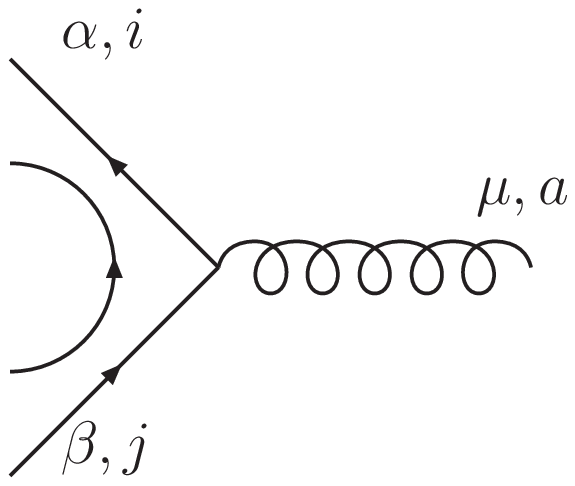}
  \end{minipage}
  =-ig_s(t^a)_{ji} (\gamma^{\mu})_{\alpha \beta},\\
  \begin{minipage}{0.24\textwidth}
      \vspace{-0.6mm}
      \includegraphics[width=\textwidth]{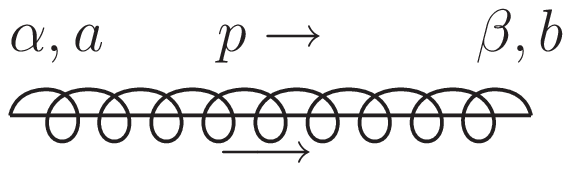}
  \end{minipage}
  &=\frac{i(\slashed{p}+m_{\tilde{g}})_{\beta \alpha} \delta_{ij}}{p^{2}-m_{\tilde{g}}^2+i\epsilon},\quad
  \begin{minipage}{0.24\textwidth}
      \vspace{-0.6mm}
      \includegraphics[width=\textwidth]{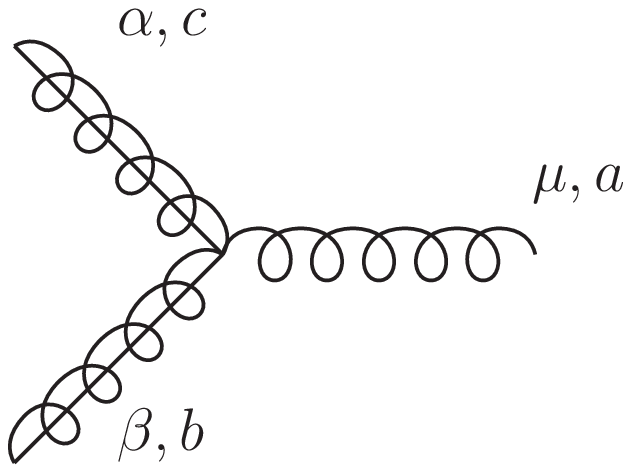}
  \end{minipage}
  =-g_s f^{abc} (\gamma^{\mu})_{\alpha \beta},\\
   \begin{minipage}{0.24\textwidth}
      \vspace{-0.9mm}
       \includegraphics[width=\textwidth]{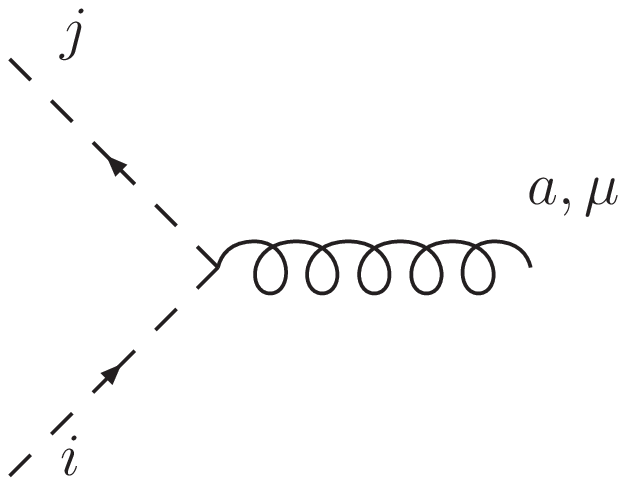}
  \end{minipage}
  &=-i g_s (t^a)_{ji} (p_i-p_j)^{\mu},\\
  \begin{minipage}{0.24\textwidth}
      \vspace{-0.6mm}
      \includegraphics[width=\textwidth]{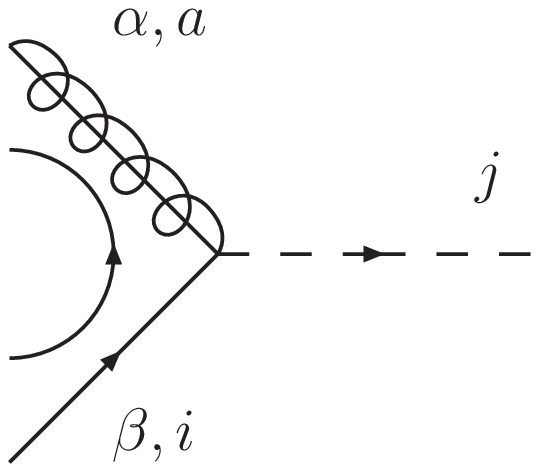}
  \end{minipage}
  &= \mp \frac{ig_s}{\sqrt{2}}(t^a)_{ji} (1 \mp \gamma_5)_{\alpha \beta},\\
  \begin{minipage}{0.24\textwidth}
      \vspace{-0.6mm}
      \includegraphics[width=\textwidth]{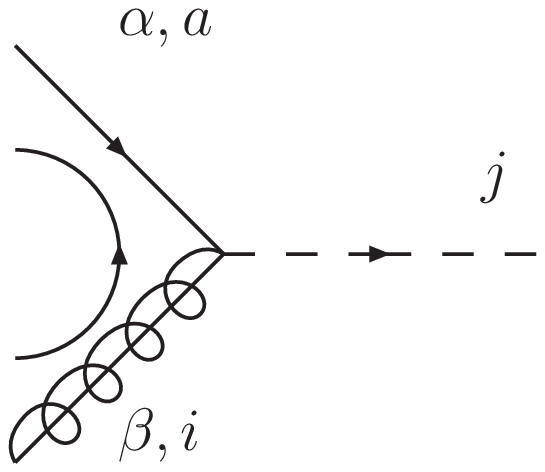}
  \end{minipage}
  &=\mp \frac{ig_s}{\sqrt{2}}(t^a)_{ji} (1 \mp \gamma_5)_{\alpha \beta},\\
  \begin{minipage}{0.24\textwidth}
      \vspace{-0.6mm}
      \includegraphics[width=\textwidth]{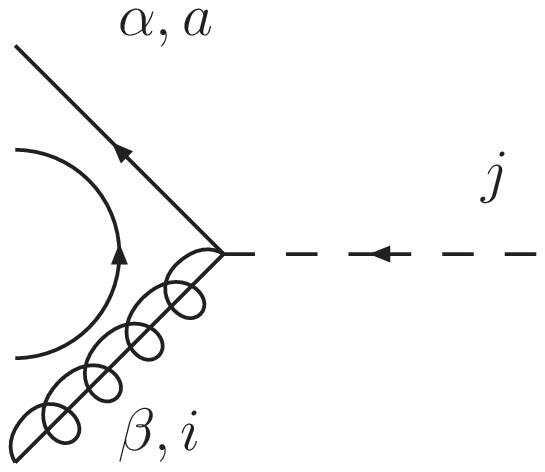}
  \end{minipage}
  &=\pm \frac{ig_s}{\sqrt{2}}(t^a)_{ji} (1 \pm \gamma_5)_{\alpha \beta} ,\\
   \begin{minipage}{0.24\textwidth}
      \vspace{-0.6mm}
      \includegraphics[width=\textwidth]{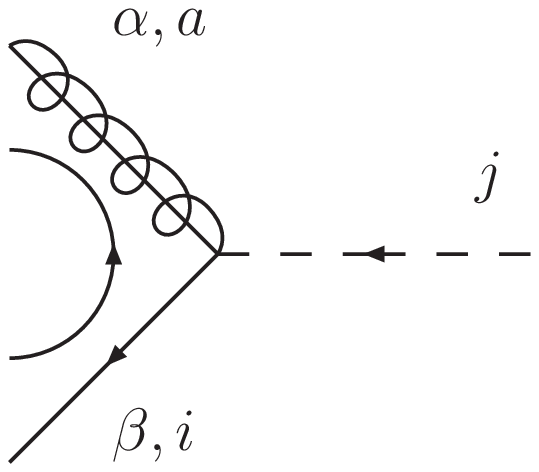}
  \end{minipage}
  &=\pm \frac{ig_s}{\sqrt{2}}(t^a)_{ji} (1 \pm \gamma_5)_{\alpha \beta},\\
\end{align*}
For the last four graphs, the \textit{upper} sign gives the Feynman rules for a \textit{left} handed squark and the \textit{lower} sign for a \textit{right} handed squark.

\newpage
\section{UED Feynman-Rules}
\subsection{UED-Interactions with Gluons and Quarks}
\begin{align*}
  \begin{minipage}{0.25\textwidth}
      \vspace{-0.6mm}
      \includegraphics[width=\textwidth]{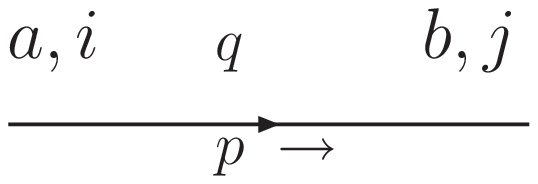}
  \end{minipage}
  &=\,\frac{i(\slashed{p}+m_q)_{b a} \delta_{ij}}{p^{2}-m_{q}^2+i\epsilon}\, ,
 &\begin{minipage}{0.25\textwidth}
      \vspace{-0.6mm}
  \hspace*{-40pt}  \qquad  \includegraphics[width=\textwidth]{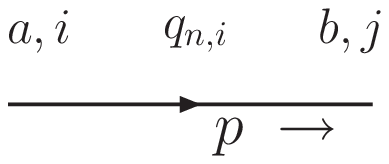}
  \end{minipage}
  \hspace*{-30pt}=\, \frac{i(\slashed{p}+M_n)_{b a} \delta_{ij}}{p^{2}-M_n^2+i M_n \Gamma_g^n}\,,\\
  \begin{minipage}{0.25\textwidth}
      \vspace{-0.6mm}
      \includegraphics[width=\textwidth]{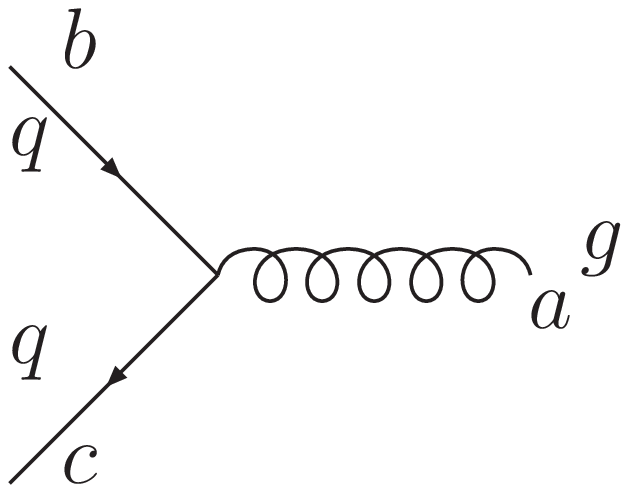}
  \end{minipage}
  &=\,-i g \gamma^{\mu} (t^a)_{cb}\, ,
 &\begin{minipage}{0.25\textwidth}
      \vspace{-0.6mm}
   \hspace*{-30pt}   \includegraphics[width=\textwidth]{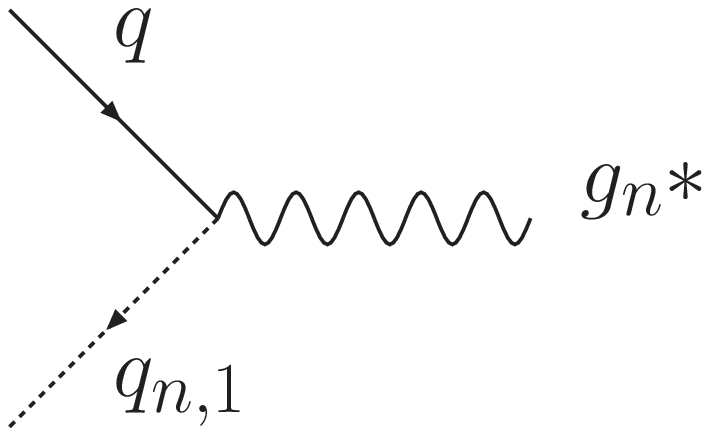}
  \end{minipage}
  \hspace*{-30pt}=\,-i g \gamma^{\mu} (t^a)_{cb} P_L\,,\\
  \begin{minipage}{0.25\textwidth}
      \vspace{-0.6mm}
      \includegraphics[width=\textwidth]{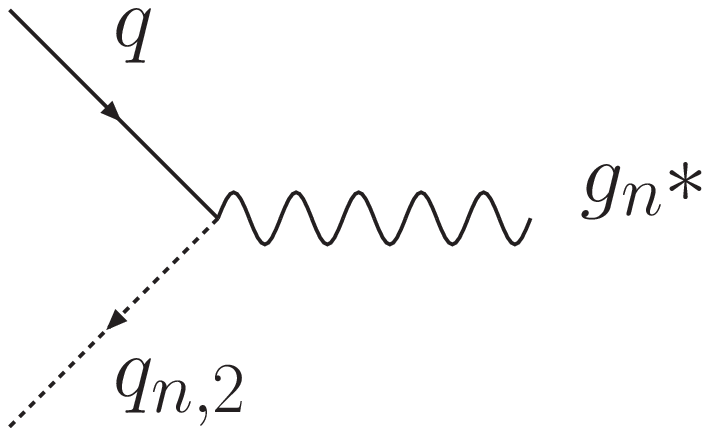}
  \end{minipage}
  &= \,-i g \gamma^{\mu} (t^a)_{cb} P_R\,,
 &\begin{minipage}{0.25\textwidth}
      \vspace{-0.6mm}
   \hspace*{-40pt}  \hspace{-17.0pt} \includegraphics[width=\textwidth]{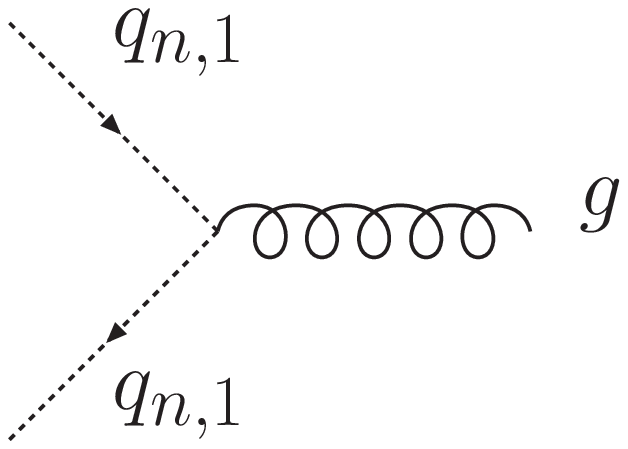}
  \end{minipage}
  \hspace*{-40pt}=\,-i g \gamma^{\mu} (t^a)_{cb}\,,\\
  \begin{minipage}{0.25\textwidth}
      \vspace{-0.6mm}
      \includegraphics[width=\textwidth]{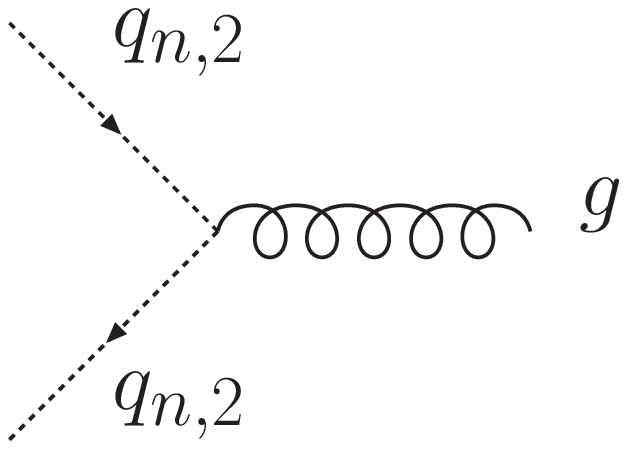}
  \end{minipage}
  &=\,-i g \gamma^{\mu} (t^a)_{cb}\,,
 &\begin{minipage}{0.25\textwidth}
      \vspace{-0.6mm}
   \hspace*{-30pt} \includegraphics[width=\textwidth]{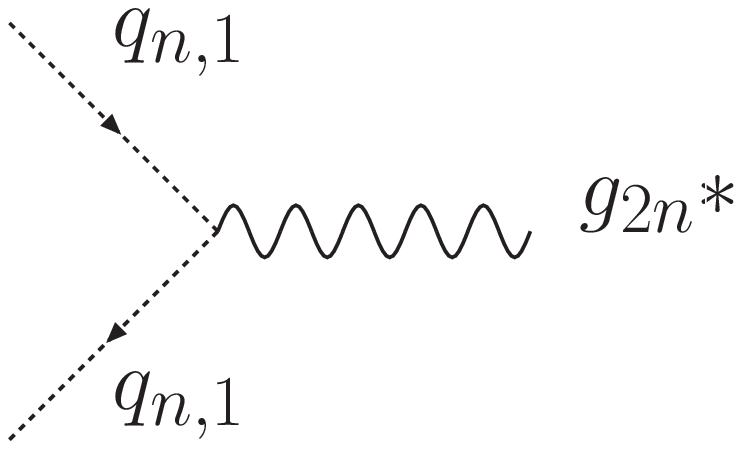}
  \end{minipage}
  \hspace*{-30pt}= ig \frac{1}{\sqrt{2}} \gamma^{\mu} \gamma_5 (t^a)_{ji},\\
  \begin{minipage}{0.25\textwidth}
      \vspace{-0.6mm}
      \includegraphics[width=\textwidth]{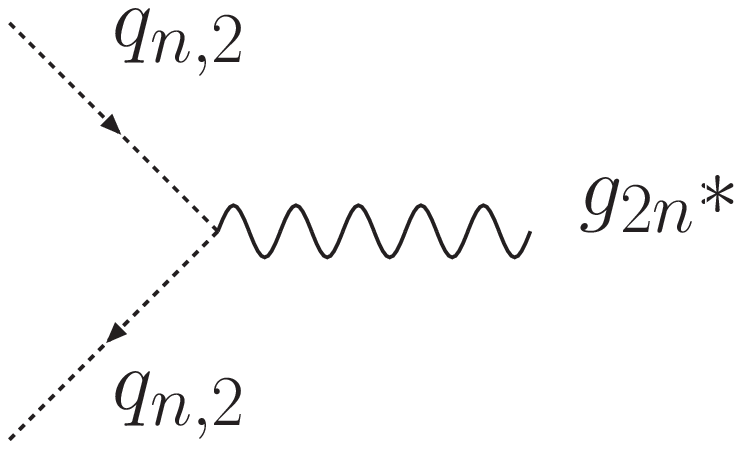}
  \end{minipage}
  &= -ig \frac{1}{\sqrt{2}} \gamma^{\mu} \gamma_5 (t^a)_{ji} ,
 &\begin{minipage}{0.25\textwidth}
      \vspace{-0.6mm}
    \hspace*{-30pt}  \includegraphics[width=\textwidth]{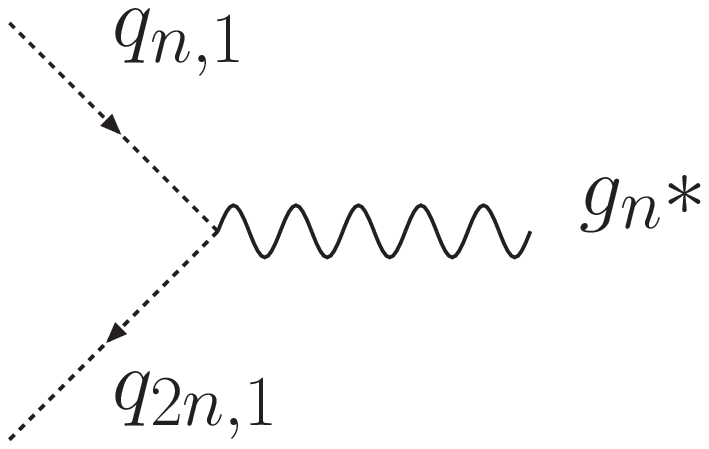}
  \end{minipage}
  \hspace*{-30pt}=-ig \frac{1}{\sqrt{2}} \gamma^{\mu} (t^a)_{ji},\\
  \begin{minipage}{0.25\textwidth}
      \vspace{-0.6mm}
      \includegraphics[width=\textwidth]{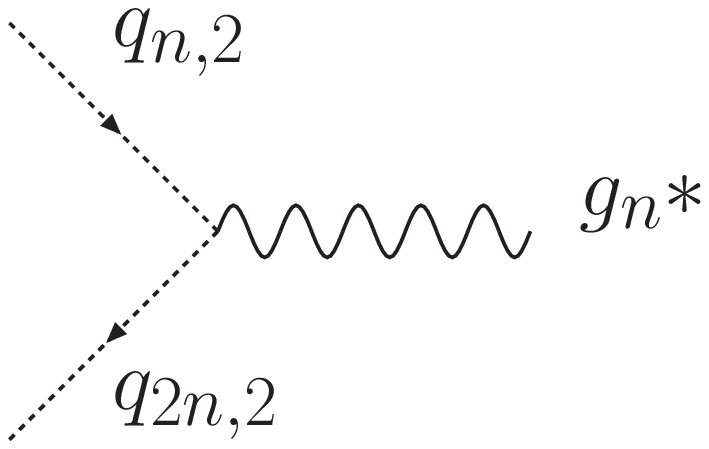}
  \end{minipage}
  &= -ig \frac{1}{\sqrt{2}} \gamma^{\mu} (t^a)_{ji},
 &\begin{minipage}{0.25\textwidth}
      \vspace{-0.6mm}
    \hspace*{-30pt}  \includegraphics[width=\textwidth]{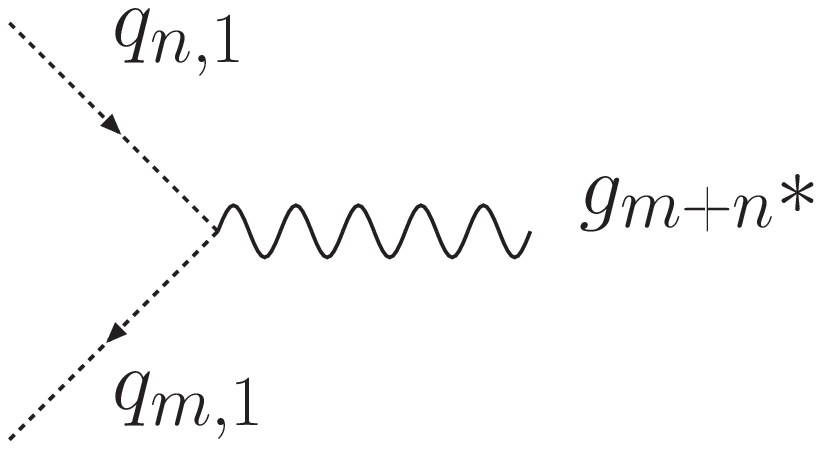}
  \end{minipage}
  \hspace*{-30pt}= ig \frac{1}{\sqrt{2}} \gamma^{\mu} \gamma_5 (t^a)_{ji},\\
  \begin{minipage}{0.25\textwidth}
      \vspace{-0.6mm}
      \includegraphics[width=\textwidth]{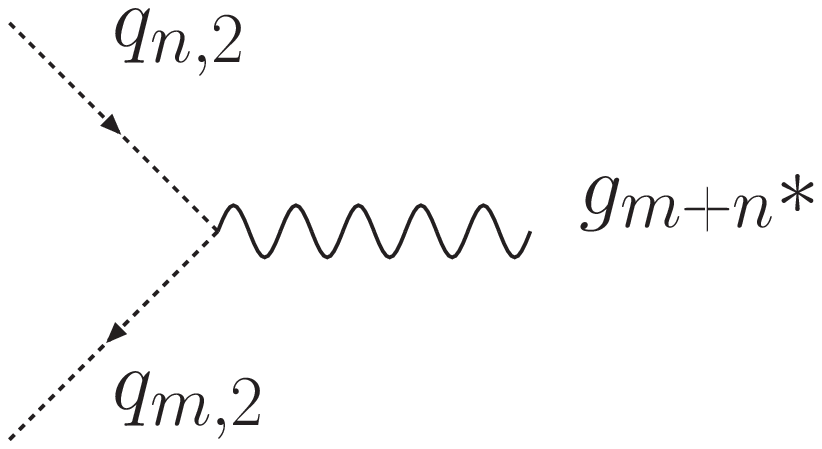}
  \end{minipage}
  &= -ig \frac{1}{\sqrt{2}} \gamma^{\mu} \gamma_5 (t^a)_{ji},
 &\begin{minipage}{0.25\textwidth}
      \vspace{-0.6mm}
    \hspace*{-30pt}  \includegraphics[width=\textwidth]{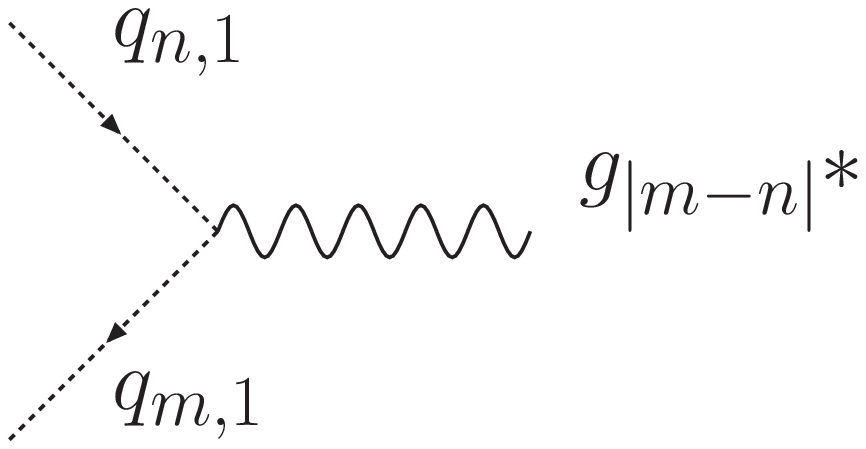}
  \end{minipage}
  \hspace*{-30pt}= -ig \frac{1}{\sqrt{2}} \gamma^{\mu} (t^a)_{ji},\\
 \begin{minipage}{0.25\textwidth}
      \vspace{-0.6mm}
      \includegraphics[width=\textwidth]{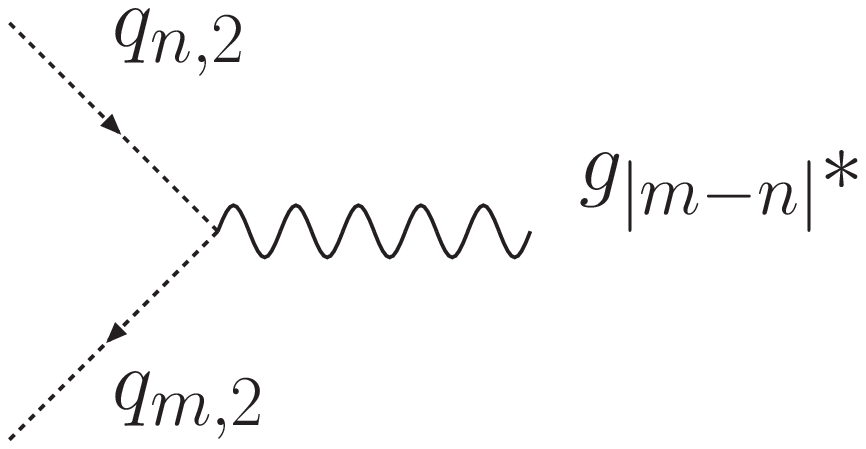}
  \end{minipage}
  &=-ig \frac{1}{\sqrt{2}} \gamma^{\mu} (t^a)_{ji},\\
\end{align*}
Indices for all vertices are equal to those at the quark-quark-gluon vertex. The coupling is given by $g=\frac{g_5}{\sqrt{\pi R}}$. For better readability we use different lines for the heavy KK-partners of the gluons $g_n^*$ and of the quarks $q_{n,i}$. The index $i=1,2$ denotes the heavy quark tower, while $n$ denotes the excitation level within the tower.
\newpage
\subsection{Purely Gluonic UED-Interactions}
\begin{align*}
 \begin{minipage}{0.22\textwidth}
  \hspace*{1mm}  \includegraphics[width=\textwidth]{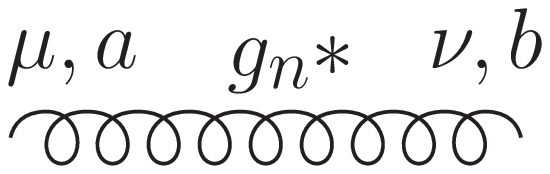}
  \end{minipage}
  \hphantom{abc} &=\left[-g^{\mu \nu}+(1-\lambda) \frac{p^{\mu} p^{\nu}}{p^2+i \epsilon} \right] \frac{i}{p^2+i\epsilon} \delta^{a b},\\
\vspace*{+0.6mm}
  \begin{minipage}{0.22\textwidth}
      \vspace*{+4.6mm}
   \hspace*{1mm}   \includegraphics[width=\textwidth]{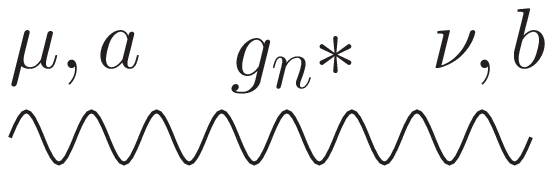}
  \end{minipage}
\bigskip
\newline
\vspace*{+4.6mm} \hphantom{abc}&={-i\;  \left(\frac{g^{\mu \nu}-\frac{p^{\mu}p^{\nu}}{M_n^2}}{p^2-M_n^2+i\;  M_n \Gamma_g^n} \right)}\delta^{ab},\\
  \begin{minipage}{0.25\textwidth}
      \vspace*{+2.6mm}
      \includegraphics[width=\textwidth]{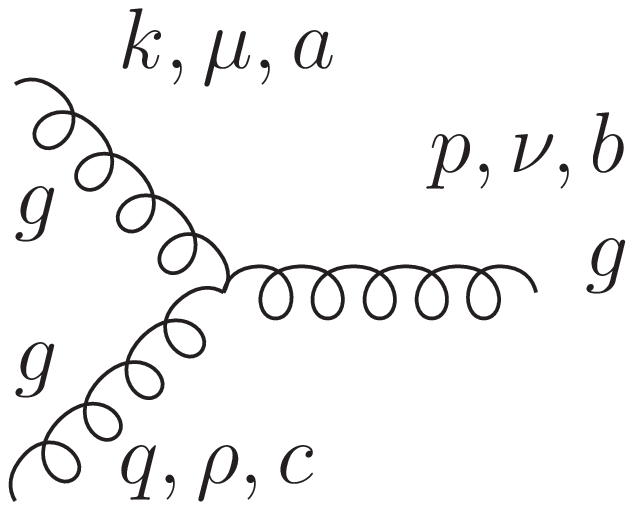}
  \end{minipage}
  &=g \; f^{abc}\;  \left[(p-k)^{\rho}\;  g^{\mu \nu}+(q-p)^{\mu}\;  g^{\nu \rho}+(k-q)^{\nu}\;  g^{\rho \mu}  \right] ,\\
 \begin{minipage}{0.27\textwidth}
      \vspace*{+2.6mm}
    \hspace*{5pt}   \includegraphics[width=\textwidth]{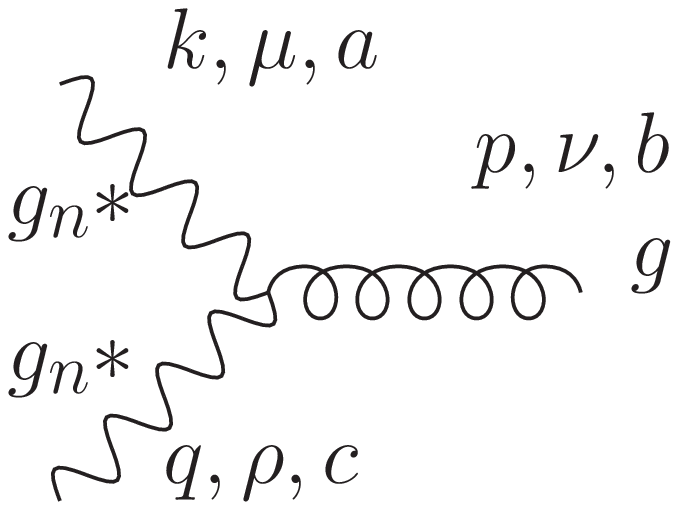}
  \end{minipage}
  &\hspace*{32pt} = g,\\
  &\hspace{-95pt} \begin{minipage}{0.27\textwidth}
      \vspace*{2.6mm}
    \includegraphics[width=\textwidth]{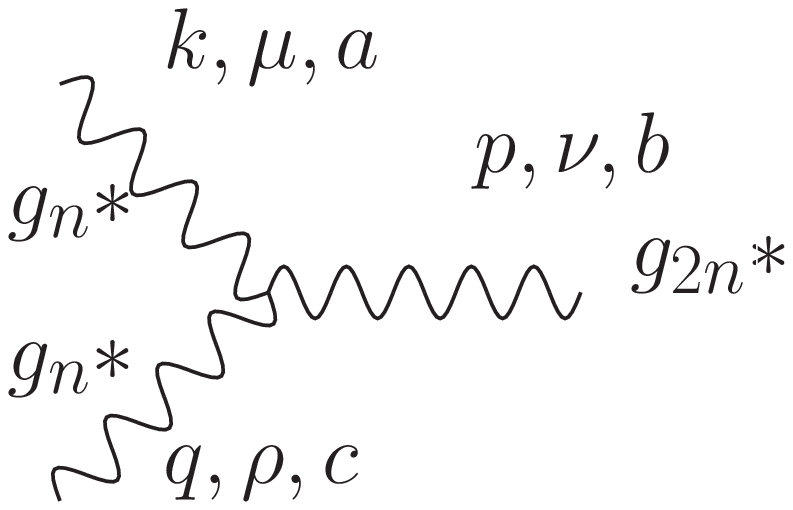}
  \end{minipage}
  \hspace*{20pt} = \frac{1}{\sqrt{2}}\;  g,\\
 & \hspace{-95pt}\begin{minipage}{0.3\textwidth}
      \vspace*{5.6mm}
      \includegraphics[width=\textwidth]{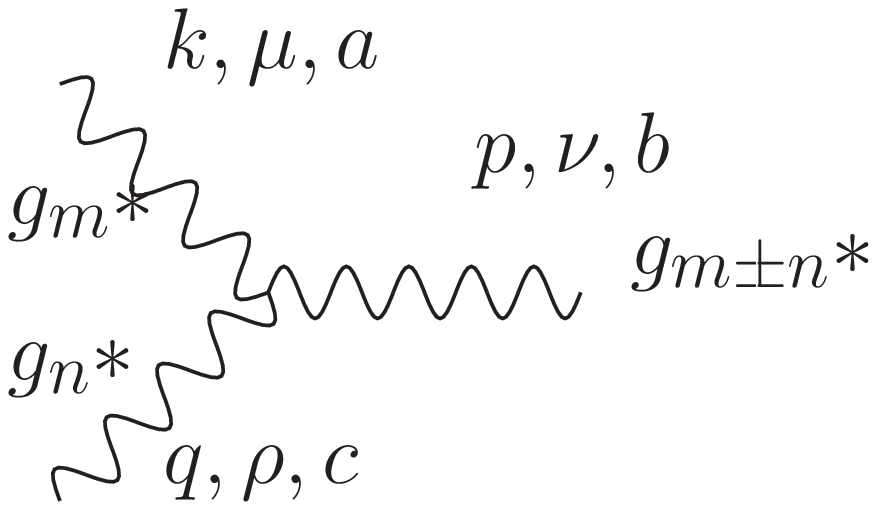}
  \end{minipage}
  \quad = \frac{1}{\sqrt{2}}\;  g,\\
  &\hspace{-97pt} \begin{minipage}{0.24\textwidth}
      \vspace*{2.6mm}
    \includegraphics[width=\textwidth]{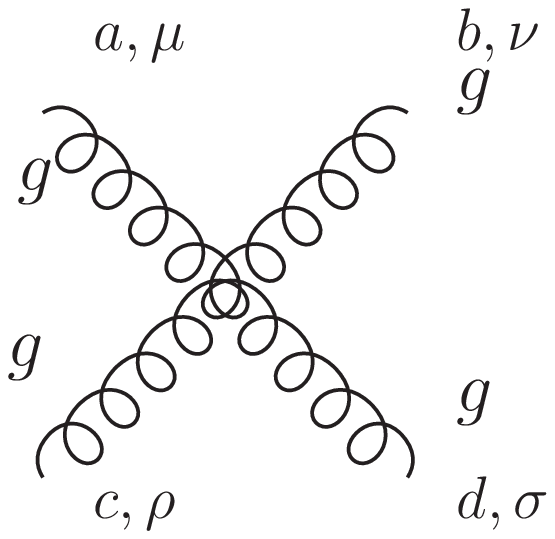}
  \end{minipage}
  \hspace{+123pt}  
  \hspace{-115pt}\begin{minipage}{0.17\textwidth}
      \vspace*{2.6mm}
     \begin{eqnarray*}
 =-i\; g^2\;  &\times&\hspace*{-3pt}[f^{xac} f^{xbd} \left(g^{\mu \nu} g^{\rho \sigma}-g^{\mu \sigma} g^{\nu \rho} \right)\\
&+&f^{xad} f^{xbc} \left(g^{\mu \nu} g^{\rho \sigma}-g^{\mu \rho} g^{\nu \sigma} \right)\\
&+&f^{xab} f^{xcd} \left(g^{\mu \rho} g^{\nu \sigma}-g^{\mu \sigma} g^{\nu \rho} \right)] ,\\
      \end{eqnarray*} 
  \end{minipage}\\
\vspace*{-45pt}
  \begin{minipage}{0.24\textwidth}
      \vspace*{2.6mm}
      \includegraphics[width=\textwidth]{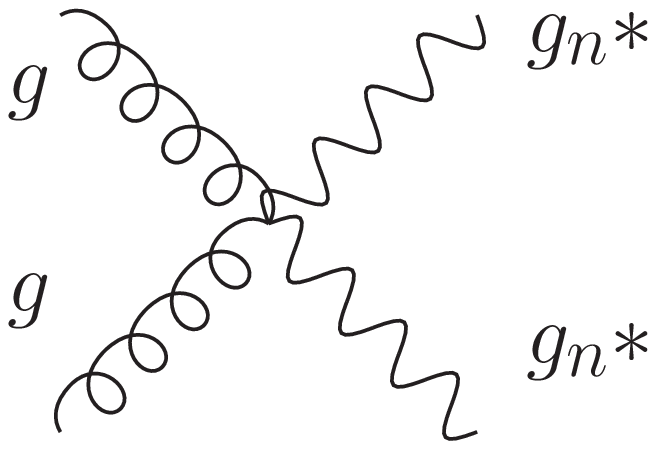}
  \end{minipage}
  &=-i\; g^2,\qquad \hspace{70pt}
  \begin{minipage}{0.24\textwidth}
      \vspace*{2.6mm}
    \hspace{-36pt}  \includegraphics[width=\textwidth]{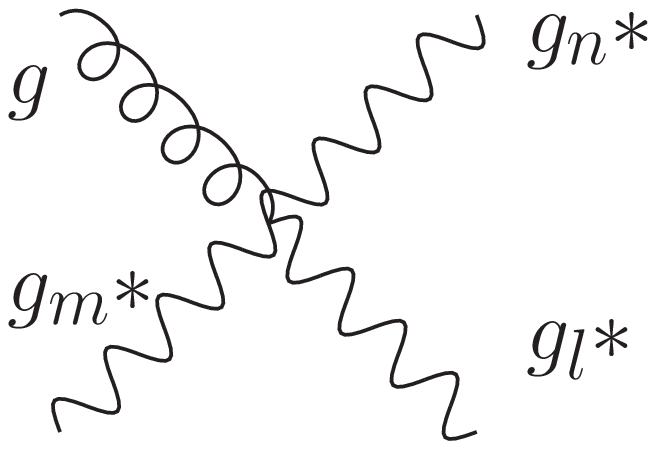}
  \end{minipage}
  &\hspace{-36pt}= -i \; \frac{1}{\sqrt{2}} \; g^2,\\
\end{align*}
\newpage
\begin{align*}
  \begin{minipage}{0.25\textwidth}
      \vspace*{3.6mm}
      \includegraphics[width=\textwidth]{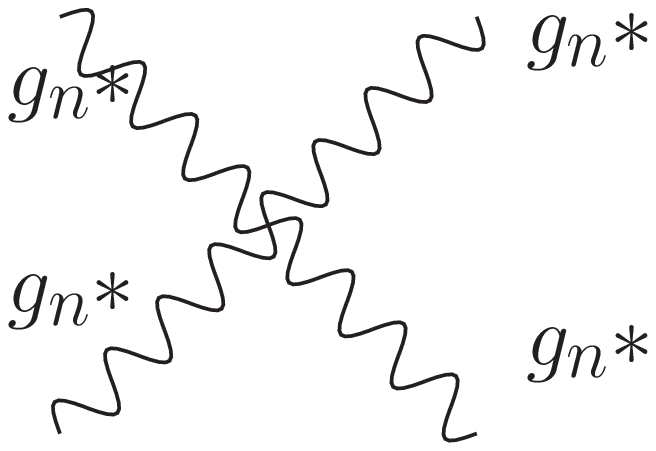}
  \end{minipage}
  &=-i\;  \frac{3}{\sqrt{2}} \;g^2,\qquad \hspace{50pt}
  \begin{minipage}{0.25\textwidth}
      \vspace*{3.6mm}
     \hspace{-36pt} \includegraphics[width=\textwidth]{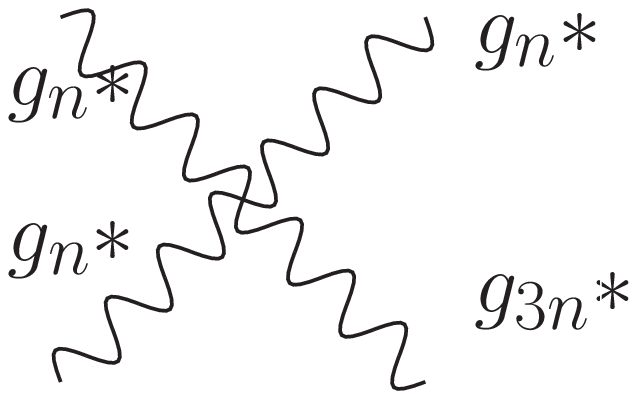}
  \end{minipage}
  &\hspace{-36pt}=-i \; \frac{1}{2}\; g^2,\\
   \begin{minipage}{0.25\textwidth}
      \vspace*{2.6mm}
      \includegraphics[width=\textwidth]{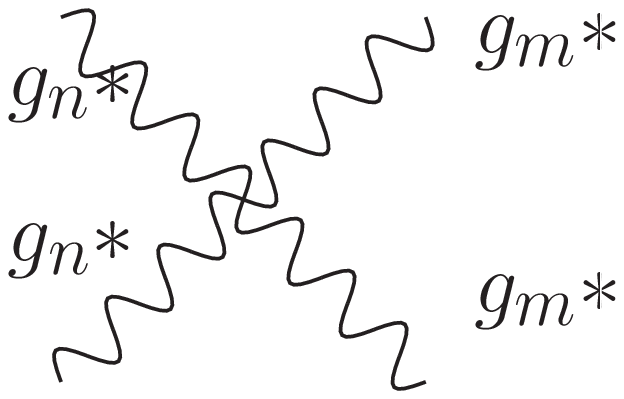}
  \end{minipage}
  &=-i \; g^2,\qquad \hspace{53pt}
  \quad \begin{minipage}{0.30\textwidth}
      \vspace*{2.6mm}
    \hspace{-50pt}  \qquad \includegraphics[width=\textwidth]{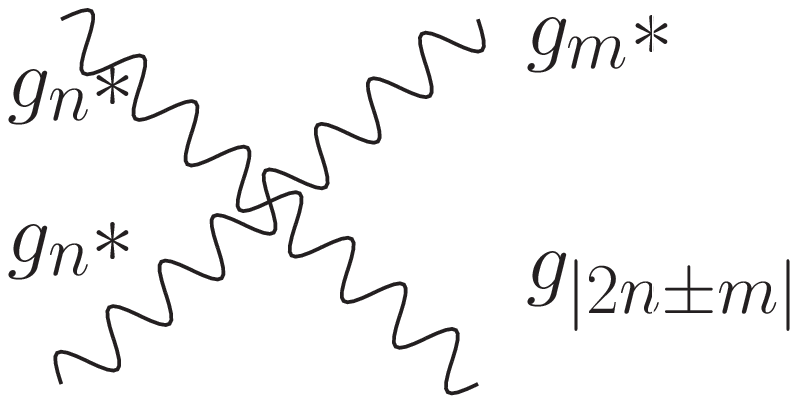}
  \end{minipage}
  &\hspace{-36pt}= -i \;  \frac{1}{2} \; g^2,\\
  & \hspace*{-95pt}\begin{minipage}{0.32\textwidth}
      \vspace*{2.6mm}
      \includegraphics[width=\textwidth]{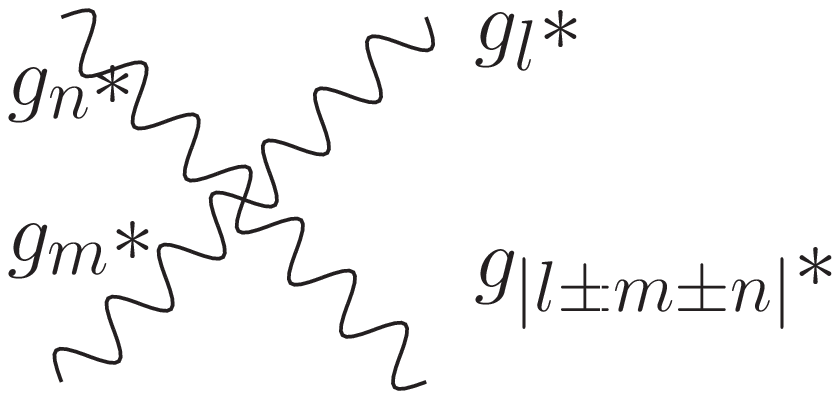}
  \end{minipage}
   =-i \; \frac{1}{2}\; g^2,\\
\end{align*}
All momenta are defined incoming. The indices belonging to the four particle interactions are the same as for the four gluon vertex. The coupling is given by \mbox{$g=\frac{g_5}{\sqrt{\pi R}}$}. For better readability we use curved lines for the heavy KK-partners of the gluons $g_n^*$. The three and four particle vertices have to be multiplied by the tensor structure term in brackets, belonging to the three and four gluon interactions.
\newpage
\subsection{Selected Feynman Rules from the Electroweak UED Sector}
\label{app:electroweak}
\hspace*{-10mm}
\begin{align*}
  \begin{minipage}{0.25\textwidth}
      \vspace*{-1.6mm}
   \hspace*{-13mm}      \includegraphics[width=\textwidth]{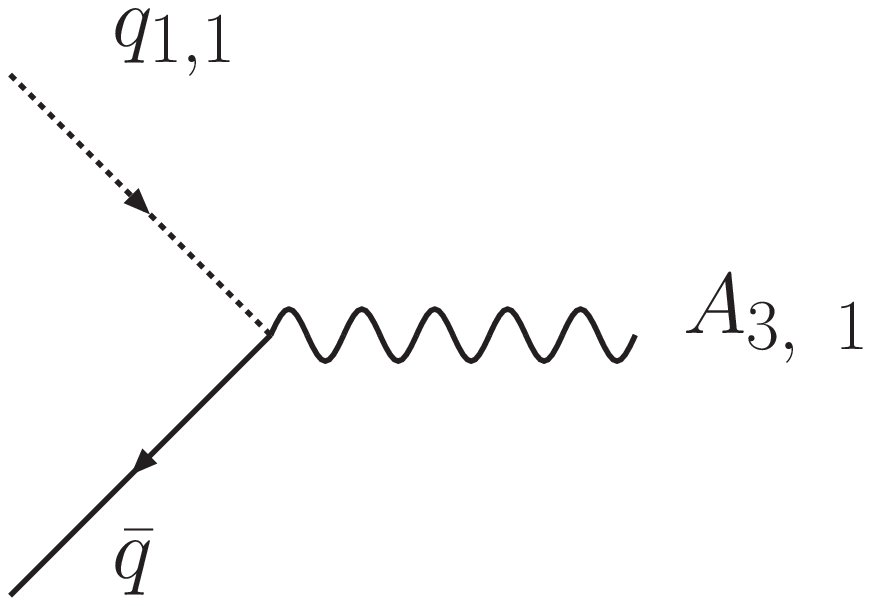}
  \end{minipage}
  \hphantom{abc}&=T_{3} \; g\; \text{cos} \alpha^{(1)} \; P_{L},\\
  \begin{minipage}{0.25\textwidth}
      \vspace*{2.6mm}
   \hspace*{-13mm}   \includegraphics[width=\textwidth]{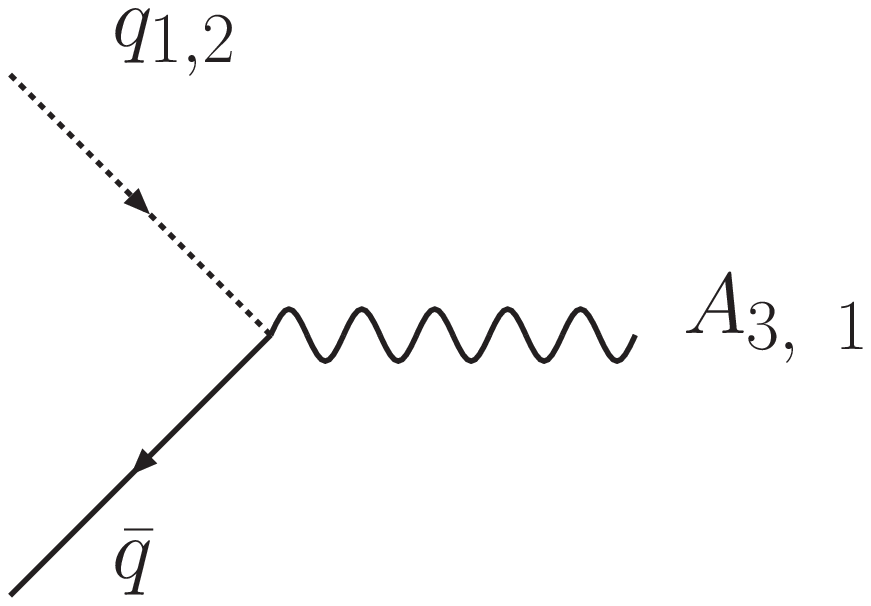}
  \end{minipage}
	\vspace*{10.6mm}
  \vspace*{10.6mm} \hphantom{abc}&=-T_{3} \; g\; \text{sin} \alpha^{(1)} \; P_{L},\\
  \begin{minipage}{0.25\textwidth}
      \vspace*{2.6mm}
   \hspace*{-13mm}   \includegraphics[width=\textwidth]{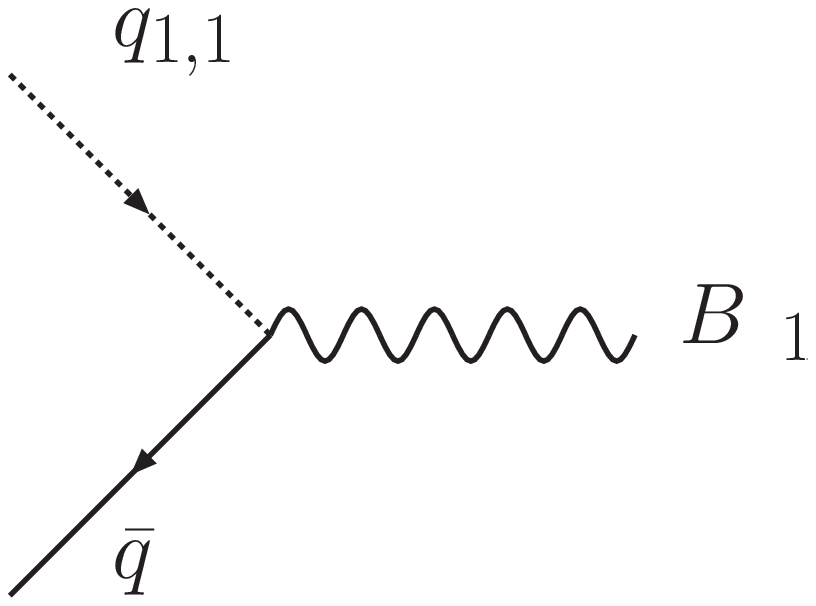}
  \end{minipage}
  \hphantom{abc}&=Y_{s} \; g_Y \; \text{sin} \alpha^{(1)} P_R + Y_{d} \; g_Y \; \text{cos} \alpha^{(1)} \; P_L ,\\
  \begin{minipage}{0.25\textwidth}
      \vspace*{2.6mm}
   \hspace*{-13mm}   \includegraphics[width=\textwidth]{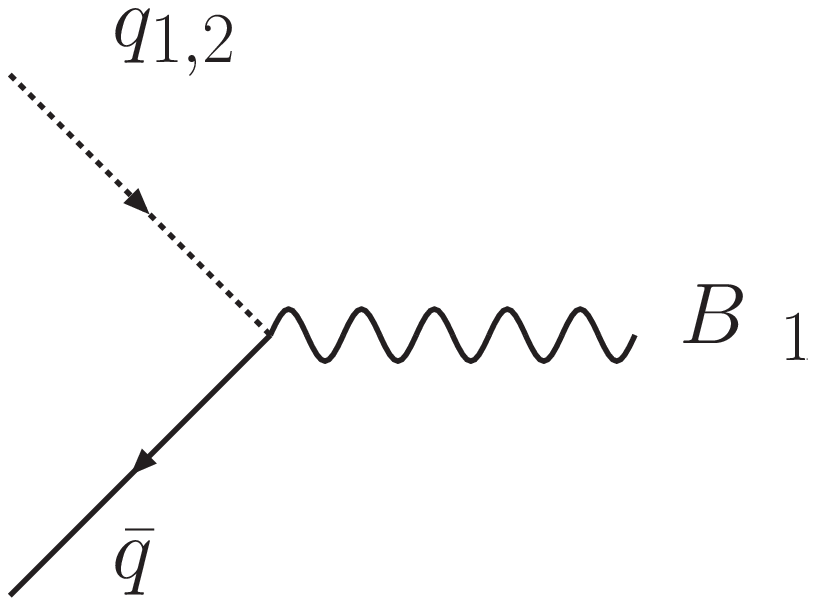}
  \end{minipage}
	\vspace*{10.6mm}
  \vspace*{10.6mm} \hphantom{abc}&=-Y_{s} \; g_Y \; \text{cos} \alpha^{(1)} P_R - Y_{d} \; g_Y \; \text{sin} \alpha^{(1)} \; P_L,\\
\end{align*}
The Lagrangian of the electroweak sector is given in chapter 3. The angle $\alpha^{(1)}$ denotes the mixing angle of singlets and doublets of the first excitation level. The particles $q_{i,n}$ here denote the mass eigenstates, derived from the eigenstates of weak interaction by mixing of the two towers. Additional electroweak Feynman rules can be found, e.g.\ , in~\cite{Bringmann}.

  \clearpage{\pagestyle{empty}\cleardoublepage}

\phantomsection 
\addcontentsline{toc}{chapter}{References}
\fancyhead[C]{References}
\bibliographystyle{bibstyles/utcaps_nott}
\bibliography{bib_reinartz}
\clearpage{\pagestyle{empty}\cleardoublepage}

  \chapter*{Acknowledgments}

First of all I would like to thank Michael Pl\"umacher for his mentoring during my year at the Max-Planck-Institute for Physics. He introduced me to a very interesting area of high energy physics.
\newline
I am grateful to Prof.\,Dr.\,Andrzej J. Buras for giving me the opportunity to conduct my diploma thesis at the Max-Planck-Institute.
\newline
For lots of discussions on the phone and his support with Madgraph, I especially thank Tilman Plehn.
\newline
Special thanks go to my roommates Daniel H\"artl, Florian Hahn-Woernle, Max Huber, Philipp Kostka and Alexandra R\"uger, to Stefan Kallweit, Tobias Kasprzik, Edoardo Mirabella, Josef Pradler and Maike Trenkel for helpful discussions and a pleasant atmosphere during my time at the institute.\\
\newline
After all I thank my parents for the continuous support and encouragement during my studies.


  \clearpage{\pagestyle{empty}\cleardoublepage}

\thispagestyle{empty}
\end{document}